\documentclass[aps,prx,onecolumn,superscriptaddress,amsmath,amssymb,10pt,epsfig,nofootinbib,eqsecnum]{revtex4-2}

\usepackage{graphicx} 
\usepackage{subfig}
\usepackage{afterpage}
\usepackage{bm}
\usepackage{amsmath}
\usepackage{caption}
\usepackage[section]{placeins}
\usepackage{color}
\usepackage{url}
\usepackage[normalem]{ulem} 
\usepackage[colorlinks=true]{hyperref}  
\hypersetup{
    unicode=false,          
    pdftoolbar=true,        
    pdfmenubar=true,        
    pdffitwindow=false,     
    pdfstartview={FitH},    
    pdftitle={HarrisSC},    
    pdfauthor={Kryhin, Lunts, Patel, Sachdev, Nosov},     
    pdfsubject={},   
    pdfcreator={},   
    pdfproducer={}, 
    pdfkeywords={} {} {}, 
    pdfnewwindow=true,      
    colorlinks=true,       
    linkcolor=blue, 
    citecolor=blue,        
    filecolor=magenta,      
    urlcolor=blue           
}

\newcommand{\p}{\partial}

\renewcommand{\vec}{\mathbf}

\begin{document}

\title{Influence of Harris disorder on quantum-critical superconductivity}

\author{Serhii Kryhin}
\affiliation{Department of Physics, Harvard University, Cambridge MA 02138, USA}
\author{Peter Lunts}
\affiliation{Episteme, Inc.}
\affiliation{Department of Physics, Harvard University, Cambridge MA 02138, USA}
\author{Aavishkar A. Patel}
\affiliation{International Centre for Theoretical Sciences, Tata Institute of Fundamental Research, Bengaluru 560089, India}
\affiliation{Center for Computational Quantum Physics, Flatiron Institute, 162 5th Avenue, New York, NY 10010, USA}
\author{Subir Sachdev}
\affiliation{Department of Physics, Harvard University, Cambridge MA 02138, USA}
\affiliation{Center for Computational Quantum Physics, Flatiron Institute, 162 5th Avenue, New York, NY 10010, USA}
\author{Pavel A. Nosov}
\affiliation{Department of Physics, Harvard University, Cambridge MA 02138, USA}
\date{\today}

\begin{abstract}
In the Hertz theory, a quantum critical metal is described by the coupling of a Fermi surface to fluctuations of a Landau-damped bosonic field $\phi$, which may represent either an order parameter or, more generally, a Higgs field for a transition without symmetry breaking. Scattering from $\phi$ produces non-Fermi-liquid behavior in the normal state, while the same fluctuations mediate enhanced Cooper pairing. By the Harris criterion, symmetry-preserving disorder couples most strongly to the coefficient of the $\phi^2$ term, locally tuning the system toward or away from criticality. This random mass (``Harris disorder'') leads to a finite density of localized low-energy $\phi$ modes even when the fermionic states remain extended and continue to provide Landau damping. We study the onset of pairing mediated by these localized overdamped bosonic modes. Starting from a real-space linearized Usadel equation for the two-particle propagator (the Cooperon), we show that the localized bosonic wave functions generate both a spatially random pairing vertex and an effective random potential for Cooper pairs. Numerical solution, a self-consistent Born approximation analysis, and single-puddle/Lifshitz-tail considerations reveal two regimes of the pairing instability. At high temperatures, pairing nucleates compact superconducting puddles on the most localized bosonic modes. At lower temperatures, an extended pairing eigenstate appears, but its transition scale and spatial structure remain strongly affected by mesoscopic correlation effects and enhanced probability of returns to favorable regions of the localized bosonic glue. The resulting distribution of local pairing scales has a power-law tail, in contrast to the stretched-exponential tails of 
disordered BCS (i.e. non-critical) superconductors. This mechanism provides a route to broad gap inhomogeneity and superconducting puddles in quantum-critical metals, and offers a natural interpretation of scanning tunneling microscopy measurements of the cuprates.
\end{abstract}

\maketitle


\section{Introduction}

The highest temperature superconductivity in the cuprates emerges upon lowering temperature ($T$) from a ``strange metal'' state without well-defined fermionic quasiparticle excitations -- a feature shared by many other correlated electron materials \cite{Keimer2015,MarginalFLPhenomenology,Hussey_foot,Hussey2011,Kirchner2020,Ayres2021,Hartnoll2022}. Itinerant quantum criticality is often postulated to be the most direct way to obtain such a metallic state, while also being experimentally plausible \cite{Sachdev_2011,Sachdev2025}. It is therefore important to understand the pairing instability of a quantum-critical metal.

In theories of itinerant quantum criticality the breakdown of quasiparticles in the normal state is induced by fluctuations of a critical boson. This boson usually represents an order parameter associated with a symmetry-breaking quantum phase transition \cite{Hertz1976,Moriya1973,Millis1993,Sachdev_2011}, but can also be a Higgs field in a `topological' transition without symmetry breaking \cite{Zhang2020,Zhang2020Decon,Mascot2022,Nikolaenko2023,Sachdev2025,Lunts2025,Bonetti_2026}. The same boson mediates the attractive interaction between the fermions, leading to pairing and the onset of superconductivity. For clean systems without disorder, many works \cite{Bonesteel1996,Son:1998uk,ChubukovFinkelstein,Abanov_review,Moon2010,Metlitski10,Wang13,Metlitski15,Raghu2015,Esterlis:2019ola,Debanjan20,Esterlis20,Chubukov20,Esterlis:2021eth,Inkof:2021ohk,Esterlis25} have made significant progress towards the understanding of the pairing instability for a variety of non-Fermi liquid parent states.

The presence of quenched disorder has a significant effect on itinerant quantum critical systems. Electronic chemical potential disorder has been extensively studied \cite{Chakravarty1998,Rosch1999,Maebashi2002,Paul2005,Nosov2020,Halbinger2021,Wu2022,Nosov2023,Wu2023,Kim2024}, but the most significant disorder for the phase transition itself is a random mass, i.e. ``Harris disorder", which is strongly relevant and locally tunes the system toward or away from the quantum critical point \cite{Harris1974}. Several recent works have considered Harris disorder and its effects on the quantum-critical normal state \cite{Aldape-2022,Hartnoll:2014gba,Patel:2014jfa,Goldman2020,Esterlis:2021eth,Aavishkar2023,Li:2024kxr,Chatterjee25,SangJin25,Boyack26,Aavishkar2024,patel2024strange}. 
\color{black}
One key finding is that Harris disorder can localize overdamped bosonic modes at low and intermediate energies across a broad region of parameter space~\cite{Aavishkar2024,patel2024strange}. This regime is related to quantum Griffiths physics near the disordered critical point, but the pairing problem considered below is not controlled solely by the asymptotic Griffiths rare-region effects. Instead, the relevant regime contains compact localized low-energy modes with an approximately constant density of states. We refer to this as the gapless-localized regime. It has been argued to describe the strange-metal ``foot''~\cite{Hussey_foot,Greene_rev,Aavishkar2024,patel2024strange,Sachdev2025} and the scaling properties of critical spin fluctuations~\cite{Hayden25} in $\mathrm{La}_{2-x}\mathrm{Sr}_x\mathrm{CuO}_4$.
\color{black}

In this paper, we focus on this extended critical region of the phase diagram, and describe the strong influence of Harris disorder on quantum-critical superconductivity. The most significant effects we find are the broadening
(in temperature) of the superconducting transition and the emergence of a power-law analog of a Lifshitz tail of localized states at temperatures well above the global superconducting transition temperature. 
This should be contrasted with conventional BCS theory, where the Lifshitz tail is expected to decay exponentially, and disorder that preserves time-reversal symmetry has little effect on $s$-wave superconductivity. At lower temperatures, the linearized gap equation develops an extended eigenstate, but the transition scale and the spatial structure of the gap remain strongly affected by mesoscopic fluctuations of the localized bosonic states. In particular, the enhanced return probability of diffusive Cooper pairs to favorable attractive regions, together with the slow dynamics of the localized bosonic modes, enhances the pairing scale and produces strong spatial gap fluctuations even when the Drude conductance is large. This behavior is qualitatively distinct from the relative insensitivity of conventional BCS pairing to time-reversal-preserving long-wavelength disorder.

An important feature of our work is the regime of localization scales that we work in: the electrons are essentially delocalized, while the overdamped boson modes are localized with a large range of length scales. The bosons thus provide an effective slowly-varying disorder potential for the electrons and Cooper pairs. This setting should be contrasted with two other, more conventional, settings: the localization problem of repulsive bosons \cite{Fisher1989} where there are no gapless fermionic excitations that can provide a dissipative bath and lead to overdamped bosonic dynamics; and Anderson localization that leads to a superconductor-to-insulator transition \cite{Sadovskii1984,Ma1985,Kapitulnik1985,Kapitulnik1986,Feigelman2007,Feigelman2010,BGM2012,BGM2021,Nosov2023} where single particle electronic wavefunctions are either completely localized \cite{Sadovskii1984,Ma1985,Kapitulnik1985,Kapitulnik1986} or remain close to the localization threshold ({\it i.e.\/} exhibit strong mesoscopic fluctuations and `multifractal' statistics) \cite{Feigelman2007,Feigelman2010,BGM2012,BGM2021,Nosov2023}.

The outline of the paper is as follows. 
In Sec.~\ref{sec:normal}, we review the normal-state bosonic properties, emphasizing the localization and spectral statistics of the low-energy modes. 
In Sec.~\ref{sec:results}, we summarize the main results for the pairing problem, combining numerical analysis with complementary analytic approaches. 
In Sec.~\ref{sec:setup}, we derive the linearized Usadel equation and discuss the assumptions and regimes of validity underlying our approach. 
In Sec.~\ref{sec:numerical}, we present additional numerical results for the pairing problem. 
Sections~\ref{sec:SCBA} and~\ref{sec:Puddles} provide detailed analytic treatments of the extended and localized pairing solutions, using the self-consistent Born approximation and the solitary-puddle approximation, respectively. 
Additional derivations and computational details are collected in the Appendices.

\section{Disordered Quantum-Critical Normal State}
\label{sec:normal}

We begin by reviewing the disordered quantum-critical normal state and the emergence of localized $\phi$ modes. 
The theory considered here is a two-dimensional Hertz-Millis-type theory \cite{Hertz1976,Millis1993,VojtaInstabilities} for an $N$-component non-conserved order-parameter field $\phi_i$ coupled via a Yukawa interaction to fermions $\psi$.
The bosons $\phi_i$ are subjected to a random mass potential, while fermions $\psi$ are subjected to a random chemical potential.
The total Lagrangian density is given by $\mathcal{L}=\mathcal{L}_{\psi}+\mathcal{L}_{\phi}+\mathcal{L}_{\psi\phi}$, where
\begin{equation}\label{eq:Model}
    \begin{aligned}
\mathcal{L}_{\psi}&=\sum_{\sigma}\bar{\psi}_{\sigma}(\partial_\tau +E(-i\nabla)+v(\mathbf{r}))\psi_{\sigma}\;,
\\
\mathcal{L}_{\phi}&= \frac{1}{2}\sum_{i = 1}^N\left[(\partial_\tau \phi_i)^2+c^2(\nabla\phi_i)^2 +(\lambda+\lambda'(\mathbf{r})) \phi_i^2 \right]+ \frac{u}{4N} \left( \sum_{i = 1}^N\phi_i^2 \right)^2\;,
\\
\mathcal{L}_{\psi\phi}&= \frac{g}{\sqrt{N}} \sum_{i = 1}^N \phi_i \bar{\psi} \hat{\mathcal{D}}_i \psi\;.
    \end{aligned}
\end{equation}
Here $\sigma{=}{\uparrow}{/}{\downarrow}$ denotes the spin component, $E(\mathbf{p})=p^2/2m -\mu$ is the fermion dispersion counted from the Fermi energy. The constant $u$ is the coupling for the order-parameter self-interaction, $\lambda$ is the bare mass, and $g$ is the translation-invariant Yukawa coupling,  with $\hat{\mathcal{D}}_i$ describing the appropriate spatial- and spin-dependent form factor. There are two sources of disorder in this problem: the random chemical potential $v(\mathbf{r})$ and the random bosonic mass $\lambda'(\mathbf{r})$, which we assume to be uncorrelated Gaussian white-noise fields 
\begin{equation}
    \overline{v(\mathbf{r})} =\overline{\lambda'(\mathbf{r})}=\overline{v(\mathbf{r})\lambda'(\mathbf{r}')} =0,\quad  \overline{v(\mathbf{r})v(\mathbf{r}')}=\bar{v}^2\delta(\mathbf{r}-\mathbf{r}'), \quad  \overline{\lambda'(\mathbf{r})\lambda'(\mathbf{r}')} =\bar{\lambda}^2\delta(\mathbf{r}-\mathbf{r}')\;.
\end{equation}
The random chemical potential $v(\mathbf r)$ produces diffusive dynamics for electrons, whereas the random mass $\lambda'(\mathbf r)$ is Harris disorder: it locally shifts the distance to the quantum critical point.

The pairing instability of the theory in Eq. (\ref{eq:Model}) is mediated by the near-critical bosons $\phi_i$, which generate an effective attractive interaction for many theories of interest (i.e. many $\hat{\mathcal{D}}_i$), such as spin-density-wave and nematic order parameters. For simplicity, we will focus on the trivial s-wave form-factor $\hat{\mathcal{D}}=\hat{1}$.
Assuming that the random chemical potential $v(\mathbf r)$ is weak (i.e. $k_Fl\gg1$, where $l$ is the fermion mean free path), the fermionic degrees of freedom remain extended and metallic. We therefore average over $v(\mathbf r)$ perturbatively, so that it enters the bosonic theory only through self-averaged metallic parameters, while keeping the random mass disorder $\lambda'(\mathbf r)$ explicit. Integrating out the disorder-averaged fermions at the one-loop level then gives an effective Landau-damped action for the bosons $\phi_i$:
\begin{align}
   \int d \tau d\vec r \,\mathcal{L}_{\phi,{\rm eff}} = \int d \tau d\vec r\, \mathcal{L}_{\phi} +
   \frac{T}{2} \sum_{\omega_m} \int d\vec r \, c_d |\omega_m| \sum_{i = 1}^N|\phi_i (i\omega_m, \vec r)|^2,
   \label{Lphieff}
\end{align}
where $\mathcal{L}_\phi$ is given in Eq.~(\ref{eq:Model}), $\omega_m$ is a bosonic Matsubara frequency, and $c_d$ is a Landau-damping coefficient determined as $c_d \propto \nu g^2$, where $\nu$ is the electron density of states at the Fermi level \cite{Aavishkar2023}. We note that Eq.~\eqref{Lphieff} assumes an overdamped but non-conserved order parameter, for which disorder-induced diffuson vertex corrections are absent in the long-wavelength theory. This should be distinguished from cases where the order parameter corresponds to a conserved density and directly couples to the soft diffusive modes of the metal: there the resulting Hertz-Millis dynamical term is nonlocal and in momentum space results in a non-analytic form $\propto |\omega_m|/q^2$ \cite{Hertz1976,Belitz2001,Brando2016,Nosov2023}. 

We solve the problem defined by Eq. \eqref{Lphieff} using the self-consistent (Hartree) approach of Refs. \cite{Maestro2008,Aavishkar2024}, which can be formally justified in the large-$N$ limit (see Appendix~\ref{sec:Large_N}). Specifically, we solve for the normal-state boson propagator $D(i \omega_m, \vec r, \vec r^\prime)$ while treating the random-mass problem for the bosons nonperturbatively. Assuming the bosonic flavor symmetry remains unbroken, the bosonic propagator can be expressed as
\begin{equation}\label{eq:D_full_sec2}
    D(i \omega_m, \vec r, \vec r^\prime) = \sum_\alpha \frac{\phi_\alpha(\vec r) \phi_\alpha^*(\vec r^\prime)}{\omega_m^2 + c_d |\omega_m| + e_\alpha},
\end{equation}
where $\omega_m$ is a bosonic Matsubara frequency, and $\mathbf r, \mathbf r'$ are the real-space coordinates on the square lattice. The real-space profiles $\phi_\alpha(\vec r)$ and auxiliary eigenvalues $e_\alpha$ are self-consistent solutions to a non-linear real-space equation
\begin{equation}\label{eq:phiaeq}
    \left[ -c^2 \nabla^2 +\lambda + \lambda'(\vec r) + u \,T\sum_{\omega_m} D(i\omega_m, \vec r, \vec r)\right]\phi_\alpha(\vec r) = e_\alpha  \phi_\alpha(\vec r).
\end{equation}
This equation is formally equivalent to a nonlinear random Schrödinger problem with a non-linearity that depends on the full bosonic eigenspectrum, not just a single eigenstate. We characterize the resulting spectrum using several quantities. The first one is the disorder-averaged local density of bosonic eigenstates (DOS),
\begin{equation}\label{eq:nE}
n(\mathcal{E})
\equiv
\left\langle \sum_\alpha |\phi_\alpha(\vec r)|^2 \delta(e_\alpha - \mathcal{E}) \right\rangle \;.
\end{equation}
After disorder averaging this quantity is spatially uniform, and summing it over lattice sites gives the corresponding global density of states. The second diagnostic is the localization length $\xi_\alpha$ of a given eigenmode $\phi_\alpha$. In practice, we extract $\xi_\alpha$ from the inverse participation ratio (IPR),
\begin{equation}\label{eq:IPR def}
    \xi_{\alpha} \equiv \frac{1}{\sqrt{2 \pi\mathcal{I}_{\alpha}}},\quad \quad \mathcal{I}_{\alpha} \equiv \sum_{\vec r} |\phi_\alpha(\vec r)|^4
    ,\quad \quad   |\phi_\alpha(\vec r)| = \sqrt{\frac{2}{\pi}} \frac{1}{\xi_\alpha} e^{- |\vec r - \vec r_\alpha|/\xi_\alpha}\;,
\end{equation}
where the last expression gives the exponentially decaying envelope used to relate the IPR to $\xi_\alpha$, and $\vec r_\alpha$ is the localization center. Another important length scale is the bosonic correlation length $l_{\rm corr}$ which characterizes the range over which the averaged propagator $\langle D(i\omega_m,\vec r,\vec r^{\prime})\rangle \sim \exp\{-|\vec r-\vec r^{\prime}|/l_{\rm corr}\}$ decays with distance. On separation scales large compared with $l_{\rm corr}$, the propagator can be coarse-grained into a local form with $\vec r\approx \vec r^{\prime}$ that depends only on the center-of-mass coordinate and varies on the scale of the typical localization length.

\begin{figure}[t!]
\centering
    \includegraphics[width=0.95\linewidth]{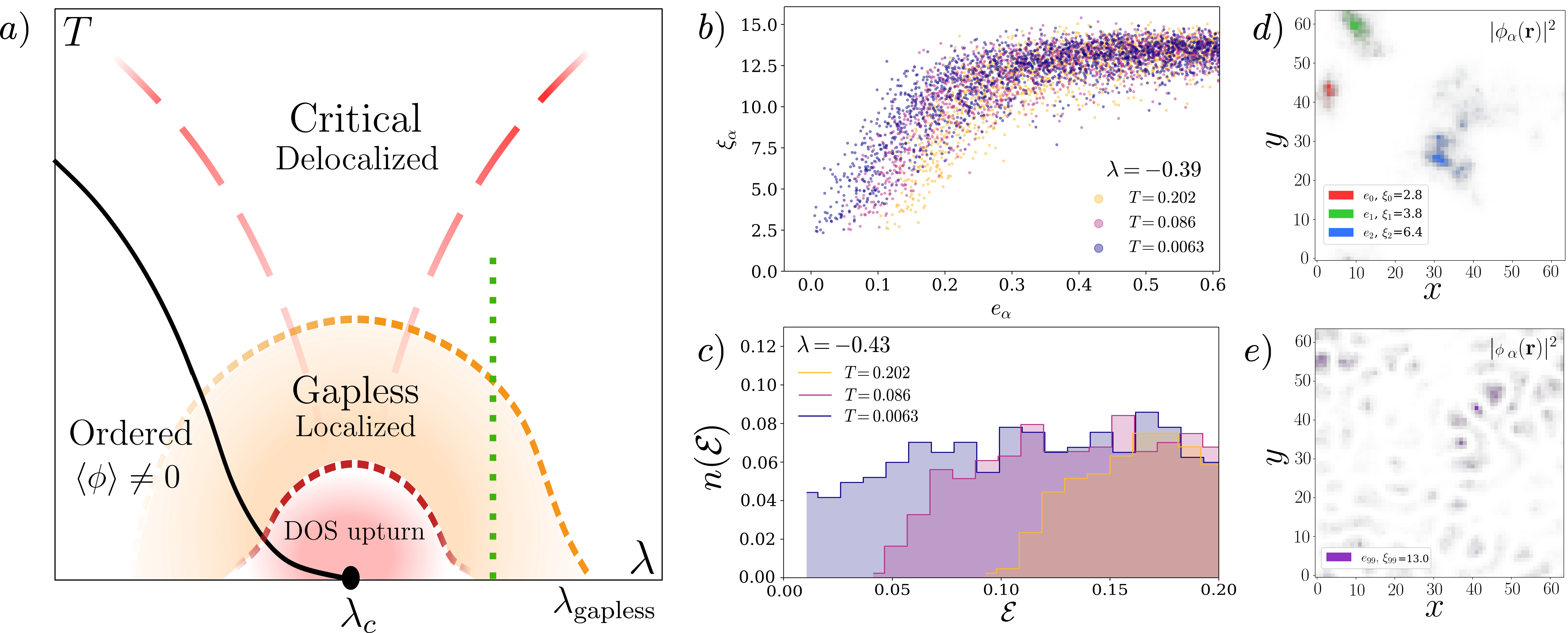}
    \captionsetup{justification=justified, singlelinecheck=false}
    \caption{
    The properties of the effective bosonic theory of the normal state are displayed. 
    (a) A schematic phase space diagram of the quantum phase transition with Harris disorder is inferred from our numerical results and Ref. \cite{Aavishkar2024}. 
    The vicinity of the quantum critical point is characterized through the behavior of the boson eigenmode DOS $n(\mathcal{E})$, defined in Eq. \eqref{eq:nE}.
    At small temperatures and large $\lambda$ the system is in a disordered paramagnetic state, and the bosonic states are generally gapped, i.e. $n(\mathcal{E} \approx 0) \approx 0$.
    As $\lambda$ is lowered, the system approaches a QCP at $\lambda_c$.
    Before the critical point is reached, the bosons become gapless at $\lambda_\mathrm{gapless} > \lambda_c$ in the sense of $n(\mathcal{E} \approx 0) = n > 0$. 
    In the immediate vicinity of the critical point, one observes accumulation of bosonic modes with extremely small energies, resulting in the appearance of an upturn at $\mathcal{E} \approx 0$ \cite{Aavishkar2024}.
    In this work, we explore the physics of a system that is still somewhat detuned from the critical point, e.g., along a representative green line in (a). 
    In this region, the low-energy modes remain compact and localized, and the bosonic DOS is approximately constant over a finite energy window.
    (b) The localization lengths $\xi_\alpha$ as a function of the energy $e_\alpha$ of the bosonic states $\phi_\alpha(\vec r)$ are small for lowest $e_\alpha$, implying strong localization present in the low energy sector.
    Three temperatures schematically correspond to three different points along the green line in panel (a).
    (c) The boson eigenmode DOS $n(\mathcal{E})$ is shown for three different temperatures along the green line in panel (a). The DOS is constant at low temperatures and develops a thermal gap as the temperature increases. 
    (d) Three bosonic eigenstates with the lowest energies $e_\alpha$ are plotted along with their localization length.
    (e) A delocalized bosonic state with a large $e_\alpha$ is plotted.
}
    \label{fig:normal_state}
\end{figure}

The action in Eq.~\eqref{Lphieff} and closely related random-mass bosonic theories have been studied extensively~\cite{vojta_rare_2006,Vojta2003,Vojta07,Vojta2009,Maestro2008,Aavishkar2024}. A central result of these works is that the random-mass bosonic theory flows toward an infinite-randomness fixed point for the class of overdamped $O(N>1)$ models considered here. The physical origin of this flow can be understood from rare extended spatial regions that support bosonic modes with anomalously small \textcolor{black}{energies} (i.e., small~$e_\alpha$). The dissipative dynamics induced by the $c_d|\omega_m|$ term in Eq.~\eqref{Lphieff} plays a crucial role. At low frequencies, the local fluctuation entering the large-$N$ constraint is logarithmically sensitive to the boson \textcolor{black}{energy}, schematically as $\int d\omega/(c_d|\omega|+e)\sim c_d^{-1}\log(\Lambda/e)$. A rare region can therefore satisfy the self-consistency condition with an exponentially small energy $e$, and hence with an exponentially long correlation time. Combining the exponentially small probability of finding a large favorable region with the exponentially long time scale of such a region gives the usual quantum Griffiths phenomenology: power-law-in-time averaged correlations and a power-law low-energy density of bosonic modes~\cite{Thill95,RSY95}. These localized modes remain dynamic for $N>1$ due to the $O(N)$ continuous symmetry with a lower critical dimension $d = 2$ (Mermin-Wagner theorem \cite{MerminWagner,Codello2013}). In terms of Eq.~\eqref{eq:nE}, the asymptotic Griffiths regime is characterized by an upturn of $n(\mathcal{E})$ at the lowest energies, which may be parameterized as $n(\mathcal{E})\approx n\mathcal{E}^{\alpha_G}$ with $\alpha_G<0$. This is the asymptotic rare-region regime indicated near $\lambda_c$ in Fig.~\ref{fig:normal_state}(a). The corresponding low-energy modes are associated with increasingly large rare regions, and their localization length grows as their energy decreases.

%
%

The pairing problem, however, is sensitive to a broader part of the bosonic spectrum, not only to its asymptotic low-energy tail. Indeed, the pairing kernel contains the full normal-state propagator in Eq.~\eqref{eq:D_full_sec2}, so modes with $e_\alpha$ up to the energy scale relevant for pairing contribute to the effective interaction. Therefore, we need the distribution of bosonic energies and wave functions over the full low- and intermediate-energy window sampled by the pairing equation, which are not attainable in analytical form. For this reason we solve Eq.~\eqref{eq:phiaeq} numerically, as was done in Ref. \cite{Aavishkar2024}, and use the resulting modes as input to the superconducting problem.

An important observation from the numerical solution is that the phase diagram summarized in Fig.~\ref{fig:normal_state}(a) contains a broad regime distinct from the immediate Griffiths-critical region. As $\lambda$ is lowered from the gapped disordered regime, the lower edge of the localized bosonic spectrum reaches zero at $\lambda_{\rm gapless}>\lambda_c$. \textcolor{black}{For $\lambda_c<\lambda<\lambda_{\rm gapless}$, the boson spectrum remains gapless and furthermore acquires a buildup of zero-energy states such that $n(\mathcal{E}\approx0) > 0$, but the corresponding low-energy modes remain strongly localized.} In this gapless-localized regime, the low-energy DOS is smooth and is well approximated by a constant over the numerical window relevant for pairing. Moreover, the bosonic correlation length $l_{\rm corr}$ remains finite and small: the gapless character comes from localized eigenmodes with small $e_\alpha$, not from a divergent spatial correlation length. 
In Fig.~\ref{fig:normal_state}(b), we plot $\xi_\alpha$ versus $e_\alpha$ for several temperatures along the representative green dashed trajectory in Fig.~\ref{fig:normal_state}(a). The localization length is shortest for the smallest $e_\alpha$, and then increases with $e_\alpha$ until reaching the finite-size cutoff. Representative low-energy localized wave functions are shown in Fig.~\ref{fig:normal_state}(d), while a high-energy extended state is shown in Fig.~\ref{fig:normal_state}(e). The corresponding density of states, shown in Fig.~\ref{fig:normal_state}(c), shifts toward lower energies as temperature is lowered and eventually becomes gapless. We emphasize that this regime should be distinguished from the asymptotic Griffiths regime closer to $\lambda_c$, where the DOS develops an upturn and the lowest-energy states are controlled by increasingly large rare regions with growing localization length. 

The origin of this gapless-localized regime can be understood directly from the nonlinear eigenvalue problem in Eq.~\eqref{eq:phiaeq}. The random mass acts as a spatially inhomogeneous potential for the bosonic modes. Regions, where the local bare potential $\lambda+\lambda'({\bf r})$ is anomalously small, form potential wells, and these wells can bind localized bosonic eigenstates. As the average tuning parameter $\lambda$ (or temperature $T$) is lowered, the distribution of such local wells shifts downward in energy (see Fig.~\ref{fig:normal_state}~(c)). In the absence of the self-consistent Hartree term,  eigenstates with negative energies eventually would appear, signaling a tendency towards local condensation of $\phi$. However, the presence of the non-linearity prevents the deepest states from simply condensing into static order by pushing their energy above zero: any anomalously soft localized mode enhances the interaction-induced part of the potential $T\sum_{\omega_m}D(i\omega_m,\mathbf r,\mathbf r)$, implying that the local well is screened by the bosonic fluctuation that it itself produces. The dissipative dynamics again plays an important role here: because a soft overdamped mode contributes to the local fluctuation with a logarithmic sensitivity to its energy, $\sim c_d^{-1}\log(\Lambda/e_\alpha)$, the feedback becomes increasingly important as $e_\alpha$ approaches zero. The nonlinearity therefore pushes would-be negative and almost condensed localized states back to small positive energies. In this sense, the self-consistent Hartree term compresses the Lifshitz tail of the underlying random-mass problem and consequently generates a finite eigenmode DOS, $n(\mathcal{E}\approx0) > 0$. We note that this mechanism is not restricted to the asymptotic Griffiths regime: it only requires random-mass wells capable of binding localized bosonic modes, a local repulsion, and dissipative dynamics that makes the local fluctuation strongly sensitive to small $e_\alpha$. The result is a broad, de-tuned regime in which compact localized overdamped modes occur with a finite, approximately constant low-energy density of states.

As we discuss in Sec. \ref{sec:results}, these modes from the gapless localized phase are what provide the inhomogeneous pairing glue for electrons. Moreover, at higher temperatures above the ``gapless" crossover (dashed orange line in Fig.~\ref{fig:normal_state}(a)), the same arguments can apply. This is because the pairing problem receives contributions from a finite window of bosonic energies, not only from the strict $e_\alpha\to0$ limit. Therefore the localized-glue description also applies on the gapped side, provided the gap is not too large relative to temperature, $e_{\rm gap}\lesssim c_d T$. For the range of temperatures and parameters we use in this work, this is indeed the case.

\section{Quantum-critical superconductivity -- results}
\label{sec:results}

\begin{figure}[t!]
        \centering
    \includegraphics[width=0.97\linewidth]{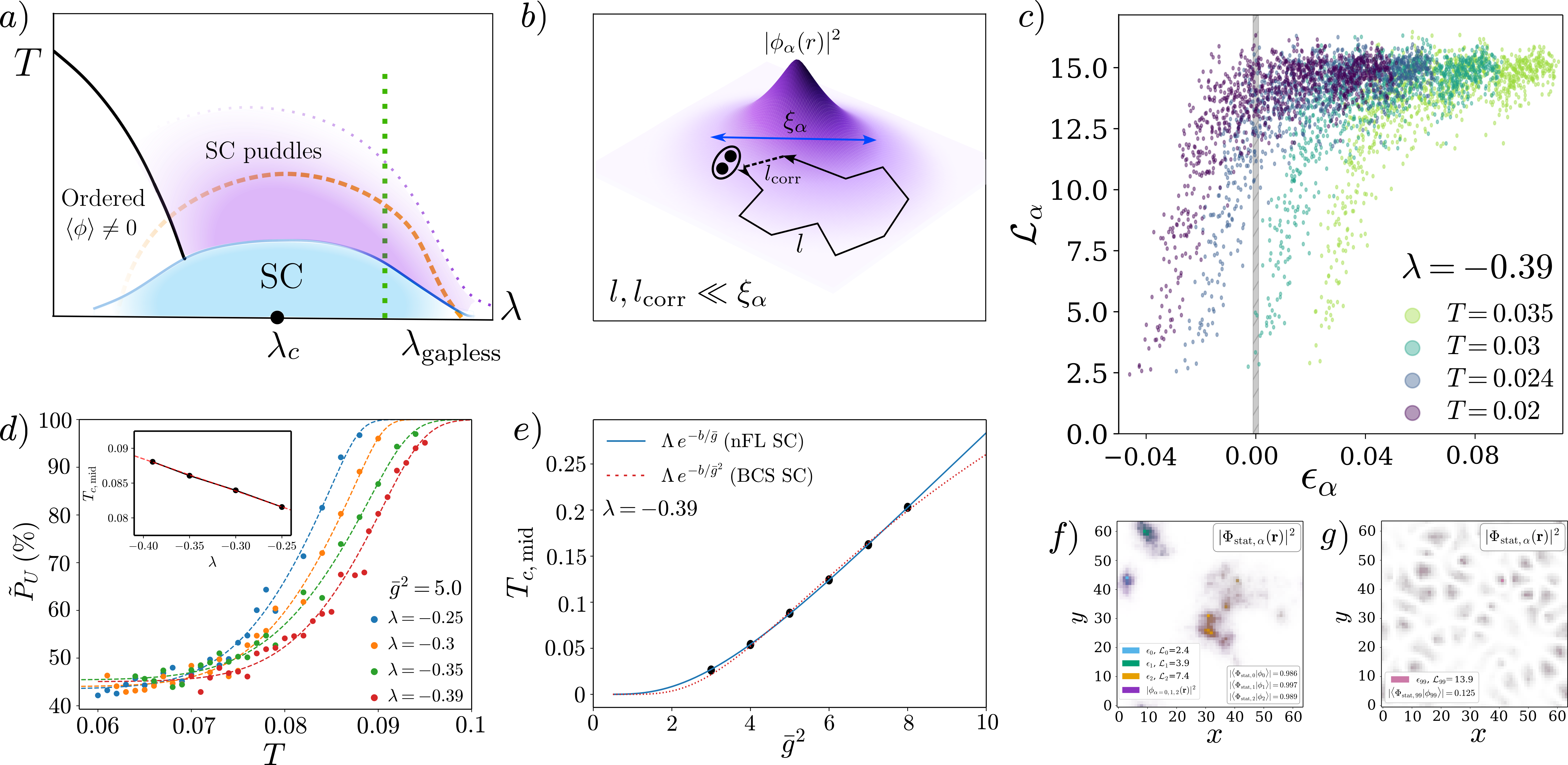}
        \caption{Properties of the superconducting state induced by the overdamped localized bosonic modes described in Sec.~\ref{sec:normal} are displayed. 
        (a) The inferred phase diagram of the superconducting state. 
       Above the dome associated with an extended macroscopic pairing instability, the system contains a region dominated by localized superconducting puddles.
        The width of the superconducting dome and the puddle regime is expected to be of the same order as the width of $n(\mathcal{E}) \approx n > 0$ region shown in Fig. \ref{fig:normal_state}.
        (b) Cartoon of a localized bosonic mode and a diffusive Cooperon loop, illustrating the hierarchy of length scales used to obtain the modified Usadel description. Here $l$ is the electronic mean free path, while $l_{\rm corr}$ is the bosonic correlation length. Both are assumed to be much smaller than the bosonic localization length $\xi_\alpha$.
        (c) The localization length $\mathcal{L}_{\alpha}$ of the eigenstates of the pairing operator in Eq. \eqref{eq:full_gap_intextSum}
        as a function of the eigenvalue $\epsilon_{\alpha}$, for several temperatures. The parameter values used to generate this data are $\lambda = -0.39, \bar g^2 = 3$. Results for all disorder configurations are plotted together. 
        All eigenstates with $\epsilon_\alpha < 0$ are unstable and therefore expected to condense (at the linearized level).
        The solutions of Eq. \eqref{eq:full_gap_intextSum} (i.e. $\epsilon_{\alpha} = 0$) are highly localized at high temperatures and delocalized below a certain temperature, which we denote as $T_{c,\text{global}}$ (here $T_{c,\text{global}} \approx 0.024$). 
        (d) The percentage of ungapped area $\tilde P_U(\%)$ is shown (cf. Eq. (\ref{eq:tilde PU}) for a precise definition) for $\bar{g}^2 = 5$. The inset shows the linear dependence of the midpoint temperature $T_{c,\text{mid}}$ on $\lambda$.
        (e) $T_{c,\text{mid}}$ as a function of the squared coupling constant $\bar g^2$ (see Eq. \eqref{eq:AvDandSigma_sec2}, \eqref{eq:TcSecIII}).
        (f), (g) The real space profile of eigenstates of the pairing operator in Eq. \eqref{eq:full_gap_intextSum} for various localization lengths. The parameter values are $\lambda = -0.39, T = 0.05, \bar{g}^2 = 4$.
        Three localized eigenstates with the highest local pairing temperatures are shown in panel (f). These superconducting puddles are pinned near the bosonic modes with the smallest energies and shortest localization lengths,  see Fig. \ref{fig:normal_state} for comparison.
        Representative extended eigenstate of Eq. \eqref{eq:full_gap_intextSum} is shown in panel (g).
        }
        \label{fig:SC_state}
    \end{figure}

With the properties of the quantum-critical normal state documented in Sec.~\ref{sec:normal}, we now turn to the superconducting instability. In this section, we summarize the main results, leaving the derivation of the effective pairing equation to Sec.~\ref{sec:setup}. The resulting phenomenology is summarized in Fig.~\ref{fig:SC_state}.

The key input from Sec.~\ref{sec:normal} is that in our regime of interest the bosonic spectrum is gapless while the spatial correlations of the bosonic propagator remain short-ranged. More precisely, the bosonic correlation length $l_{\rm corr}$ stays finite and small, whereas the low-energy bosonic eigenmodes have localized envelopes whose length scale is set by their localization length $\xi_\alpha$. This separation allows us to coarse-grain the bosonic propagator over its short relative-coordinate structure. After this coarse-graining, the relative coordinate of the Cooper pair is integrated out, and the remaining spatial dependence of the linearized Cooper problem is through the slow center-of-mass coordinate $\vec r$, as in the standard theory of disordered superconducting thin films. The length-scale hierarchy underlying this description is illustrated schematically in Fig.~\ref{fig:SC_state}(b).

The normal-state bosonic modes then enter the pairing problem through two local, spatially inhomogeneous quantities:
\begin{equation}\label{eq:InteractionFull}
\bar D(i \omega_m, \vec r) = \sum\limits_{\alpha}\frac{ |\phi_\alpha(\vec{r})|^2 }{\omega_m^2+c_d|\omega_m|+e_\alpha},
\quad\quad
\bar \Sigma(i {\varepsilon_n}, \vec r) = \frac{\bar g^2}{\pi c_d} \sum_\alpha |\phi_\alpha(\vec r)|^2 \log \left( \frac{c_d |{\varepsilon_n}| + e_\alpha}{\pi c_d T +e_\alpha}\right).
\end{equation}
Here $\bar D(i\omega_m,\vec r)$ is the local bosonic propagator entering the pairing vertex, while $\bar\Sigma(i\varepsilon_n,\vec r)$ is the corresponding local normal-state fermion self-energy (already integrated over the internal frequency). Both quantities vary in space through the weights $|\phi_\alpha(\vec r)|^2$, and hence their spatial variation is controlled by the localization lengths $\xi_\alpha$ of the bosonic modes that dominate at the relevant frequency. The effective local coupling is $\bar g^2=8\pi^2 l_{\rm corr}^2\nu g^2$.

The linearized pairing problem is therefore an eigenvalue problem in Matsubara-frequency and center-of-mass coordinate space. We define pairing eigenstates $\Phi_\alpha(i\varepsilon_n,\vec r)$ and eigenvalues $\epsilon_\alpha$ by
\begin{equation}\label{eq:full_gap_intextSum}
\Big[-D\nabla^2+2|\varepsilon_n|+2 \bar \Sigma(i\varepsilon_n,\vec{r})\Big]\Phi_\alpha(i\varepsilon_n , \vec{r})
-2\bar{g}^2 T \sum\limits_{\varepsilon_m} \bar D(i \varepsilon_m - i \varepsilon_n, \vec r) \Phi_\alpha(i\varepsilon_m,\vec{r})
=
\epsilon_\alpha\Phi_\alpha(i\varepsilon_n,\vec r).
\end{equation}
Here $D=v_F^2\tau/2$ is the electron diffusion coefficient and $\tau^{-1}$ is the elastic scattering rate induced by the random potential $v(\vec r)$.
We note that Eq.~\eqref{eq:full_gap_intextSum} is the modified version of the standard linearized Usadel equation ~\cite{Usadel1970,Meyer2001,Tikhonov2020,larkin1972density} 
for the anomalous pair amplitude.
For instance, in case of a random BCS coupling $\lambda_{\rm BCS}(\mathbf{r})$ (which was first studied in \cite{larkin1972density}), this equation becomes 
\begin{equation}\label{eq:random_BCS}
    \Big[-D\nabla^2+2|\varepsilon_n|\Big]\Phi_\alpha(i\varepsilon_n , \vec{r})-2\lambda_{\rm BCS}(\mathbf{r}) T \sum\limits_{\varepsilon_m} \Phi_\alpha(i\varepsilon_m,\vec{r}) = \epsilon_\alpha\Phi_\alpha(i\varepsilon_n,\vec r)\;.
\end{equation}
 When tuned away from the critical point, all bosonic energies $e_\alpha$ in Eq.~\eqref{eq:InteractionFull} become large, and the boson eigenmode DOS opens a gap.
 With the low-energy bosonic states absent, the frequency dependence in the denominator of Eq. \eqref{eq:InteractionFull} can be neglected, leading to the random-BCS case described by Eq.~\eqref{eq:random_BCS} with $\lambda_{\mathrm{BCS}}(\vec r) = \bar{g}^2\sum_\alpha |\phi_\alpha(\vec r)|^2 / e_\alpha$.

The superconducting instability occurs when the lowest pairing eigenvalue crosses zero. For $\epsilon_\alpha<0$, the pair mode is unstable towards condensation (at the linearized level). If the corresponding eigenfunction is localized in space, then we interpret this mode as a nucleating  superconducting puddle, corresponding to the upper inhomogeneous regime in Fig.~\ref{fig:SC_state}(a). If the eigenfunction is extended, then the transition corresponds to the onset of a macroscopic pairing state, schematically indicated by the dome in Fig.~\ref{fig:SC_state}(a). The numerical evolution from localized pairing zero modes at high temperature to an extended zero mode at lower temperature is shown in Fig.~\ref{fig:SC_state}(c), with representative localized and extended eigenfunctions shown in Figs.~\ref{fig:SC_state}(f) and~\ref{fig:SC_state}(g), respectively.

Equation~\eqref{eq:full_gap_intextSum} also makes clear the physical competition controlling the transition. The local propagator $\bar D(i\omega_m,\vec r)$ enhances pairing through a retarded attractive interaction. The same localized bosonic modes also generate $\bar\Sigma(i\varepsilon_n,\vec r)$, which enters as a dephasing term. Thus favorable regions with strong bosonic glue also tend to have enhanced normal-state scattering. The structure of the pairing eigenstates is determined by the balance between this local attraction, local dephasing, and diffusive spreading of Cooper pairs through the kinetic term $-D\nabla^2$.

Let us finally comment on the self-consistency of using the normal-state bosonic propagator as input to Eq.~\eqref{eq:full_gap_intextSum}. First, we treat the superconducting correlations as a linear instability out of the normal state. The anomalous amplitude is infinitesimal at the transition, so its feedback on the bosonic spectrum, including the depletion of low-energy fermionic states, is beyond the linearized approximation. Second, in the regime considered here, the fermionic polarization bubble is self-averaging and remains local on the scale of the electron mean free path, $l$. Since the inhomogeneous self-energy $\bar\Sigma(i\varepsilon_n,\vec r)$ varies only on the longer bosonic localization scale, it is approximately constant over the region that controls the local polarization bubble. Thus, within the coarse-grained approximation used here, the Landau-damping term $c_d|\omega_m|$ is not affected by the inhomogeneous self-energy background. The effect of $\bar\Sigma(i\varepsilon_n,\vec r)$ is retained in the pairing problem only as a dephasing term. Finally, we note that the same coarse-grained pairing equation \eqref{eq:full_gap_intextSum} can be derived in a large-$N$ limit of a modified version of Eq.~\eqref{eq:Model}, in which the Yukawa coupling is taken to be spatially random with zero mean (see Appendix~\ref{sec:Large_N}).

In the rest of this section, we summarize the main properties obtained from  Eq.~\eqref{eq:full_gap_intextSum} in three complementary ways. In Sec.~\ref{sec:numerics_results}, we obtain the spectrum of Eq.~\eqref{eq:full_gap_intextSum} numerically using the exact bosonic modes obtained from the normal-state problem of Sec.~\ref{sec:normal}. In Sec.~\ref{sec:SCBA_results}, we develop a self-consistent Born approximation that describes the extended (but inhomogeneous) pairing solution and the macroscopic superconducting transition by treating the bosonic statistics phenomenologically. In Sec.~\ref{sec:puddles_results}, we analyze the localized high-temperature tail of the pairing spectrum, corresponding to isolated superconducting puddles nucleated by the most favorable low-energy localized bosonic modes.

\subsection{Numerical analysis}\label{sec:numerics_results}

We first analyze Eq.~\eqref{eq:full_gap_intextSum} numerically, using the exact normal state boson eigenmodes described in Fig. \ref{fig:normal_state}. All simulations are done on a square lattice of linear size $L$ with periodic boundary conditions and with the following fixed parameters: $L = 64, D = 0.5, \bar{\lambda} = 0.5, c_d = 1, c^2 = 10, u = 1$. The number of disorder configurations is $N_d = 10$. The results of the numerical analysis are summarized in Fig. \ref{fig:SC_state}, and more details are given in Sec. \ref{sec:numerical}.

The spatial structure of the superconducting state is contained in the spectral analysis of the linear pairing operator defined in Eq.~\eqref{eq:full_gap_intextSum}. Diagonalizing this operator yields eigenvectors $\Phi_{\alpha} (i \varepsilon_m, \vec r) $ and corresponding eigenvalues $\epsilon_\alpha$. \color{black}
In order to characterize the spatial structure of these emergent pairing eigenmodes and their relation to the underlying localized bosonic states $\phi_\alpha$, we focus on the static component $ \Phi_\mathrm{stat, \alpha}(\mathbf r) \equiv T\sum_{m} \Phi_{\alpha} (i\varepsilon_m, \vec r)$, which enters the static pair susceptibility. Indeed, for a frequency-independent local pairing source (cf. Eq.~\eqref{eq:full_gap_intext}), the non-local zero-frequency pair susceptibility has the spectral representation
\begin{equation}
\Phi_\mathrm{stat}(\mathbf r,\mathbf r')  =\sum\limits_\alpha \frac{\Phi_\mathrm{stat, \alpha}(\mathbf r) \Phi_\mathrm{stat, \alpha}(\mathbf r')}{\epsilon_\alpha}\;.
\end{equation}
Thus the divergence of the pair susceptibility is controlled by the eigenmodes with $\epsilon_\alpha \approx 0$, while the spatial structure of the pair response is encoded in $\Phi_{\rm stat,\alpha}(\mathbf r)$.
\color{black}

The relationship between the spatial structure of $\Phi_{\rm stat,\alpha}$ and the eigenvalue $\epsilon_\alpha$ closely resembles that of the localized bosonic modes $\phi_\alpha(\mathbf r)$. Defining its localization length $\mathcal L_{\alpha}$ exactly as done in Eq. (\ref{eq:IPR def}) (making sure to normalize $\Phi_\mathrm{stat, \alpha}(\mathbf r)$ beforehand), we can see in Fig. \ref{fig:SC_state}(c) that $\mathcal L_\alpha$ is $\mathcal O(1)$ for the lowest pairing eigenstates, and then increases linearly to $\mathcal O(L/2)$. As the temperature decreases from well above $T_c$, the solutions of the gap equation, i.e. $\epsilon_{\alpha'} = 0$, have a monotonically increasing $\mathcal L_{\alpha'}$, from highly localized puddles to a fully delocalized solution. From the solution's localization length we can extract the percentage of area that is ungapped, $\tilde P_U$ (defined in Eq. (\ref{eq:tilde PU})), which is shown as a function of $T$ in Fig. \ref{fig:SC_state}(d). 

The shape of these curves and their behavior with $\lambda$ is consistent with the same curves extracted from scanning tunneling microscopy (STM) measurements of $\text{Bi}_2\text{Sr}_2\text{CaCu}_2\text{O}_{8+\delta}$ from Ref. \cite{Yazdani_puddles_2007} (c.f. Fig. \ref{fig:Yazdani plots percentage} and Sec. \ref{sec:numerical}). Both the experimental and numerical simulation curves show a very gradual onset of  superconductivity as temperature is lowered: the interval of temperatures in which the sample changes from mostly ungapped to mostly gapped is comparable to the macroscopic $T_c$ of the sample. Moreover, both sets of data are well fit by skew-normal distributions, and
their midpoints shift linearly with a tuning parameter (i.e. doping in experiment and detuning from the critical point in our numerics).

The transition temperature of delocalized solutions has a dependence on the coupling strength $\bar g^2$ that is $T_{c} \sim \Lambda e^{-b/\bar g}$, which is a known result in the theory of non-Fermi liquid superconductivity \cite{Son:1998uk}. This is shown in Fig. \ref{fig:SC_state}(e), where it is compared to the BCS expectation of $T_{c} \sim \Lambda e^{-b/\bar g^2}$. This result is derived in Sec. \ref{sec:SCBA_results} and \ref{sec:SCBA} using the self-consistent Born approximation. 

Highly localized states with $\epsilon_\alpha \leq 0$ correspond to isolated condensed puddles that will not result in global phase coherence. Some examples of these are shown in Fig. \ref{fig:SC_state}(f). One important feature is that the most localized superconducting puddles sit almost entirely on the most localized bosonic modes that form the pairing glue (in contrast to the delocalized state in Fig. \ref{fig:SC_state}(g)). In Sec. \ref{sec:numerical} we show that this feature is generic across $\lambda, \bar g, T$. Focusing on higher temperature, we analyze these localized solutions in detail in Sec. \ref{sec:puddles_results} and \ref{sec:Puddles}, where we manage to reproduce this high overlap of localized states, and make a prediction for the high $T$ behavior of $\tilde P_U(T)$ (which we currently lack the numerical precision to confirm).

\subsection{Extended pairing instability and mesoscopic gap inhomogeneity}
\label{sec:SCBA_results}

The numerical solution of Eq.~\eqref{eq:full_gap_intextSum}, summarized in Fig.~\ref{fig:SC_state}, shows that the pairing equation develops extended eigenstates at sufficiently low temperature. To analyze these extended solutions analytically, we employ a self-consistent Born approximation (SCBA), analogous to the approximation commonly used to study disorder-averaged spectral properties in disordered conductors~\cite{Lee1985}. The strategy is to separate the spatially averaged pairing problem from fluctuations around it. More specifically, we replace $\bar\Sigma(i\varepsilon_n,\mathbf r)$ and $\bar D(i\omega_m,\mathbf r)$ by their spatial averages plus their leading spatial cumulants, and treat the latter as weak disorder in the pairing kernel. Building on the phenomenology of Sec.~\ref{sec:normal}, we assume that the low-energy bosonic density of states is approximately constant, $n(\mathcal{E}) = n = \mathrm{const}$ (see Eq. \eqref{eq:nE}), from the lowest energies up to a cutoff $\mathcal E\sim \Lambda_B^2$, while the bosonic localization length $\xi$ is roughly energy independent.

Averaging over the bosonic spectrum gives the following low-frequency forms for the spatially averaged pairing interaction and fermion self-energy:
\begin{equation}\label{eq:AvDandSigma_sec2}
    \langle \bar D \rangle(i \omega_m) \approx n
    \ln\left(\frac{\Lambda}{|\omega_m|}\right),
    \quad\quad
    \langle \bar \Sigma \rangle(i \varepsilon_n) \approx \tilde g^2
    |\varepsilon_n| \ln \left(\frac{\Lambda}{|\varepsilon_n|}\right), \quad\quad \tilde g^2 = \frac{\bar g^2 n}{\pi},
\end{equation}
where $\Lambda = \Lambda_B^2/c_d$.
The electron self-energy $\langle\bar \Sigma\rangle(i \varepsilon_n)$ has a marginal Fermi-liquid form \cite{MarginalFLPhenomenology}, implying that electrons form a strange metal with a $T$-linear resistivity and $T\ln T$ contribution to the specific heat in a normal state.
The derivation of the averaged expressions for $\langle \bar D\rangle(i \omega_m)$ and $\langle \bar \Sigma \rangle(i \varepsilon_n)$ can be found in Section \ref{sec:SCBA} with the result being Eq. \eqref{eq:AvDandSigma}.

Spatial fluctuations of $\bar D(i\omega_m,\mathbf r)$ generate an additional  effective pairing kernel of the form
\begin{equation}\label{eq:KDDFin_sec2}
    K_{DD}(\varepsilon_n, \varepsilon_m) \simeq \frac{\vartheta n}{2} \, \Theta\left(E_\mathrm{Th} - \eta\right) \ln^2 \left( \frac{E_\mathrm{Th}}{\eta} \right), \quad \quad \eta = \max(|\varepsilon_n|, |\varepsilon_m|), 
    \quad\quad \vartheta = \frac{3 \bar g^2}{2 \pi^2 c_d D}.
\end{equation}
In the expression above, $E_\mathrm{Th} = 2 D/\xi^2$ is the Thouless energy of a Cooper pair confined in a region of size $\xi^2$.
In this case, $\xi = \xi(\pi c_d T)$ is a characteristic bosonic localization length.
The total pairing vertex in Eq.~\eqref{eq:full_gap_intextSum} in this case takes a form $\bar D(i \varepsilon_n - i \varepsilon_m, \vec r) \rightarrow \langle \bar D \rangle(i \varepsilon_n - i \varepsilon_m) + K_{DD}(\varepsilon_n, \varepsilon_m)$.
\textcolor{black}{The double logarithm $\ln^2(E_\mathrm{Th}/\eta)$ in Eq.~\eqref{eq:KDDFin_sec2} has a simple physical origin.
One logarithm comes} from the static spatial redistribution of the Cooper pairs away from the homogeneous solution.
Since $\bar D(i\omega_m, \vec r)$ plays the role of an attractive potential, Cooper pairs naturally concentrate in the regions of the stronger attraction, leading to higher values of the Gor'kov function $\Phi(i\varepsilon_n, \vec r)$ in those regions. In turn, higher values of $\Phi(i\varepsilon_n, \vec r)$ imply that the Cooper pairs experience, on average, a larger potential than $\langle \bar D \rangle(i \omega_m)$.
\textcolor{black}{The second logarithm originates from the enhanced return probability of diffusive Cooper pairs in two dimensions. Cooper pairs that enter a favorable attractive region are likely to return to it many times before escaping, further increasing the effective attraction.}
\textcolor{black}{It also turns out that spatial fluctuations of $\bar\Sigma(i\varepsilon_n,\mathbf r)$ produce only subleading corrections, which mainly renormalize the effective coupling constant $\bar{g}^2$ and do not change the qualitative structure of the pairing kernel.}
The derivation of Eq. \eqref{eq:KDDFin_sec2} is given in Sec.~\ref{sec:PairingPotCorr}.

It is important to note that the explicit dependence of $\vartheta$ on the Yukawa coupling constant $\bar g$ in Eq.~\eqref{eq:KDDFin_sec2} should not be interpreted as an additional small parameter associated with the interaction strength. In a self-consistent treatment, the Landau-damping coefficient $c_d$ is itself generated by the same electron-boson coupling (e.g., at the one-loop level, we have  $c_d\sim \nu \bar g^2$). With this scaling, $\vartheta
\propto \bar g^2/c_d D
\sim
1/\nu D$, which is proportional to the inverse dimensionless Drude conductance $\mathcal G^{-1}\sim 1/k_F l\ll 1$. We should therefore regard $\vartheta$ as an electronic parameter (independent of $\bar g$) controlling the strength of mesoscopic fluctuation effects.

Dimensionless constants $\tilde g$ and $\vartheta$, along with two energy scales $\Lambda$ and $E_\mathrm{Th}$, fully define the behavior of the averaged theory.
With the self-energy and the pairing potential given by Eqs. \eqref{eq:AvDandSigma_sec2} and \eqref{eq:KDDFin_sec2}, Eq. \eqref{eq:full_gap_intextSum} attains a superconducting solution at $T_c$ given by
\begin{equation}\label{eq:TcSecIII}
    T_c \simeq \begin{cases}
        E_\mathrm{Th} \exp \left[- \#/(\tilde g^2 \vartheta)^{1/3} \right], \quad \tilde g \lesssim \tilde g_*,
        \\
        \Lambda \exp \left[ - \#/ \tilde g \right], \quad\quad\quad\quad\quad\;\, \tilde g \gtrsim \tilde g_*,
    \end{cases}
    \quad\quad\quad \tilde g_* = \min[\vartheta, \; \ln^{-1}(\Lambda/E_\mathrm{Th})].
\end{equation}
Note that above it is assumed that $E_\mathrm{Th} \lesssim \Lambda$.
In case of the opposite, one should set $E_\mathrm{Th} = \Lambda$ in all the equations that involve $E_\mathrm{Th}$.
The calculation supporting Eq. \eqref{eq:TcSecIII}, along with details of the structure of spatially averaged Gor'kov function $\langle\Phi\rangle(i\varepsilon_n)$, can be found in Section \ref{sec:AvGapSCBA}.

\textcolor{black}{The two lines of Eq.~\eqref{eq:TcSecIII} describe two distinct regimes.} When $\tilde g$ is large, the effects of spatial inhomogeneity provided by Eq. \eqref{eq:KDDFin_sec2} are negligible. 
In this scenario, the Cooper pairs \textcolor{black}{diffuse over many bosonic localization volumes} and probe numerous bosons. 
This behavior efficiently averages over the inhomogeneity, resulting in the naively expected answer shown in the second line of Eq. \eqref{eq:TcSecIII}. 
For small $\tilde g$, the inhomogeneity correction from Eq. \eqref{eq:KDDFin_sec2} dominates over the average effects, given by Eq. \eqref{eq:AvDandSigma_sec2}.
This signals the importance of the Cooper pair diffusive return physics and therefore predicts a significant gap inhomogeneity.

To further support the described physical picture, we estimate the degree of inhomogeneity of the pairing amplitude in the same formalism.
We evaluate the structure factor of the spatial fluctuations of the Gor'kov function $\delta \Phi(i \varepsilon_n, \vec r)$ and the pairing amplitude $\delta \Phi_\mathrm{stat}(\vec r)$, where 
\begin{align}
\Phi(i \varepsilon, \vec r) = \langle \Phi\rangle(i \varepsilon) + \delta \Phi(i \varepsilon, \vec r), 
\quad\quad
\Phi_\mathrm{stat}(\vec r) = \int \frac{d\varepsilon}{2 \pi} \Phi(i \varepsilon, \vec r) = \langle \Phi_\mathrm{stat} \rangle + \delta \Phi_\mathrm{stat}(\vec r).
\end{align}
We seek $\delta \Phi(i\varepsilon_n, \vec r)$ as a solution to Eq. \eqref{eq:full_gap_intextSum}, with the fluctuation effects of $\delta \bar \Sigma(i\varepsilon_n, \vec r) = \bar \Sigma(i\varepsilon_n, \vec r) - \langle \bar\Sigma\rangle(i\varepsilon_n)$ and $\delta \bar D(i \omega_m, \vec r) = \bar D (i\omega_m, \vec r) - \langle \bar D\rangle(i \omega_m)$ accounted up to the leading order. \textcolor{black}{The correlation function of $\delta\Phi_\mathrm{stat}(\vec r)$ for momenta above the thermal diffusive scale, $q\gtrsim \sqrt{T_c/D}$, takes the form (cf. Eq.~(\ref{dLdLres}))}
\begin{equation}
      \frac{\langle \delta \Phi_\mathrm{stat} \, \delta \Phi_\mathrm{stat}\rangle(\vec q)}{\langle\Phi_\mathrm{stat}\rangle^2} \simeq \frac{32 \pi^4 \vartheta}{3 \vec q^2} \frac{\ln \left( \frac{D \vec q^2}{\pi T_c} \right)}{\left(1 + \vec q^2 \xi^2/4 \right)^{3}}, \quad D \vec q^2 \gtrsim \pi T_c.\label{eq:LL_corr}
\end{equation}
The details of this calculation can be found in Section \ref{sec:GapFluctSCBA}.
The non-integrable divergence of Eq. \eqref{eq:LL_corr} at small momenta implies that the scale of relative spatial pairing amplitude fluctuations  \textcolor{black}{have a logarithmic  enhancement at large distances.}
These fluctuations become of the order of the average quantity at the distance $\xi_\mathrm{SC}$, where 
\begin{equation}\label{eq:theta_star1}
    \ln \left( \frac{\xi_\mathrm{SC}}{\xi} \right) \simeq \begin{cases}
        1 \quad \tilde g \lesssim \tilde g_*,
        \\
        \tilde g/\vartheta, \quad \tilde g_* \lesssim \tilde g \lesssim \vartheta/\sqrt{\vartheta_*},
        \\
        1/\sqrt{\vartheta_*}, \quad \vartheta/\sqrt{\vartheta_*} \lesssim \tilde g ,
    \end{cases}
    \quad\quad
    \vartheta_* = \max(\vartheta, \vartheta^2 \ln^2(\Lambda/E_\mathrm{Th})).
\end{equation}
At the scale $\xi_\mathrm{SC}$, the impact of spatial fluctuations becomes significant, hence the SCBA approximation breaks down in favor of the strongly inhomogeneous profile of the superconducting gap.
For large values of $\tilde g$, the inhomogeneity length scale of the superconducting gap is exponentially larger than a typical inhomogeneity length scale of bosons $\xi$.
This conclusion is consistent with the intuition of an extended gap with freely propagating Cooper pairs that efficiently average out the bosonic inhomogeneity.
For small values of $\tilde g$, the length scales $\xi_\mathrm{SC}$ and $\xi$ are comparable.
This implies that the Cooper pairs prefer to occupy the regions of higher attraction and strongly prefer to remain in their vicinity due to the increased probability of return.

The effects of bosonic disorder described above are qualitatively different from the effects of potential disorder in BCS in multiple aspects.
First, the BCS problem with the UV scale $\Lambda$ and effective BCS coupling $\tilde g^2$ necessarily results in $T_{\mathrm{BCS}} \sim \Lambda \exp[- \#/ \tilde g^2]$, which is parametrically much smaller than $T_c$ in Eq. \eqref{eq:TcSecIII}. 
The effects of bosonic inhomogeneity should also be contrasted with the role of disorder in conventional BCS mean-field analyses. 
For an $s$-wave order parameter, time-reversal-preserving potential disorder is largely inconsequential due to Anderson's theorem, and substantial spatial variations of the pairing amplitude arise primarily from rare-region effects \cite{Larkin1981,Meyer2001,Dodaro2018}. For a $d$-wave order parameter, short-wavelength disorder can additionally act as a pair breaker by scattering quasiparticles between regions of the Fermi surface with opposite signs of the gap, leading to stronger gap inhomogeneity already at the BCS level \cite{Atkinson2000,Nunner2005,KivelsonLee21}. In our problem, the origin of the gap inhomogeneity is different: the overdamped dynamics of localized bosonic modes produces strong long-wavelength variations of the pairing vertex, with a nontrivial dependence on the  Matsubara frequency of the Gor'kov function. Rather than acting as a conventional pair-breaking impurity potential, this vertex locally favors pairing, but the presence of a finite density of low-energy localized modes strongly fragments the pairing response in space, as evidenced by Eq.~\eqref{eq:LL_corr}. We further corroborate the role of such finite-density low-energy localized modes below, where we analyze pairing solutions associated with individual superconducting puddles.

\subsection{Statistics of localized superconducting puddles}
\label{sec:puddles_results}

The large number of localized superconducting solutions displayed in Fig. \ref{fig:SC_state} and the breakdown of SCBA treatment of inhomogeneity calls for a complementary approach to the pairing problem.
Evidently, these superconducting puddles emerge on top of the most localized bosons, which act as strong local mediators of pairing (compare, for example, the low-energy bosonic eigenstates in  Fig.~\ref{fig:normal_state}(d) with the localized pairing eigenstates in Fig.~\ref{fig:SC_state}(f) obtained for the same realization of disorder).
We consider a single strongly localized boson $\phi_\alpha(\vec r)$ with a localization length $\xi_\alpha$. 
Since the strongly localized bosons are relatively rare, we assume that this boson is embedded in the background of other localized bosonic states $\phi_\beta(\vec r)$ with localization lengths $\xi_\beta \gg \xi_\alpha$.
If a superconducting puddle nucleates on top of $\phi_\alpha(\mathbf r)$ with local pairing scale $T_{c,\alpha}$, its spatial extent is controlled by the thermal diffusion length of the Cooperon, $L_{T_{c, \alpha}} \sim \sqrt{D/T_{c,\alpha}}$.
We focus on the regime suggested by the numerical solutions (Fig. \ref{fig:SC_state}), in which different puddles do not overlap and the localized pairing solution is comparable in size to the parent bosonic mode. In terms of length scales, this corresponds to
$\xi_\alpha \lesssim L_{T_{c, \alpha}} \ll \xi_\beta$ for $\beta \neq \alpha$.
Hence, we are interested in a solution to Eq. \eqref{eq:full_gap_intextSum} with the self-energy and pairing potential given by
\begin{equation}\label{eq:Sec2_puddle_setup}
    \bar\Sigma(i \varepsilon_n, \vec r) = \langle \bar \Sigma\rangle(i \varepsilon_n) + \frac{\bar g^2}{\pi c_d} |\phi_\alpha(\vec r)|^2 \ln\left( \frac{c_d |\varepsilon_n| + e_\alpha}{\pi c_d T + e_\alpha} \right),
    \quad\quad
    \bar D(i\omega_m, \vec r) = \langle \bar D \rangle(i \omega_m) + \frac{|\phi_\alpha(\vec r)|^2}{\omega_m^2 + c_d |\omega_m| + e_\alpha}.
\end{equation}
We solve Eq. \eqref{eq:full_gap_intextSum} for the highest $T_{c,\alpha}$ that corresponds to an exponentially localized solution for a simplified bosonic mode profile  $|\phi_\alpha(\vec r)|^2 \sim \Theta(\xi_\alpha - |\vec r - \vec r_\alpha|)$, with details of the calculation displayed in Section \ref{sec:Puddles}.

The ``tail" of the distribution of the localized states is of most interest. 
As suggested by Fig. \ref{fig:SC_state}, the local pairing scales $T_{c,\alpha}$ associated with the localized bosonic modes can be significantly larger than the pairing temperature of the extended eigenstate, $T_c$.
Therefore, assuming $T_{c, \alpha} \gg T_c$ we can neglect the effects of pairing coming from the bulk, and focus on the pairing from the strongly localized part of the bosonic spectrum.
This allows to neglect $\langle \bar \Sigma\rangle(i\varepsilon_n)$ and $\langle\bar D\rangle(i\omega_m)$ in Eq. \eqref{eq:Sec2_puddle_setup}, leaving only terms proportional to $|\phi_\alpha(\vec r)|^2$.
The important consequences of such ``puddle" physics are numerous.
Eq. \eqref{eq:Sec2_puddle_setup} with $\langle \bar \Sigma\rangle(i\varepsilon_n)$ and $\langle \bar D\rangle(i\omega_m)$ neglected immediately implies that $T_{c,\alpha}$ is independent of $n$ and the energy scale $\Lambda$.
Instead, the values of $D$ and $\xi_\alpha$ play an important role in determining $T_{c, \alpha}$ through a strong inverse proximity effect from the surrounding metallic state and diffusive Cooperon physics of a disordered metal -- a consequence of the compact spatial support of the pairing kernel.
As a result, the pairing temperature $T_{c, \alpha}$ is controlled by $\vartheta$ and $E_{\mathrm{Th}, \alpha} = 2 D/\xi_\alpha^2$ rather than $\tilde g$ and $\Lambda$ -- quantities controlling $T_c$ in SCBA-type averaged theory, Section \ref{sec:SCBA}.
In Section \ref{sec:Puddles} we solve Eq. \eqref{eq:full_gap_intextSum} with the pairing kernel given in Eq. \eqref{eq:Sec2_puddle_setup} and show that the temperature $T_{c, \alpha}$ obeys the following equation
\begin{equation}\label{eq:TcPuddleSecIII}
    \pi T_{c, \alpha} + e_\alpha/c_d = E_{\mathrm{Th},\alpha} \exp \left[ - \frac{3}{4 \vartheta} \right], \quad\quad
    E_{\mathrm{Th}, \alpha} = \frac{2 D}{\xi_\alpha^2}.
\end{equation}
Note that $T_{c,\alpha}$ depends only on $E_{\mathrm{Th},\alpha}$, $e_\alpha$, and $\vartheta$, supporting the intuition presented above.

There are several conclusions to be drawn from Eq. \eqref{eq:TcPuddleSecIII}. 
First, only bosonic states that satisfy $e_\alpha \xi_\alpha^2 < 2 D c_d e^{-3/4\vartheta}$ are capable of hosting a localized superconducting solution at finite temperature.
Second, the largest values of $T_{c,\alpha}$ correspond to the smallest values of $e_\alpha$, assuming that $\xi(e_\alpha)$ grows with $e_\alpha$ (see Fig. \ref{fig:normal_state}).
Such behavior is consistent with the spectrum of Eq. \eqref{eq:full_gap_intextSum} in Fig. \ref{fig:SC_state}.
Whenever $\xi_\alpha \sim e_\alpha^\gamma$ with $\gamma > 0$
and $n(e_\alpha) = \mathrm{const}$ for the smallest $e_\alpha$ (Fig.~\ref{fig:normal_state} suggests $\gamma \approx 1$), the probability density $p(T_{c, \alpha})$ of the local pairing scale distribution, and the fraction of gapped area $1-\tilde P_U(T)$ of the sample at $T \gg T_c$ scale as
\begin{equation}\label{eq:PuddleResult}
    p(T_{c, \alpha}) \sim T_{c, \alpha}^{- 1 - 1/2\gamma}, 
    \quad\quad
    1-\tilde P_U(T) \sim T^{-1 - 1/2 \gamma}.
\end{equation}
The details of the calculation are shown in Section \ref{sec:Puddles}, Eqs. \eqref{eq:PTcPuddle} and \eqref{eq:PUTheory}.

The power-law scaling of $p(T_{c, \alpha})$ and $P_U(T)$ originates from the Harris disorder that leads to the localization of bosonic glue.
The superconducting puddles with the highest transition temperatures are centered on the most localized bosonic states with the lowest values of $e_\alpha$, according to Eq. \eqref{eq:TcPuddleSecIII}.
Therefore, the proliferation of strongly localized bosonic states near the critical point causes the appearance of a wide distribution of strongly localized superconducting states.

The behavior described above is qualitatively distinct from the Lifshitz-tail physics of superconducting puddles in a conventional random-BCS problem~\cite{bulaevskii1987,Larkin1981,vojta_rare_2006,Dodaro2018}.
Eq. \eqref{eq:PuddleResult} predicts a power-law-like pairing temperature distribution of superconducting puddles, while the BCS problem conventionally predicts a stretched exponential scaling.
It is worth repeating the argument of \cite{vojta_rare_2006,Dodaro2018} for the stretched-exponential Lifshitz tails of the random-BCS problem to assess its robustness.
Assuming that the inverse BCS coupling constant $\lambda_{\mathrm{BCS}}^{-1}(\vec r) = \langle\lambda_\mathrm{BCS}^{-1} \rangle - \delta \gamma_\alpha$ with $\delta \gamma_\alpha$ being Gaussian-distributed, the probability to find a region of size $R_\alpha$ with coupling enhancement $\delta \gamma$ is $p(R_\alpha, \delta \gamma_\alpha) \sim \exp[-A R^d_\alpha \delta\gamma^2_\alpha]$ for $d = 2$.
To ensure the condensation of the puddle of size $R_\alpha$ at a temperature $T_{c, \alpha} > T_c$, the coupling enhancement $\delta \gamma_\alpha$ should be sufficient to overcome both the increased thermal fluctuations and the inverse proximity effect.
It follows from Eq.~\eqref{eq:random_BCS} that $\delta \gamma_\alpha$ can be estimated by
\begin{equation}\label{eq:BCSLog}
    \delta \gamma_\alpha \sim \ln \left( \frac{D/R_\alpha^2 + T_{c, \alpha}}{T_c} \right),
\end{equation}
which should be considered the BCS analog of Eq. \eqref{eq:TcPuddleSecIII}.
The temperature $T_c$ is the superconducting onset temperature that corresponds to the averaged BCS coupling $\langle \lambda^{-1}_{\mathrm{BCS}}\rangle$.
Eq. \eqref{eq:BCSLog} implies two distinct scaling regimes for the local pairing scale distribution of puddles $p_\mathrm{BCS}(T_{c, \alpha})$:

\begin{equation}\label{eq:BCSpuddleDist}
    p_\mathrm{BCS}(T_{c, \alpha}) \sim 
    \begin{cases}
        \exp\left[- A' \mathcal{T}_\alpha\right], \quad\quad\quad\quad\quad\quad\;\; \mathcal{T}_\alpha \lesssim 1,
        \\
        \exp\left[-A a'^2 \ln^2(T_{c, \alpha}/T_c)\right],  \quad \mathcal{T}_\alpha \gtrsim 1,
    \end{cases}
    \quad 
    \mathcal{T}_\alpha = (T_{c, \alpha} - T_c)/T_c.
\end{equation}

Constants $A$ and $A'$ are independent of $T_{c, \alpha}$, and $a'$ is a UV length scale defined below.
To obtain the first line of Eq. \eqref{eq:BCSpuddleDist}, Eq. \eqref{eq:BCSLog} is linearized, leading to $\delta \gamma_\alpha = \mathcal T_\alpha + L_{T_{c, \alpha}}^2/R^2$, where $L_T = \sqrt{D/T}$.
For a fixed $T_{c, \alpha}$, probability density $p(R_\alpha, \delta \gamma_\alpha)$ is strongly peaked for $R_\alpha \sim L_{T_{c, \alpha}} \mathcal{T}_\alpha^{-1/2}$.
The existence of the characteristic length scale $R_\alpha$ for a given $T_{c, \alpha}$ is what results in the exponential scaling $p_\mathrm{BCS}(T_{c, \alpha}) \sim \exp[-A'\mathcal{T}_\alpha]$.
In the regime that describes the second line of Eq. \eqref{eq:BCSpuddleDist}, $\mathcal{T}_\alpha \gtrsim 1$, the size of a puddle is reduced to its minimum allowed by thermal diffusion: $R_\alpha \sim L_{T_{c, \alpha}}$.
For such short length scales, the coupling enhancement that is required for the puddle condensation is $\delta \gamma_\alpha \sim \ln(T_{c, \alpha} /T_c)$.
For the small puddles, the size of the strongly attractive core of the puddle is no longer important for $T_{c, \alpha}$, as the effect of the inverse proximity effect has reached its maximum.
Therefore, as far as the UV cutoff $\Lambda$ of the BCS problem is dominant, the cores of all sizes in the range $a' \lesssim R \lesssim L_{T_{c, \alpha}}$ contribute to the puddles of similar $T_{c, \alpha}$.
The UV cutoff $a'$ can be estimated by $a' = \max(L_\Lambda, a)$, where $a$ is a characteristic length scale of $\lambda_{\mathrm{BCS}}(\vec r)$ disorder.
As a result, for $\mathcal{T}_\alpha \gtrsim 1$ one expects log-exponential scaling $p_\mathrm{BCS}(T_{c, \alpha}) \sim \exp[- A a'^2 \ln^2(T_{c, \alpha}/T_c)]$.

The BCS argument presented above would fail if the distribution of $\delta \gamma_\alpha$ was non-Gaussian, which would require a failure of the central limit theorem. 
The value $\delta \gamma_\alpha$ is an effective quantity of a large puddle composed of a large number of microscopic fluctuations.
The central limit theorem can only fail if the microscopic fluctuations are strongly correlated at large, mesoscopic distances comparable to the puddle, or if the microscopic fluctuations are drawn from a distribution with infinite variance \cite{gnedenko1968limit}. 
The former option is the signature of long-range non-Fermi-liquid correlations, and the latter option only occurs in the vicinity of the infinite randomness fixed point.
The importance of the central limit theorem can be seen from a simple example: if the distribution of $\delta\gamma_\alpha$ is a power-law instead of a Gaussian, it is easy to find relatively small impurities of size $L_T \lesssim R \lesssim L_T \tau^{-1/2}$.
This, in turn, would lead to a proliferation of small, power-law-distributed puddles with sufficiently large $\delta \gamma_\alpha$ to overcome the inverse proximity effect, as described in the quantum-critical scenario.


\section{Linearized Usadel Equation}
\label{sec:setup}

\color{black}
In this section, we provide further details regarding the origin and underlying assumptions of the linearized pairing equation, Eq.~\eqref{eq:full_gap_intextSum}, used in Sec.~\ref{sec:results}. The key point is that the random-mass disorder must be treated nonperturbatively in the bosonic sector, since it localizes the low-energy modes that mediate pairing. By contrast, the fermionic states at the Fermi level remain extended and diffusive (provided $k_F l\gg 1$), and can be treated within a self-averaging approximation. To capture this regime, we first coarse-grain the non-local structure of both the electronic and bosonic propagators, controlled by $l$ and $l_{\rm corr}$, respectively. We then derive a real-space Usadel equation for the anomalous Cooper-pair amplitude, whose remaining spatial dependence is in the center-of-mass coordinate and varies on the scale of the bosonic localization length $\xi$.
\color{black}

We begin by specifying the hierarchy of length scales, which is schematically illustrated in Fig.~\ref{fig:SC_state}(b). The disorder-averaged fermionic Green's function decays on the scale of the elastic mean free path $l$, so $\langle G(i\varepsilon_n,\mathbf{r},\mathbf{r}')\rangle \propto \exp\left\{-|\mathbf{r}-\mathbf{r}'|/2l\right\}$. 
Within the self-consistent Born approximation (SCBA), the random chemical potential $v(\mathbf{r})$ alone leads to $l=v_F\tau$, where $v_F$ is the Fermi velocity, and $\tau=1/m\bar{v}^2$ is the mean free time. 
Length scale $l$ is unrelated to localization and only reflects the fact that the phases of the wave-functions become uncorrelated at long distances. 
Similarly, the spatial decay of the averaged bosonic propagator $\langle D(i\omega_n,\mathbf{r},\mathbf{r}') \rangle\propto \exp\left\{-|\mathbf{r}-\mathbf{r}'|/2l_{\rm corr}\right\}$ is characterized by the correlation length $l_{\rm corr}$. 
Finally, anticipating localization of the bosonic wave-functions due to random mass disorder, we characterize them by a (low-energy) localization length $\xi$. 
Our main physical assumption will be the following hierarchy of scales
\begin{equation}\label{eq:scales}
  k_F^{-1} \; \lesssim\;\; l_{\rm corr}\;\ll \;l \;\;\ll \xi\;.
\end{equation}
Let us clarify the physical significance of this hierarchy.
First, the condition $k_F^{-1}\ll l$ is standard for weakly-disordered metals and translates to a large dimensionless Drude conductivity $\mathcal{G}\gg 1$ in the normal state.
In 2D, the fermion localization length is exponentially longer than the mean free path, i.e. $\sim l e^{\# \mathcal{G}}$, and thus we can disregard it and treat fermions as delocalized.
In combination with the fact that the order parameter is non-conserved (e.g., Ising-nematic or AFM),  the condition $k_F^{-1}\ll l$ also leads to the absence of disorder-induced vertex corrections (i.e., the bosonic fluctuations are not enhanced by diffusion directly), which otherwise would lead to strong temperature-dependent Altshuler-Aronov corrections to conductivity (for the discussion of these issues see \cite{Nosov2020,Nosov2023}). 

The condition $l_{\rm corr}\ll l$ allows us to treat the bosonic propagator as local on the scale over which the electronic motion becomes diffusive. Thus the short-distance relative-coordinate structure of $D(i\omega_m,\mathbf r,\mathbf r')$ can be integrated out, leaving a local pairing kernel $D(i\omega_m,\mathbf r,\mathbf r')
\simeq
8\pi l_{\rm corr}^2
\bar D(i\omega_m,\mathbf r)
\delta(\mathbf r-\mathbf r')$,
where $\bar D(i\omega_m,\mathbf r)$ is given in Eq.~\eqref{eq:InteractionFull}. Finally, the condition $l\ll\xi$ implies that the spatial fluctuations of the boson-induced self-energy and pairing interaction occur on scales much longer than the fermionic mean free path. The localized bosons therefore generate random long-wavelength dephasing and pairing potentials for diffusive Cooper pairs. This last assumption is important for our derivation, but is also physically plausible: the electronic mean free path in strongly correlated metals rarely exceeds tens of lattice spacings, while the relevant localization lengths in our model mostly exceed that scale; see Fig.~\ref{fig:normal_state}(b).

Making use of the essentially local nature of the bosonic Green's function on the scale of $l$ and $l_{\rm corr}$, we can also approximate the contribution to a fermionic self-energy from scattering off bosons. The full one-loop expression, 
\begin{equation}
    \Sigma(i\varepsilon_n, \vec r, \vec r^\prime) = g^2 T\sum_{\omega_m} D(i\varepsilon_n+i\omega_m, \vec r, \vec r^\prime) G(i\omega_m, \vec r, \vec r^\prime),
\end{equation}
is coarse-grained to 
\begin{equation}
\label{eq:Sigma_full_eq}
    \Sigma(i\varepsilon_n, \vec{r},\vec{r}')  =-i \, \mathrm{sgn}(\varepsilon_n) \, \bar{\Sigma}(i\varepsilon_n, \vec{r})\delta(\vec{r}-\vec{r}'),\quad\quad
    \bar{\Sigma}(i\varepsilon_n, \vec{r})= \bar{g}^2 \, \mathrm{sgn}(\varepsilon_n) \, T\sum\limits_{\omega_m}  \operatorname{sgn}(\varepsilon_n+\omega_m) \bar D(i \omega_m, \vec r), 
\end{equation}
where $\bar{g}^2=8\pi^2 l^2_\mathrm{corr} \nu g^2$ is an effective coupling constant, and $\nu=  m/2\pi$ is a fermionic density of states at the Fermi level (see Appendix ~\ref{sec:AppendixGreensFunction} for details).
At this level of coarse-graining, the local value of the fermionic propagator only depends on the density of states and not on the quasi-particle dynamics.
Hence, the approximation is self-consistent: the electronic part of the system is self-averaging in the $k_Fl\gg 1$ limit.

\begin{figure}
    \centering
    \includegraphics[width=0.7\linewidth]{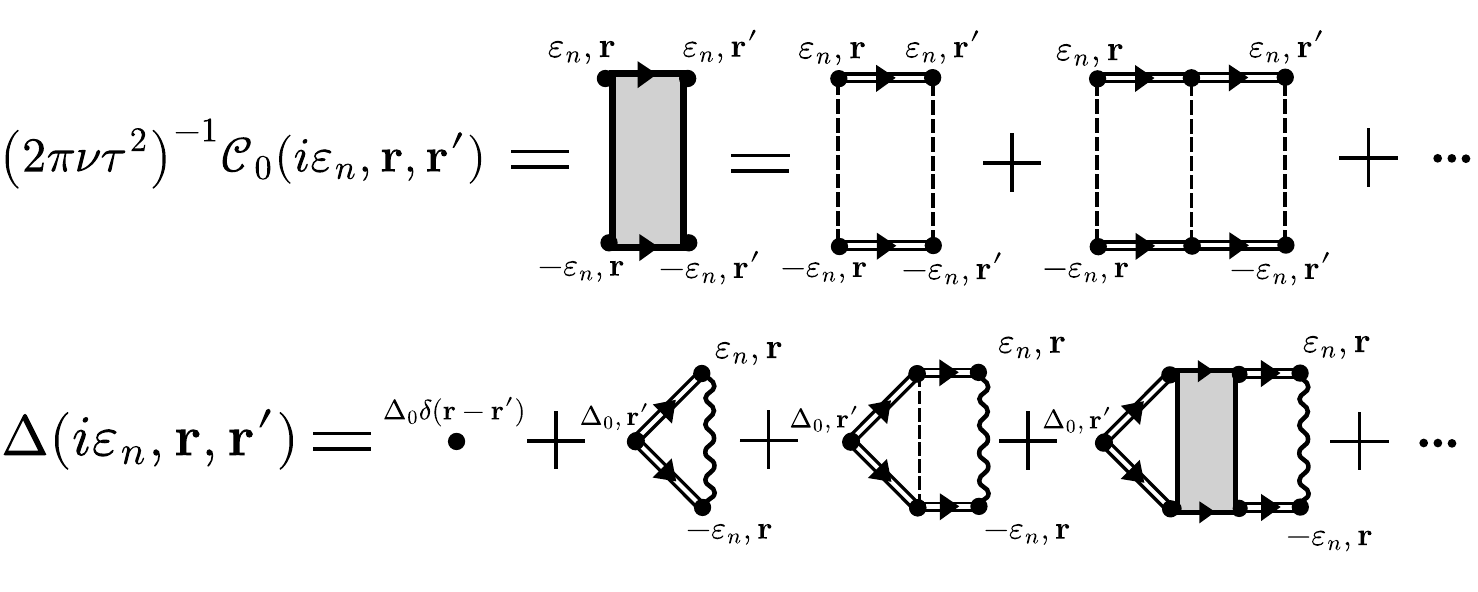}
    \caption{The first line is a definition for $\mathcal{C}_0$ -- a ``bare" Cooperon -- as diagrammatic series. The dashed line represents a conventional impurity scattering induced by a random chemical potential $v(\bm{r})$.
    The second line is a diagrammatic series definition for pairing vertex $\Delta$. The wavy line denotes the position-dependent bosonic propagator $\bar{D}(i\omega_m,\bm{r})$.}
    \label{fig:DiagSeries}
\end{figure}

The pairing problem in this setting is reduced to the calculation of the pairing vertex and the pairing susceptibility, in a conventional way \cite{AGD,Levitov}. 
The construction of a real-space pairing vertex $\Delta(i\varepsilon_n, \vec r, \vec r^\prime)$ induced by the presence of a local source $\Delta_0\delta(\vec r - \vec r')$ is performed with the use of a real-space Cooperon propagator $\mathcal{C}_0(i \varepsilon_n, \vec r, \vec r^\prime)$. This propagator excludes the vertex corrections originating from the inelastic interactions with a boson, and only includes the potential disorder ladder and the self-energy corrections (see Fig. \ref{fig:DiagSeries} for a diagrammatic representation). The remaining pairing vertex corrections originating from $\bar{D}(i\omega_m, \vec r)$ can then be conveniently expressed by using $\mathcal{C}_0(i \varepsilon_n, \vec r, \vec r^\prime)$, as also shown in Fig. \ref{fig:DiagSeries}. The diagrammatic series for $\mathcal{C}_0(i\varepsilon_n, \vec r, \vec r^\prime)$ and $\Delta(i \varepsilon_n, \vec r, \vec r^\prime)$ can be resummed (see \cite{Levitov} and Sec.~\ref{sec:AppendixUsadel} for the details) by solving the following equations 
\begin{equation}
\begin{aligned}
    &\Big(-D\nabla^2+2|\varepsilon_n|+2\bar{\Sigma}(i\varepsilon_n, \vec{r})\Big)\mathcal{C}_0(i\varepsilon_n , \vec{r},\vec{r}')=\delta(\vec{r}-\vec{r}'),\\
    &\Delta(i\varepsilon_n,\vec{r},\vec{r}
')-2\bar{g}^2 T \sum\limits_{\varepsilon_m} \bar D(i \varepsilon_n - i \varepsilon_m, \vec r) \int d\bar{\vec{r}} \;\mathcal{C}_0(i\varepsilon_m , \vec{r},\bar{\vec{r}}) \Delta(i\varepsilon_m,\bar{\vec{r}} ,\vec{r}') = \Delta_0 \delta(\mathbf{r}-\vec{r}')\;.\label{eq:Usadel}
    \end{aligned}
\end{equation}
Here $D=v_F^2\tau/2$ is the diffusion coefficient, and $\Delta_0$ is an infinitesimal local pairing source. One can further define the real-space Gor'kov function, which obeys the following equivalent equations
\begin{equation}
   \Phi(i\varepsilon_n , \vec{r},\vec{r}')= \int d\bar{\vec{r}}\;\mathcal{C}_0(i\varepsilon_n , \vec{r},\bar{\vec{r}}) \Delta(i\varepsilon_n , \bar{\vec{r}},\vec{r}')\;,\quad \quad  \left[-D\nabla^2+2|\varepsilon_n|+2\bar{\Sigma}(i\varepsilon_n, \vec{r})\right]\Phi(i\varepsilon_n , \vec{r},\vec{r}')=\Delta(i\varepsilon_n ,\vec{r},\vec{r}')\;.
\end{equation}
After substituting the second expression above into the second equation in Eq.~\eqref{eq:Usadel}, we find
\begin{equation}\label{eq:full_gap_intext}
    \Big(-D\nabla^2+2|\varepsilon_n|+2 \bar \Sigma(i\varepsilon_n, \vec{r})\Big)\Phi(i\varepsilon_n , \vec{r},\vec{r}')-2\bar{g}^2 T \sum\limits_{\varepsilon_m} \bar D(i \varepsilon_m - i \varepsilon_n, \vec r) \Phi(i\varepsilon_m,\vec{r},\vec{r}')=\Delta_0\delta(\vec{r}-\vec{r}')\;.
\end{equation}
Note that this equation is completely local in space, and the second coordinate $\vec r'$ only reflects the position of the source term on the right-hand side.
If we now set $\Delta_0=0$, we find
\begin{equation}\label{eq:full_gap}
\Big(-D\nabla^2+2|\varepsilon_n|+2\bar{\Sigma}(i\varepsilon_n, \vec{r})\Big)\Phi(i\varepsilon_n , \vec{r})-2\bar{g}^2 T \sum\limits_{\varepsilon_m} \bar D(i \varepsilon_m - i \varepsilon_n, \vec r) \Phi(i\varepsilon_m,\vec{r})=0\;.
\end{equation}
This equation is a special zero mode case, $\epsilon_\alpha=0$, of the general eigenvalue problem defined in Eq.~\eqref{eq:full_gap_intextSum}. 
 Lastly, we note that the `static' terms in the self-energy and the pairing vertex in Eq.~\eqref{eq:full_gap} cancel each other identically -- manifestation of Anderson's theorem.


\section{Numerical Results}
\label{sec:numerical}

We first solve Eq.~\eqref{eq:full_gap_intextSum} numerically, using the boson wavefunctions $\phi_{\alpha}$ obtained from the procedure outlined in Sec. \ref{sec:normal}. More precisely, we find the entire spectrum of the operator of the pairing equation, rather than solving only for its zero eigenvalue. This allows us to do a more thorough analysis of the inhomogeneous superconducting state. The technical details of how we obtain the numerical solution can be found in Appendix \ref{sec:Numerical_details}. 

The full spectrum of eigenvalues $\epsilon_{\alpha}$ behaves as follows. At high temperatures above the pairing instability, all $\epsilon_{\alpha} > 0$. As the temperature is lowered beyond a critical value, the smallest $\epsilon_{\alpha}$ changes sign. For a translationally invariant system, this temperature denotes the mean-field $T_c$. For a strongly disordered system, this is less straightforward, since the smallest $\epsilon_{\alpha}$ correspond to gap functions $\Phi_{\alpha}(i\varepsilon_m,\vec{r})$ that are highly localized \cite{Dodaro2018} and therefore cannot support a phase with global phase coherence. 

As in the SCBA analysis of Sec. \ref{sec:SCBA}, we can quantify the degree of localization of each eigenvector of Eq. (\ref{eq:full_gap}) by looking at the static pair density, $\Phi_\mathrm{stat, \alpha}(\mathbf r) = T \sum_n \Phi_{\alpha}(i \varepsilon_n,\vec{r})$. The localization length $\mathcal L_{\alpha}$ is then defined from the inverse participation ratio, analogous to Eq. (\ref{eq:IPR def}). Similar to the boson localization length $\xi_{\alpha}$, $\mathcal L_{\alpha}$ increases roughly linearly as a function of the eigenvalue $\varepsilon_\alpha$ from tightly localized states with $\mathcal L_{\alpha} \sim \mathcal{O}(1)$ to completely delocalized states that have a maximum of $\mathcal L_{\alpha} \sim \mathcal{O}(L/2)$, as shown in Fig. \ref{fig:SC_state}c. To illustrate the localization length of the eigenstates, we show $|\Phi_\mathrm{stat, \alpha}(\mathbf r)|^2$ for $\alpha = 0,1,2,3$ (highly localized) and $\alpha = 99$ (delocalized) in Fig. \ref{fig:SC_state}f-g, for the parameter values $\lambda = -0.39, T = 0.05, \bar{g}^2 = 4$.   

The temperature dependence of the spectrum is illustrated in 
Fig. \ref{fig:SC_state}c. At high temperatures, all $\epsilon_\alpha > 0$. Upon lowering the temperature, eventually the most localized eigenstate, $\Phi_{\alpha=0}$, will acquire a zero eigenvalue, $\epsilon_0 = 0$. In order to define the localization length of the zero eigenvalue, $\mathcal{L}_{\text{SC}}$, we choose a small averaging window of $e = 0 \pm 0.001$ (gray strip in Fig. \ref{fig:SC_state}c), within which we average $\mathcal{L}_{\alpha}$. Due to the small gap coverage of this state, this does not yet signal global superconductivity (at the mean field level). Upon lowering the temperature further, gap functions with larger and larger values of $\mathcal L_{\alpha}$ will satisfy Eq. (\ref{eq:full_gap}), down until the first fully delocalized solution, which defines the transition temperature $T_{c,\text{global}}$ with `global' gap coverage. Operationally, we define $T_{c,\text{global}}$ as the temperature at which the localization length of the solution to Eq. (\ref{eq:full_gap}), denoted as $\mathcal{L}_{\text{SC}}$, reaches $95\%$ of its maximum value for that set of parameter values (slightly different across parameters). $T_{c,\text{global}}$ acts as a minimum $T_c$ in this mean-field theory.  

\begin{figure}[h]
    \centering
    \includegraphics[width=0.4\linewidth]{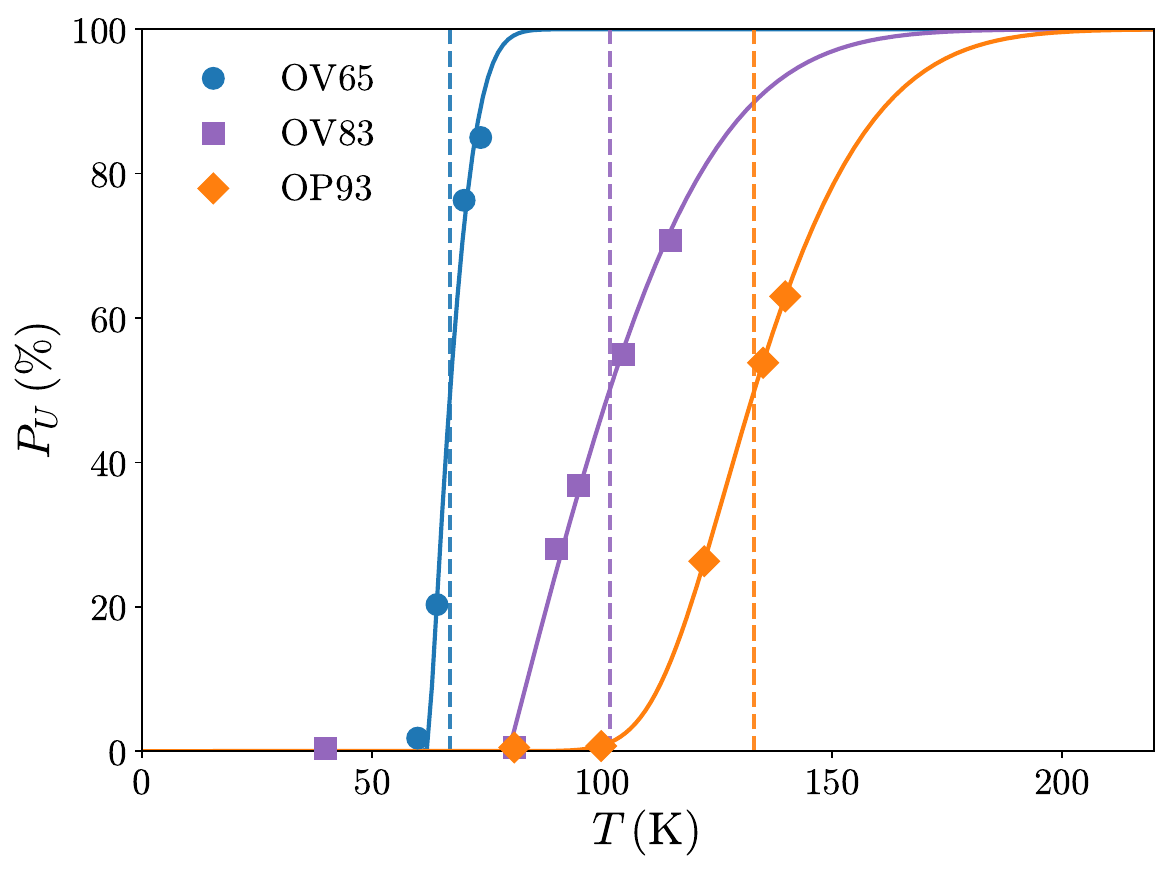}
    \includegraphics[width=0.4\linewidth]{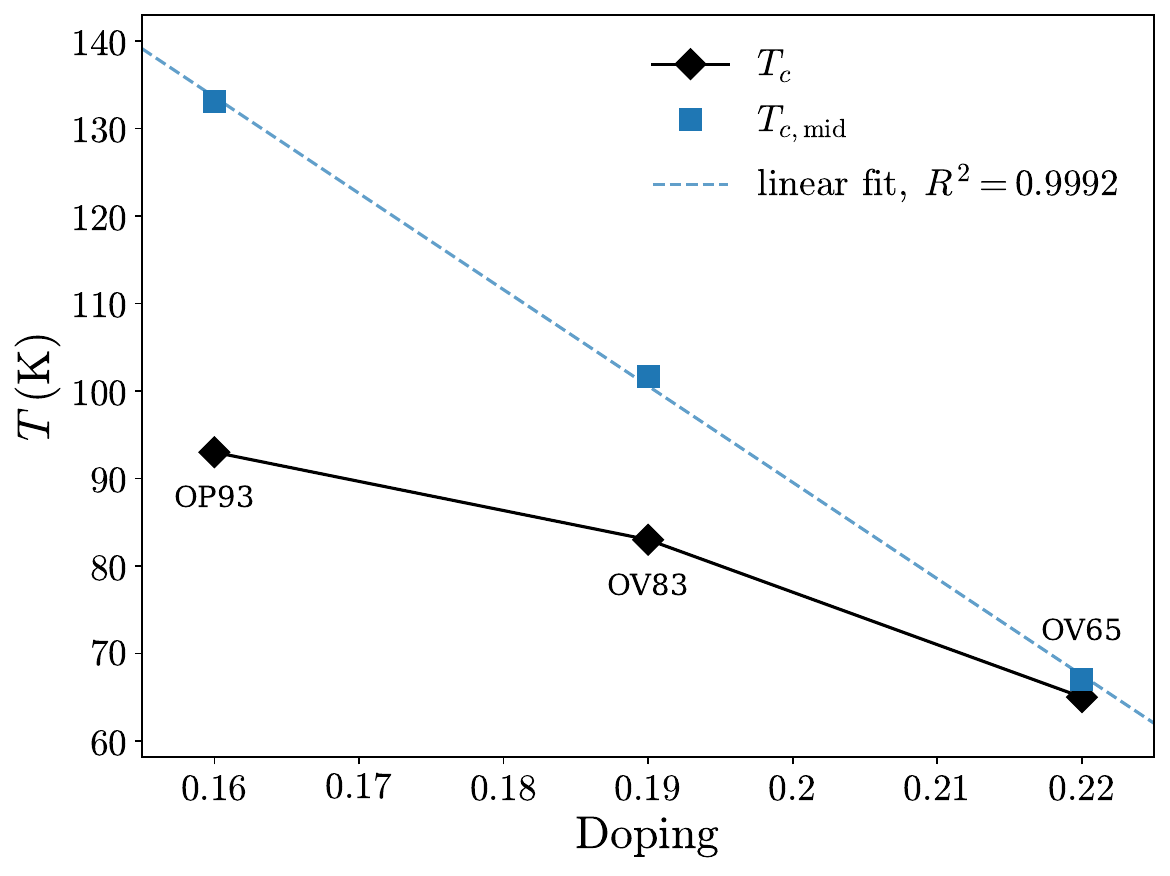}
    \caption{Analysis of STM data on $\text{Bi}_2\text{Sr}_2\text{CaCu}_2\text{O}_{8+\delta}$ from Ref. \cite{Yazdani_puddles_2007} from the Yazdani group. (a) Data points from Figure 3e of Ref. \cite{Yazdani_puddles_2007} (reproduced with the corresponding author's permission). The points measure the percentage of the area (measured with STM) that is ungapped, i.e. in the normal state, versus temperature. The three colors correspond to three different samples at different doping concentrations. The curves are fits to a skew-normal function. (b) $T_{c,\text{mid}}$ (with a linear fit) and $T_c$ extracted from plot (a) versus doping.}
    \label{fig:Yazdani plots percentage}
\end{figure}

We can also define another proxy for the superconducting transition temperature from the partial gap coverage. For this, we look to the scanning tunneling microscopy (STM) experiments of Ref. \cite{Yazdani_puddles_2007} on $\text{Bi}_2\text{Sr}_2\text{CaCu}_2\text{O}_{8+\delta}$. In Fig. \ref{fig:Yazdani plots percentage} we reproduce data points from Ref. \cite{Yazdani_puddles_2007} that show the percentage of the scanned area that is ungapped, as a function of temperature, along with our fit of the data to a skew-normal distribution. \footnote{Specifically, we use the skewnorm.cdf(x, a, loc, scale) function from \url{https://docs.scipy.org/doc/scipy/reference/generated/scipy.stats.skewnorm.html}, allowing all parameters to vary.}
The finite widths of the distributions are a sharp measure of the superconducting inhomogeneity. From the fitted distribution functions we can define $T_{c,\text{mid}}$ as the temperature at which the spatial gap coverage reaches $50\%$ of the area. The reason for focusing on this precise percentage is that Ref. \cite{Yazdani_puddles_2007} notes that $T_{c,\text{mid}}$ empirically corresponds to the onset of well-defined vortex responses in $\text{Bi}_2\text{Sr}_2\text{CaCu}_2\text{O}_{8+\delta}$, including the Nernst effect. One interesting feature of $T_{c,\text{mid}}$ is that it follows a very linear dependence on doping value. 

To make contact with the experimental data in Fig. \ref{fig:Yazdani plots percentage}, we can similarly define the percentage of spatial area that is \textit{ungapped} using spatial support of the static pair density $\Phi_{\text{stat}, \bar{\alpha}}(\mathbf r)$. This is done through $\mathcal{L}_{\text{SC}}$ as
\begin{equation}\label{eq:tilde PU}
    \tilde P_{U} \equiv 1 - \left(\frac{\mathcal{L}_{\text{SC}}}{L/(2\sqrt{\pi})}\right)^2.
\end{equation} 
The denominator $L/(2\sqrt{\pi})$ roughly represents the maximal possible localization length of the system with periodic boundary conditions, but its precise numerical value is not important for the shape of the curve.  

We distinguish $\tilde P_{U}$ from the experimentally measured $P_{U}$, since at lower temperatures they saturate to different values owing to our ignorance of the precise $\mathcal{O}(L)$ numerical value of the maximum localization length. However, aside from the different offset, the shape of the curves $\tilde P_U(T)$ can be directly compared with the experimental data for $P_U(T)$ in Fig. \ref{fig:Yazdani plots percentage}. These curves are shown in Fig. \ref{fig:SC_state}d for $\bar{g}^2 = 5$ across several $\lambda$ values. Additional plots for more parameter values are presented in Appendix \ref{sec:Numerical_details}. All the data fits very well to the same skew-normal form. Extracting $T_{c,\text{mid}}$ as the $50\%$ point between the minimum and maximum values of $\tilde P_U(T)$, we can see that it has a very linear dependence on the tuning parameter $\lambda$, which acts as the doping concentration in our model. This reproduces, at a qualitative level, the roughly linear evolution of $T_{c,\mathrm{mid}}$ with doping observed in the STM data, despite the simplicity of the model and the absence of a microscopic mapping between $\lambda$ and doping.

We can also use $\tilde P_U(T)$ and $T_{c,\text{mid}}$ to compare the numerical simulation with the analytical results of Sec. \ref{sec:SCBA} and \ref{sec:Puddles}. To make contact with the SCBA, we focus on $T_{c,\text{mid}}$, as the superconducting state at this temperature is rather delocalized, so here the system is in the regime of effectively weak inhomogeneity. We can attempt to reproduce the scaling of $T_{c,\text{mid}}$ with $g^2$ from Eq. (\ref{eq:TcSecIII}), specifically for $\tilde g > \tilde g_*$ where a comparison can be made. This scaling is shown in Fig. \ref{fig:SC_state}e, where we can indeed see a very good fit to the form predicted by SCBA, $T_{c,\text{mid}} \approx \Lambda \, e^{-b/\bar{g}}$ (plots for other $\lambda$ values are shown in Appendix \ref{sec:Numerical_details}).
We note that such an analysis is not entirely consistent: since the Landau damping parameter $c_d$ and $\bar{g}^2$ both depend on $g^2$, the boson wavefunctions need to be re-solved for each new $\bar{g}^2$. However, here we assume that the new wavefunctions $\phi_{\alpha}$ would not change significantly enough to qualitatively modify the above result. 

To compare with Sec. \ref{sec:Puddles}, one could attempt to use the $T \gg T_{c,\text{mid}}$ behavior of $\tilde P_U(T)$ predicted by the behavior of very rare puddles in Eq. (\ref{eq:PUTheory}). This would imply that the skew-symmetric normal is not a good fit in this large temperature regime. However, neither the experimental data points, nor the numerical simulation has such resolution to be able to compare the precise way that $\tilde P_U(T)$ (or $P_U(T)$) deviates from $\tilde P_U(T \rightarrow \infty) = 1$ at finite $T$. 

\begin{figure}[t!]
    \centering
    \includegraphics[width=0.45\linewidth]{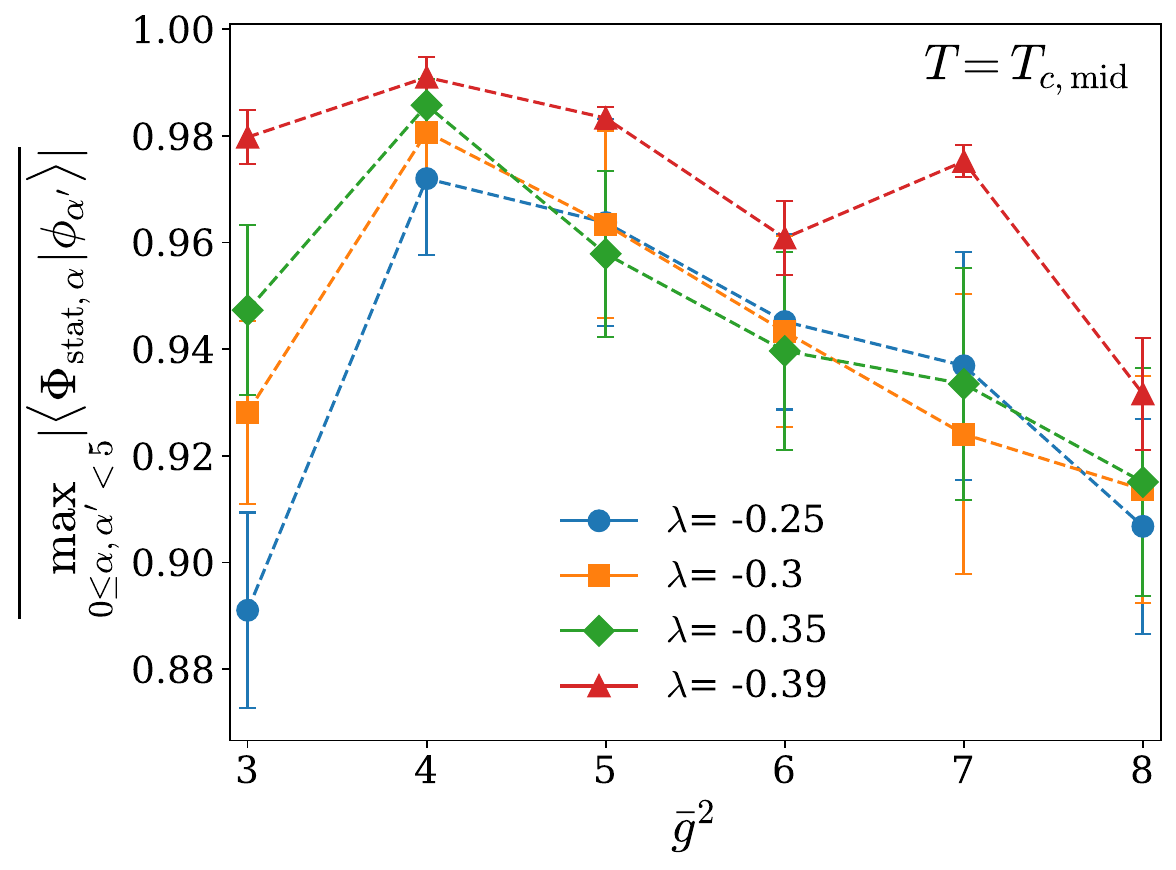}
    \caption{Overlap of the most localized superconducting puddles $\Phi_{\text{stat},\alpha}(\mathbf r)$ and localized bosons $\phi_{\alpha'}(\mathbf r)$. This plot demonstrates that for each disorder configuration, among the most localized puddles and most localized bosons there are pairs which essentially sit on top of each other (cf. Fig. \ref{fig:normal_state}(d) and Fig. \ref{fig:SC_state}(f)). These pairs need not have the same eigenvalue index.}
    \label{fig:overlap_general}
\end{figure}

Alternatively, we can focus in detail on the structure of the most localized puddles themselves. One prediction from Sec. \ref{sec:Puddles} is that the most localized puddles will sit on top of the most localized bosons. In Fig. \ref{fig:SC_state}e we illustrate that this is indeed true. There we plot $\Phi_{\text{stat},\alpha}(\mathbf r)$ on top of the highly localized boson states $\phi_{\alpha'}(\mathbf r)$ for $\alpha, \alpha' = 0,1,2$. We can see that, among the most localized bosons and most localized superconducting puddles, there is a very large (close to one) overlap. To show how general this is, in Fig. \ref{fig:overlap_general} we plot the maximal overlap among $\alpha, \alpha' = 0,\dots,4$, averaged over disorder seeds, $\overline{\max_{0 \leq \alpha, \alpha' < 5} |\langle \Phi_{\text{stat},\alpha}(\mathbf r) | \phi_{\alpha'}(\mathbf r) \rangle|}$, for a few different temperatures $T$ and $\lambda$ values. We can see that the overlap is very large for nearly all $\lambda$ and $T$. We thus numerically corroborate parts of both types of analytical analyses from Sec. \ref{sec:SCBA} and \ref{sec:Puddles}.  

\section{Self-consistent Born approximation}
\label{sec:SCBA}

We use Eq.~\eqref{eq:full_gap} to analyze how spatial inhomogeneity modifies the linearized Gor'kov function.
We begin by considering the analog of the Born approximation for the spatially inhomogeneous boson-mediated pairing interaction. 

To perform disorder averaging, we use the phenomenological assumptions consistent with Section \ref{sec:normal}.
In particular, we assume that all the bosonic eigenstates $\phi_\alpha(\vec r)$ with eigenvalues $e_\alpha$ are localized and their localization length $\xi_\alpha = \xi(e_\alpha)$ is only a function of the eigenvalue.
Adding the assumption of disorder-averaged translation invariance allows to express the average electron self-energy $\langle\bar \Sigma\rangle(i \varepsilon)$ and the effective pairing potential $\langle \bar D \rangle(i\omega)$ through $n(\mathcal{E}) = \left\langle \sum_\alpha |\phi_\alpha(\vec r)|^2 \delta(e_\alpha - \mathcal{E}) \right\rangle$ (see Eq. \eqref{eq:nE}). 
Averaging of Eqs. \eqref{eq:InteractionFull} leads to
\begin{equation}
    \langle \bar D\rangle(i\omega) \approx  \int d\mathcal{E} \, n(\mathcal{E}) \, \frac{ 1}{\omega^2 + c_d |\omega| + \mathcal{E}},
    \quad\quad
    \langle \bar \Sigma\rangle(i\varepsilon) \approx 2 \bar g^2 \int_{\pi T}^{\varepsilon} \frac{d\omega}{2\pi} \langle \bar D \rangle (i \omega).  
\end{equation}

For low-energy bosonic states distributed according to the power law $n(\mathcal{E}) = n \mathcal{E}^{\alpha_G}$, as is expected in the Griffiths regime, the average self-energy and the pairing potential scale as

\begin{equation}\label{eq:AvDandSigma}
    \langle \bar D \rangle(i \omega) \approx n
    \begin{cases}
        \pi \csc(\pi |\alpha|) |c_d \omega|^{\alpha_G}, \quad \alpha_G < 0,
        \\
        \ln(\Lambda/|\omega|), \quad\quad\quad\quad\;\;\;\, \alpha_G = 0,
    \end{cases}
    \langle \bar \Sigma \rangle(i \varepsilon) \approx \frac{\bar g^2 n}{\pi}
    \begin{cases}
        \pi \csc(\pi |\alpha_G|)(1+\alpha_G) c_d^{\alpha_G} |\varepsilon|^{\alpha_G+1}, \quad \alpha_G < 0,
        \\
        |\varepsilon| \ln (\Lambda/|\varepsilon|), \quad\quad\quad\quad\quad\quad\quad\quad\quad\quad \alpha_G = 0.
    \end{cases}
\end{equation}
In case of $\alpha_G = 0$, the cutoff $\Lambda$ is defined as $\Lambda = \Lambda_B^2/c_d$, where $\Lambda_B^2$ is the UV cutoff of the bosonic eigenvalues that span $n(\mathcal{E})$.
The derivation above assumes $c_d \gg T$, $\varepsilon_n, \omega_n \lesssim \Lambda$, which allows to neglect $\omega_n^2$ term in Eq.~\eqref{eq:InteractionFull}.
From Eq. \eqref{eq:AvDandSigma}, one observes that different values of $\alpha$ reproduce different limits of ``$\gamma$-model" \cite{Gamma1,Gamma2,Gamma3,Gamma4,Gamma5,Gamma6}, including it into the perspective of current work.

For a moment, we focus on the case of non-vanishing density of states $n(\mathcal{E}) = n = \mathrm{const}$, as it will turn out to be the most physically interesting regime, which is also supported by numerical simulation (see Sec. \ref{sec:normal}).
For a moment, we assume that spatial fluctuations of $\bar \Sigma(i \varepsilon_n, \vec r)$ and $\bar D(i \omega_n, \vec r)$ are neglected, so $\langle \bar \Sigma\rangle(i \varepsilon_n)$ and $\langle\bar D\rangle(i\omega_n)$ can be used instead in Eq. \eqref{eq:full_gap}.
With the notation $\Phi_0(i \varepsilon_n, \vec r)$ reserved for the solution of Eq. \eqref{eq:full_gap} with the inhomogeneity neglected, the linearized Usadel equation can be written as
\begin{equation}\label{eq:HomGapEq}
    \Big(-D\nabla^2+2|\varepsilon_n|+2 \langle\bar{\Sigma}\rangle(i\varepsilon_n)\Big)\Phi_0(i\varepsilon_n , \vec{r})-2\bar{g}^2 T \sum\limits_{\varepsilon_m} \langle \bar D \rangle(i \varepsilon_m - i \varepsilon_n) \Phi_0(i\varepsilon_m,\vec{r})= \Delta_0.
\end{equation}
The leading instability at temperature $T_{c0}$ in this translationally invariant problem corresponds to a homogeneous solution.
To tackle the problem analytically, we adopt the approach presented in \cite{Gamma1}.
First, we approximate the Matsubara frequency sums with integrals by employing the Euler-Maclaurin approximation \cite{Apostol01051999} under the assumption of a small $T_{c0}$. 
Then, we estimate the contribution from $\langle \bar D \rangle$ to Eq. \eqref{eq:HomGapEq}, by approximating the kernel ${\langle \bar D \rangle (i |\varepsilon_n -  \varepsilon_m|) \approx \langle \bar D \rangle(i \max(|\varepsilon_n|, |\varepsilon_m|))}$, leading to 
\begin{equation}\label{eq:HomIntegralEq}
    \varepsilon\left[ 1  +  \tilde g^2 \ln \left( \frac{\Lambda}{\varepsilon} \right) \right] \Phi_0(i \varepsilon) = \tilde g^2 \int_{\pi T}^{\Lambda} d\varepsilon^\prime \, \ln \left( \frac{\Lambda}{\max (\varepsilon, \varepsilon^\prime)} \right) \Phi_0(i \varepsilon^\prime) + \Delta_0,
\end{equation}
for $\varepsilon >0$, and $\tilde g^2 = \bar g^2 n / \pi$. Eq. \eqref{eq:HomIntegralEq} can be solved by reducing it to a differential equation. The solution to the integro-differential equation takes the form $\Phi_0(i \varepsilon) =[\varepsilon+ \tilde g^2 \varepsilon \ln(\Lambda / \varepsilon)]^{-1} F(\varepsilon)$,
and 
\begin{equation}\label{eq:DiffExSolNoInh}
    F(\varepsilon) = c_1 \sqrt{x} \, J_1\left( 2\sqrt{x}\right) + c_2 \sqrt{x} \, Y_1\left( 2 \sqrt x\right), \quad x(\varepsilon) = \log\left( \frac{\Lambda}{\varepsilon} \right) + \frac{1}{\tilde g^2}.
\end{equation}
The substitution of Eq. \eqref{eq:DiffExSolNoInh} back into Eq. \eqref{eq:HomIntegralEq} fixes $c_1$ and $c_2$ up to a multiplicative constant, resulting in 
the pair density that for $\tilde g \ll 1$ can be written as
\begin{equation}\label{eq:PairingSusc}
    \Phi_\mathrm{0,stat}(\omega = 0) = \int \frac{d\varepsilon}{2 \pi} \Phi_0(i\varepsilon) \approx \frac{\Delta_0}{2 \pi \tilde g} \tan \left( \tilde g\, \ln \left( \frac{\Lambda}{\pi T} \right)\right).
\end{equation}
For algebraic details of solving Eq. \eqref{eq:HomIntegralEq} that lead to Eq. \eqref{eq:PairingSusc} see Appendix \ref{sec:NoInhomogeneitySolution}.
The pairing susceptibility in Eq. \eqref{eq:PairingSusc} diverges at
\begin{equation}\label{eq:Tc0}
    T_{c0} = \frac{\Lambda}{\pi} \exp \left[ - \frac{\pi}{2 \tilde g}\right] \gg T_\mathrm{NFL} \sim \frac{\Lambda}{\pi} \exp \left[ - \frac{1}{\tilde g^2} \right],
\end{equation}
which is a well-known result in the theory of non-Fermi liquid superconductivity \cite{Son:1998uk}.
By comparing it against the temperature $T_\mathrm{NFL}$, at which the effects of $\langle \bar \Sigma \rangle(\varepsilon)$ become comparable to $|\varepsilon|$ on the left side of Eq. \eqref{eq:HomIntegralEq}, one concludes that $T_{c0} \gg T_\mathrm{NFL}$.
This implies that the effects of the self-energy $\bar \Sigma(i \varepsilon, \vec r)$ are negligible in Eq. \eqref{eq:full_gap} for temperatures comparable to and above $T_{c0}$.
This motivates neglecting the $\bar \Sigma(i \varepsilon_n, \vec r)$ in Eq. \eqref{eq:full_gap} and considering a simplified problem. 
For a more formal treatment and estimates, see Appendix \ref{sec:AppendixSCBA}.

\subsection{Effects of boson localization on the pairing interaction}
\label{sec:PairingPotCorr}

We treat the corrections from spatial fluctuations of $\bar D(i \omega_n, \vec r)$ by accounting for them perturbatively at the level of the Usadel equation in Eq. \eqref{eq:full_gap}.
The effects of spatial fluctuations of $\bar D(i \omega_n, \vec r)$ are encoded in $\delta \bar D(i\omega_n, \vec r)$, where
\begin{equation}\label{eq:DeltaDdef}
    \bar D(i \omega_n, \vec r) = \langle \bar D\rangle(i \omega_n) + \delta \bar D(i\omega_n, \vec r)
\end{equation}
To capture the effects of $\delta \bar D(i \omega_n, \vec r)$ up to second order, we convolve Eq. \eqref{eq:full_gap} with $\bar{\mathcal C}_0(i\varepsilon_n, \vec r- \vec r')$, which is given by Eq. \eqref{eq:Usadel}, but with self-energy dropped. 
This allows to rewrite Eq. \eqref{eq:full_gap} as
\begin{equation}\label{eq:PhiReduced}
        \Phi(i \varepsilon_n, \vec r, \vec r^\prime) = 2 \bar g^2 T \int d^2 \vec x \; \bar{\mathcal C}_0(i \varepsilon_n, \vec r- \vec x) \sum_{\varepsilon_m}  \Big( \langle\bar D\rangle (i \varepsilon_n - i \varepsilon_m) + \delta \bar D (i \varepsilon_n - i \varepsilon_m, \vec x)  \Big)  \Phi(i \varepsilon_m, \vec x, \vec r^\prime).
\end{equation}
After substitution of Eq. \eqref{eq:PhiReduced} into itself, averaging over the bosonic disorder ensemble, and acting with $(-D\nabla^2 +2 |\varepsilon_n|)$ from the left on the resulting equation, we recover
\begin{equation}\label{eq:GapEqCorrected}
        \Big(- D \nabla^2 + 2 |\varepsilon_n|  \Big) \langle\Phi \rangle(i\varepsilon_n, \vec r, \vec r^\prime) = 
    2 \bar g^2   T \sum_{\varepsilon_m} \Big( \langle  \bar D\rangle (i\varepsilon_n - i\varepsilon_m)+\bar K_{DD}(\varepsilon_n, \varepsilon_m) \Big)\langle\Phi \rangle(i\varepsilon_m, \vec r , \vec r^\prime),
\end{equation}
where $\bar K_{DD}$ is
the inhomogeneity correction kernel, which can be expressed as 
\begin{equation}\label{eq:barKDDDef}
    \bar K_{DD}(\varepsilon_n, \varepsilon_n + \omega_m) = 2 \bar g^2T \sum_{\substack{l \neq 0\\l \neq m}} \int d^2 \vec r \, \langle \delta \bar D(i\omega_m - i \omega_l, \vec r) \delta \bar D(i\omega_l, 0) \rangle  \bar{\mathcal C}_0(i \varepsilon_n + i \omega_m - i \omega_l, \vec r).
\end{equation}
In the process of deriving Eq. \eqref{eq:GapEqCorrected}, we neglected all $\delta \bar D-\delta \bar D$ correlations beyond second order. 
In particular, we approximated $\langle \delta \bar D \, \delta \bar D \, \Phi\rangle \approx \langle \delta \bar D \, \delta \bar D \rangle \langle \Phi \rangle$.  
The algebraic details that lead to the Eqs. \eqref{eq:GapEqCorrected} and \eqref{eq:barKDDDef} are rather technical and can be found in
Appendix \ref{sec:AppendixSCBA}.

The structure of $\bar K_{DD}$ is determined from the spatial correlations of $\bar D(i \omega_n, \vec r)$ by Eq. \eqref{eq:barKDDDef}.
To evaluate the expression for $\bar K_{DD}$, we first evaluate $\langle \delta \bar D(i \omega_n, \vec r) \delta \bar D(i \omega_m, 0) \rangle$.
This correlation function is most conveniently expressed through $P_{2,2}(\mathcal{E}, \vec r) = \left\langle \sum_\alpha |\phi_\alpha(\vec r)|^2 |\phi_\alpha(0)|^2 \, \delta(\mathcal{E} - e_\alpha) \right\rangle$ \cite{RevModPhys.80.1355,Cuevas2007}:
\begin{equation}\label{eq:delDdelD}
    \langle \delta \bar D(i \omega_n, \vec r) \delta \bar D(i \omega_m, 0) \rangle \approx
    \int d\mathcal{E} \, P_{2,2}(\mathcal{E}, \vec r) W(\omega_n, \mathcal{E}) W(\omega_m, \mathcal{E}),
    \quad\quad
    W(\omega_n, \mathcal{E}) \approx (c_d |\omega_n| + \mathcal{E})^{-1}.
\end{equation}
The bosonic states correlation function $P_{2,2}(\mathcal{E}, \vec r)$ in the momentum space takes form $P_{2,2}(\mathcal{E}, \vec q) = \int d^2 \vec r \, e^{-i \vec r \cdot \vec q} P_{2,2}(\mathcal{E}, \vec r) = n(\mathcal{E})(1 + \vec q^2 \xi^2/4)^{-3}$, where $n(\mathcal{E}) = n \mathcal{E}^{\alpha_G}$.
Finally, in a diffusive metal, the bare Cooperon in momentum space is of the form ${\bar{\mathcal{ C}}_0(\varepsilon_n, \vec q) = (D \vec q^2 + 2 |\varepsilon_n|)^{-1}}$. 
For the scaling estimates below, we replace $\xi(\mathcal E)$ by a representative low-energy value $\xi = \xi(\pi c_d T)$.
Substitution of Eq. \eqref{eq:delDdelD} into Eq. \eqref{eq:barKDDDef} along with the expressions above results in 
\begin{equation}\label{eq:KDDalpha}
    \bar K_{DD}(\varepsilon, \varepsilon') \simeq 
    \frac{\bar g^2 n}{4 \pi^2 c_d D} \frac{\pi \alpha_G c_d^{\alpha_G} }{\sin \pi \alpha_G}
    \begin{cases}
            \eta^{\alpha_G} R(\eta), \quad \quad\quad\quad  \alpha_G < 0,
            \\
            \left|\ln\left( \frac{E_\mathrm{Th}}{\eta} \right)\right| R(\eta), \quad  \alpha_G = 0,
    \end{cases}
    \quad\quad
    R(\eta) = 
    \begin{cases}
        \ln \left( \frac{E_\mathrm{Th}}{\eta} \right), \quad \eta < E_\mathrm{Th},
        \\
        \frac{E_\mathrm{Th}}{2 \eta}, \quad\quad\quad\;\; \eta > E_\mathrm{Th},
    \end{cases}
\end{equation}
where $\eta = \max(|\varepsilon|, |\varepsilon'|)$ and $E_\mathrm{Th} = 2 D / \xi^2$.
The algebraic details of the calculation outlined in this section can be found in the Appendices \ref{sec:AppendixSCBA} and \ref{sec:AppKenelEstimation}.

It is interesting to note that the inhomogeneity corrections are rather mild for $\alpha_G < 0$: the power-law scaling of $\langle \bar D \rangle(\varepsilon - \varepsilon')$ is only logarithmically amplified in $\bar K_{DD}(\varepsilon, \varepsilon')$.
To leading order, the effects of these corrections can be absorbed into the homogeneous theory of Eq. \eqref{eq:HomGapEq} by adjusting the coupling constant.
%
%
The status of the spatial-inhomogeneity corrections is qualitatively different for $\alpha_G = 0$.
The logarithmic scaling of $\langle \bar D \rangle(\varepsilon - \varepsilon')$ is amplified by the additional square-logarithmic $\bar K_{DD}(\varepsilon, \varepsilon')$. 
Thus, inhomogeneity provides a much stronger correction to the effective pairing potential of Eq. \eqref{eq:HomGapEq} at low energies.

Once more, we narrow our scope to $n(\mathcal{E}) = n= \mathrm{const}$ ($\alpha_G = 0$ case), as it is most relevant according to the argument above and is supported by numerical results (Section \ref{sec:normal}).
Following the structure of Eq. \eqref{eq:KDDalpha}, we approximate the inhomogeneity kernel as in Eq.~\eqref{eq:KDDFin_sec2}, 
where $\vartheta = 3 \bar g^2 /2 \pi^2 c_d D$ and $\Theta(\epsilon)$ is a Heaviside step-function.

The physical meaning of $\vartheta$ can be understood in the context of the Y-SYK model \cite{Aavishkar2023}.
The combination $\bar g^2/\nu$ should be regarded as a spatially disordered coupling constant in the notation of \cite{Aavishkar2023}, leading to $c_d \sim \nu \bar g^2$ in the framework of Y-SYK.
Consequently, $\vartheta \sim (k_F l)^{-1}$, where $l$ is an electron mean free path.
The quantity $k_F l$ is proportional to the number of bosonic states that an electron visits before its phase decoheres due to disorder.
Therefore, the larger $k_Fl$, the more bosonic states the electron coherently probes, smearing the effects of inhomogeneity and decreasing the effects of $\bar K_{DD}$.
The electronic nature of the parameter $\vartheta$ does not permit treating $\vartheta$ as a bosonic parameter that reproduces the homogeneous limit in the limit $\vartheta \rightarrow 0$.
Instead, one should consider $\xi \rightarrow \infty$ as a homogeneous limit: it leads to $E_\mathrm{Th} \rightarrow 0$, so the effects of $\bar K_{DD}(\varepsilon, \varepsilon')$ are negligible when $E_\mathrm{Th} < T_{c0}$.
As a reminder, $T_{c0}$ is a superconducting onset temperature in the homogeneous limit, which is given by Eq. \eqref{eq:Tc0}.

\subsection{Spatially averaged pairing amplitude and superconducting transition temperature}
\label{sec:AvGapSCBA}

\begin{figure}
 \subfloat{
    \includegraphics[width=0.38\linewidth]{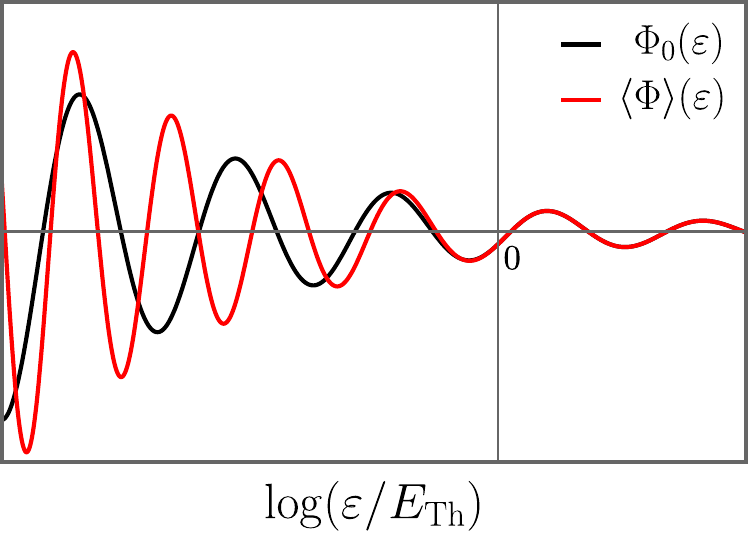}
    \label{fig:a}
  }
 \subfloat{
    \includegraphics[width=0.26\linewidth]{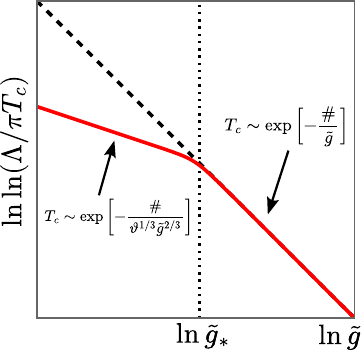}
    \label{fig:b}
  }
  \subfloat{\includegraphics[width=0.29\linewidth]{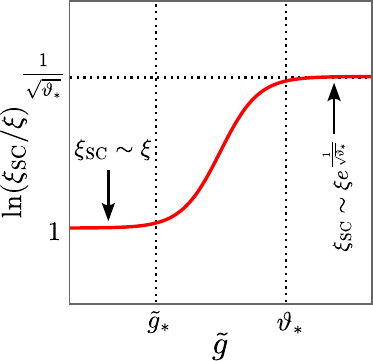}
  \label{fig:c}
  }
\caption{The figure shows the behavior of $T_c$ as a function of $\tilde g$, as described by Eq. \eqref{eq:GapEqInhom}. The lower value on the plot implies a higher $T_c$. The constant $\tilde g_* = \min(\vartheta, \ln^{-1}(\Lambda/E_\mathrm{Th}))$ determines the crossover between the two regimes. In the regime $\tilde g \geq \tilde g_*$, the $T_c$ follows a rule similar to homogeneous solution: $\pi T_c = \Lambda \exp[- \#\tilde g^{-1}]$. In the opposite regime, $\tilde g \lesssim \tilde g_*$, $T_c$ follows $\pi T_c = E_\mathrm{Th} \exp [- \# \vartheta^{-1/3} \tilde g^{-2/3}]$. In both regimes, numerical values $\#$ depend on $\vartheta$ and $\ln (E_\mathrm{Th}/T_{c0})$, but always remain of the order of unity.}
    \label{fig:TcBehavior}
\end{figure}

To estimate the effects of $\bar K_{DD}(\varepsilon, \varepsilon')$ on the structure of the spatially averaged Gor'kov function $\langle\Phi\rangle(i \varepsilon_n)$, we solve
\begin{equation}\label{eq:GapEqInhom}
        |\varepsilon|  \langle\Phi \rangle(i\varepsilon) 
    = 2 \bar g^2   \int_{\pi T}^{\Lambda} \frac{d\varepsilon^\prime}{2\pi} \Big( \langle  \bar D\rangle (i\max(|\varepsilon|, |\varepsilon^\prime|))+ \bar K_{DD}(\varepsilon, \varepsilon^\prime) \Big) \langle\Phi \rangle(i\varepsilon^\prime),
\end{equation}
where $\bar K_{DD}(\varepsilon, \varepsilon')$ is given by Eq. \eqref{eq:KDDFin_sec2} and, as before, we approximated $\langle \bar D \rangle(\varepsilon - \varepsilon')$ by $\langle \bar D \rangle(\max(|\varepsilon|, |\varepsilon'|))$.
Eq. \eqref{eq:GapEqInhom} can be solved exactly by
\begin{align}\label{eq:SCBAGapInhom}
    \langle\Phi \rangle(i\varepsilon) =
    \begin{cases}
        \left[ A_+ \cos\left( u(\varepsilon) \right) + B_+ \sin\left( u(\varepsilon) \right)\right]/\varepsilon,
        \quad
        u(\varepsilon) = \tilde g \ln(\Lambda/\varepsilon), \quad \quad\quad\quad\quad\quad\quad\quad\quad\quad\quad\quad\quad\quad\quad
        \varepsilon \geq E_\mathrm{Th}
        \\
        \left[ A_- \, \mathrm{Ai}^\prime\left( h(\varepsilon)\right) + B_- \, \mathrm{Bi}^\prime\left(h(\varepsilon)\right)\right]/\varepsilon,
        \quad
        h(\varepsilon) = - \left( \frac{\tilde g}{\vartheta}\right)^{\frac{2}{3}}  \left( 1 - \vartheta\ln \left( \Lambda/E_\mathrm{Th}\right) +(\vartheta/\tilde g) u(\varepsilon) \right), 
        \quad \;
        \varepsilon \leq E_\mathrm{Th}
    \end{cases}
\end{align}
The functions $\mathrm{Ai}\,(u)$ and $\mathrm{Bi}\,(u)$ are the Airy functions of the first and second kind, respectively, and the constants $A_\pm$ and $B_\pm$ are determined self-consistently.
The smoothness conditions implied by Eq. \eqref{eq:GapEqInhom} at $\varepsilon = E_\mathrm{Th}$ ensure that the values of 
$\langle\Phi \rangle(i \varepsilon)$ and 
$\p_\varepsilon\langle\Phi \rangle(i \varepsilon)$ are continuous across $\varepsilon = E_\mathrm{Th}$.
The self-consistency conditions at $\varepsilon = \Lambda$ ensure that
$\langle\Phi \rangle(i\Lambda) =0$.
Substitution of Eq. \eqref{eq:SCBAGapInhom} into Eq. \eqref{eq:GapEqInhom} with the three conditions mentioned above allows to fully determine all the unknown coefficients in Eq. \eqref{eq:SCBAGapInhom}.
The resulting behavior of the Gor'kov function is displayed in Fig. \ref{fig:a} and $T_c$ is shown in Fig. \ref{fig:b}.
The exact solution to Eq. \eqref{eq:GapEqInhom} that results in Fig. \ref{fig:a} is not physically illuminative, so it can be found in Appendix \ref{sec:AppendixGapEqCorrections}.

The behavior of $T_c$ shown in Fig. \ref{fig:a} also follows from the exact solution and can be easily understood from the Rayleigh-Ritz method applied to  \eqref{eq:GapEqInhom}.
Eq. \eqref{eq:GapEqInhom} can be understood as the eigenvalue equation $\langle\Phi \rangle = V[\langle\Phi \rangle]$ for a linear operator $V$ that has a form
\begin{equation}\label{eq:Koperator}
    V[\langle\Phi \rangle](\varepsilon) = \frac{\tilde g^2}{\varepsilon}  \int_{\pi T}^{\Lambda} d \varepsilon^\prime \left[ \ln \left( \frac{\Lambda}{\max(\varepsilon, \varepsilon^\prime)} \right) 
     + \frac{\vartheta }{2}  \; \Theta\left(E_\mathrm{Th} - \varepsilon\right) \Theta\left(E_\mathrm{Th} - \varepsilon'\right) \ln^2 \left( \frac{E_\mathrm{Th}}{\max(\varepsilon, \varepsilon^\prime)} \right) \right]\langle\Phi \rangle(i\varepsilon^\prime).
\end{equation}
The system goes through the superconducting transition at the temperature  $T_c$ when the largest eigenvalue of the operator $V$ equals unity.
The largest eigenvalue can be estimated from below using the Rayleigh-Ritz method by evaluating
$( \Phi_T, V[\Phi_T]) / ( \Phi_T, \Phi_T )$ for a test function
$\Phi_T(i\varepsilon)$.
The brackets $(f, g)$ for functions $f$ and $g$ denote the dot product of the form  $\langle f, g \rangle = \int_{\pi T}^{\Lambda} d\varepsilon \, \varepsilon f(\varepsilon) g(\varepsilon)$.
The solution to the gap equation with no inhomogeneity effects provides a good candidate for a test gap function, so we set $\Phi_T(i \varepsilon) \approx \varepsilon^{-1} \cos(\tilde g \ln(\varepsilon/\pi T))$.
The lower estimate of the largest eigenvalue of $V$ then evaluates to
\begin{equation}\label{eq:ReleighRitz}
    \frac{( \Phi_T, V[\Phi_T])}{(\Phi_T, \Phi_T)} = \frac{2 x - \sin (2 x)}{2 x + \sin (2x)} + \frac{\vartheta}{\tilde g} \Theta(x-g_\mathrm{Th}) \frac{2 (x - g_\mathrm{Th})^2 - 2 \sin^2(x-g_\mathrm{Th})}{2 x + \sin (2x)}, \quad x = \tilde g \ln \left( \frac{\Lambda}{\pi T_c}\right), \quad g_\mathrm{Th} = \tilde g\ln \left( \frac{\Lambda}{E_\mathrm{Th}}\right).
\end{equation}
The lower bound on $T_c$ is determined by the value of $x$ that equates to unity the right side of Eq. \eqref{eq:ReleighRitz}.
The expression in Eq. \eqref{eq:ReleighRitz} contains two regimes that are controlled solely by the ratio $\vartheta/\tilde g$.
When $\vartheta/\tilde g$ is small, the second term in Eq. \eqref{eq:ReleighRitz} can be neglected.
This corresponds to the regime where the disorder-averaged interaction produced by $\langle D \rangle(i \varepsilon)$ dominates over $\bar K_{DD}(\varepsilon, \varepsilon')$ in Eq. \eqref{eq:GapEqInhom}.
The smallest $x$ that equates the right side of Eq. \eqref{eq:ReleighRitz} to unity is $x = \pi/2$, leading to 
\begin{equation}\label{eq:DisAver1}
    T_c \sim \Lambda \exp \left[-\frac{\pi}{2 \tilde g}\right], 
\end{equation}
as expected from Eq. \eqref{eq:Tc0}.
This regime is valid when $\tilde g \gtrsim \vartheta$ and the effects of $\tilde g_\mathrm{Th}$ in the second term of Eq. \eqref{eq:ReleighRitz} are negligible.
The latter occurs when $\tilde g \gtrsim \ln^{-1}(\Lambda/E_\mathrm{Th})$, leading to $\tilde g_* = \min(\vartheta, \ln^{-1}(\Lambda/E_\mathrm{Th}))$, as shown in Fig. \ref{fig:b}.

When $\vartheta/\tilde g$ is large, the first term on the right of Eq. \eqref{eq:ReleighRitz} can be neglected, so $\bar K_{DD}(\varepsilon, \varepsilon')$ dominates over $\langle \bar D \rangle(i\varepsilon)$ in Eq. \eqref{eq:GapEqInhom}, resulting in the eigenvalue estimate reaching unity for $\vartheta x^3 /\tilde g = 1$.
The superconducting transition temperature then behaves as
\begin{equation}\label{eq:DisAver2}
    T_c \sim E_\mathrm{Th} \exp \left[-\frac{\#}{(\vartheta\tilde g^2)^{1/3}} \right],
\end{equation} 
which is also indicated in Fig. \ref{fig:b}.
This behavior is in agreement with the exact solution displayed in Appendix \ref{sec:AppendixGapEqCorrections}.

\subsection{Spatial fluctuations of the pairing amplitude}
\label{sec:GapFluctSCBA}

Besides the access to the disorder-averaged prediction of $T_c$, our approach additionally allows limited access to the inhomogeneity structure of the Gor'kov function $\Phi(i \varepsilon_n, \vec r)$.
In particular, we consider the statistics of the spatial fluctuations of $\Phi(i \varepsilon_n, \vec r)$, which is defined as $\delta \Phi(i \varepsilon_n, \vec r) = \Phi(i\varepsilon_n, \vec r) - \langle \Phi\rangle(i \varepsilon_n)$.
The spatial fluctuation $\delta \Phi(i \varepsilon_n, \vec r)$ obeys 
\begin{equation}\label{eq:DeltaPhiGapEq}
    \Big( - D \nabla^2 + 2 |\varepsilon_n| \Big) \delta \Phi(i \varepsilon_n, \vec r) - 2 \bar g^2 T \sum_{\varepsilon_m} \left[ \langle \bar D \rangle(\varepsilon_n - \varepsilon_m) + \bar K_{DD}(\varepsilon_n, \varepsilon_m) \right] \delta \Phi(i \varepsilon_m, \vec r) = 2 \bar g^2 T\sum_{\varepsilon_m} \delta \bar D(\varepsilon_n - \varepsilon_m, \vec r) \langle \Phi\rangle(i \varepsilon_m).
\end{equation}
Eq. \eqref{eq:DeltaPhiGapEq} is a result of linearization of Eq. \eqref{eq:full_gap} in 
$\delta \Phi(i \varepsilon_n, \vec r)$ and $\delta \bar D(i \omega_n, \vec r)$ under the assumption of
$ n(\mathcal{E}) = \mathrm{const}$.
The expressions for $\langle \bar D \rangle(i \omega)$ and $\bar K_{DD}(\varepsilon_n, \varepsilon_m)$ remain unchanged and are given by Eqs. \eqref{eq:AvDandSigma} and \eqref{eq:KDDFin_sec2} correspondingly. 

Eq. \eqref{eq:DeltaPhiGapEq} can be solved by approximating $\langle\bar D \rangle(i \varepsilon_n - i \varepsilon_m) \approx \langle \bar D \rangle(i \max (\varepsilon_n, \varepsilon_m))$ in the limit of $T_c \ll \Lambda$.
We are interested in the solution $\delta \Phi(i \varepsilon, \vec q) = \int d^2 \vec r \, e^{- i \vec q \cdot \vec r} \delta \Phi(i \varepsilon, \vec r)$ at finite momentum $\vec q$, such that $D \vec q^2 \gtrsim \pi T_c$.
Such momenta correspond to microscopic gap fluctuations that cannot be reduced to an adiabatic in-space change in the system's parameters.
Neglecting the effects of interaction in Eq. \eqref{eq:DeltaPhiGapEq} allows to write its solution as
\begin{equation}\label{eq:DeltaPhiExp}
    \delta \Phi(i \varepsilon_n, \vec q) \approx \frac{2 \bar g^2T \sum_{\varepsilon_m}  \delta \bar D(i\varepsilon_n - i\varepsilon_m, \vec q) \langle \Phi \rangle(i\varepsilon_m) }{D \vec q^2 + 2 |\varepsilon_n|}.
\end{equation}
The effects of the pairing $\langle\bar D \rangle(\varepsilon_n - \varepsilon_m)$ and $\bar K_{DD}(\varepsilon_n, \varepsilon_m)$ are to be discarded in this case, since they result in corrections to $\delta \Phi(i \varepsilon_n, \vec r)$ sub-leading in $\bar g^2$
(see Appendix \ref{sec:SCBAGapFluctuations} for algebraic details).

The strength of the inhomogeneity of the pairing density is characterized by $\delta \Phi_\mathrm{stat}(\omega = 0, \vec q) = (2 \pi)^{-1} \int d\varepsilon \, \delta \Phi(i \varepsilon, \vec q)$.
In particular, we evaluate the static structure factor $\langle \delta \Phi_\mathrm{stat}(\vec r) \, \delta \Phi_\mathrm{stat}(\vec r')\rangle / \langle\Phi_\mathrm{stat}\rangle^2$ of the pair density fluctuations.
The quantity can be evaluated from Eq. \eqref{eq:DeltaPhiExp}, given that statistics of $\langle \delta \bar D \delta \bar D\rangle$ and the spatially averaged Gor'kov function $\langle \Phi \rangle(i \varepsilon)$ were given in Eqs. \eqref{eq:delDdelD} and  Eq.~\eqref{eq:SCBAGapInhom} correspondingly.
The resulting expression in momentum space, ${\langle \Phi_\mathrm{stat}\,\delta\Phi_\mathrm{stat}\rangle(\vec q) = \int d^2 \vec r \, e^{- i \vec r \cdot \vec q} \langle \delta \Phi_\mathrm{stat}(\vec r) \, \delta \Phi_\mathrm{stat}(0) \rangle}$, takes a form (see Appendix \ref{sec:SCBAGapFluctuations} for algebraic details):
\begin{equation}\label{dLdLres}
      \frac{\langle \delta \Phi_\mathrm{stat} \, \delta \Phi_\mathrm{stat}\rangle(\vec q)}{\langle\Phi_\mathrm{stat}\rangle^2} \simeq \frac{32 \pi^4 \vartheta}{3 \vec q^2} \frac{\ln \left( \frac{D \vec q^2}{\pi T_c} \right)}{\left(1 + \vec q^2 \xi^2/4 \right)^{3}}, \quad D \vec q^2 \gtrsim \pi T_c.
\end{equation}
Divergence of $\langle \delta \Phi_\mathrm{stat} \, \delta \Phi_\mathrm{stat}\rangle$ at small momenta $\vec q$ implies that the spatial fluctuations of the pair amplitude grow at large distances.
The fluctuations grows comparable to $\langle \Phi_\mathrm{stat}\rangle^2$ at the length scale $|\vec r - \vec r'| \sim \xi_{\mathrm{SC}}$,
which is estimated by
\begin{equation}\label{eq:XiSCCondition}
    1 \sim \int_{|\vec q| > 2 \pi/\xi_{\mathrm{SC}}} \frac{d^2 \vec q}{(2 \pi)^2}  \frac{\langle \delta \Phi_\mathrm{stat} \, \delta \Phi_\mathrm{stat}\rangle(\vec q)}{\langle\Phi_\mathrm{stat}\rangle^2} \simeq \frac{8\pi^3 \vartheta}{3} \ln \left( \frac{\xi_{\mathrm{SC}}} {\xi} \right) \left[ \ln\left( \frac{E_\mathrm{Th}}{T_c} \frac{\xi}{\xi_{\mathrm{SC}}}\right) \right].
\end{equation}

Eq. \eqref{eq:XiSCCondition} allows two limiting physical regimes separated by transient behavior:
\begin{equation}\label{eq:XiSC}
    \ln \left( \frac{\xi_\mathrm{SC}}{\xi} \right) \simeq \begin{cases}
        1 \quad \tilde g \lesssim \tilde g_*
        \\
        \tilde g/\vartheta, \quad \tilde g_* \lesssim \tilde g \lesssim \vartheta/\sqrt{\vartheta_*}
        \\
        1/\sqrt{\vartheta_*}, \quad \vartheta/\sqrt{\vartheta_*} \lesssim \tilde g 
    \end{cases}
\end{equation}
where $\vartheta_*$ is given in Eq.~\eqref{eq:theta_star1}. Schematic behavior of $\xi_\mathrm{SC}$ with the change of $\tilde g$ is shown in Fig. \ref{fig:c}.
Thus, the pairing amplitude is approximately homogeneous for $\tilde g \gtrsim \tilde g_*$ and fragments for $\tilde g \lesssim \tilde g_*$.

\section{Single-Puddle Approximation}
\label{sec:Puddles}

Sec. \ref{sec:numerical} demonstrated numerically the presence of a large number of localized superconducting puddles.
We explore the structure of localized solutions analytically by considering a single-puddle model. 
In particular, we consider a pairing problem induced by a single strongly localized bosonic state $\phi_\alpha(\vec r)$ with center at $\vec r_\alpha$, localization length $\xi_\alpha$, and eigenvalue $e_\alpha$. 
We assume that nearby bosonic states have localization lengths much larger than $\xi_\alpha$.
Each localized boson $\phi_\beta(\vec r)$ contributes with amplitude proportional to $|\phi_\beta(\vec r)|^2 \sim \xi_\beta^{-2}$ into self-energy $\bar \Sigma(i \varepsilon_n, \vec r)$ and pairing vertex $\bar D(i \omega_m,\vec r)$.
Since $\xi_\alpha \ll \xi_\beta$ for all bosons $\phi_\beta(\vec r)$ in the vicinity of $\vec r_\alpha$, the local intensity of $\phi_\alpha$ dominates near $\vec r_\alpha$.

In our model we assume that $|\phi_\alpha(\vec r)|^2 = \Theta(\xi_\alpha - |\vec r- \vec r_\alpha|)/ \pi \xi_\alpha^2$, and neglect the contributions into $\bar \Sigma(i \varepsilon_n, \vec r)$ and $\bar D(i \omega_m, \vec r)$ from $\phi_\beta(\vec r)$ in the vicinity of $\vec r_\alpha$.
Then,  $\bar \Sigma(i \omega_m, \vec r)$ and $\bar D(i \varepsilon_n, \vec r)$ in the linearized Usadel equation  Eq. \eqref{eq:full_gap} for a Gor'kov function $\Phi(i \varepsilon_n ,\vec r)$ can be approximated as
\begin{align}\label{eq:SinglePuddle}
    \bar \Sigma(i \varepsilon_n, \vec r) 
    = 
    \begin{cases}
    \frac{\bar g^2}{\pi^2 \xi^2_\alpha c_d} \log \left( \frac{e_\alpha + c_d |\varepsilon_n|}{e_\alpha + \pi c_d T} \right),\quad |\vec{r}|\leq\xi_\alpha,
    \\
    0 ,\quad\quad\quad\quad\quad\quad\quad\quad\quad\quad\, |\vec{r}|>\xi_\alpha,
    \end{cases}
    \quad\quad\quad
    \bar D(i \omega_m, \vec r) 
    = 
    \begin{cases} 
    \frac{1}{\pi \xi^2_\alpha}\frac{1}{c_d |\omega_m| + e_\alpha},\quad |\vec{r}|\leq\xi_\alpha,
    \\
    n  \log \left( \frac{\Lambda}{|\omega_m|} \right) ,\quad\; |\vec{r}|>\xi_\alpha.
    \end{cases}
\end{align}
Above, we assumed that $\vec r_\alpha = 0$. 
The pairing for $|\vec r| > \xi_\alpha$ is the same as in the averaged problem considered in Section~\ref{sec:SCBA}.

First, we seek to simplify Eq. \eqref{eq:full_gap} by reducing the $\bar D(i \varepsilon_n - i \varepsilon_m, \vec r)$ term to $\bar D(i \max( \varepsilon_n, \varepsilon_m), \vec r)$.
Justification for this step for $|\vec r| > \xi_\alpha$ can be found in Sec. \ref{sec:SCBA}.
An additional complication arises for $|\vec r|< \xi_\alpha$:
the contribution of $(c_d |\varepsilon_n - \varepsilon_m| + e_\alpha)^{-1}$ diverges logarithmically near $\varepsilon_n \approx \varepsilon_m$.
Therefore, an additional contribution is present in the right side of Eq. \eqref{eq:full_gap} for $\varepsilon_m = \varepsilon_n$:
\begin{equation}\label{eq:SingularityRemove}
      T\sum_m \frac{\Phi(i \varepsilon_m, \vec r)}{c_d |\varepsilon_n - \varepsilon_m| + e_\alpha} \approx  T \sum_m \frac{\Phi(i \varepsilon_m, \vec r)}{c_d\max(|\varepsilon_m|, |\varepsilon_n|) + e_\alpha} +  \frac{\Phi(i \varepsilon_n, \vec r)}{\pi c_d} \ln \frac{e_\alpha+ c_d |\varepsilon_n|}{e_\alpha + \pi c_d T}.
\end{equation}

Substitution of Eq. \eqref{eq:SingularityRemove} into Eq. \eqref{eq:full_gap} leads to a cancellation of the logarithmic divergences between the self-energy and the second term of Eq. \eqref{eq:SingularityRemove}, resulting in
\begin{equation}\label{eq:gap_puddle}
    \Big(-D\nabla^2+2|\varepsilon_n|\Big)\Phi(i\varepsilon_n , \vec{r})-2\bar{g}^2 T \sum\limits_{\varepsilon_m} \bar D (i \max (\varepsilon_n, \varepsilon_m), \vec r) \Phi(i\varepsilon_m,\vec{r})=0.
\end{equation}

We seek a non-trivial solution $\Phi(i\varepsilon, \vec r)$ of Eq. \eqref{eq:gap_puddle} with the highest superconducting transition temperature $T_{c, \alpha} > T_c$.
This solution is expected to be exponentially localized, thus requiring solving for $\Phi(i\varepsilon, \vec r)$ as a function of both Matsubara frequency and coordinate.
The problem defined by Eq. \eqref{eq:gap_puddle} is too involved in a general setting, so we study two limiting cases described below.
The only length scale in the problem that explicitly appears in Eq. \eqref{eq:SinglePuddle} is $\xi_\alpha$.
Therefore, it is natural to assume that $D\nabla^2 \Phi \sim E_{\mathrm{Th}, \alpha} \Phi$ up to logarithmic corrections, where $E_{\mathrm{Th}, \alpha} = 2D/\xi_\alpha^2$ is the electronic Thouless energy associated with characteristic length $\xi_\alpha$.
Behavior of $T_{c, \alpha}$ changes significantly as the ratio between $E_{\mathrm{Th}, \alpha}$ and $T_{c, \alpha}$ changes, which leads to the following two regimes. 

The first scenario to be considered is when $\xi_\alpha$ is large enough for $E_{\mathrm{Th}, \alpha} \lesssim T_{c,\alpha}$.
Then the gradient term in Eq. \eqref{eq:gap_puddle} can be neglected, resulting in the Gor'kov function of the form $\Phi(i \varepsilon_n, \vec r) = \Phi(i \varepsilon_n) \Theta(\xi_\alpha - |\vec r|)$.
Solving Eq. \eqref{eq:gap_puddle} for $\Phi(i \varepsilon_n)$ leads to $T_{c, \alpha}$ being 
\begin{equation}\label{eq:SinglePuddleNoGradTc}
    \frac{E_{\mathrm{Th}, \alpha}}{\pi T_{c, \alpha}} \sim  \frac{g_\alpha}{\vartheta} \left( \exp \left[\frac{1}{g_\alpha}\right] -1 \right), 
    \quad\quad
    g_\alpha = \frac{\bar g^2}{\pi^2 \xi^2_\alpha e_\alpha}.
\end{equation}
Since we require $E_{\mathrm{Th}, \alpha} \lesssim T_{c, \alpha}$, it follows that $\vartheta$ must satisfy
$\vartheta \gtrsim g_\alpha (\exp[g_\alpha^{-1}]  - 1) \gtrsim 1$.
Therefore, this scenario is only valid in the presence of strong electronic disorder outside the metallic regime.
The derivation of Eq.~\eqref{eq:SinglePuddleNoGradTc} is given in Appendix \ref{sec:AppBigSolitaryPuddle}.

The second scenario we consider corresponds to $E_{\mathrm{Th}, \alpha} \gg T_{c, \alpha}$, where the gradient term in Eq. \eqref{eq:gap_puddle} dominates over $|\varepsilon|$ term. 
To approach the problem analytically, we assume that $\xi_\alpha$ is the smallest length scale in the problem, particularly $\xi_\alpha \lesssim \sqrt{D/T_{c, \alpha}}$. 
We seek a solution to Eq. \eqref{eq:gap_puddle} by first constructing the eigenfunctions that satisfy Eq. \eqref{eq:gap_puddle} for all $|\vec r| > \xi_\alpha$ and ``stitch'' them at the boundary $|\vec r| = \xi_\alpha$ with a proper boundary condition.
The eigenfunctions at $|\vec r| \geq \xi_\alpha$ have a form $C_{\varkappa_m}(\varepsilon_n) K_0\left(k_m r\right)$, so the solution to  Eq. \eqref{eq:gap_puddle} has a form 
\begin{equation}\label{eq:SinglePuddleForm}
    \Phi(i \varepsilon_n, \vec{r}) = 
    \sum\limits_{\varkappa_m} \tilde\Psi(\varkappa_m) \,C_{\varkappa_m}(\varepsilon_n) \, K_0\left(k_m r\right).
\end{equation}
The $k_m$ values $k_m = \sqrt{2\varkappa_m/D}$ are defined by the eigenvalues $\varkappa_m$ of the eigenfunctions $C_{\varkappa_m}(\varepsilon_n)$, 
and coefficients $\tilde \Psi(\varkappa_m)$ are determined by the behavior of $\Phi(i \varepsilon_n, \vec r)$ at $|\vec r| < \xi_\alpha$.
Substitution of Eq. \eqref{eq:SinglePuddleForm} into Eq. \eqref{eq:gap_puddle} determines the equation that eigenfunctions $C_{\varkappa_m}(\varepsilon_n)$ and their respective eigenvalues $\varkappa_m$ must satisfy.
In the limit of $T, T_{c0} \ll \Lambda$, the equation for $C_\varkappa(\varepsilon)$ reduces to
\begin{equation}\label{eq:Ckeq}
    (|\varepsilon|- \varkappa) C_{\varkappa}(\varepsilon) = \tilde g^2 \int_{\pi T}^\Lambda d\varepsilon' \, \ln \left( \frac{\Lambda}{ \max (|\varepsilon|, \varepsilon')}\right)  C_{\varkappa}(\varepsilon').
\end{equation}
Eq. \eqref{eq:Ckeq} always attains a non-trivial solution of the form
\begin{equation}\label{eq:Csolution}
    C_\varkappa(\varepsilon) = \delta(|\varepsilon| - \varkappa) + A_\varkappa F_1 \left( \frac{|\varepsilon|}{\varkappa} \right) + B_\varkappa F_2\left(1- \frac{|\varepsilon|}{\varkappa}\right),
\end{equation}
where $B_\varkappa = - \tilde g^2/\varkappa$ and $A_\varkappa = B_\varkappa \tilde g^{-1} \cot (\tilde g \ln (\varkappa/\pi T_{c0}))$, and $\pi T_{c0} = \Lambda e^{-\pi/2\tilde g}$ is superconducting transition temperature of a homogeneous problem.
Functions $F_1$ and $F_2$ are  $F_1(z) = \mathrm{Re} \, F\left(1 + i \tilde g, 1- i \tilde g ; \,  1 ; \, z\right)$, and $F_2(z) = \Theta(-z)\mathrm{Re} \, F\left(1 + i \tilde g, 1- i \tilde g ; \,  2 ; \, z\right)$, where $F(a,b,c; z)$ is a Gauss hypergeometric function.
Eq. \eqref{eq:Csolution} is valid, provided $\pi T<|\varkappa|<\Lambda$ and $\tilde g$ is small.
Verifying that Eq. \eqref{eq:Csolution} solves Eq. \eqref{eq:Ckeq} is a task similar to solving Eqs. \eqref{eq:HomIntegralEq},
so its details are displayed in Appendix \ref{sec:SmallPuddleEigenvalueEq}.

It is important to note that the regular, non-delta-function parts of Eq.~\eqref{eq:Csolution} are proportional to $A_\varkappa$ and $B_\varkappa$, and therefore are of the order $\tilde g^2$ for $\varepsilon \gtrsim \sqrt{T_{c0} \Lambda}$.
This allows to approximate $k_m$ in Eq. \eqref{eq:SinglePuddleForm} by $k_n \approx \sqrt{2 |\varepsilon_n|/D}$.
With new coefficients defined as $\Psi(\varepsilon_n) = \sum_{\varkappa_m} \tilde \Psi(\varkappa_m) C_{\varkappa_m}(\varepsilon_n)$, the Gor'kov function $\Phi(i \varepsilon_n, \vec r)$ can be approximated by $\Phi(i \varepsilon, \vec r) \approx \Psi(\varepsilon_n) K_0(k_n r)$.

To determine the coefficients $\Psi(\varkappa_m)$ in Eq. \eqref{eq:SinglePuddleForm}, the boundary conditions for $\Phi(i \varepsilon_n, \vec r)$ at $|\vec r| = \xi_\alpha$ have to be considered.
We assume that $\xi_\alpha$ is the smallest scale in the problem, so $\Phi(i \varepsilon_n, \vec r)$ does not vary significantly for $|\vec r| < \xi_\alpha$. 
Under this assumption, one can integrate 
Eq. \eqref{eq:gap_puddle} over $\vec r$ over a ball of radius $\xi_\alpha$ centered at the bosonic localized state, leading to
\begin{equation}\label{eq:BoundaryCon}
    - 2 \pi D \xi_\alpha \left( \p_{\vec r} \Phi(i\varepsilon_n, \xi_\alpha) \right)  - 2 \bar g^2 T  \sum_{\varepsilon_m}  \frac{\Phi(i\varepsilon_m, \xi_\alpha)}{c_d \max(|\varepsilon_n|, |\varepsilon_m|) + e_\alpha}  =0.
\end{equation}
Eq. \eqref{eq:BoundaryCon} can be rewritten through $\Psi(\varepsilon_n)$, $k_n = \sqrt{2 \varepsilon_n/D}$, and $\vartheta = 3 \bar g^2 /2 \pi^2 c_d D$ (see Sec. \ref{sec:SCBA}) as 
\begin{equation}\label{eq:PuddleBoundary}
    \Psi(\varepsilon_n) = \frac{2 \pi T \vartheta}{3} \sum_{\varepsilon_m} \frac{\Psi(\varepsilon_m)K_0(k_m \xi_\alpha)}{\max(|\varepsilon_n|, |\varepsilon_m|) + e_\alpha/c_d}.
\end{equation}
In the limit of small $T$, Eq. \eqref{eq:PuddleBoundary} can be solved in a manner similar to Eqs. \eqref{eq:HomIntegralEq} 
(see Appendix \ref{sec:PuddleBoundaryCond}).
The resulting expression for $\Psi(\varepsilon)$ takes a form 
\begin{align}\label{eq:PuddleBoundSol}
    \Psi(\varepsilon) = \sqrt{\frac{E_{\mathrm{Th}, \alpha}}{ \varepsilon}} \left[ a \, \mathrm{Ai} \,(z(\varepsilon)) + b \, \mathrm{Bi} \, (z(\varepsilon)) \right], \quad\quad
    z(\varepsilon) = \left(\frac{\vartheta}{3}\right)^{-2/3} \left( \frac{1}{4} - \frac{\vartheta}{3} \ln \left(\frac{E_{\mathrm{Th}, \alpha}}{\varepsilon} \right) \right).
\end{align}
The coefficients $a$ and $b$ are determined self-consistently by substituting Eq. \eqref{eq:PuddleBoundSol} back into Eq. \eqref{eq:PuddleBoundSol}.

The behavior of the Airy functions in Eq. \eqref{eq:PuddleBoundSol} for large positive and large negative arguments is principally different.
For a positive arguments the asymptotics of Airy functions is exponential: $Ai(z)\sim (2 \sqrt{\pi} z^{1/4})^{-1} \exp[-2 z^{3/2}/3]$, $Bi(z)\sim (2 \sqrt{\pi} z^{1/4})^{-1} \exp[2 z^{3/2}/3]$. 
For large negative arguments the Airy functions are oscillatory $Ai(-z) \sim (\sqrt \pi z^{1/4})^{-1} \sin( 2 z^{3/2}/3 + \pi/4)$, $Bi(-z)\sim (\sqrt \pi z^{1/4})^{-1} \cos (2 z^{3/2}/3 + \pi/4)$.
The lowest energies to appear in the argument $z(\varepsilon)$ that enters the Airy functions in Eq. \eqref{eq:PuddleBoundSol} are of the order of $\pi T + e_\alpha/c_d$. 
Therefore, if $z(\pi T + e_\alpha/c_d) < 0$, the behavior of $\Psi(\varepsilon)$ at the smallest energies is highly oscillatory, signaling the onset of superconductivity at $z(\pi T + e_\alpha/c_d) = 0$.
Therefore, the superconducting transition temperature $T_{c, \alpha}$ for the puddle satisfies 
\begin{equation}\label{eq:TcPuddleRel}
    \pi T_{c, \alpha} + e_\alpha/c_d = E_{\mathrm{Th},\alpha} \exp \left[ - \frac{3}{4 \vartheta} \right], \quad\quad
    E_{\mathrm{Th}, \alpha} = \frac{2 D}{\xi_\alpha^2}.
\end{equation}

There are several conclusions to be drawn from Eq. \eqref{eq:TcPuddleRel}.
First, it is important to note that only the localized bosons that satisfy
\begin{equation}\label{eq:PuddleBound}
    e_\alpha \xi_\alpha^2 < 2 D c_d e^{-3/4\vartheta}    
\end{equation}
can form localized superconducting solutions at finite temperature.
If $\xi_\alpha$ decreases as $e_\alpha$ decreases, as suggested by the numerical data, one expects the existence of an upper cutoff for the values of $e_\alpha$ that allows for the formation of superconducting puddles on top of the bosons.
The second conclusion concerns the high-temperature behavior of the distribution of superconducting puddles.
It follows from Eq. \eqref{eq:TcPuddleRel} that superconducting puddles with the highest $T_{c, \alpha}$ correspond to the localized bosons with the smallest $e_\alpha$ and $\xi_\alpha$.
Therefore, the second term on the left side of Eq. \eqref{eq:TcPuddleRel} can be neglected, leading to $T_{c,\alpha} = 2D e^{-3/4 \vartheta}/\xi_\alpha^2$.
Assuming that $\xi_\alpha \sim e_\alpha^\gamma$, it follows that $T_{c,\alpha} \sim e_\alpha^{- 2 \gamma}$. 
Since $e_\alpha$ is distributed homogeneously at smallest values, it follows that $T_{c,\alpha}$ is distributed with a power-law probability density 
\begin{equation}\label{eq:PTcPuddle}
    p(T_{c, \alpha}) \sim T_{c, \alpha}^{- 1 - 1/2\gamma}    
\end{equation}
until a microscopic cutoff defined by the smallest $e_\alpha$ is reached.
Each localized state with superconducting transition temperature $T_{c, \alpha}$ is expected to have a size of the order of $\sqrt{D/T_{c, \alpha}}$.
Multiplying the density of active puddles by their typical area, of order $D/T_{c,\alpha}$, gives the same power law for the fraction of space covered by puddles, $1 -\tilde P_U(T)$:
\begin{equation}\label{eq:PUTheory}
    1-\tilde P_U(T) \sim T^{-1 - 1/2 \gamma}, \quad \quad T \gg T_c.
\end{equation}
The power law behavior in Eqs. \eqref{eq:PTcPuddle} and \eqref{eq:PUTheory} contrasts with the exponential behavior of superconducting puddles in BCS superconductivity \cite{Dodaro2018,bulaevskii1987,Larkin1981}.

\section{Discussion and outlook}

In this work we have shown that disorder affects quantum-critical superconductivity in a way that is qualitatively distinct from its role in conventional BCS theory. The central reason is that the disorder couples directly to the local distance from the quantum critical point (so-called ``Harris disorder''), and non-perturbatively reshapes the spectrum and wave functions of the bosonic modes that both mediate pairing and produce non-Fermi-liquid behavior in the normal state. In the regime considered here, the fermionic states remain extended and diffusive, while the low-energy bosonic modes become spatially localized~\cite{patel2024strange,Aavishkar2024}. The superconducting problem is consequently controlled by Cooper-pair diffusion through a highly inhomogeneous, strongly overdamped pairing glue. The main results supporting this picture are summarized in Sec.~\ref{sec:results}.

This pairing mechanism produces a broad hierarchy of pairing scales. The most localized low-energy bosonic modes act as compact pairing centers and nucleate superconducting puddles at temperatures above the onset of an extended pairing eigenstate. These puddles are not the conventional rare large regions of enhanced BCS attraction. Rather, they are tied to localized overdamped collective modes, and their pairing scales are controlled by the competition between the singular local glue and the inverse proximity effect from the surrounding metal. This leads to a power-law distribution of local pairing scales, in contrast to the stretched-exponential Lifshitz tails expected in weakly disordered BCS superconductors. At lower temperatures, the pairing eigenstate becomes extended, but the localized structure of the bosonic glue continues to affect the transition through the mesoscopic correlation effects and enhanced return probability of Cooper pairs to favorable regions.

Several assumptions underlie this description. We have focused on pairing from bosonic modes with a localization length that is large compared to the single-particle mean free path. \textcolor{black}{In this regime, the boson-induced pairing vertex and self-energy act as long-wavelength inhomogeneities for diffusive Cooper pairs.} Relaxing this restriction would allow us to include effective short-wavelength scattering processes that can act as conventional pair breakers, especially for sign-changing order parameters. We have also assumed a regime in which the electronic states remain metallic on the length scales relevant for pairing, while the bosonic modes are already localized. Finally, the transition studied here is the linearized pairing instability. In two dimensions, the ultimate thermodynamic transition into a phase-coherent superconducting state will also depend on phase fluctuations and Josephson coupling between puddles. Thus, the extended zero mode of the pairing problem should be viewed as a mean-field criterion for macroscopic pairing, while the true transition temperature may be reduced by phase-ordering physics.

The broader implication of our results is that strong superconducting inhomogeneity does not require localized electrons or microscopic granularity. It can arise naturally in a metallic quantum-critical system from the localization of the collective modes responsible for pairing. This provides a route by which strange-metal behavior, broad gap distributions, and superconducting puddles can emerge from a common origin: Harris disorder acting on the quantum-critical bosonic pairing glue. 

The evolution of the inhomogeneous superconducting states that we find is broadly consistent with STM observations of temperature-dependent gap inhomogeneity in the cuprates~\cite{Yazdani_puddles_2007,Yazdani_puddles_2008}. However, the existing data is too sparse to discern the high-temperature tails of these distributions and confirm our prediction. We hope that future experimental works will revisit this important problem of high-temperature gap inhomogeneity in the cuprates with STM. 

It is also useful to compare our results with recent theoretical work on superconducting puddles in the overdoped cuprates~\cite{EliashbergLinear,Tulipman24,Tulipman24B,Tulipman26}. Those works postulate a set of ``two-level systems'' that provide local Andreev scattering centers for itinerant electrons. The closest analog in our theory is the solitary-puddle regime discussed in Sec.~\ref{sec:Puddles}, where pairing is nucleated by an exponentially localized $\phi$ mode. The important difference is that in our case the puddles are not introduced phenomenologically; they emerge from the generic effect of Harris disorder on a quantum-critical metal.

The theory developed here suggests several directions for future work, aside from simply relaxing the assumptions about length scales that we used. One important extension is to derive an effective phase-only theory for the superconducting puddles and calculate the resulting superfluid stiffness. This would clarify the relation between the local pairing scale, the onset of an extended pairing eigenstate, and the eventual phase-ordering transition. Another important direction is to treat the bosonic localization problem and the superconducting gap equation fully self-consistently, allowing the pairing gap to feed back on the Landau damping by depleting the low-energy fermionic spectral weight.

\color{black}

\acknowledgements

We thank Erez Berg, Andrey Chubukov, Steven Kivelson, Alex Kamenev, Alex Levchenko, Charlie Marcus, Srinivas Raghu, Joerg Schmalian, and Evyatar Tulipman for valuable discussions. We thank Ali Yazdani for valuable discussions and for permission to extract data from the figures of Ref. \cite{Yazdani_puddles_2007}. This research was supported by NSF Grant DMR-2245246, by the
Harvard Quantum Initiative Postdoctoral Fellowship in
Science and Engineering, and by the Simons Collaboration on Ultra-Quantum Matter which is a grant from the Simons Foundation (651440, SS). The Flatiron Institute is a division of the Simons Foundation.

\appendix


\section{Contribution of localized bosonic modes  into electronic self-energy}\label{sec:AppendixGreensFunction}

The inhomogeneous Dyson equation for the electronic self-energy can be derived using the Wigner transform expansion technique. 
First, we consider the bosonic Green's function expanded in localized eigenmodes $\phi_\alpha(\vec r)$ with eigenvalues $e_\alpha$.
Such an ansatz phenomenologically accounts for bosonic localization in the presence of electrons: 
\begin{equation}
    D(i \omega_m, \vec r, \vec r^\prime) = \sum_\alpha \frac{\phi_\alpha(\vec r) \phi^*_\alpha(\vec r^\prime)}{\omega_m^2 + c_d |\omega_m| + e_\alpha}.
\end{equation}
We assume that each mode consists of a smooth envelope $|\phi_\alpha(\vec r)| $ are localized at $\vec r_\alpha$ with a localization length $\xi_\alpha$, while its short-distance phase correlations decay at the length scale $l_{\rm corr}$.
We model the disorder-averaged short-distance phase correlations by
$\langle e^{i[\theta_\alpha(\vec r)-\theta_\alpha(\vec r^\prime)]}\rangle
\sim e^{-|\vec r-\vec r^\prime|/2l_{\rm corr}}$, which gives
\begin{equation}
    D(i \omega_m, \vec r, \vec r^\prime) = \sum_\alpha \frac{|\phi_\alpha(\vec r) ||\phi_\alpha(\vec r^\prime)| e^{- |\vec r - \vec r^\prime|/2l_{\rm corr}}}{\omega_m^2 + c_d |\omega_m| + e_\alpha}.
\end{equation}
It is convenient to redefine the bosonic Green's function as a function of the center of mass $\vec R = (\vec r + \vec r^\prime)/2$ and the difference $\vec y = \vec r - \vec r^\prime$: 
\begin{equation}
    D(i \omega_m, \vec R, \vec y) = \sum_\alpha \frac{\phi_\alpha(\vec R+ \vec y/2) \phi_\alpha(\vec R- \vec y/2) e^{- |\vec y|/2l_{\rm corr}}}{\omega_m^2 + c_d |\omega_m| + e_\alpha}.
\end{equation}
 Above and in the following, we absorb the absolute value of $\phi_\alpha$ into the definition of the envelope for convenience.
 If $l_{\rm corr} \ll \xi_\alpha$ is assumed, the contribution of $\vec y$ into the argument of $\phi_\alpha$ can be neglected.
This is equivalent to neglecting the corrections of the order of $(l_{\rm corr}/\xi_\alpha)^2$. 
After performing a Wigner transform over the coordinate $\vec y$, one obtains
\begin{equation}
    D(i \omega_m, \vec R, \vec q) = \int d^2 y \, e^{- i \vec q \cdot \vec y} D(i \omega_m, \vec R, \vec y) =\sum_\alpha \frac{8 \pi l^2_\mathrm{corr} \phi^2_\alpha(\vec R) f(\vec q)}{\omega_m^2 + c_d |\omega_m| + e_\alpha},
\end{equation}
where $f(\vec q) = (1 + 4 \vec q^2 l_{\mathrm{corr}}^2)^{-3/2}$. 
Assuming that $|\vec q| \ll l^{-1}_{\mathrm{corr}}\sim k_F$ (see the main text), we can approximate $f(\vec q)$ as $f(\vec q) \approx 1$.
As a result, the bosonic Green's function is local after coarse-graining over length scales large compared with $l_\mathrm{corr}$. 
Hence we define $\bar D(i \omega_m, \vec r)$ as $D(i \omega_m, \vec r, \vec r^\prime) \approx 8 \pi l^2_\mathrm{corr} \bar D(i\omega_m, \vec r) \delta(\vec r - \vec r^\prime)$, where
\begin{equation}
    \bar D(i\omega_m, \vec r) = \sum_\alpha \frac{\phi_\alpha^2(\vec r)}{\omega_m^2 + c_d |\omega_m| + e_\alpha}.
\end{equation}

We proceed by considering the Dyson equation for the electron Green's function.
We assume that the electron self-energy contribution due to the chemical potential disorder has already been self-consistently accounted for in a ``bare" electron Green's function
\begin{equation}
    G_0(i \varepsilon, \vec k) = \frac{1}{i \varepsilon - E(\vec k) + i \Gamma \, \mathrm{sgn} \, \varepsilon},
\end{equation}
where $\Gamma=1/2\tau$ is the elastic scattering rate.
The Dyson equation for the electron position-dependent Green's function $G(i\varepsilon, \vec r, \vec k)$ can be written with the use of the Wigner transform exactly:
\begin{equation}\label{eq:DysonG}
    \left[ G_0^{-1}(i \varepsilon, \vec k) - \Sigma(i \varepsilon, \vec r, \vec k) \right] \circ G(i \varepsilon, \vec r, \vec k) = 1, 
\end{equation}
where $A(\vec x, \vec p)\circ B(\vec x, \vec p) \equiv A(\vec x, \vec p) e^{\frac{i}{2}( \overleftarrow{\p}_{\vec x}  \overrightarrow{\p}_{\vec p} - \overleftarrow{\p}_{\vec p}  \overrightarrow{\p}_{\vec x})} B(\vec x, \vec p)$ is a Moyal product. 
For details regarding the Moyal product and Wigner transform, see \cite{Kamenev} or \cite{Kryhin2025}. 
In the Dyson equation given by Eq. \eqref{eq:DysonG}, the electron self-energy $\Sigma(i\varepsilon, \vec r, \vec k)$ after the Wigner transform is given by
\begin{equation}\label{eq:Dyson}
    \Sigma(i \varepsilon_n, \vec r, \vec k) = g^2 T \sum_{\omega_m} \int \frac{d^2 q}{(2 \pi)^2} G(i \varepsilon_n + i \omega_m, \vec r, \vec k - \vec q) D(i \omega_m, \vec r, \vec q). 
\end{equation}

To access the effects of spatial inhomogeneity on $G(i\varepsilon_n, \vec r, \vec k)$ and $\Sigma(i\varepsilon_n, \vec r, \vec k)$, we consider the gradient corrections to Eq. \eqref{eq:DysonG} induced by the Wigner transform.
Assume that the electronic Green's function can be approximately expressed as
\begin{equation}
    G(i\varepsilon_n, \vec r, \vec k) = \frac{1}{i \varepsilon_n - E(\vec k) + i \Gamma \, \mathrm{sgn} \, \varepsilon_n - \Sigma(i \varepsilon_n, \vec r, \vec k) - \delta \Sigma (i\varepsilon_n, \vec r, \vec k)}.
\end{equation}
The additional term $\delta \Sigma(i \varepsilon_n, \vec r, \vec k)$ is a small correction to $\Sigma$ due to spatial inhomogeneity terms arising from the Moyal product.
Indeed, $G(i\varepsilon_n, \vec r, \vec k)$ for $\delta \Sigma = 0$ not an exact Moyal inverse of $G_0^{-1} - \Sigma$ due to the presence of derivative terms. 
One can compute the correction $\delta \Sigma$ by expanding Eq. \eqref{eq:Dyson} up to the second-order corrections in $\p_{\vec r}$ and $\p_{\vec k}$. 
The self-energy correction at energies close to the Fermi surface takes the form
\begin{equation}
    \delta \Sigma \sim \frac{(\vec v_F \p_{\vec r})^2}{\Gamma^2} \Sigma + \frac{\p_{k_i}\p_{k_j} E(\vec k)}{\Gamma} \p_{r_i} \p_{r_j}\Sigma.
\end{equation}
The expression above is a non-singular small correction when $\xi_\alpha \gg v_F/\Gamma = 2l$, where $l$ is an electron mean free path. 
Note that the leading correction to $G(i\varepsilon_n, \vec r, \vec k)$ is on the order of $l^2/\xi_\alpha^2$ and therefore can be neglected or reabsorbed into $\Sigma$ by defining $\Sigma^\prime(i \varepsilon_n, \vec r) \equiv \Sigma(i\varepsilon_n, \vec r, \vec k) + \delta \Sigma(i\varepsilon_n, \vec r, \vec k)$.
With the form of the Green's function confirmed to be approximately described by 
\begin{equation}\label{eq:GApprox}
    G(i\varepsilon_n, \vec r, \vec k) = \frac{1}{i \varepsilon_n - \xi(\vec k) + i \Gamma \, \mathrm{sgn} \, \varepsilon_n - \Sigma(i \varepsilon_n, \vec r, \vec k)}, 
\end{equation}
one can compute the position-dependent self-energy of the Green's function.
Substituting Eq. \eqref{eq:GApprox} into Eq. \eqref{eq:Dyson} results in 
\begin{equation}\label{eq:AppFullSigma}
    \Sigma(i\varepsilon_n, \vec r) = - i \bar g^2 T \sum_\alpha \sum_m \frac{\phi_\alpha^2(\vec r) \, \mathrm{sgn} \, (\varepsilon_n + \omega_m)}{\omega_m^2 + c_d |\omega_m| + e_\alpha} = - 2 i \bar g^2  T \, \mathrm{sgn} \, (\varepsilon_n) \sum_\alpha \left[ \sum_{m = 1}^{|n| - \Theta(-n - 1/2)} \frac{\phi_\alpha^2(\vec r)}{\omega_m^2 + c_d |\omega_m| + e_\alpha} + \frac{\phi_\alpha^2(\vec r)}{2 e_\alpha}\right],
\end{equation}
where $\bar{g}^2=8\pi^2 l^2_\mathrm{corr} \nu g^2$ and $\nu = k_F/2 \pi v_F$. 
Note that $\Sigma(i\varepsilon_n, \vec r) = \Sigma(i\varepsilon_n, \vec r, \vec k)$ is approximately independent of momentum $\vec k$, which immediately implies that the electron self-energy $\Sigma(i\varepsilon_n, \vec r, \vec r^\prime)$ is local in real space at the scales greater than $k_F^{-1}$. 
Therefore, one can define
\begin{equation}
    \Sigma(i\varepsilon_n, \vec r, \vec r^\prime) \approx - i \, \mathrm{sgn}(\varepsilon_n) \, \bar \Sigma(i\varepsilon_n, \vec r) \delta(\vec r - \vec r^\prime),
\end{equation}
which reproduces the Eq. \eqref{eq:Sigma_full_eq} in the main text.
In the overdamped regime, where the relevant Matsubara frequencies satisfy $|\omega_m|\ll c_d$ so that the $\omega_m^2$ term can be neglected, the sum over $\omega_m$ in Eq. \eqref{eq:AppFullSigma} gives
\begin{equation}\label{eq:AppSigmaBar}
    \bar \Sigma(i\varepsilon_n, \vec r) = \frac{ \bar g^2}{\pi c_d}  \sum_\alpha \phi_\alpha^2(\vec r) \left[ \psi\left(\frac{1}{2} + \frac{|\varepsilon_n|}{2 \pi T} + \frac{e_\alpha}{2 \pi c_d  T} \right) - \psi \left( \frac{1}{2} + \frac{e_\alpha}{2 \pi c_d  T} \right) + \frac{\pi c_d T}{e_\alpha}\right],
\end{equation}
which at $T = 0$ reduces to
\begin{equation}
   \bar \Sigma(i \varepsilon, \vec r) = \frac{\bar g^2}{\pi c_d} \sum_\alpha \phi_\alpha^2(\vec r) \log \left( \frac{c_d |\varepsilon| + e_\alpha}{e_\alpha}\right).
\end{equation}


\section{Derivation of the linearized Usadel equation}
\label{sec:AppendixUsadel}

We now derive the linearized gap equation using the coarse-grained boson propagator and the electronic Green's function obtained in Appendix~\ref{sec:AppendixGreensFunction}.
We begin by evaluating the Cooperon propagator $\mathcal{C}_0(i \varepsilon_n, i \omega_m, \vec r, \vec r^\prime)$, given in Fig. \ref{fig:DiagSeries}, which is a central quantity to the theory of disordered superconductivity \cite{AGD,Levitov}.
Here ``bare'' means that the Cooperon includes potential-disorder ladders and the local electronic self-energy, but not the boson-mediated pairing interaction.
The ``bare" Cooperon propagator satisfies the Bethe-Salpeter equation described by the series in Fig. \ref{fig:DiagSeries}, which can be written explicitly as
\begin{multline}
    \left(2 \pi \nu \tau^2\right)^{-1}\mathcal{C}_0(i \varepsilon, i \omega, \vec r, \vec r^\prime) =\left(2 \pi \nu \tau\right)^{-2} G\left(i\varepsilon + i\frac{\omega}{2}, \vec r, \vec r^\prime\right) G\left(- i \varepsilon + i \frac{\omega}{2}, \vec r, \vec r^\prime\right)
    +
    \\
    + \left( 2 \pi \nu \tau \right)^{-1} \int d^2 \vec x \; G\left(i \varepsilon + i \frac{\omega}{2}, \vec r, \vec x\right) G\left(-i \varepsilon + i \frac{\omega}{2}, \vec r, \vec x \right) \left( 2 \pi \nu \tau^2 \right)^{-1} \mathcal{C}_0(i \varepsilon, i \omega, \vec x, \vec r^\prime).
\end{multline}
The ``bare'' Cooperon depends on two spatial coordinates, so the Wigner transform in the real space can be constructed ordinarily. 
After applying the Wigner transform, the Bethe-Salpeter equation reduces to
\begin{equation} \label{eq:CooperonSalpeter}
     \mathcal{C}_0(i\varepsilon, i \omega, \vec r, \vec p) =  \tau K_C(i \varepsilon, i \omega, \vec r, \vec p) + K_C(i \varepsilon, i \omega, \vec r, \vec p) \circ \mathcal{C}_0(i\varepsilon, i \omega, \vec r, \vec p),
\end{equation}
where
\begin{equation}\label{eq:AppKCDef}
    K_C(i \varepsilon, i \omega, \vec r, \vec p) = \left(2 \pi \nu \tau \right)^{-1} \int \frac{d^2 \vec w}{(2 \pi)^2} G\left(i\varepsilon + i\frac{\omega}{2}, \vec r, \vec p + \vec w\right) G\left(- i \varepsilon + i \frac{\omega}{2}, \vec r,  - \vec w\right).
\end{equation}
Therefore, the equation for the ``bare" Cooperon propagator can be expressed as
\begin{equation}\label{eq:AppShortCooperon}
    \tau^{-1} K_C^{-1} \circ (1 - K_C) \circ\mathcal{C}_0(i \varepsilon, i \omega, \vec r, \vec p) = 1,
\end{equation}
where $K_C^{-1}$ denotes the Moyal inverse of $K_C$.
Note that Eq. \eqref{eq:AppShortCooperon} reduces to the corresponding homogeneous expression with ordinary multiplication instead of the Moyal product, if the system is assumed to be translation invariant \cite{Levitov}. 
Within the usual Fermi-surface approximation, the integral in $K_C$ can be evaluated analytically by substituting Eq. \eqref{eq:GApprox} into Eq. \eqref{eq:AppKCDef}.
This leads to
\begin{equation}\label{eq:NonEqKC}
    K_C(i \varepsilon, i \omega, \vec r, \vec p) = \frac{\theta\left(\varepsilon^2 - \frac{\omega^2}{4}\right)}{\sqrt{\frac{v_F^2 p^2}{4 \Gamma^2} + \left( 1 + \frac{|\varepsilon|}{\Gamma} + \frac{|\bar \Sigma(\varepsilon + \omega/2, \vec r) + \bar \Sigma(-\varepsilon + \omega/2, \vec r)|}{2 \Gamma}\right)^2}},
\end{equation}
where $\bar \Sigma(\varepsilon, \vec r)$ is a local electron self-energy defined in Eq. \eqref{eq:AppSigmaBar}.
At $\omega = 0$ and low momentum $p v_F \ll \Gamma$, one can expand $K_C$ near $K_C = 1$:
\begin{equation}\label{eq:AppCopperonAprox}
    1 - K_C \approx \frac{\frac{|\varepsilon|}{\Gamma} + \frac{\bar \Sigma(\varepsilon, \vec r)}{\Gamma}}{1 + \frac{|\varepsilon|}{\Gamma} + \frac{\bar \Sigma(\varepsilon, \vec r)}{\Gamma}} + \frac{p^2 v_F^2 / 8 \Gamma^2}{\left( 1 + \frac{|\varepsilon|}{\Gamma} + \frac{\bar \Sigma(\varepsilon, \vec r)}{\Gamma}  \right)^3}.
\end{equation}
For $|\varepsilon|,\bar\Sigma(\varepsilon, \vec r) \ll \Gamma$, Eq. \eqref{eq:AppCopperonAprox} further simplifies to 
\begin{equation}
    1 - K_C \approx \frac{|\varepsilon|}{\Gamma} + \frac{\bar \Sigma(\varepsilon, \vec r)}{\Gamma} + \frac{p^2 v_F^2}{8 \Gamma^2}.
\end{equation}
The inverse of $K_C$ under the Moyal transform can be written as
\begin{equation}
    K_C^{-1}(i \varepsilon, i \omega, \vec r, \vec p) = \theta \left(\varepsilon^2 - \frac{\omega^2}{4}\right) \sqrt{\frac{v_F^2 p^2}{4 \Gamma^2} + \left( 1 + \frac{|\varepsilon|}{\Gamma} + \frac{|\bar \Sigma(\varepsilon + \omega/2, \vec r) + \bar \Sigma(-\varepsilon + \omega/2, \vec r)|}{2 \Gamma}\right)^2} + \delta K_C^{(-1)},
\end{equation}
where the correction $\delta K_C^{(-1)}$ can be estimated in the leading order as
\begin{equation}
    \delta K_C^{(-1)} \approx \frac{v_F^2}{64 \Gamma^2} K_C^3 \p_{\vec r}^2 \bar \Sigma(\varepsilon, \vec r).
\end{equation}
Since these corrections are parametrically of order $(l/\xi_\alpha)^2$, they are negligible under the assumed hierarchy $l\ll \xi_\alpha$.
The approximate Cooperon equation in real space can now be recovered by applying an inverse Wigner transform to Eq. \eqref{eq:AppShortCooperon} with $K_C$ approximated by Eq. \eqref{eq:AppCopperonAprox}. 
For small pair momentum and arbitrary $|\varepsilon|+\bar\Sigma$ within the diffusive approximation, the ``bare" Cooperon equation takes a form
\begin{equation}\label{eq:NonEqCooperonDiffEq}
    \left[- \frac{v_F^2 \tau}{2} \frac{1}{1 + 2 \tau |\varepsilon| + 2 \tau \bar \Sigma(\varepsilon, \vec r)} \p_{\vec r}^2 + 2 |\varepsilon| + 2 \, \bar \Sigma(\varepsilon, \vec r) \right] \mathcal{C}_0(i \varepsilon, \vec r, \vec r^\prime) =  \delta(\vec r - \vec r^\prime).
\end{equation}
When $2\tau(|\varepsilon|+\bar\Sigma)\ll1$, the expression above simplifies to
\begin{equation} \label{eq:AppCooperonEq}
    \left( - D\p_{\vec r}^2 + 2 |\varepsilon| +2 \, \bar \Sigma(\varepsilon, \vec r) \right) \mathcal{C}_0(i \varepsilon, \vec r, \vec r^\prime) =  \delta(\vec r - \vec r^\prime),
\end{equation}
where $D = v_F^2 \tau / 2$ is the electronic diffusion constant. 
Eq. \eqref{eq:AppCooperonEq} is the ``bare" Cooperon equation used in the main text.

We proceed to evaluate the equation that governs the pairing vertex $\Delta(i\varepsilon_n, \vec r, \vec r^\prime)$, as defined by the diagrammatic series of Fig. \ref{fig:DiagSeries}. 
We consider the response of the system to a bare pairing vertex $\Delta_0(\vec r, \vec r^\prime) = \Delta_0\delta(\vec r - \vec r^\prime)$. 
The dressed vertex $\Delta(i \varepsilon, \vec r, \vec r^\prime)$ defined by the diagrammatic series in Fig. \ref{fig:DiagSeries} satisfies a Bethe-Salpeter equation of the form
\begin{multline}\label{eq:AppOrigVertex}
    \Delta(i\varepsilon_n, \vec r, \vec r^\prime) = \Delta_0  \delta(\vec r - \vec r^\prime)
    + g^2  T \sum_{\omega_m} \int d^2 \vec x \, \Delta(i \varepsilon_n + i \omega_m,  \vec x, \vec r^\prime) (2 \pi \nu \tau) K_C(i \varepsilon_n + i \omega_m, 0, \vec x, \vec r) 
    (8 \pi l^2_\mathrm{corr})\bar D( i \omega_m, \vec r)
    \\
    + g^2 T \sum_{\omega_m} \int d^2 \vec x \, d^2 \vec y \, \Delta(i \varepsilon_n + i \omega_m, \vec x, \vec r^\prime) (2 \pi \nu \tau) K_C(i \varepsilon_n + i \omega_m, 0, \vec x, \vec r) K_C(i \varepsilon_n + i \omega_m, 0,\vec y, \vec r) (8 \pi l^2_\mathrm{corr}) \bar D( i \omega_m, \vec r)
    \\
    + g^2 T \sum_{\omega_m} \int d^2 x \, d^2 x^\prime \, d^2 y \, \Delta(i \varepsilon_n + i \omega_m, \vec x) 
    (2 \pi \nu \tau)^2 K_C(i \varepsilon_n + i \omega_m, 0, \vec x, \vec x^\prime) \times
    \\
    \times (2 \pi \nu \tau^2)^{-1} \mathcal{C}_0(i \varepsilon_n + i\omega_m, \vec x^\prime, \vec y)  
    K_C(i \varepsilon_n + i \omega_m, 0, \vec y, \vec r) (8 \pi l^2_\mathrm{corr}) \bar D( i \omega_m, \vec r). 
\end{multline}
Summing the impurity ladder using Eq.~\eqref{eq:CooperonSalpeter} gives
\begin{equation}
    \Delta(i\varepsilon_n, \vec r, \vec r^\prime) = \Delta_0 \delta(\vec r - \vec r^\prime) + 16 \pi^2 l^2_\mathrm{corr} \nu g^2  T \sum_{\omega_m} \int d^2 x \, \Delta(i \varepsilon_n + i \omega_m, \vec x, \vec r^\prime) \mathcal{C}_0(i \varepsilon_n + i\omega_m, \vec x, \vec r) \bar D( i \omega_m, \vec r).
\end{equation}
With the substitution of $\bar{g}^2=8\pi^2 l^2_\mathrm{corr} \nu g^2$ into the equation above, the pairing vertex equation is
\begin{equation}
    \Delta(i\varepsilon_n, \vec r, \vec r^\prime) = \Delta_0 \delta(\vec r - \vec r^\prime) + 2 \bar g^2  T \sum_{\omega_m} \int d^2 x \, \Delta(i \varepsilon_n + i \omega_m, \vec x, \vec r^\prime) \mathcal{C}_0(i \varepsilon_n + i\omega_m, \vec x, \vec r) \bar D( i \omega_m, \vec r), 
\end{equation}
which is the pairing-vertex equation used in the main text.


\section{Large-$N$ Yukawa-SYK realization of the model} \label{sec:Large_N}

In this section, we extend our model in Eq.~\eqref{eq:Model} to a large-$N$ limit, which will formally replace some of the physical assumptions that we made before. The Lagrangian density reads as
\begin{equation}\label{eq:Model_large_N}
    \begin{aligned}
\mathcal{L}_{\psi}&=\bar{\psi}_{j\sigma}(\partial_\tau +\xi(i\nabla))\psi_{j\sigma} +\frac{1}{\sqrt{N}}v_{ij}(\mathbf{r})\bar{\psi}_{i\sigma}\psi_{j\sigma} \;\\
\mathcal{L}_{\phi}&= \frac{1}{2}\left[(\partial_\tau \phi_l)^2+c^2(\nabla\phi_l)^2 +(\lambda+\lambda'(\mathbf{r})) \phi_l^2 \right]+ \frac{u}{4 N}(\phi_l^2)^2\\
\mathcal{L}_{\psi\phi}&=g_{ijl}(\mathbf{r}) \phi_l \bar{\psi}_i  \psi_j
    \end{aligned}
\end{equation}
where $i,j,l=1,..,N$, and $g_{ijl}(\mathbf{r})$ is a random coupling with zero mean and variance $\overline{g_{ijl}(\mathbf{r})g_{i'j'l'}(\mathbf{r}')}=g^2\delta(\mathbf{r}-\mathbf{r}')\delta_{ii'}\delta_{jj'}\delta_{ll'}$. In the large-$N$ limit, the normal state saddle point equations read as
\begin{equation}\label{eq:Problem}
    \begin{aligned}
     G^{-1}(i\varepsilon_n, \vec{r},\vec{r}')&=\Big(i\varepsilon_n - \xi(i\nabla)+\frac{i}{2\tau}\operatorname{sgn}(\varepsilon_n) +i \operatorname{sgn}(\varepsilon_n) \bar{\Sigma}(i\varepsilon_n, \vec{r}) \Big)\delta(\vec{r}-\vec{r}')\\
     D^{-1}(i\omega_n, \vec{r},\vec{r}')&=\Big(\omega_n^2 -c^2\nabla^2 +\lambda+\lambda'(\vec{r}) -\bar{\Pi}(i\varepsilon_n, \vec{r})\Big)\delta(\vec{r}-\vec{r}') \\
       \bar{\Pi}(i\varepsilon_n, \vec{r})&=u T\sum\limits_{m}D(i\omega_m, \vec{r},\vec{r})- g^2 T\sum\limits_{m} G(i\omega_n+i\varepsilon_m, \vec{r},\vec{r})G(i\varepsilon_m, \vec{r},\vec{r})\;,\\
        \bar{\Sigma}(i\varepsilon_n, \vec{r}) &=g^2 T\sum\limits_{m} G(i\omega_m+i\varepsilon_n, \vec{r},\vec{r})D(i\omega_m, \vec{r},\vec{r})
    \end{aligned}
\end{equation}
Here $G$ ($D$) is the fermionic (bosonic) Green's function. The equations in \eqref{eq:Problem} are assumed to be exact in a given realization of the random mass  $\lambda(\vec{r})$, thus capturing certain disorder-induced effects (such as localization of bosonic modes) in a non-perturbative manner. Meanwhile, the averaging over regular potential disorder for fermions has already been carried out exactly due to the particular flavor structure of the random potential $v_{ij}(\mathbf{r})$. In particular, the disorder-induced vertex corrections are $1/N$ suppressed in this model. This simplification is important in order to keep the dissipative form of the Landau damping term in the bosonic self-energy $\bar{\Pi}$, and to rule out strong temperature-dependent Altshuler-Aronov corrections. It thus appears that the large-$N$ limit and the delta-correlated random coupling constant combined can mimic the physical separation of scales assumed in Eq.~\eqref{eq:scales}, and thus lead to Eq.~\eqref{eq:D_full_sec2}, Eq.~\eqref{eq:Sigma_full_eq}, and Eq.~\eqref{eq:Usadel}.


\section{Self-consistent Born approximation for the gap equation}
\label{sec:AppendixSCBA}

We analyze Eq.~\eqref{eq:full_gap} by expanding around the disorder-averaged problem and retaining disorder correlations to second order self-consistently. 
After separating $\bar \Sigma(i \varepsilon_n, \vec r)$ and $\bar D(i \omega_m, \vec r)$ into disorder-averaged and fluctuating parts, the gap equation can be written as
\begin{multline}\label{eq:AppAlmEq}
  \Big(-D\nabla^2+2|\varepsilon_n|+2\langle\bar{\Sigma}\rangle(i\varepsilon_n) +2 \delta \bar \Sigma(i \varepsilon_n, \vec r)\Big)\Phi(i\varepsilon_n , \vec{r}, \vec r^\prime)=
  \\
  = 
  2\bar{g}^2 T \sum\limits_{\varepsilon_m} \Big( \langle\bar D\rangle (i \varepsilon_m - i \varepsilon_n) + \delta \bar D(i\varepsilon_m - i \varepsilon_n, \vec r) \Big) \Phi(i\varepsilon_m,\vec{r}, \vec r^\prime) + \Delta_0 \delta(\vec r - \vec r^\prime).
\end{multline}
In the equation above, $\langle\cdots\rangle$ denotes disorder-averaged quantities and $\delta$ denotes a fluctuation, so $\langle \delta \bar \Sigma \rangle = \langle \delta \bar D\rangle = 0$. 
If the mixed averages $\langle \delta\bar\Sigma\,\Phi\rangle$ and $\langle \delta\bar D\,\Phi\rangle$ are neglected, the averaged equation reduces to Eq.~\eqref{eq:HomGapEq}.
To account for the spatial correlations between $\delta \bar \Sigma$, $\delta\bar D$, and $\Phi$, we expand Eq. \eqref{eq:AppAlmEq} in fluctuating quantities. 
One can formally express the solution of Eq. \eqref{eq:AppAlmEq} as
\begin{multline}\label{eq:AppPhi}
    \Phi(i \varepsilon_n, \vec r, \vec r^\prime) = \Delta_0 \bar{\mathcal{C}}_0(i \varepsilon_n, \vec r - \vec r^\prime) - 2 \int d^2 \vec x \; \bar{\mathcal{C}}_0(i \varepsilon_n, \vec r - \vec x) \delta\bar\Sigma(i \varepsilon_n, \vec x) \Phi(i \varepsilon_n, \vec x, \vec r^\prime)
    +
    \\
    + 2 \bar g^2 T \int d^2 \vec x \; \bar{\mathcal{C}}_0(i \varepsilon_n, \vec r - \vec x) \sum_{\varepsilon_m} \Big( \langle \bar D \rangle(i \varepsilon_n - i \varepsilon_m) + \delta \bar D(i \varepsilon_m - i \varepsilon_n, \vec x) \Big) \Phi(i \varepsilon_m, \vec x, \vec r^\prime),
\end{multline}
where $\bar{\mathcal{C}}_0(i \varepsilon_n, \vec r)$ is a disorder-averaged Cooperon propagator defined by
\begin{equation}\label{eq:AppCoopProp}
    \Big(-D\nabla^2+2|\varepsilon_n|+2\langle\bar{\Sigma}\rangle(i\varepsilon_n) \Big)\bar{\mathcal{C}}_0(i\varepsilon_n , \vec{r} - \vec r^\prime) = \delta(\vec r - \vec r^\prime).
\end{equation}
Substituting Eq. \eqref{eq:AppPhi} back into Eq. \eqref{eq:AppAlmEq}, averaging over the disorder fluctuations, and neglecting fluctuation cumulants above the second order, leads to
\begin{multline}\label{eq:AppLongPhi}
    \langle\Phi\rangle(i \varepsilon_n, \vec r, \vec r^\prime) = \Delta_0 \bar{\mathcal{C}}_0(i \varepsilon_n, \vec r - \vec r^\prime) + 2 \bar g^2 T \int d^2 \vec x \; \bar{\mathcal{C}}_0(i \varepsilon_n, \vec r - \vec x) \sum_{\omega_m \neq 0} \langle\Phi\rangle(i \varepsilon_n + i \omega_m, \vec x, \vec r^\prime)
     \langle \bar D \rangle(i \omega_m)
    \\
    + 4 \int d^2 \vec x \, d^2 \vec y \; \bar{\mathcal{C}}_0(i\varepsilon_n, \vec r - \vec x) \langle  \delta \bar \Sigma(i \varepsilon_n, \vec x) \delta \bar \Sigma(i \varepsilon_n, \vec y) \rangle \bar{\mathcal{C}}_0(i\varepsilon_n, \vec x - \vec y)  \langle\Phi\rangle(i \varepsilon_n, \vec y, \vec r^\prime)
    \\
     -4 \bar g^2 T \int d^2 \vec x \, d^2 \vec y  \; \sum_{\omega_m \neq 0} \bar{\mathcal{C}}_0(i\varepsilon_n, \vec r - \vec x) \langle \delta \bar D(i\omega_m, \vec x)  \delta \bar \Sigma(i \varepsilon_n + i \omega_m, \vec y) \rangle \bar{\mathcal{C}}_0(i\varepsilon_n + i\omega_m, \vec x - \vec y)\langle\Phi\rangle(i \varepsilon_n + i\omega_m, \vec y, \vec r^\prime)
     \\
    - 4 \bar g^2  T \int d^2 \vec x \, d^2 \vec y \; \sum_{\omega_m \neq 0} \bar{\mathcal{C}}_0(i \varepsilon_n, \vec r - \vec x) \langle \delta \bar \Sigma (i \varepsilon_n, \vec x) \delta \bar D(i\omega_m, \vec y) \rangle \bar{\mathcal{C}}_0(i \varepsilon_n, \vec x - \vec y) \langle\Phi\rangle(i \varepsilon_n + i \omega_m, \vec y, \vec r ^\prime)
     \\
     + (2 \bar g^2 T)^2 \int d^2 \vec x \, d^2 \vec y \; \sum_{\omega_l, \omega_m \neq 0} \bar{\mathcal{C}}_0(i \varepsilon_n, \vec r - \vec x) \langle \delta \bar D(i\omega_m, \vec x) \delta \bar D(i \omega_l, \vec y) \rangle \bar{\mathcal{C}}_0(i\varepsilon_n + i \omega_m, \vec x - \vec y) \langle\Phi\rangle(i \varepsilon_n + i \omega_m + i \omega_l, \vec y, \vec r^\prime).
\end{multline}
In the equation above $\langle\Phi\rangle(i \varepsilon_n, \vec r, \vec r^\prime)$ is a disorder-averaged Gor'kov function.
After disorder averaging, statistical translational invariance implies that the correlation functions $\langle \delta \bar \Sigma(i \varepsilon_n, \vec r) \delta \bar \Sigma(i \varepsilon_m, \vec r^\prime)\rangle$, $\langle \delta \bar \Sigma(i \varepsilon_n, \vec r) \delta \bar D(i \omega_m, \vec r^\prime)\rangle$, and $\langle \delta \bar D(i \omega_n, \vec r) \delta \bar D(i \omega_m, \vec r^\prime)\rangle$ are only functions of $\vec r - \vec r^\prime$.
Moreover, these correlation functions are significantly large only in the region $|\vec r - \vec r^\prime| \lesssim \xi$, where $\xi$ is a typical fluctuation length scale of $\delta\bar\Sigma$ and $\delta \bar D$.
Under this approximation of slow variation, we approximate the terms $\Phi(i \varepsilon_n, \vec y, \vec r^\prime)$ by $\Phi(i \varepsilon_n, \vec x, \vec r^\prime)$, as far as $\Phi(i \varepsilon_n, \vec y, \vec r')$ varies slowly at the length scales of the order of the bosonic inhomogeneity scale. 
The approximate form of Eq. \eqref{eq:AppLongPhi} reduces to 
\begin{multline}\label{eq:AppSCBA}
    \Big(- D \nabla^2 + 2 |\varepsilon_n| + 2\langle \bar \Sigma\rangle(i \varepsilon_n) - 2 \bar K_{\Sigma\Sigma}( \varepsilon_n) \Big) \langle\Phi\rangle(i\varepsilon_n, \vec r, \vec r^\prime)   - \Delta_0 \delta(\vec r - \vec r^\prime)=
    \\
    = 2 \bar g^2 T \sum_{\omega_m \neq 0} \left[ \langle  \bar D\rangle (i\omega_m)+  \bar K_{\Sigma D}(\varepsilon_n, \omega_m)  + \bar K_{DD}(\varepsilon_n, \varepsilon_n +\omega_m) \right] \langle\Phi\rangle(i\varepsilon_n + i \omega_m, \vec r , \vec r^\prime). 
\end{multline}
The $\omega_m=0$ contribution is omitted below because the static part of the pairing kernel cancels the corresponding static contribution from the self-energy.
Kernels $\bar K_{\Sigma \Sigma}$, $\bar K_{\Sigma D}$, and $\bar K_{D D}$ are corrections arising from the spatial fluctuations and are given by
\begin{align}\label{eq:KSigmaSigma}
    \bar K_{\Sigma \Sigma}(\varepsilon_n) &= 2 \int d^2 \vec r \, \langle \delta \bar \Sigma(i \varepsilon_n, \vec r) \delta \bar \Sigma(i \varepsilon_n, 0)\rangle  \bar{\mathcal{C}}_0(i \varepsilon_n, \vec r),
    \\\nonumber
    \bar K_{\Sigma D}(\varepsilon_n, \omega_m) &= -2 \int d^2 \vec r \, \langle \delta \bar D(i \omega_m, \vec r) \delta \bar \Sigma(i \varepsilon_n + i \omega_m, 0) \rangle  \bar{\mathcal{C}}_0(i \varepsilon_n + i \omega_m, \vec r) +
    \\\label{eq:KSigmaD}
    &\left. \right. \quad \quad\quad\quad\quad\quad\quad\quad\quad\quad\quad\quad - 2 \int d^2 \vec r \, \langle \delta \bar \Sigma(i \varepsilon_n, \vec r) \delta \bar D(i \omega_m, 0)\rangle  \bar{\mathcal{C}}_0(i \varepsilon_n, \vec r),
    \\\label{eq:KDD}
    \bar K_{D D}(\varepsilon_n, \varepsilon_n + \omega_m) &= 2 \bar g^2 T\sum_{\substack{l \neq 0\\l \neq m}}
    \int d^2 \vec r \, \langle \delta \bar D(i\omega_m - i \omega_l, \vec r) \delta \bar D(i\omega_l, 0) \rangle  \bar{\mathcal C}_0(i \varepsilon_n + i \omega_m - i \omega_l, \vec r).
\end{align}
Equation~\eqref{eq:AppSCBA} is the disorder-averaged gap equation that includes second-order corrections from spatial fluctuations. 
Neglecting $\bar K_{\Sigma\Sigma}$ and $\bar K_{\Sigma D}$, which are subleading in the regime considered in the main text, gives Eq.~\eqref{eq:GapEqCorrected}.


\section{Estimates of disorder corrections to the Usadel equation}
\label{sec:AppKenelEstimation}

To determine how spatial fluctuations affect the superconducting transition temperature $T_c$, we estimate the kernels $\bar K_{\Sigma \Sigma}$, $\bar K_{\Sigma D}$, and $\bar K_{DD}$. 
We assume that the spatial structure of the wavefunctions is $|\phi_\alpha(\vec r)| = \sqrt{\frac{2}{\pi}} \frac{1}{\xi_\alpha} e^{- |\vec r - \vec r_\alpha|/\xi_\alpha}$.
A useful quantity to describe the spatial fluctuations of $|\phi_\alpha(\vec r)|^2$ is
\begin{equation}\label{eq:P4function}
    P_{2,2}(\mathcal{E}, \vec r - \vec r^\prime)\equiv \left\langle \sum_\alpha |\phi_\alpha(\vec r)|^2 |\phi_\alpha(\vec r^\prime)|^2 \, \delta(\mathcal{E} - e_\alpha) \right\rangle 
    = n(\mathcal{E})\int d^2 \vec x \, \phi^2(\mathcal{E},\vec r - \vec x) \phi^2(\mathcal{E}, \vec r^\prime - \vec x) 
    = \frac{n(\mathcal{E}) r^2}{\pi \xi^4} K_2 \left( \frac{2r}{\xi} \right),
\end{equation}
where $r = |\vec r - \vec r^\prime|$, $\xi = \xi(\mathcal{E})$ and the function $\phi(\mathcal{E}, \vec r)$ is given by $\phi(\mathcal{E}, \vec r) = \sqrt{\frac{2}{\pi}} \frac{1}{\xi} e^{-|\vec r|/\xi}$. 
In writing $P_{2,2}$, we retain only the same-mode contributions: the disconnected part is removed by the definition of the fluctuations.
With the use of Eq.~\eqref{eq:P4function}, the correlation functions of $\delta \bar \Sigma$ and $\delta \bar D$ can be written as:
\begin{align}\label{eq:AppDeltaDDeltaDCalc}
    \langle \delta \bar D(i \omega_n, \vec r) \delta \bar D(i \omega_m, \vec r^\prime) \rangle &=  B(\omega_n, \omega_m, \vec r - \vec r^\prime),    
    \\
    \langle \delta \bar \Sigma(i \varepsilon_n, \vec r) \delta \bar D(i \omega_m, \vec r^\prime) \rangle &= 2 \bar g^2 T \sum_{a = 1}^{|n| - \Theta(-n - 1/2)} B(\omega_a, \omega_m, \vec r - \vec r^\prime),
    \\
    \langle \delta \bar \Sigma(i\varepsilon_n, \vec r) \delta \bar \Sigma(i \varepsilon_m, \vec r^\prime) \rangle &=
    4 \bar g^4 T^2 \sum_{a = 1}^{|n| - \Theta(-n - 1/2)} \sum_{b = 1}^{|m| - \Theta(-m - 1/2)} B(\omega_a, \omega_b, \vec r - \vec r^\prime),
\end{align}
where the kernel $B(\omega_n, \omega_m, \vec r)$ has a form
\begin{equation}\label{eq:AppB}
    B(\omega_n, \omega_m, \vec r) =
    \int d\mathcal{E} \,  P_{2,2}(\mathcal{E}, \vec r) \left( \frac{1}{\omega_n^2 + c_d |\omega_n| + \mathcal{E}} \right) \left( \frac{1}{\omega_m^2 + c_d |\omega_m| + \mathcal{E}}  \right).
\end{equation}
The kernel $B(\omega_n, \omega_m, \vec r)$ is most conveniently evaluated in momentum space. 
In the estimates below, we work in the overdamped regime, where
$\omega^2\ll c_d|\omega|$ for the relevant frequencies.
As a result, application of Parseval's identity to Eqs. \eqref{eq:KSigmaSigma}-\eqref{eq:KDD} allows us to express $\bar K_{\Sigma\Sigma}$, $\bar K_{\Sigma D}$, and $\bar K_{DD}$ through $B(\omega_n, \omega_m, \vec q) =\int d^2 \vec r \, e^{-i \vec k \vec r} B(\omega_n, \omega_m, \vec r)$:
\begin{align}
    \label{eq:KDDSimp}
    \bar K_{DD}(\varepsilon_n, \varepsilon_n + \omega_m) &= 2 \bar g^2 T \sum_{\substack{l \neq 0\\l \neq m}} \mathcal{K}(\varepsilon_n + \omega_m - \omega_l, \omega_m - \omega_l, \omega_l),
    \\
    \label{eq:KSigmaDSimp}
    \bar K_{\Sigma D}(\varepsilon_n, \omega_m) &= -4 \bar g^2 T \sum_{a = 1}^{|n+m| - \Theta(-n-m - 1/2)} \mathcal{K}(\varepsilon_n + \omega_m, \omega_m, \omega_a) 
    - 4 \bar g^2 T \sum_{a = 1}^{|n| - \Theta(-n - 1/2)} \mathcal{K}(\varepsilon_n, \omega_a, \omega_m),
    \\
    \label{eq:KSigmaSigmaSimp}
    \bar K_{\Sigma \Sigma}(\varepsilon_n) &= 8 \bar g^4 T^2 \sum_{a = 1}^{|n| - \Theta(-n - 1/2)} \sum_{b = 1}^{|n| - \Theta(-n - 1/2)} \mathcal{K}(\varepsilon_n, \omega_a, \omega_b),
\end{align}
where the kernel $\mathcal{K}$ is given by
\begin{equation}\label{eq:appK}
    \mathcal{K}(\varepsilon_n, \omega_m, \omega_l) = \int \frac{d^2 \vec q}{(2 \pi)^2} B(\omega_m, \omega_l, \vec q) \bar{\mathcal{C}}_0(i \varepsilon_n, \vec q).
\end{equation}
The correlation function $P_{2,2}(\mathcal{E}, \vec q)$ in momentum space has a form $P_{2,2}(\mathcal{E}, \vec q) = n(\mathcal{E}) (1 + \vec q^2 \xi^2/4)^{-3}$, where ${n(\mathcal{E}) = n \mathcal{E}^{\alpha_G}}$.
The expression for the bare Cooperon propagator in momentum space is $\bar{\mathcal{C}}_0(i \varepsilon_n, \vec q) = [D \vec q^2 + 2 |\varepsilon_n| + 2 \langle\bar \Sigma\rangle(i \varepsilon_n)]^{-1}$.
For the analytic estimates, we approximate $\xi(\mathcal E)$ by a typical value $\xi =\xi(c_d T)$ in the relevant low-energy window.
Substitution of Eq. \eqref{eq:AppB} into Eq. \eqref{eq:appK} results in
\begin{equation}\label{eq:AppKerK}
    \mathcal{K}(\varepsilon_n, \omega_m, \omega_l) \approx \frac{n}{4 \pi c_d D} \frac{\pi c_d^{\alpha_G}}{\sin \pi \alpha_G} \frac{|\omega_m|^{\alpha_G} - |\omega_l|^{\alpha_G}}{|\omega_m| - |\omega_l|} R(\varepsilon_n),
    \quad
    R(\varepsilon_n) \approx
    \begin{cases}
        \ln \left( \frac{E_\mathrm{Th}}{|\varepsilon_n| + \langle\bar \Sigma\rangle(\varepsilon_n)}\right), \quad E_\mathrm{Th} \gtrsim |\varepsilon_n| + \langle\bar \Sigma \rangle(\varepsilon_n),
        \\
        \frac{1}{2} \frac{E_\mathrm{Th}}{|\varepsilon_n| + \langle\bar \Sigma\rangle(\varepsilon_n)},
        \quad\quad\;\;\; 
        E_\mathrm{Th} \lesssim |\varepsilon_n| + \langle\bar \Sigma \rangle(\varepsilon_n),
    \end{cases}
\end{equation}
where $E_\mathrm{Th} = 2 D/\xi^2$ and $-1< \alpha_G \leq 0$.
Eq. \eqref{eq:AppKerK} along with Eqs. \eqref{eq:KDDSimp}-\eqref{eq:KSigmaSigmaSimp} fully describes the impact of inhomogeneity correlations on the linearized gap equations in Eq. \eqref{eq:AppSCBA}.

We now estimate the magnitudes of $\bar K_{\Sigma\Sigma}$, $\bar K_{\Sigma D}$ and $\bar K_{DD}$.
The Euler-Maclaurin approximation, applied to $\bar K_{\Sigma\Sigma}$ in Eq. \eqref{eq:KSigmaSigmaSimp}, results in
\begin{equation}
     \bar K_{\Sigma \Sigma}(\varepsilon) = 8 \bar g^4 \int_{2 \pi T}^{|\varepsilon|} \frac{d\omega}{2 \pi} \frac{d\omega^\prime}{2 \pi}  \mathcal{K}(\varepsilon, \omega, \omega^\prime) 
     \approx
    \frac{\bar g^4 n}{6 \pi c_d D} f_\Sigma(\alpha_G)
    \begin{cases}
         |\varepsilon|^{1+\alpha_G}\ln\left( \frac{E_\mathrm{Th}}{|\varepsilon|} \right), \quad |\varepsilon| \lesssim E_\mathrm{Th},
        \\
        |\varepsilon|^{\alpha_G}\frac{E_\mathrm{Th}}{2}, \quad\quad\quad\quad \;\; |\varepsilon| \gtrsim E_\mathrm{Th},
        \end{cases}
\end{equation}
where
\begin{equation}
    f_\Sigma(\alpha_G) = \frac{3}{\pi^2}\frac{\pi c_d^{\alpha_G}}{\sin \pi \alpha_G} \frac{2 + 2 \alpha_G (\psi(\alpha_G)+ \gamma)}{\alpha_G(1+\alpha_G)}, \quad \quad
    f_\Sigma(0) = 1. 
\end{equation}

For $\alpha_G \neq 0$, $\bar K_{\Sigma\Sigma}$ is a parametrically small logarithmic correction to a power-law scaling of $\langle\bar \Sigma\rangle(i \varepsilon)$ that is suppressed by $\bar g^2/c_d D$. 
For $\alpha_G = 0$, $\bar K_{\Sigma\Sigma}$ has the same frequency dependence as $\langle\bar \Sigma\rangle(i \varepsilon)$ and therefore only renormalizes effective coupling.
In both cases, the effects of $\bar K_{\Sigma\Sigma}$ can be neglected.

The mixed correction $\bar K_{\Sigma D}$ is 
\begin{equation}\label{eq:AppKSigmaDLowT}
    \bar K_{\Sigma D}(\varepsilon, \omega) = -4 \bar g^2 \int_{2 \pi T}^{|\varepsilon + \omega|} \frac{d\omega^\prime}{2 \pi}  \mathcal{K}(\varepsilon + \omega, \omega, \omega^\prime) 
    - 4 \bar g^2 \int_{2 \pi T}^{|\varepsilon|} \frac{d\omega^\prime}{2\pi} \mathcal{K}(\varepsilon, \omega^\prime, \omega),
\end{equation}
after the Euler-Maclaurin approximation is applied to Eq. \eqref{eq:KSigmaDSimp}.
The leading order behavior of $K_{\Sigma D}$ can be summarized by
\begin{equation}
    \bar K_{\Sigma D}(\varepsilon, \varepsilon^\prime - \varepsilon) \sim 
    - \frac{\bar g^4 n}{12 c_d D}
    \begin{cases}
         \eta^{\alpha_G} \ln\left( \frac{E_\mathrm{Th}}{\eta} \right), \quad\quad\quad E_\mathrm{Th} \gtrsim \max(|\varepsilon|, |\varepsilon'|),
        \\
         \eta^{\alpha_G - 1}E_\mathrm{Th}, \quad\quad\quad\quad\;\; E_\mathrm{Th} \lesssim \max(|\varepsilon|, |\varepsilon'|),
    \end{cases}
\end{equation}
where $\eta = \max(|\varepsilon|, |\varepsilon'|)$.
The effects of $\bar K_{\Sigma D}$ are small logarithmic corrections to $\langle \bar D \rangle(i \varepsilon-i \varepsilon')$ in the case of $\alpha_G \neq 0$ and can be absorbed into a coupling constant for $\alpha_G = 0$.
We therefore neglect $\bar K_{\Sigma D}$ in the main analysis.

Evaluation of $\bar K_{DD}(\varepsilon, \varepsilon')$ requires slightly more care.
The internal Matsubara frequency sum in Eq. \eqref{eq:KDDSimp} has a non-trivial convolution structure.
To estimate it, we approximate the integrand as
\begin{equation}\label{eq:AppKOptions}
    \mathcal{K}(\omega, \omega - \varepsilon, \varepsilon^\prime - \omega) \simeq 
    \frac{n}{4 \pi c_d D} \frac{\pi c_d^{\alpha_G}}{\sin \pi \alpha_G}
    \begin{cases}
         \frac{ |\varepsilon|^{\alpha_G} - |\varepsilon'|^{\alpha_G}}{|\varepsilon| - |\varepsilon^\prime|} \ln \left( \frac{E_\mathrm{Th}}{|\omega|} \right), \quad \quad E_\mathrm{Th} \gtrsim |\omega|, \quad  \eta  \gtrsim |\omega|.
        \\
        \frac{|\varepsilon|^{\alpha_G} - |\varepsilon'|^{\alpha_G}}{|\varepsilon| - |\varepsilon^\prime|} \frac{E_\mathrm{Th}}{2|\omega|}, \quad\quad\quad\quad\;\;  E_\mathrm{Th} \lesssim |\omega|; \quad \eta \gtrsim |\omega|.
        \\
        \frac{\alpha_G}{|\omega|^{1 - \alpha_G}} \ln \left( \frac{E_\mathrm{Th}}{|\omega|} \right), \quad \quad \quad \;\;\; E_\mathrm{Th} \gtrsim |\omega|; \quad \eta \lesssim |\omega|.
        \\
        \frac{\alpha_G}{|\omega|^{1-\alpha_G}} \frac{E_\mathrm{Th}}{2|\omega|}, \quad \quad \quad\quad\quad \quad\; E_\mathrm{Th} \lesssim |\omega|; \quad \eta \lesssim |\omega|.
    \end{cases}
\end{equation}
where $\eta = \max(|\varepsilon|, |\varepsilon'|)$.
The integral kernel $\bar K_{DD}$ in the regime $E_\mathrm{Th} \lesssim \eta$ involves integrating options (1), (2), and (4) of Eq. \eqref{eq:AppKOptions}, which results in 
\begin{equation}\label{eq:AppKDD1}
    \bar K_{DD}(\varepsilon, \varepsilon^\prime) = 2 \bar g^2 \int \frac{d \omega}{2 \pi} \mathcal{K}(\omega, \omega - \varepsilon, \varepsilon^\prime - \omega) \simeq \frac{\bar g^2 n}{4 \pi^2 c_d D} \frac{\pi c_d^{\alpha_G}}{\sin \pi \alpha_G} \left[ 
    E_\mathrm{Th}   \frac{|\varepsilon|^{\alpha_G} - |\varepsilon'|^{\alpha_G}}{|\varepsilon| - |\varepsilon^\prime|} + \frac{\alpha_G}{1-\alpha_G} \frac{E_\mathrm{Th}}{2 \eta^{1- \alpha_G}}\right].
\end{equation}
Above $\omega$ denotes the internal Cooperon frequency after the change of variables in the convolution.
In the regime, where $E_\mathrm{Th} \gtrsim \eta$, the kernel $\bar K_{DD}$ is evaluated by integrating options (1), (3), and (4) from Eq. \eqref{eq:AppKOptions}, which results in
\begin{multline}\label{eq:AppKDD2}
    \bar K_{DD}(\varepsilon, \varepsilon^\prime) = 2 \bar g^2 \int \frac{d \omega}{2 \pi} \mathcal{K}(\omega, \omega - \varepsilon, \varepsilon^\prime - \omega) \simeq 
    \frac{\bar g^2 n}{4 \pi^2 c_d D} \frac{\pi c_d^{\alpha_G}}{\sin \pi \alpha_G} \left[
     2\eta \, \ln \left(  \frac{E_\mathrm{Th}}{\eta} \right)  \frac{|\varepsilon|^{\alpha_G} - |\varepsilon'|^{\alpha_G}}{|\varepsilon| - |\varepsilon^\prime|} \right. + 
     \\
    +2 \left.\alpha_G^{-1} \left(E_\mathrm{Th}^{\alpha_G} - \eta^{\alpha_G} \left(1 + \alpha_G \ln \left(\frac{E_\mathrm{Th}}{\eta} \right) \right) \right) + \frac{\alpha_G \, E^{\alpha_G}_\mathrm{Th}}{1 - \alpha_G} \right].
\end{multline}
For $\alpha_G \neq 0$, the leading order behavior of $\bar K_{DD}$ is
\begin{equation}
    \bar K_{DD}(\varepsilon, \varepsilon^\prime) \simeq \frac{\bar g^2 n}{4 \pi^2 c_d D} \frac{\pi \alpha c_d^{\alpha_G} }{\sin \pi \alpha_G}
    \begin{cases}
         \eta^{\alpha_G} \ln \left( \frac{E_\mathrm{Th}}{\eta} \right), \quad E_\mathrm{Th} \gtrsim \eta,
    \\
    \eta^{\alpha_G - 1} E_\mathrm{Th}, \quad\quad \;\; E_\mathrm{Th} \lesssim \eta.
    \end{cases}
\end{equation}
Value $\alpha_G = 0$ is a special point of Eqs. \eqref{eq:AppKDD1} and \eqref{eq:AppKDD2}, and up to nonlogarithmic terms and order-one constants produces $\bar K_{DD}$ of the form
\begin{equation}
    \bar K_{DD}(\varepsilon, \varepsilon^\prime) \simeq \frac{\bar g^2 n}{4 \pi^2 c_d D} 
    \begin{cases}
         3 \ln^2 \left( \frac{E_\mathrm{Th}}{\eta} \right), \quad E_\mathrm{Th} \gtrsim \eta
    \\
    \frac{E_\mathrm{Th}}{\eta}, \quad E_\mathrm{Th} \lesssim \eta.
    \end{cases}
\end{equation}
For energies $\eta \gtrsim E_\mathrm{Th}$, the scaling of $\bar K_{DD}(\varepsilon, \varepsilon')$ is subleading to $\langle\bar D \rangle(i \varepsilon - i \varepsilon')$,
so the effects of $\bar K_{DD}$ can be neglected at energies larger than $E_\mathrm{Th}$.
When $\eta \lesssim E_\mathrm{Th}$, the kernel is square-logarithmic and therefore dominates over $\langle \bar D\rangle(\varepsilon - \varepsilon')$ at lowest energies.
Therefore, this effect cannot be absorbed into $\langle \bar D \rangle$ by redefinition of the theory parameters.
This leads to a final form of the inhomogeneity-induced correction:
\begin{equation}\label{eq:AppBarKDDFin}
    \bar K_{DD}(\varepsilon, \varepsilon^\prime) \simeq \frac{\vartheta n}{2} \, \theta\left(E_\mathrm{Th} - \eta\right) \ln^2 \left( \frac{E_\mathrm{Th}}{\eta} \right),
\end{equation}
where $\theta(x)$ is the Heaviside step function, $E_\mathrm{Th} = 2 D/\xi^2$ is the Thouless energy, $\vartheta = 3 \bar g^2 /2 \pi^2 c_d D$, and $\eta = \max(|\varepsilon|, |\varepsilon^\prime|)$.


\section{Solution to Usadel equation without spatial inhomogeneity} \label{sec:NoInhomogeneitySolution}

For $\varepsilon>0$, Eq. \eqref{eq:HomIntegralEq} in the main text becomes 
\begin{equation}\label{eq:AppIntEqApprox}
         \Big( \varepsilon + \langle \bar \Sigma\rangle(i \varepsilon) \Big) \Phi_0(i \varepsilon) = \Delta_0 + \frac{\bar g^2}{\pi} \int_{\pi T}^{\Lambda} d \varepsilon^\prime \, \Phi_0(i \varepsilon^\prime)
     \langle\bar D \rangle (i \max(\varepsilon, \varepsilon^\prime)).
\end{equation}
This integral equation can be reduced to the ordinary differential equation
\begin{equation}\label{eq:AppDiffEq0}
    \p_\varepsilon \left( \frac{\p_\varepsilon \left((\varepsilon + \langle \bar \Sigma \rangle (\varepsilon)) \Phi_0(i\varepsilon)\right)}{\p_\varepsilon \langle\bar D \rangle(i \varepsilon)} \right) = \frac{\bar g^2}{\pi} \Phi_0(i\varepsilon).
\end{equation}
Substituting 
 $F(\varepsilon) = \left( \varepsilon + \langle \bar \Sigma \rangle(i \varepsilon) \right) \Phi_0(i\varepsilon)$ along with $\partial_\varepsilon\langle\bar D\rangle(i\varepsilon)=-n/\varepsilon$ into Eq. \eqref{eq:AppDiffEq0} leads to
 \begin{equation}
     \p_\varepsilon (\varepsilon \p_\varepsilon F(\varepsilon)) = -\tilde g^2  \frac{F(\varepsilon)}{\varepsilon + \langle \bar \Sigma \rangle(i \varepsilon)},
 \end{equation}
where
$\tilde g^2 = \bar g^2 n / \pi$ is an effective dimensionless coupling constant. 
It is convenient to define $x = \ln(\Lambda'/\varepsilon)$, $\ln(\Lambda'/\Lambda) = \tilde g^{-2}$
leads to
\begin{equation}
    \p^2_x F = - \frac{F}{x}
\end{equation}
with an exact solution given by
\begin{equation}\label{eq:AppDiffEqSol}
    F(x) = c_1 \sqrt{x} \, J_1\left( 2\sqrt{x}\right) + c_2 \sqrt{x} \, Y_1\left( 2 \sqrt x\right).
\end{equation}
We substitute Eq. \eqref{eq:AppDiffEqSol} into  Eq. \eqref{eq:AppIntEqApprox} to determine constants $c_1$ and $c_2$.
In terms of $F(\varepsilon)$ the integro-differential equation can be written as
\begin{equation}\label{eq:F_int_eq}
    F(\varepsilon) = \Delta_0 + \frac{\bar g^2}{\pi}  \left( \int_{\pi T}^{\varepsilon} d \varepsilon^\prime \frac{ F(\varepsilon^\prime) \bar D(i \varepsilon
    )}{\varepsilon^\prime + \langle \bar \Sigma \rangle(i \varepsilon^\prime)}  + \int_\varepsilon^{\Lambda} d\varepsilon^\prime \frac{ F(\varepsilon^\prime) \bar D(i \varepsilon^\prime
    )}{\varepsilon^\prime + \langle \bar \Sigma \rangle(i \varepsilon^\prime)}\right).
\end{equation}
The integrals on the right-hand side of Eq. \eqref{eq:F_int_eq} evaluate to
\begin{equation}\label{eq:AppFInt}
   \frac{\bar g^2}{\pi}  \left( \int_{\pi T}^{\varepsilon} d \varepsilon^\prime \frac{ F(\varepsilon^\prime) \bar D(i \varepsilon
    )}{\varepsilon^\prime + \langle \bar \Sigma \rangle(\varepsilon^\prime)}  + \int_\varepsilon^{\Lambda} d\varepsilon^\prime \frac{ F(\varepsilon^\prime) \bar D(i \varepsilon^\prime
    )}{\varepsilon^\prime + \langle \bar \Sigma \rangle(\varepsilon^\prime)}\right) = F(\varepsilon) - F\left( \Lambda \right) -  
     (x_\varepsilon -\tilde g^{-2})  \left[ c_1 J_0 \left( 2\sqrt{x_T} \right) + c_2 Y_0 \left( 2\sqrt{ x_T} \right) \right],
\end{equation}
where $x_\varepsilon = \log(\Lambda'/\varepsilon)$ and $x_T = \log(\Lambda'/\pi T)$.
Substituting Eq.~\eqref{eq:AppFInt} into Eq.~\eqref{eq:F_int_eq} imposes the two boundary conditions
\begin{align}\label{eq:c1c2eq}
    \Delta_0 = F\left( \Lambda\right) , \quad\quad
    0  = c_1 J_0 \left( 2\sqrt{ \ln \left( \frac{\Lambda^\prime}{\pi T} \right)} \right) + c_2 Y_0 \left( 2 \sqrt{\ln \left( \frac{\Lambda^\prime}{\pi T} \right)} \right),
\end{align}
which fix $c_1$ and $c_2$ for finite $\Delta_0$.
The system of linear equations in Eq. \eqref{eq:c1c2eq} can be solved exactly and leads to
\begin{equation}
    F(\varepsilon) = \Delta_0 \frac{y(\varepsilon)}{y\left( \Lambda \right)} \frac{Y_0(y(\pi T)) J_1(y(\varepsilon)) - J_0(y(\pi T)) Y_1(y(\varepsilon))}{Y_0(y(\pi T)) J_1\left( y \left( \Lambda \right) \right) - J_0(y(\pi T)) Y_1\left( y \left( \Lambda \right) \right)}, 
    \quad \quad 
    y(\varepsilon) = 2 \sqrt{\ln \left( \frac{\Lambda^\prime}{\varepsilon} \right)}.
\end{equation}
To evaluate the static superconducting susceptibility $\Phi_\mathrm{stat}(\omega = 0)$, we compute
\begin{equation}\label{eq:AppExactGapFunction}
    \Phi_\mathrm{0,stat}(\omega = 0) = \int \frac{d \varepsilon}{2 \pi} \Phi_0(i\varepsilon) = \frac{\Delta_0}{2 \pi \tilde g} \frac{Y_0(y(\pi T)) J_0\left( y \left( \Lambda \right) \right) - J_0(y(\pi T)) Y_0\left( y \left( \Lambda \right) \right)}{Y_0(y(\pi T)) J_1\left( y \left( \Lambda \right) \right) - J_0(y(\pi T)) Y_1\left( y \left(\Lambda \right)\right)},
\end{equation}
which is the static pair density response to $\Delta_0$ source.
For $\tilde g\ll1$, using the large-argument asymptotics of the Bessel functions gives
\begin{equation}\label{eq:SupSuscSmallg}
    \Phi_\mathrm{0,stat}(\omega = 0) \approx \frac{\Delta_0}{2 \pi \tilde g} \tan \left( \tilde g\, \ln \left( \frac{\Lambda}{\pi T} \right)\right),
    \quad \quad \tilde g^2 \ll 1.
\end{equation}
Eq.~\eqref{eq:SupSuscSmallg} shows that the static susceptibility diverges at $T=T_{c0}$,, where  
\begin{equation}\label{eq:AppTcNoInhomogeneity}
    T_{c0} = \frac{\Lambda}{\pi} \exp \left[ - \frac{\pi}{2 \tilde g}\right] \gg T_\mathrm{NFL} \sim \frac{\Lambda}{\pi} \exp \left[ - \frac{1}{\tilde g^2} \right].
\end{equation}
For later use, it is useful to compare $T_c$ to $T_\mathrm{NFL}$ --
the scale at which $\langle\bar\Sigma\rangle(i\varepsilon)$ becomes comparable to $|\varepsilon|$.
Comparing $T_{c0}$ and $T_\mathrm{NFL}$ shows that $T_\mathrm{NFL} \ll T_{c0}$, which implies the electronic self-energy is parametrically small at the onset of superconductivity.
Finally, it is interesting to note that neglecting electron self-energy effects in Eq. \eqref{eq:AppIntEqApprox} leads to the same effect as the small $\tilde g$ asymptotic expansion of the exact gap function in Eq. \eqref{eq:AppExactGapFunction}.


\section{Solution of the Usadel equation with inhomogeneity corrections}\label{sec:AppendixGapEqCorrections}

In this section, we determine how the inhomogeneity corrections generated by $\delta\bar\Sigma$ and $\delta\bar D$ modify $T_c$.
The relevant gap equation has been derived in the main text as Eq. \eqref{eq:GapEqInhom}:
\begin{equation}\label{eq:SCBAGapApprox}
     |\varepsilon| \langle \Phi \rangle (i\varepsilon) =  2 \bar g^2 \int_{\pi T}^{\Lambda} \frac{d\varepsilon^\prime}{2\pi} \left[  \langle \bar D \rangle(i \max(\varepsilon, \varepsilon^\prime)) + \bar K_{DD}(\varepsilon, \varepsilon^\prime) \right] \langle \Phi\rangle(i\varepsilon^\prime) + \Delta_0.
\end{equation}
As argued in Appendix~\ref{sec:AppKenelEstimation}, the $\bar K_{\Sigma\Sigma}$ and $\bar K_{\Sigma D}$ terms only renormalize parameters in the regimes of interest and therefore are dropped. 
Eq. \eqref{eq:SCBAGapApprox} for $\varepsilon > 0$ can be written out explicitly as
\begin{equation}    
\label{eq:DisoerderCorrectedGapEq}
     \varepsilon \langle\Phi\rangle(i\varepsilon) = \tilde g^2 \int_{\pi T}^{\Lambda} d \varepsilon^\prime \ln \left( \frac{\Lambda}{\max(\varepsilon, \varepsilon^\prime)} \right) \langle\Phi\rangle(i \varepsilon^\prime) 
     + \frac{\vartheta \tilde g^2}{2}  \; \Theta\left(E_\mathrm{Th} - \varepsilon\right) \int_{\pi T}^{E_\mathrm{Th}} d\varepsilon^\prime \ln^2 \left( \frac{E_\mathrm{Th}}{\max(\varepsilon, \varepsilon^\prime)} \right) \langle\Phi\rangle(i\varepsilon^\prime)  + \Delta_0.
\end{equation}
The integral equation above separates into two frequency regimes: $\varepsilon \geq E_\mathrm{Th}$, and $\varepsilon \leq E_\mathrm{Th}$. 
We write 
\begin{equation}
    \langle\Phi\rangle(i\varepsilon) = 
    \begin{cases}
        \Phi_-(i\varepsilon), \quad \varepsilon \leq E_\mathrm{Th},
        \\
        \Phi_+(i\varepsilon), \quad \varepsilon \geq E_\mathrm{Th},
    \end{cases}
\end{equation}
so Eq. \eqref{eq:DisoerderCorrectedGapEq} is equivalent to a system of equations
\begin{align}
    - \p_\varepsilon \left( \varepsilon \p_\varepsilon(\varepsilon \Phi_+(i\varepsilon)) \right) &= \tilde g^2 \Phi_+(i\varepsilon), \quad \quad \varepsilon > E_\mathrm{Th},
    \\
    - \p_\varepsilon\left( \frac{\varepsilon \p_\varepsilon(\varepsilon \Phi_-(i\varepsilon))}{1 + \vartheta \ln \left( \frac{E_\mathrm{Th}}{\varepsilon} \right)} \right) &= \tilde g^2 \Phi_-(i\varepsilon), \quad \quad \varepsilon < E_\mathrm{Th}.
\end{align}
The differential equations above are solved by
\begin{align}\label{eq:SCBACorGapPlus}
    \Phi_+(i\varepsilon) &= \frac{1}{\varepsilon} \left[ A_+ \cos\left( u(\varepsilon) \right) + B_+ \sin\left( u(\varepsilon) \right)\right],
    \quad\quad u(\varepsilon) = \tilde g \ln\left( \frac{\Lambda}{\varepsilon} \right),
    \\\label{eq:SCBACorGapMinus}
    \Phi_-(i\varepsilon) &= \frac{1}{\varepsilon} \left[ A_- \, \mathrm{Ai}^\prime\left( h(\varepsilon)\right) + B_- \, \mathrm{Bi}^\prime\left(h(\varepsilon)\right)\right],
    \quad\quad
    h(\varepsilon) = - \left( \frac{\tilde g}{\vartheta
    } \right)^{\frac{2}{3}}  \left( 1 + \vartheta\ln \left( \frac{E_\mathrm{Th}}{\varepsilon} \right) \right),
\end{align}
where $\mathrm{Ai}\,(x)$ and $\mathrm{Bi}\,(x)$ are the Airy functions.

Coefficients $A_\pm$ and $B_\pm$ are fully determined from the continuity properties of $\langle \Phi\rangle(i\varepsilon)$.
At $\varepsilon = E_\mathrm{Th}$, Eq. \eqref{eq:DisoerderCorrectedGapEq} dictates
$\Phi_+(iE_\mathrm{Th}) = \Phi_-(iE_\mathrm{Th})$, and $\p_\varepsilon(\varepsilon \Phi_+(i\varepsilon))|_{\varepsilon =E_\mathrm{Th}} = \p_\varepsilon(\varepsilon \Phi_-(i\varepsilon))|_{\varepsilon = E_\mathrm{Th}}$. 
For the eigenvalue problem with $\Delta_0=0$, Eq.~\eqref{eq:DisoerderCorrectedGapEq} gives $\Phi_+(i\Lambda)=0$, implying $A_+ = 0$.
The continuity conditions immediately imply 
\begin{align}\label{eq:SCBACorGapAMinus}
    A_-&= - \pi B_+ \sin\left( u(E_\mathrm{Th}) \right) \mathrm{Bi}(h(E_\mathrm{Th})) + \pi B_+ \left( \frac{\vartheta}{\tilde g} \right)^{\frac{1}{3}} \cos\left( u(E_\mathrm{Th}) \right) \mathrm{Bi}^\prime(h(E_\mathrm{Th})),
    \\\label{eq:SCBACorGapBMinus}
    B_- &= \pi B_+ \sin\left( u(E_\mathrm{Th}) \right) \mathrm{Ai}(h(E_\mathrm{Th}))- \pi B_+ \left( \frac{\vartheta}{\tilde g} \right)^{\frac{1}{3}} \cos\left( u(E_\mathrm{Th}) \right) \mathrm{Ai}^\prime(h(E_\mathrm{Th})).
\end{align}
Substitution of Eqs. \eqref{eq:SCBACorGapPlus} -- 
\eqref{eq:SCBACorGapBMinus} back into Eq. \eqref{eq:SCBAGapApprox} results in the following self-consistency condition for $T_c$:
\begin{equation}\label{eq:AppTcEqInhomogeneous}
     A_- \, \mathrm{Ai}\,(h(\pi T_c)) + B_- \, \mathrm{Bi}\, (h(\pi T_c))  = 0.
\end{equation}

We first analyze Eq. \eqref{eq:AppTcEqInhomogeneous} in the simplest scenario. 
We examine the solution in the regime of a fixed $\tilde g$ and $\gamma = \vartheta/\tilde g \lesssim 1$. 
In this regime, one can perform the asymptotic expansion of the Airy functions, leading to
\begin{align}\label{eq:SCBAExpansionsAm}
    A_- &= \sqrt{\pi} B_+ \left( \frac{\tilde g}{\vartheta} \right)^{-\frac{1}{6}} \sin \left( \frac{2}{3\gamma}  - \tilde g \ln \left( \frac{\Lambda}{E_\mathrm{Th}} \right) + \frac{\pi}{4} \right),
    \\\label{eq:SCBAExpansionsAp}
    B_- &=  \sqrt{\pi} B_+ \left( \frac{\tilde g}{\vartheta} \right)^{-\frac{1}{6}} \cos \left( \frac{2}{3\gamma}  - \tilde g \ln \left( \frac{\Lambda}{E_\mathrm{Th}} \right) + \frac{\pi}{4} \right).
\end{align}
Substitution of Eqs. \eqref{eq:SCBAExpansionsAm} and \eqref{eq:SCBAExpansionsAp} into Eq. \eqref{eq:AppTcEqInhomogeneous} and the asymptotic expansion of the Airy functions leads to
\begin{equation}
    \frac{2}{3\gamma} \left[ \left(1 + \gamma \tilde g \ln \left( \frac{E_\mathrm{Th}}{\pi T_c}\right) \right)^\frac{3}{2} - 1 \right] + \tilde g \ln \frac{\Lambda}{
    E_\mathrm{Th}} = \frac{\pi}{2}.
\end{equation}
This equation can be solved with the following notation:
$\mathcal A = \tilde g \ln (\Lambda/\pi T_c)$, and $\tilde g_\mathrm{Th} = \tilde g \ln (\Lambda/E_\mathrm{Th})$, which allows to express $T_c$ as
\begin{equation}\label{eq:AppTcMod}
    T_c = \frac{\Lambda}{\pi} \exp\left[ - \frac{\mathcal A}{\tilde g}\right], \quad \mathcal A = \tilde g_\mathrm{Th} + \frac{1}{\gamma} \left[ \left( 1 + \frac{3 \gamma}{2}\left( \frac{\pi}{2} - \tilde g_\mathrm{Th}\right) \right)^{\frac{2}{3}} -1 \right].
\end{equation}

We next consider $\gamma\gtrsim1$, which itself contains two sub-cases: $|h(\pi T_c)| \lesssim 1$ and $|h(\pi T_c)| \gtrsim 1$.
We begin by considering the first sub-case, $|h(\pi T_c)| \lesssim 1$, where we use the Airy-function expansion around $h=0$:
\begin{equation}\label{eq:AppAiryExp}
    \mathrm{Ai}\, (h) \approx \frac{1}{3^\frac{2}{3} \Gamma \left( \frac{2}{3} \right)} - \frac{h}{3^{\frac{1}{3}} \Gamma \left( \frac{1}{3}
    \right)},
    \quad\quad
    \mathrm{Bi}(h) \approx \frac{1}{3^{\frac{1}{6}} \Gamma\left( \frac{2
    }{3} \right)} + \frac{3^{\frac{1}{6}} h}{\Gamma \left( \frac{1}{3}\right)},
\end{equation}
which is a good approximation as far as $|h| \lesssim 1$.
Using the approximation given in Eq. \eqref{eq:AppAiryExp} leads to
\begin{equation}
    A_- = 1 - \frac{\Gamma\left( \frac{1}{3} \right)  \tan (\tilde g_\mathrm{Th})}{3^{\frac{1}{3}} \Gamma \left( \frac{2}{3} \right) \gamma^{\frac{1}{3}}},
    \quad\quad
    B_- = \frac{1}{3^{\frac{1}{2}}} + \frac{\Gamma\left( \frac{1}{3} \right) \tan \left( \tilde g_\mathrm{Th} \right)}{3^\frac{5}{6} \Gamma \left( \frac{2}{3} \right) \gamma^{\frac{1}{3}}}
\end{equation}
after dividing both coefficients by a common nonzero factor.
From Eq. \eqref{eq:AppTcEqInhomogeneous} and the approximation in Eq. \eqref{eq:AppAiryExp}, the value of $h(\pi T_c)$ can be deduced, resulting in 
\begin{equation}\label{eq:ApphTcNearZero}
    h(\pi T_c) \approx - \frac{\gamma^{\frac{1}{3}}}{\tan(\tilde g_\mathrm{Th})}, \quad\quad \gamma^{\frac{1}{3}} \lesssim \tan(\tilde g_\mathrm{Th}).
\end{equation}
The inequality in Eq. \eqref{eq:ApphTcNearZero} follows from $|h(\pi T_c)| \lesssim 1$.
Using the definition of $h(\varepsilon)$ then gives
\begin{equation}\label{eq:AppTcTransient}
    T_c  = \frac{E_\mathrm{Th}}{\pi} \exp\left[-\frac{\cot(\tilde g_\mathrm{Th})}{\tilde g}  + \frac{1}{\vartheta} \right], \quad \quad 1 \lesssim \left(\frac{\vartheta}{\tilde g}\right)^\frac{1}{3} \lesssim \tan(\tilde g_\mathrm{Th}).
\end{equation}

Proceeding to the second sub-case, $|h(\pi T_c)| \gtrsim 1$, we use the asymptotic expansion of the Airy functions, simplifying Eq. \eqref{eq:AppTcEqInhomogeneous} to 
\begin{equation}
    \sqrt{3} (1- x) \sin \left( \frac{\pi}{4} + \frac{2}{3} |h(\pi T_c)|^\frac{3}{2} \right) + (1+ x) \cos \left( \frac{\pi}{4} + \frac{2}{3} |h(\pi T_c)|^\frac{3}{2} \right) = 0, \quad x= \frac{\Gamma\left( \frac{1}{3} \right) \tan(\tilde g_\mathrm{Th})}{3^\frac{1}{3} \Gamma\left( \frac{2}{3} \right) \gamma^\frac{1}{3}} \approx 1.37 \; \frac{\tan(\tilde g_\mathrm{Th})}{\gamma^{\frac{1}{3}}}.
\end{equation}
The solution with minimal $|h(\pi T_c)|$ can be expressed as
\begin{equation}
    h(\pi T_c) = - \left[ \frac{3\pi}{8} - \frac{3 \alpha(x)}{2}  \right]^\frac{2}{3},
    \quad
    \quad
    \alpha(x) = \tan^{-1}\left(\sqrt{3} \frac{x-1}{1+x} \right), \quad \quad x \lesssim 1 .
\end{equation}
In the range $0 < x < 1$ the value of $h(\pi T_c)$ changes insignificantly: $h(\pi T_c) \approx - 1.12$ for $x = 1$ and $h(\pi T_c) \approx - 1.96$ for $x = 0$.
Therefore, the superconducting transition temperature can be expressed as 
\begin{equation}\label{eq:AppTc23}
    T_c = \frac{E_\mathrm{Th}}{\pi} \exp\left[ -  \frac{\mathcal{ B}}{\vartheta^{\frac{1}{3}} \tilde g^{\frac{2}{3}} } + \frac{1}{\vartheta} \right],
    \quad
    \tilde g \lesssim \vartheta, \quad \tan(\tilde g_\mathrm{Th}) \lesssim (\vartheta/\tilde g)^{\frac{1}{3}},
\end{equation}
where $\mathcal B$ is an order-one constant determined by $x$; in the $x\to0$ limit, $\mathcal B=(7\pi/8)^{2/3}$..

Physically, Eq. \eqref{eq:AppTcMod} and Eq. \eqref{eq:AppTcTransient} correspond to the same $T_c \sim \Lambda e^{- \#/\tilde g}$ regime, where the numerical constant and the prefactor slowly flow with $\tilde g$. 
Eq. \eqref{eq:AppTc23} corresponds to a distinct regime, in which $T_c \sim E_\mathrm{Th} e^{-\#/ \vartheta^{1/3} \tilde g^{2/3}}$. 
Therefore, the crossover occurs when $\tan (\tilde g_\mathrm{Th}) \sim \gamma^\frac{1}{3}$, which occurs for $\vartheta \gtrsim \ln^{-1} (\Lambda/E_\mathrm{Th})$ and when $\vartheta \lesssim \tilde g \lesssim 1$. 
Thus, the behavior of $T_c$ as a function of $\tilde g$ can be summarized as
\begin{equation}
    T_c \sim \begin{cases}
        \Lambda \exp  \left[ - \frac{\#}{\tilde g} \right], \quad\quad\quad \tilde g \gtrsim \tilde g_*,
        \\
        E_\mathrm{Th} \exp\left[ - \frac{\#}{\vartheta^{1/3} \tilde g^{2/3}} \right], \quad \tilde g \lesssim \tilde g_*,
    \end{cases}
    \quad\quad
    \tilde g_* = \min(\vartheta, \pi \ln^{-1}(\Lambda/E_\mathrm{Th})/2).
\end{equation}


\section{Solution of the Usadel equation for gap fluctuation}
\label{sec:SCBAGapFluctuations}

We solve Eq.~\eqref{eq:DeltaPhiGapEq}, which for $\varepsilon > 0$ is
\begin{multline}\label{eq:AppDeltaPhiGapEqDetail}
    \Big( - D \nabla^2 + 2 \varepsilon \Big) \delta \Phi(i \varepsilon, \vec r) - 2 \tilde g^2 \int_{\pi T}^\Lambda d\varepsilon' \, \left[ \ln \left( \frac{\Lambda}{\max(\varepsilon, \varepsilon')} \right) + \frac{\vartheta}{2} \Theta(E_\mathrm{Th} - \max(\varepsilon, \varepsilon'))  \ln^2 \left( \frac{E_\mathrm{Th}}{\max(\varepsilon, \varepsilon')} \right) \right] \delta \Phi(i \varepsilon', \vec r) =
    \\
    = 4 \bar g^2 \int_{\pi T}^{\Lambda} d\varepsilon' \, \delta\bar D (i\varepsilon - i\varepsilon', \vec r) \langle \Phi\rangle(i\varepsilon').
\end{multline}
We seek a solution in the form
\begin{equation}\label{eq:AppDeltaPhiExpr}
    \delta \Phi(i \varepsilon, \vec r) = \int_{\pi T}^\Lambda d\varepsilon' \int d^2\vec r' S(\varepsilon', \vec r') \mathcal{F}(\varepsilon, \varepsilon', \vec r - \vec r'), \quad \quad  S(\varepsilon, \vec r) = 4 \bar g^2 \int_{\pi T}^\Lambda d\varepsilon' \, \delta \bar D(i\varepsilon - i\varepsilon', \vec r) \langle \Phi\rangle(i\varepsilon'),
\end{equation}
and the Green's function $\mathcal{F}(\varepsilon, \varepsilon', \vec r - \vec r')$ satisfies:
\begin{multline}
    (- D \nabla^2 + 2 \varepsilon)\mathcal{F}(\varepsilon, \varepsilon_0, \vec r - \vec r') - \delta(\varepsilon - \varepsilon_0) \delta(\vec r - \vec r')=
    \\
    = 2 \tilde g^2 \int_{\pi T}^\Lambda d \varepsilon' \, \left[ \ln \left( \frac{\Lambda}{\max(\varepsilon, \varepsilon')} \right) + \frac{\vartheta}{2} \Theta(E_\mathrm{Th} - \max(\varepsilon, \varepsilon'))  \ln^2 \left( \frac{E_\mathrm{Th}}{\max(\varepsilon, \varepsilon')} \right) \right] \mathcal{F}( \varepsilon', \varepsilon_0, \vec r - \vec r') .
\end{multline}
In the momentum space, the equation for $\mathcal{F}(\varepsilon, \varepsilon', \vec q) = \int d^2 \vec r \, e^{-i \vec r \cdot \vec q} \mathcal{F}(\varepsilon, \varepsilon', \vec r)$ takes the form
\begin{equation}\label{eq:AppEqForFq}
        (D \vec q^2 + 2 \varepsilon)\mathcal{F}(\varepsilon, \varepsilon_0, \vec q) - 2 \tilde g^2 \int_{\pi T}^\Lambda d \varepsilon' \, \left[ \ln \left( \frac{\Lambda}{\max(\varepsilon, \varepsilon')} \right) + \frac{\vartheta}{2} \Theta(E_\mathrm{Th} - \max(\varepsilon, \varepsilon'))  \ln^2 \left( \frac{E_\mathrm{Th}}{\max(\varepsilon, \varepsilon')} \right) \right] \mathcal{F}( \varepsilon', \varepsilon_0, \vec q) = \delta(\varepsilon - \varepsilon_0).
\end{equation}
We write $\mathcal{F}(\varepsilon, \varepsilon_0, \vec q)$ in the form $\mathcal{F}(\varepsilon, \varepsilon_0, \vec q) = \mathcal{F}'(\varepsilon, \varepsilon_0, \vec q) + \delta(\varepsilon - \varepsilon_0)(D \vec q^2 + 2 \varepsilon_0)^{-1}$.
The function $\mathcal{F'(\varepsilon, \varepsilon_0, \vec q)}$ satisfies the following equation:
\begin{multline}\label{eq:AppEqForFprime}
    (D \vec q^2 + 2 \varepsilon)\mathcal{F}'(\varepsilon, \varepsilon_0, \vec q) - 2 \tilde g^2 \int_{\pi T}^\Lambda d \varepsilon' \, \left[ \ln \left( \frac{\Lambda}{\max(\varepsilon, \varepsilon')} \right) + \frac{\vartheta}{2} \Theta(E_\mathrm{Th} - \max(\varepsilon, \varepsilon'))  \ln^2 \left( \frac{E_\mathrm{Th}}{\max(\varepsilon, \varepsilon')} \right) \right] \mathcal{F}'( \varepsilon', \varepsilon_0, \vec q) =
    \\
    = \frac{2 \tilde g^2}{D \vec q^2 + 2 \varepsilon_0} \left[ \ln \left( \frac{\Lambda}{\max(\varepsilon, \varepsilon_0)} \right) + \frac{\vartheta}{2} \Theta(E_\mathrm{Th} - \max(\varepsilon, \varepsilon_0))  \ln^2 \left( \frac{E_\mathrm{Th}}{\max(\varepsilon, \varepsilon_0)} \right) \right].
\end{multline}
Eq. \eqref{eq:AppEqForFprime} is devoid of $\delta$-functions, therefore $\mathcal{F}'(\varepsilon, \varepsilon_0, \vec q)$ is a continuous function of $\varepsilon$ in the vicinity of $\varepsilon = \varepsilon_0$.
Differentiating Eq.~\eqref{eq:AppEqForFprime} with respect to $\varepsilon$ across $\varepsilon=\varepsilon_0$ gives
\begin{equation}
    \p_\varepsilon((D \vec q^2 + 2 \varepsilon)\mathcal{F}'(\varepsilon, \varepsilon_0, \vec q))|_{\varepsilon = \varepsilon_0-0} - \frac{2\tilde g^2} {D\vec q^2 + 2 \varepsilon_0} \left[ \frac{1}{\varepsilon_0} + \frac{\vartheta}{\varepsilon_0} \Theta(E_\mathrm{Th} - \varepsilon_0) \ln \left( \frac{E_\mathrm{Th}}{\varepsilon_0} \right) \right]=\p_\varepsilon((D \vec q^2 + 2 \varepsilon)\mathcal{F}'(\varepsilon, \varepsilon_0, \vec q))|_{\varepsilon = \varepsilon_0+0}.  
\end{equation}
One expects $\mathcal{F}'(\varepsilon, \varepsilon_0, \vec q) \sim ( D \vec q^2 + 2 \varepsilon)^{-1}$ from the structure of Eq. \eqref{eq:AppEqForFprime}.
Therefore, the divergence of the integral in the second term of Eq. \eqref{eq:AppEqForFprime} is expected to be cut off at $\varepsilon \sim D\vec q^2$ instead of $\varepsilon  = \pi T$, leading to
\begin{align}
    \tilde g^2\int_{\pi T}^\Lambda d\varepsilon'\, \ln \left( \frac{\Lambda}{\max(\varepsilon, \varepsilon')} \right) (D \vec q^2 + 2 \varepsilon')^{-1} &\sim \tilde g^2 \ln \left( \frac{\Lambda}{\varepsilon}\right) \ln \left( \frac{\Lambda}{D \vec q^2} \right) \ll 1,
    \\
    \frac{\tilde g^2 \vartheta}{2} \int_{\pi T}^{E_\mathrm{Th}} d\varepsilon' \, \ln^2 \left( \frac{E_\mathrm{Th}}{\max(\varepsilon, \varepsilon')} \right) (D \vec q^2 + 2 \varepsilon')^{-1} &\sim \frac{\tilde g^2 \vartheta}{2} \ln^2 \left( \frac{E_\mathrm{Th}}{\varepsilon}\right) \ln \left( \frac{E_\mathrm{Th}}{D \vec q^2} \right) \ll 1,
\end{align}
as far as $D \vec q^2 \gg \pi T_c$, so the logarithmic terms remain small.
Therefore,
\begin{equation}
    \mathcal{F}'(\varepsilon, \varepsilon_0, \vec q) \approx \frac{2 \tilde g^2}{(D \vec q^2 + 2 \varepsilon_0)(D \vec q^2 + 2 \varepsilon)} \left[ \ln \left( \frac{\Lambda}{\max(\varepsilon, \varepsilon_0)} \right) + \frac{\vartheta}{2} \Theta(E_\mathrm{Th} - \max(\varepsilon, \varepsilon_0))  \ln^2 \left( \frac{E_\mathrm{Th}}{\max(\varepsilon, \varepsilon_0)} \right) \right].
\end{equation}
Since $\mathcal{F}'(\varepsilon, \varepsilon_0, \vec q) = O( \tilde g^2$) with no additional divergences, $\mathcal{F}'$ is a subleading contribution.
The structure of $\mathcal{F}(\varepsilon, \varepsilon_0, \vec q)$ for $D \vec q^2 \gtrsim \pi T_c$ and $T \approx T_c$ is therefore $\mathcal{F}(\varepsilon, \varepsilon_0, \vec q) = \delta(\varepsilon - \varepsilon_0)(D \vec q^2 + 2 \varepsilon_0)^{-1}$ to the leading order in $\tilde g$.

From Eq. \eqref{eq:AppDeltaPhiExpr}, the correlation function $\langle \delta \Phi(i \varepsilon, \vec r) \, \delta \Phi(i \varepsilon', \vec r')\rangle$ for $D \vec q^2 \gtrsim \pi T_c$ takes the form
\begin{equation}\label{eq:AppDeltaPhiDeltaPhiCorrelator}
    \langle \delta \Phi(i \varepsilon, \vec r) \delta \Phi(i \varepsilon', \vec r')\rangle = \int_{\pi T}^\Lambda d\varepsilon_0 \, d\varepsilon_0' \int d^2\vec x \, d^2\vec x' \; \langle S(\varepsilon_0, \vec x) S(\varepsilon_0', \vec x')\rangle  \mathcal{F}(\varepsilon, \varepsilon_0, \vec r - \vec x) \mathcal{F}(\varepsilon', \varepsilon_0', \vec r' - \vec x').
\end{equation}
The ``source" term correlation function $\langle S(\varepsilon_0, \vec x) S(\varepsilon_0', \vec x') \rangle$ follows from Eq. \eqref{eq:AppDeltaPhiExpr}:
\begin{equation}
    \langle S(\varepsilon, \vec x) S(\varepsilon', \vec x') \rangle = 16 \bar g^4  \int_{\pi T}
^\Lambda d \varepsilon_0 d\varepsilon_0' \; \langle \delta \bar D(i\varepsilon - i\varepsilon_0, \vec x) \, \delta \bar D( i\varepsilon' - i\varepsilon'_0, \vec x') \rangle \langle\Phi\rangle(i\varepsilon_0) \langle\Phi\rangle(i\varepsilon_0').
\end{equation} 
With the help of Eq. \eqref{eq:delDdelD}, the above expressions in momentum space reduce to
\begin{align}\label{eq:AppDPhidPhiDef}
        \langle \delta \Phi \delta \Phi\rangle (i \varepsilon, i\varepsilon', \vec q) &= \int_{\pi T}^\Lambda d\varepsilon_0 \, d\varepsilon_0' \; \langle SS\rangle(\varepsilon_0, \varepsilon_0', \vec q)  \mathcal{F}(\varepsilon, \varepsilon_0, \vec q) \mathcal{F}(\varepsilon', \varepsilon_0', \vec q),
        \\\label{eq:AppSSDef}
        \langle S S\rangle(\varepsilon, \varepsilon', \vec q) &= 16 \bar g^4  \int_{\pi T}
^\Lambda d \varepsilon_0 d\varepsilon_0' \; B(\varepsilon- \varepsilon_0, \varepsilon' - \varepsilon'_0, \vec q) 
\langle\Phi\rangle(i\varepsilon_0) \langle\Phi\rangle(i\varepsilon_0').
\end{align}
where $B(\omega, \omega', \vec q) =\int d^2 \vec r \, e^{-i \vec q \vec r} B(\omega, \omega', \vec r)$ and $B(\omega, \omega', \vec r)$ is given by Eq. \eqref{eq:AppB}.
Assuming $n(\mathcal{E}) = n = \mathrm{const}$ and $\xi(\mathcal{E}) = \xi = \mathrm{const}$,
Eq. \eqref{eq:AppSSDef} simplifies to
\begin{equation}\label{eq:AppSSExpression2}
    \langle S S\rangle(\varepsilon, \varepsilon', \vec q) =  \frac{16\bar g^4 n}{c_d} (1 + \vec q^2 \xi^2/4)^{-3}  \int_{\pi T}
^\Lambda d \varepsilon_0 d\varepsilon_0' \;  
\frac{\ln(|\varepsilon - \varepsilon_0|/|\varepsilon' - \varepsilon_0'|)}{|\varepsilon - \varepsilon_0| - |\varepsilon' - \varepsilon_0'|}\langle\Phi\rangle(i\varepsilon_0) \langle\Phi\rangle(i\varepsilon_0').
\end{equation}
The main contributions to the integrals over the energies in Eq. \eqref{eq:AppSSExpression2} come from small energies $\varepsilon_0, \varepsilon_0' \lesssim \varepsilon, \varepsilon'$, since $\langle\Phi\rangle(i \varepsilon) \sim \varepsilon^{-1}$ up to logarithmic corrections (see Eqs. \eqref{eq:SCBACorGapPlus} and \eqref{eq:SCBACorGapMinus}).
Therefore, Eq. \eqref{eq:AppSSExpression2} is estimated as
\begin{equation}\label{eq:AppSourceExpression}
    \frac{\langle S S\rangle(\varepsilon, \varepsilon', \vec q)}{\langle \Phi_\mathrm{stat} \rangle^2(\omega = 0)} \simeq  \frac{16\bar g^4 n}{c_d} (1 + \vec q^2 \xi^2/4)^{-3}  
    \frac{\ln(\varepsilon/\varepsilon')}{\varepsilon -\varepsilon'}
\ell(\varepsilon)\ell(\varepsilon'),
\quad \quad
\ell(\varepsilon) = \int_{\pi T}^\varepsilon
 d \varepsilon_0  \;  
 \frac{\langle\Phi\rangle(i\varepsilon_0)}{\langle \Phi_\mathrm{stat} \rangle(\omega = 0)}.
\end{equation}
Substitution of Eq. \eqref{eq:AppSourceExpression} into Eq. \eqref{eq:AppDPhidPhiDef} results in
\begin{equation}\label{eq:AppDeltaPhiDeltaPhiFin}
    \frac{\langle \delta \Phi \, \delta \Phi\rangle(i \varepsilon, i \varepsilon', \vec q)}{\langle \Phi_\mathrm{stat}\rangle^2(\omega = 0)} \simeq \frac{16 \bar g^4 n}{c_d} \frac{\ell(\varepsilon)\ell(\varepsilon')}{(D \vec q^2 + 2 \varepsilon) (D \vec q^2 + 2 \varepsilon') (1 + \vec q^2 \xi^2 /4)^{3}} \frac{\ln (\varepsilon/ \varepsilon')}{\varepsilon - \varepsilon'}, \quad D \vec q^2 \gg \pi T_c, 
\end{equation}
A useful quantity that can be evaluated from Eq. \eqref{eq:AppDeltaPhiDeltaPhiFin} is the structure factor of the fluctuation of the pairing amplitude $\delta \Phi_\mathrm{stat}(\omega = 0, \vec r) = \int d\varepsilon \, \delta \Phi(i \varepsilon, \vec r)$. In momentum space $\langle \delta \Phi_\mathrm{stat} \, \delta \Phi_\mathrm{stat}\rangle (\omega = 0, \vec q) = \int d^2 \vec r \, e^{- i \vec r \cdot \vec q} \langle \delta \Phi_\mathrm{stat}(\omega = 0, \vec r) \, \delta \Phi_\mathrm{stat}(\omega = 0, 0)\rangle$, which estimates to
\begin{equation}\label{eq:AppDeltaLDeltaL}
      \frac{\langle \delta \Phi_\mathrm{stat} \, \delta \Phi_\mathrm{stat}\rangle(\omega = 0, \vec q)}{\langle\Phi_\mathrm{stat}\rangle^2(\omega = 0)} \simeq \frac{16 \pi^2 \bar g^2}{c_d D \vec q^2} \frac{\ln \left( \frac{D \vec q^2}{\pi T_c} \right)}{\left(1 + \vec q^2 \xi^2/4 \right)^{3}}, \quad D \vec q^2 \gg \pi T_c,
\end{equation}
which is the Eq. \eqref{dLdLres} of the main text.


\section{Details of the numerical solution and additional figures} \label{sec:Numerical_details}

\subsection{Numerical solution}

Here we provide the technical details of the numerical solution to Eq. (\ref{eq:full_gap}), along with some additional data plots that did not fit into the main text. 

The dynamical Usadel equation in Eq. (\ref{eq:full_gap}) is an eigenvalue equation, and can be solved efficiently using iterative methods. We write it again below, to make this more transparent, 
\begin{equation}\label{eq:full_gap_matrix_vector}
  \sum\limits_{m = -m_{\text{max}}}^{m_{\text{max}}} \left[\Big(-D\nabla^2+2|\varepsilon_n|+2\bar{\Sigma}(i\varepsilon_n, \vec{r})\Big) \delta_{n,m} - 2\bar{g}^2 T \bar D(i \varepsilon_m - i \varepsilon_n, \vec r) \right]\Phi(i\varepsilon_m,\vec{r})=0
\end{equation}
Here, $m_{\text{max}}$ is a necessary ``frequency cutoff". The size of the vector is $N_{\omega}\times L^2$, where $N_{\omega} = 2 \, m_{\text{max}} + 1$. As discussed in the main text, due to the strong disorder in the problem, we are interested in the entire low-lying spectrum (tens to hundreds of eigenvalues), beyond just the zero eigenvalue. We therefore use the locally optimal block preconditioned conjugate gradient (LOBPCG) algorithm, since it is particularly well equipped to find these simultaneously, running on graphical processing units (GPUs) (a similar approach was used in Ref. \cite{Huang2024} to solve an exciton Schr\"odinger equation).

In order to implement LOBPCG we simply need to efficiently apply the matrix-vector (matvec) multiplication of the operator in Eq. (\ref{eq:full_gap_matrix_vector}). We apply the gradient term by a fast Fourier transform (FFT): $\nabla^2 \Phi(i\varepsilon_m,\vec{r}) = \mathcal F^{-1} (2 - \cos(k_x) - \cos(k_y))\mathcal F \, \Phi(i\varepsilon_m,\vec{r})$. The attractive part of the pairing operator, which contains the boson Green's function, is Toeplitz in the frequency index, and can therefore be fast multiplied by FFT with a standard procedure. All other matvecs are elements-wise multiplication. LOBPCG is at its best when there is a fast preconditioner $P$ available for acceleration. In this case, we simply use the free kernel with a constant shift $\alpha$: $P^{-1} = \mathcal F^{-1} (2 - \cos(k_x) - \cos(k_y))\mathcal F + 2 |\varepsilon_n| + \alpha$. 

As the temperature $T$ is lowered, we keep the actual frequency cutoff in the Matsubara sum fixed, i.e. $T \times N_{\omega} = C$, where $C$ is the same constant throughout all simulations.

\subsection{Additional plots}

\begin{figure}[h]
    \centering
    \includegraphics[width=0.3\linewidth]{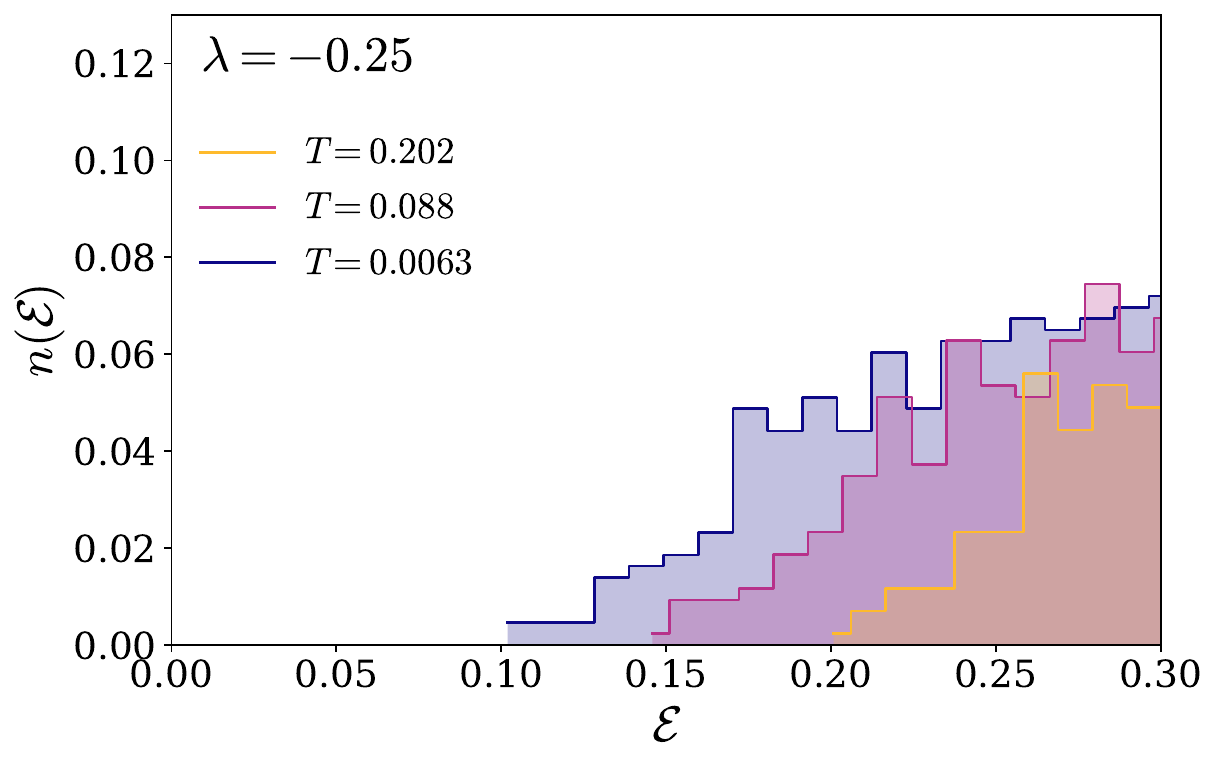}
    \includegraphics[width=0.3\linewidth]{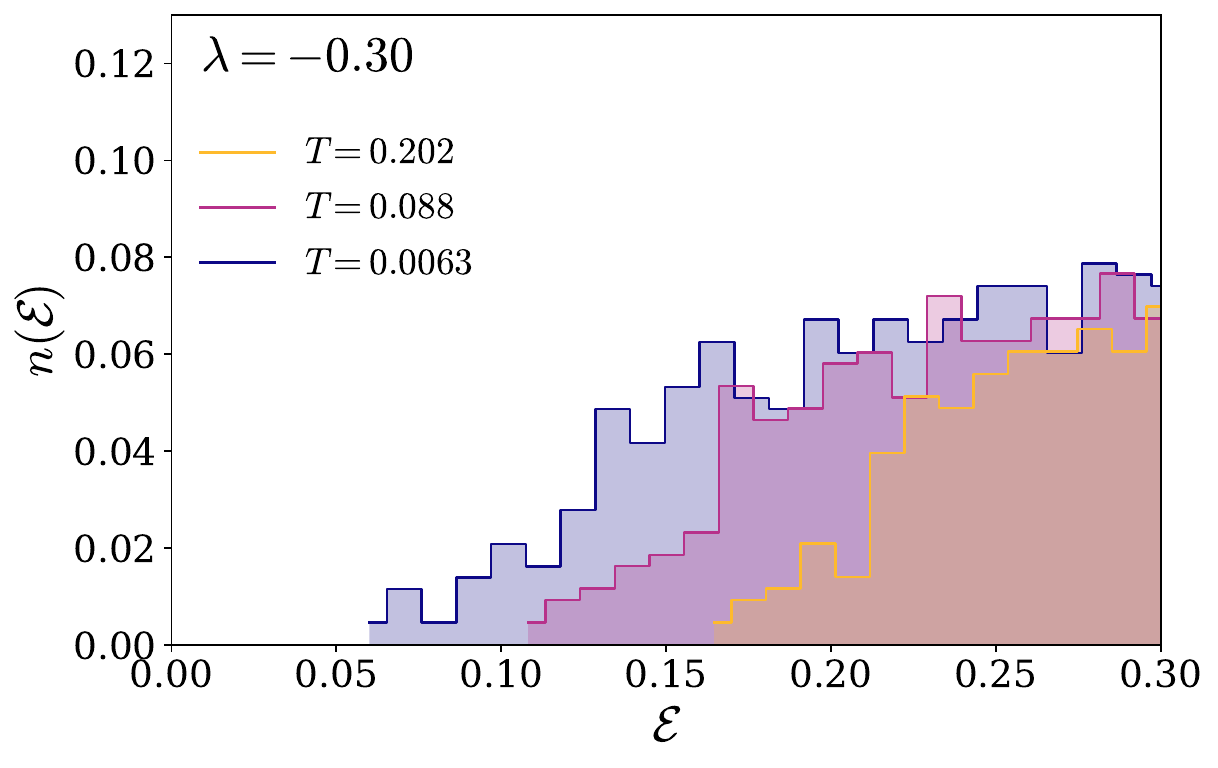}
    \includegraphics[width=0.3\linewidth]{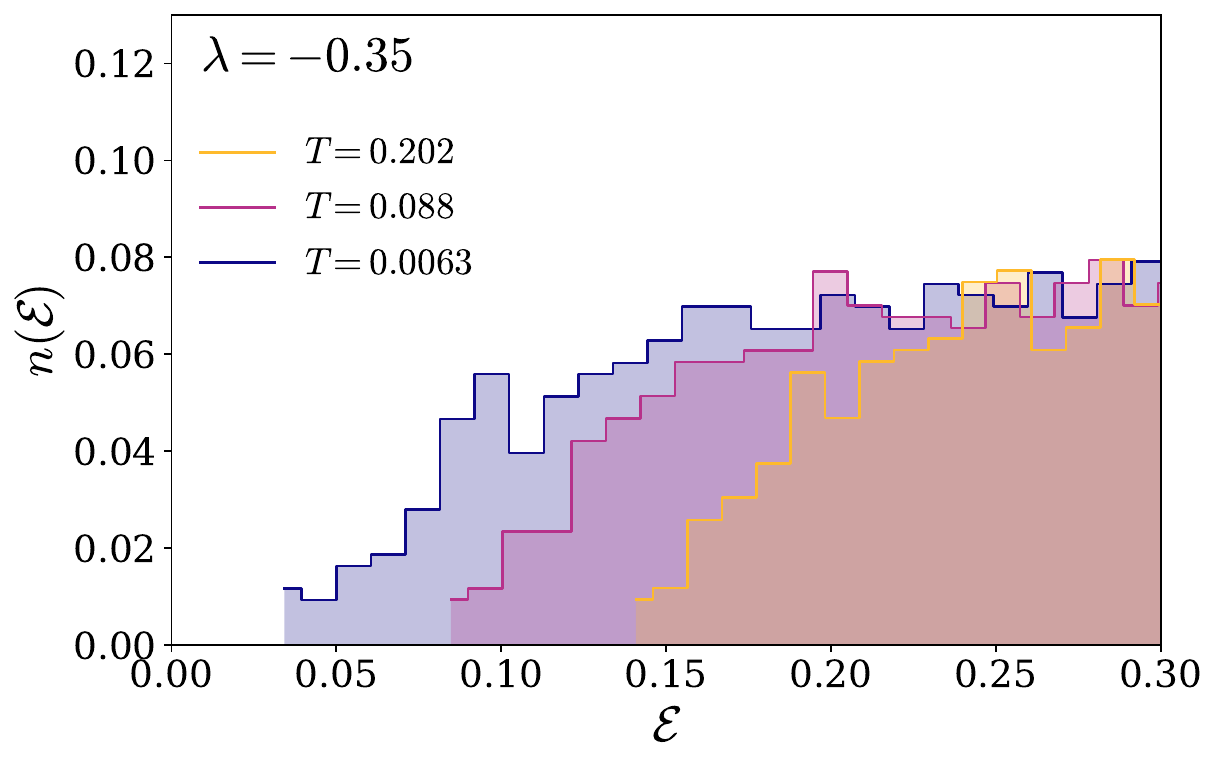}
    \includegraphics[width=0.3\linewidth]{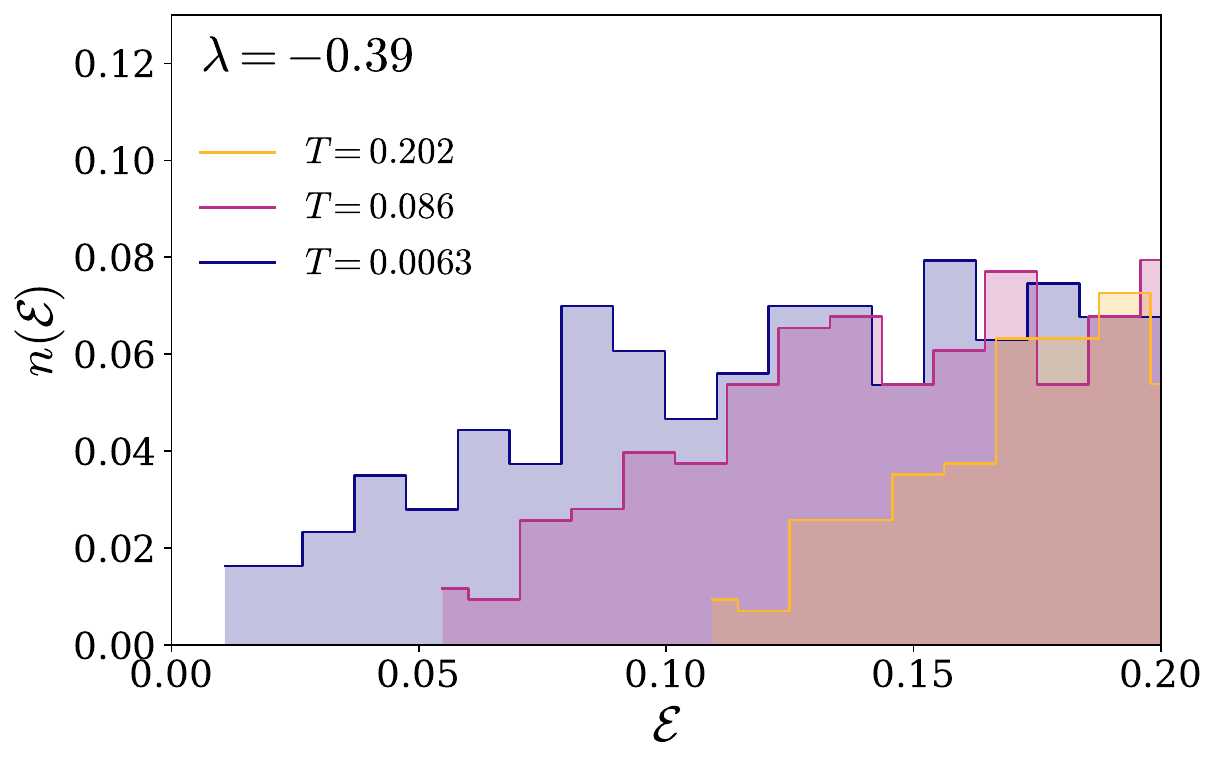}
    \includegraphics[width=0.3\linewidth]{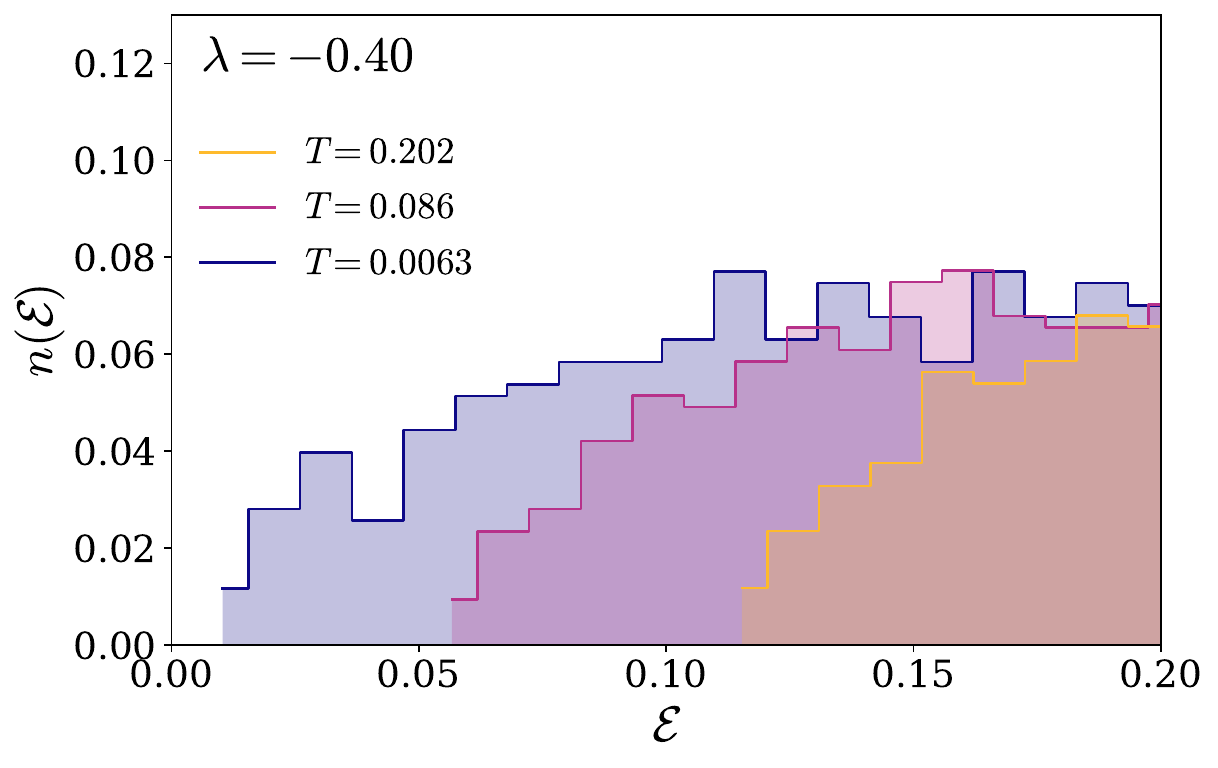}
    \includegraphics[width=0.3\linewidth]{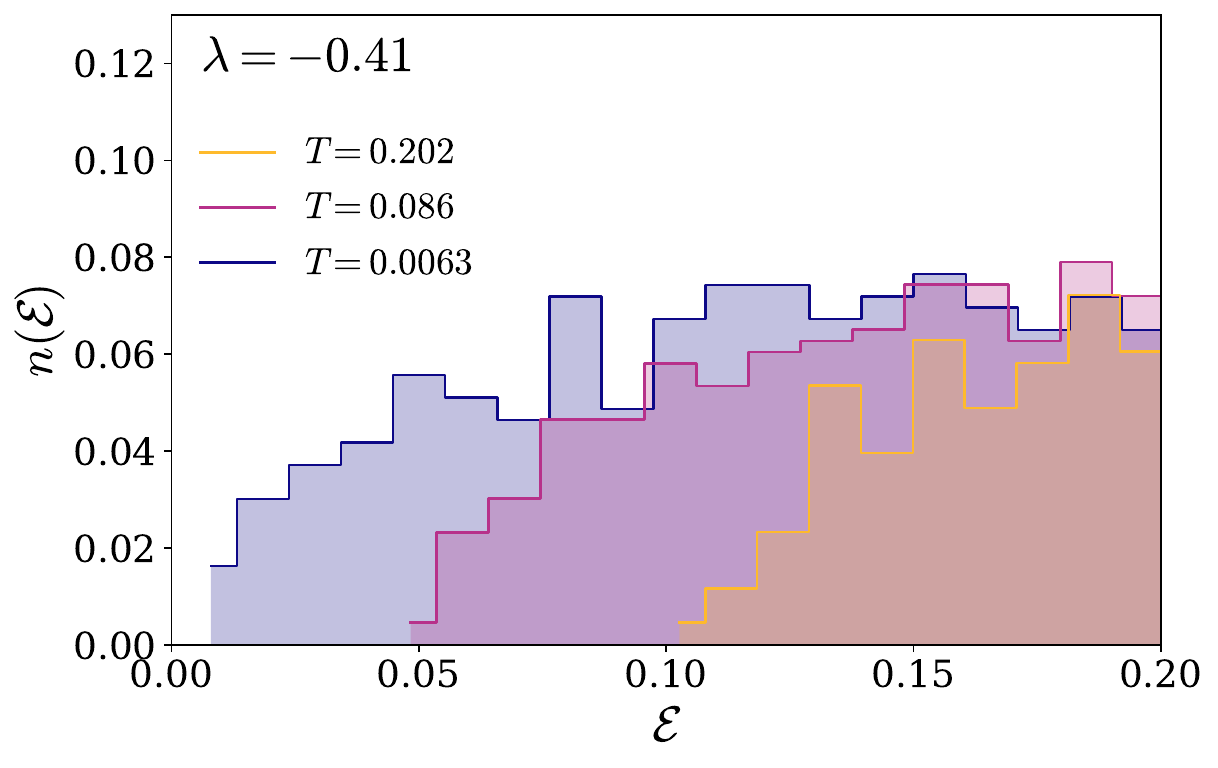}
    \includegraphics[width=0.3\linewidth]{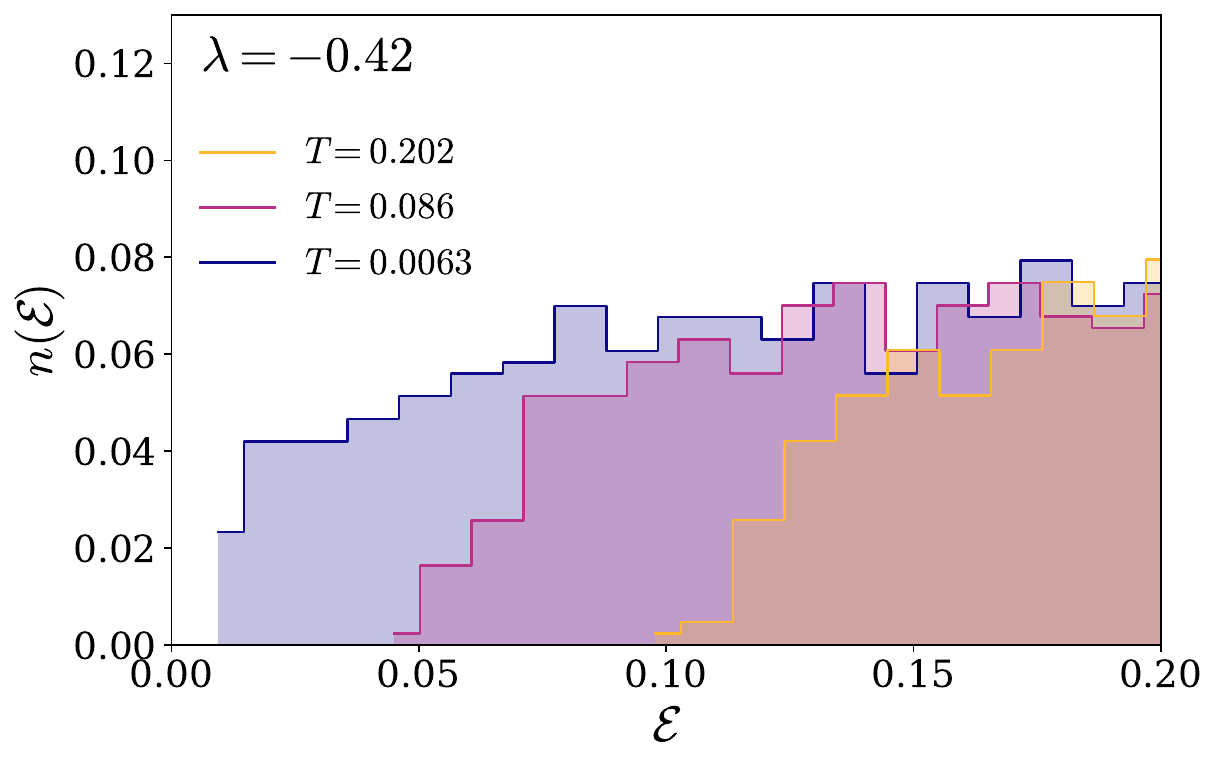}
    \caption{The boson eigenmode density of states, across the phase diagram.}
    \label{fig:DOS_extra}
\end{figure}

Now we provide some additional figures that were too voluminous to include in the main text. In Fig. \ref{fig:DOS_extra} we show the boson eigenmode DOS for many values of $\lambda$, complementing Fig. \ref{fig:normal_state}c.

\begin{figure}[h]
    \centering
    \includegraphics[width=0.32\linewidth]{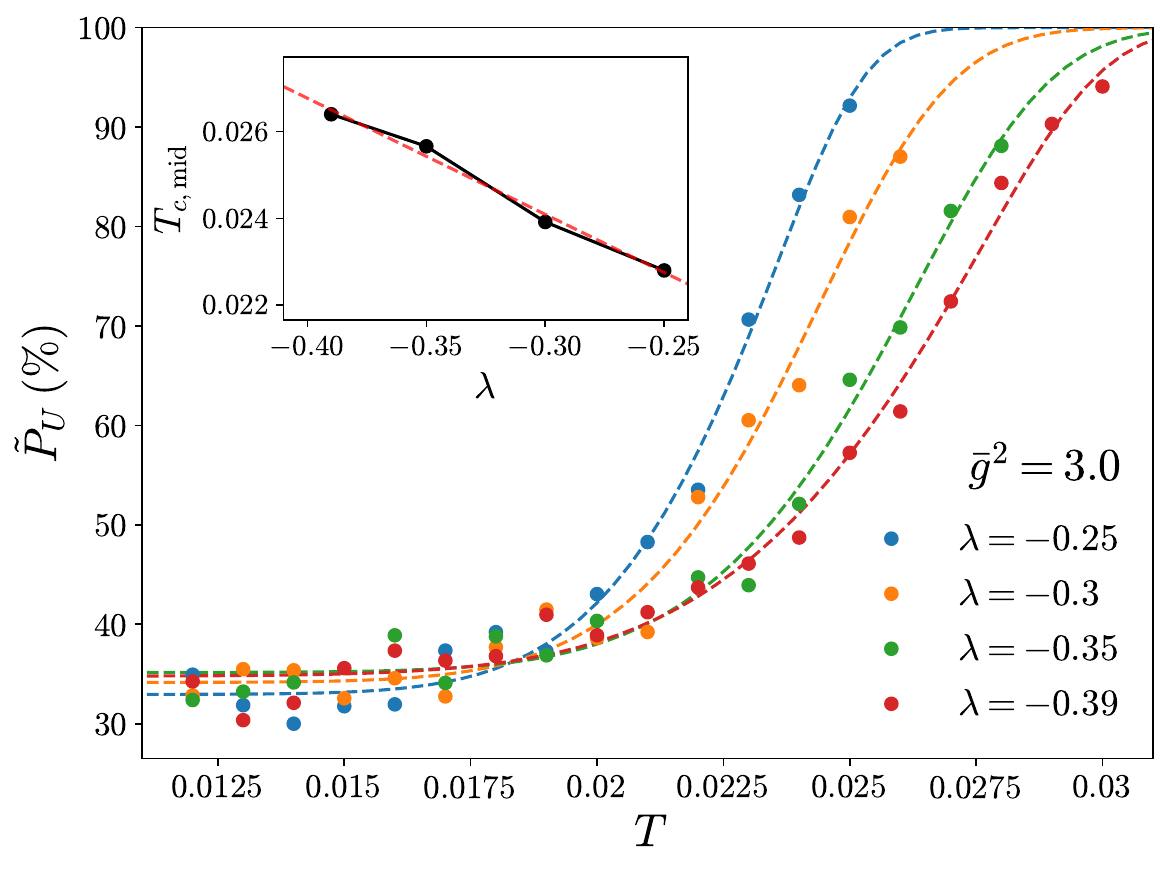}
    \includegraphics[width=0.32\linewidth]{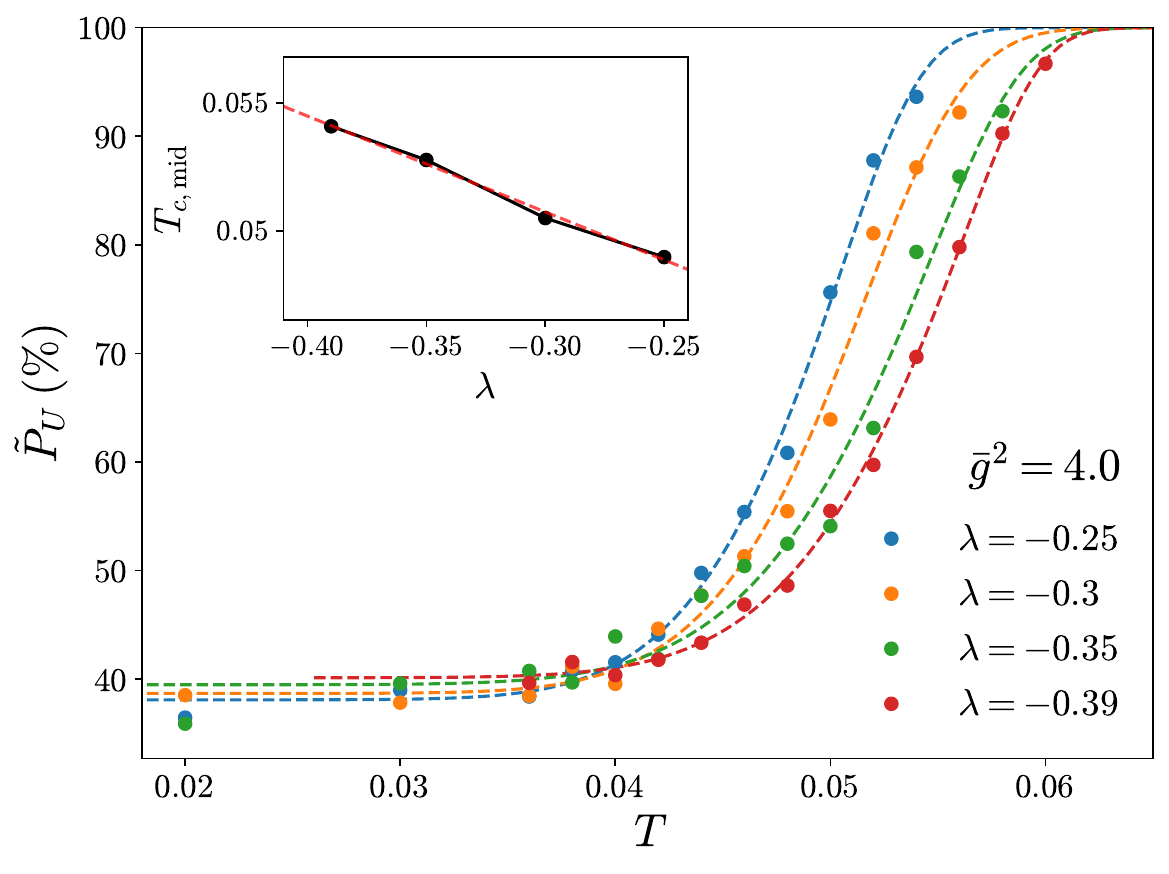}
    \includegraphics[width=0.32\linewidth]{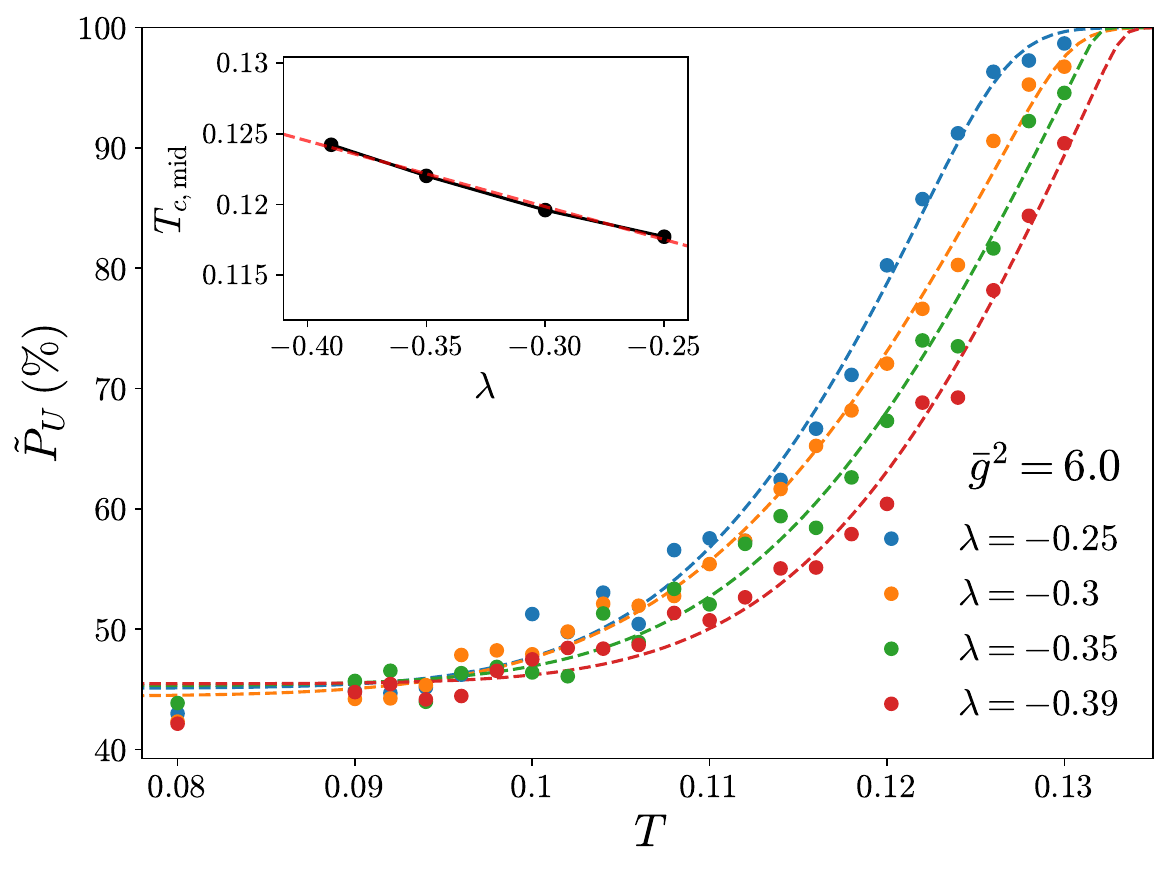}
    \includegraphics[width=0.32\linewidth]{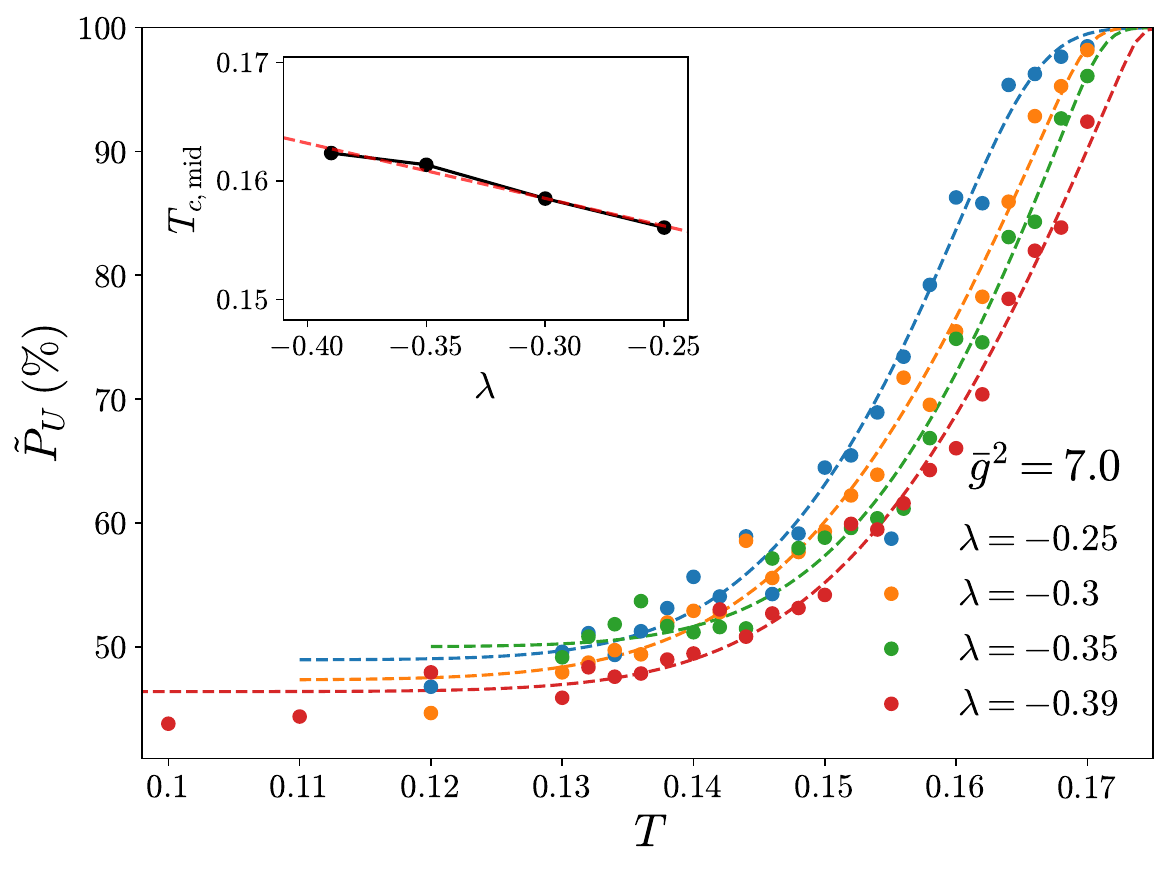}
    \includegraphics[width=0.32\linewidth]{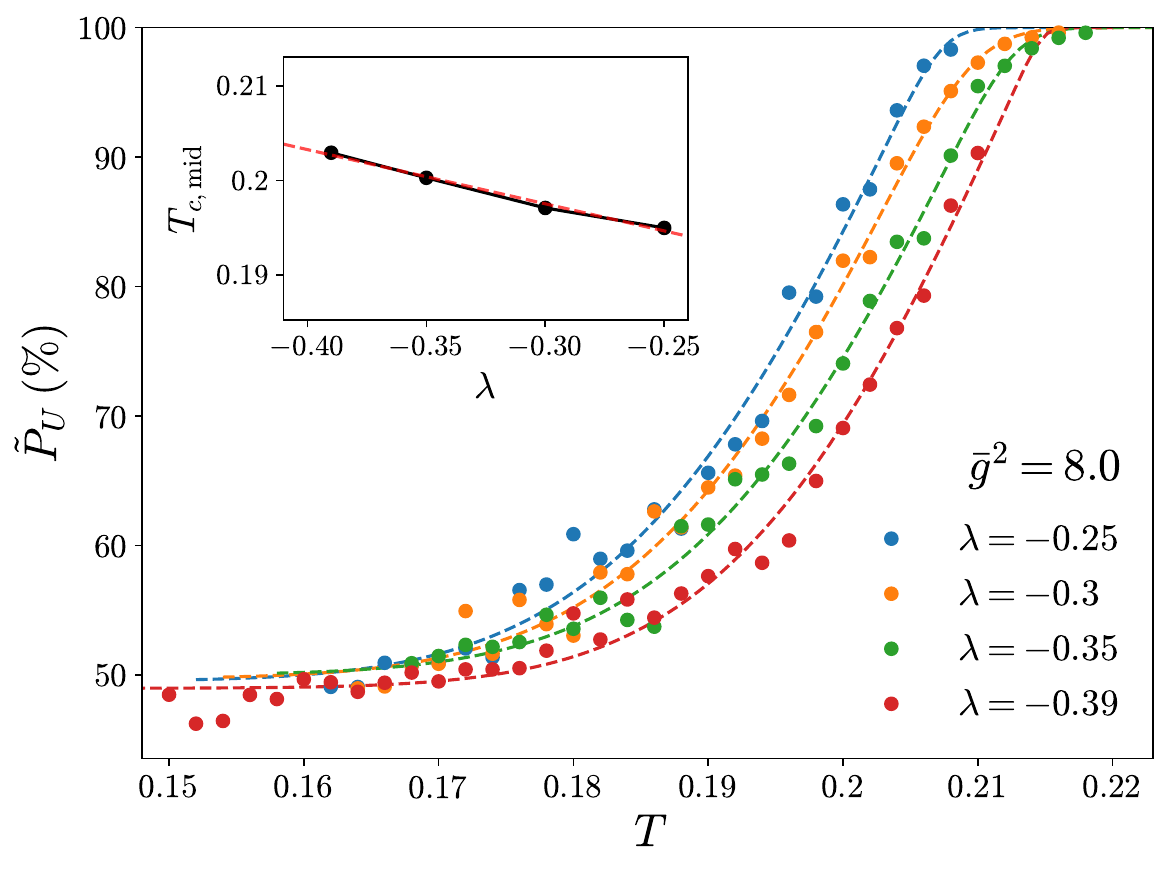}
    \caption{$\tilde P_U$ plots for $\bar{g}^2 = 3, 4, 6, 7, 8$. Insets show $T_{c,\text{mid}}$ versus $\lambda$ and the linear fits.}
    \label{fig:tilde P_U plots extra}
\end{figure}

In Fig. \ref{fig:tilde P_U plots extra} we show complementary plots to Fig. \ref{fig:SC_state}c for more values of $\bar{g}^2$.

\begin{figure}[h]
    \centering
    \includegraphics[width=0.45\linewidth]{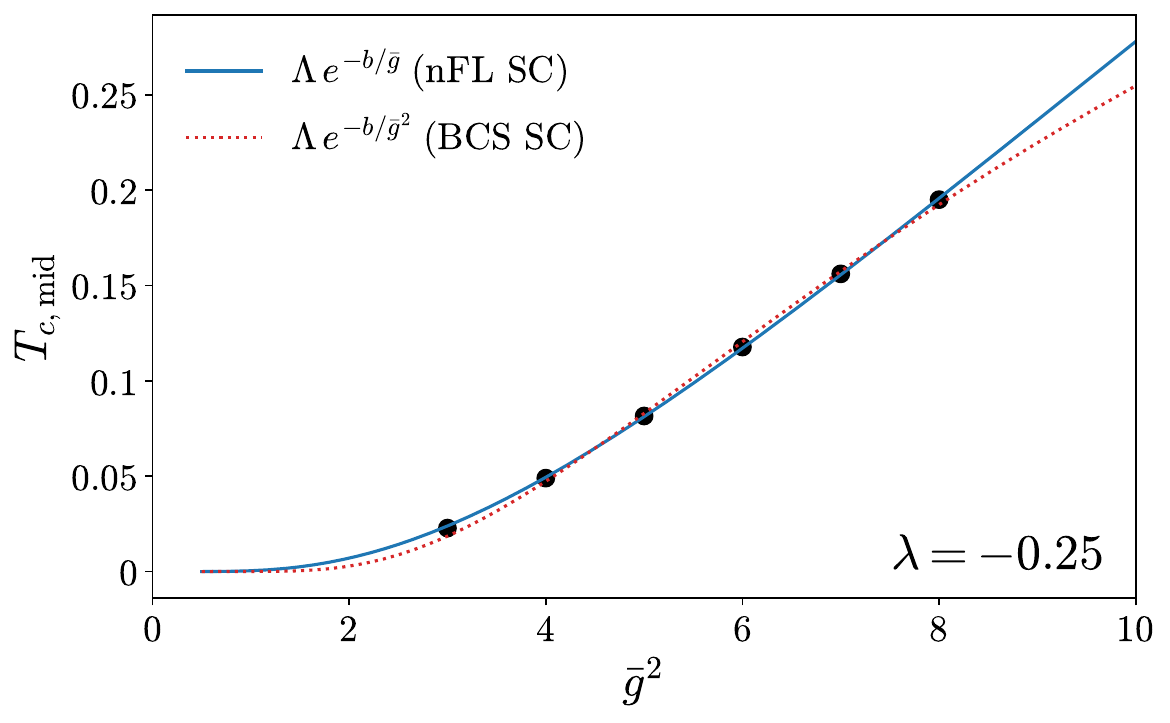}
    \includegraphics[width=0.45\linewidth]{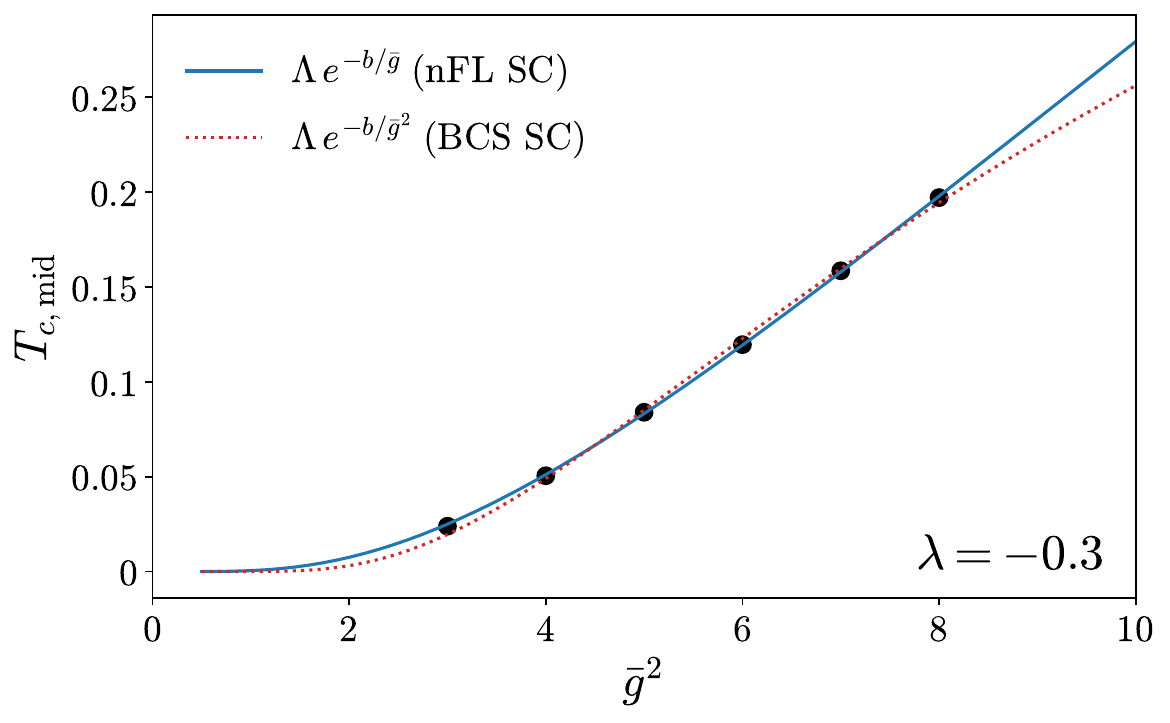}
    \includegraphics[width=0.45\linewidth]{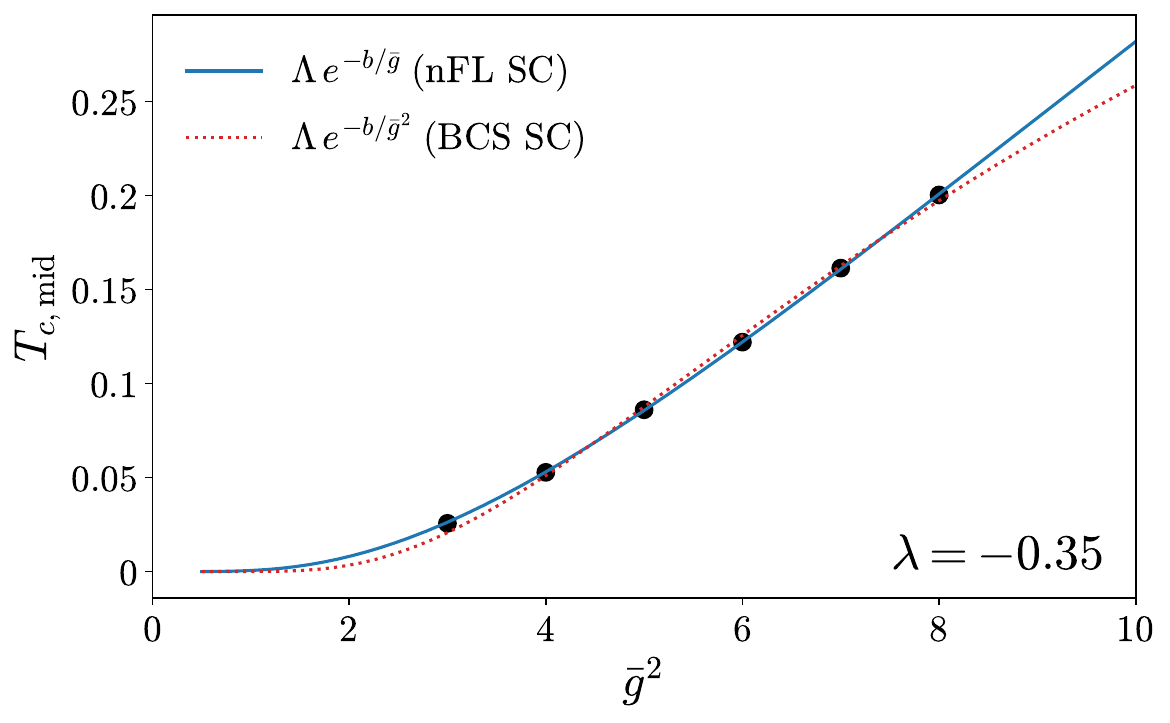}
    \caption{Comparison of the nFL and BCS form of the transition temperature as a function of coupling strength, across $\lambda$ values.}
    \label{fig:T_c_mid plots all lambda}
\end{figure}

In Fig. \ref{fig:T_c_mid plots all lambda} we show the scaling of $T_{c,\text{mid}}$ with $\bar{g}^2$ for more values of $\lambda$, complementary to Fig. \ref{fig:SC_state}d. All curves show equally good fits to the non-Fermi liquid form.


\section{Large solitary puddle}
\label{sec:AppBigSolitaryPuddle}

We consider the regime in which the local pairing kernel is dominated by a single bosonic eigenstate $\phi_\alpha(\vec r)$.
As was determined in the main text, in the limit of $T_{c, \mathrm{loc}} \gtrsim E_{\mathrm{Th}, \alpha}$, the gradient term in Eq. \eqref{eq:gap_puddle} can be neglected, leading to
\begin{equation}\label{eq:AppBigPuddle}
    |\varepsilon_n| \Phi(i \varepsilon_n) =  \bar g^2 T \sum_{\varepsilon_m} \Phi(i \varepsilon_m)\bar D(i \max(\varepsilon_n, \varepsilon_m)),
\end{equation}
where
\begin{align}
    \bar D(i \omega_n) 
    = 
    \frac{1}{\pi \xi^2_\alpha}\frac{1}{c_d |\omega_n| + e_\alpha},
\end{align}
and the Gor'kov function is given by $\Phi(i \varepsilon_n, \vec r) = \Phi(i \varepsilon_n) \Theta(\xi_\alpha - |\vec r|)$.
A standard approximation
$\bar D(i\varepsilon_n-i\varepsilon_m) \approx
\bar D(i\max(|\varepsilon_n|,|\varepsilon_m|))$ was used in Eq. \eqref{eq:AppBigPuddle}.
For this large-puddle estimate, we approximated the mode as roughly uniform over an area $\pi\xi_\alpha^2$.
Applying the Euler-Maclaurin approximation to the right side of  Eq. \eqref{eq:AppBigPuddle} 
 simplifies the equation to 
\begin{equation}\label{eq:AppReducedEq}
    \varepsilon \Phi(i \varepsilon) = \frac{2 \bar g^2}{\pi \xi^2_\alpha} \int_{\pi T}^{\Lambda^\prime} \frac{d \varepsilon^\prime}{2 \pi} \frac{\Phi(i \varepsilon^\prime)}{c_d \max (\varepsilon, \varepsilon^\prime) + e_\alpha}.
\end{equation}
With a substitution $g_\alpha = \bar g^2/\pi^2 \xi^2_\alpha e_\alpha$ and $x = c_d \varepsilon/e_\alpha$, Eq. \eqref{eq:AppReducedEq} becomes dimensionless:
\begin{equation}\label{app:AppBigPuddleIntEq}
    x \Phi(i x) = g_\alpha \int_{x_T}^{x_\Lambda} d x^\prime \frac{\Phi(i x^\prime)}{\max(x, x^\prime) + 1},
\end{equation}
where $x_T = \pi c_d T/e_\alpha$ and $x_\Lambda = c_d \Lambda/e_\alpha$.
The solution of Eq.~\eqref{app:AppBigPuddleIntEq} satisfies the differential equation
\begin{equation}\label{eq:DiffEqBigPuddle}
    - \p_x \left[ (1+x)^2 \p_x (x \Phi) \right] = g_\alpha \Phi.
\end{equation}
The solution to Eq. \eqref{eq:DiffEqBigPuddle} has a form
\begin{equation}\label{eq:BigPuddleSol}
    \Phi(i x) = A C_1(x) + B C_2(x),
\end{equation}
where
\begin{equation}
    C_1(x) = \frac{1}{x (1+x)} F\left(\alpha_+,\alpha_-,2; \frac{1}{1+x}\right), \quad\quad C_2(x) = \frac{1}{(1+x)^2}F\left(1+\alpha_+,1+\alpha_-,2; \frac{x}{1+x}\right).
\end{equation}
The constants $\alpha_\pm$ are given by $\alpha_\pm = 1/2 \pm \sqrt{1 + 4 g_\alpha}/2$, and $F(a, b, c; z)$ is a Gauss ${}_2F_1$ hypergeometric function.
Using standard hypergeometric identities \cite{bateman1953}, one can show that $C_1(x)$ satisfies the following identity:
\begin{multline}\label{eq:C1FinEq}
    \frac{g_\alpha}{1+x} \int_{x_T}^x C_1(x^\prime) \, dx^\prime + g_\alpha \int_{x}^{x_\Lambda} \frac{C_1(x^\prime)}{1+x^\prime} \, dx^\prime - x C_1(x) = 
    \\
    -\frac{1}{(1 + x)(1+x_T)} \left[ x_T F\left( 1+\alpha_+, 1+\alpha_-, 1, \frac{1}{1+x_T}\right) - F\left( 1+\alpha_+, 1+\alpha_-, 2, \frac{1}{1+x_T} \right) \right]
    +
    \\
    +
    \frac{1}{1+x_\Lambda} \left[\frac{x_\Lambda}{1+ x_\Lambda} F\left( 1+ \alpha_+, 1+ \alpha_-, 1, \frac{1}{1+x_\Lambda}\right) -  F\left( 1+\alpha_+, 1+\alpha_-, 2, \frac{1}{1+x_\Lambda} \right) \right].
\end{multline}
The second solution $C_2(x)$ satisfies an analogous identity:
\begin{multline}\label{eq:C2FinEq}
    \frac{g_\alpha}{1+x} \int_{x_T}^x C_2(x^\prime)\, dx^\prime + g_\alpha \int_x^{x_\Lambda} \frac{C_2(x^\prime)}{1+x^\prime}\, dx^\prime - x C_2(x) = 
    \\
    - \frac{1}{1+x} \left[ \frac{x_T}{(x_T+1)} F\left( 1+\alpha_+, 1+\alpha_-, 2, \frac{x_T}{1+x_T} \right) - \frac{1}{1+x_T} F\left( 1 + \alpha_+, 1+ \alpha_-, 1, \frac{x_T}{1+x_T} \right) \right] +
    \\
    + \frac{1}{(1+ x_\Lambda)^2}  F\left( 1+ \alpha_+, 1+ \alpha_-, 1, \frac{x_\Lambda}{1+x_\Lambda} \right).
\end{multline}
Substituting Eq. \eqref{eq:C1FinEq} and \eqref{eq:C2FinEq} into Eq. \eqref{app:AppBigPuddleIntEq} reduces to the self-consistency condition for $T = T_{c, \mathrm{loc}}$ of the form
\begin{equation}\label{eq:BigPuddleFullCondition}
    \left[ (1-z_T) \tilde F_1(z_T) - z_T \tilde F_2(z_T) \right]\left[ z_\Lambda \tilde F_1(1- z_\Lambda) \right] - \left[ (1 - z_\Lambda) \tilde F_1(z_\Lambda) - \tilde F_2(z_\Lambda) \right] \left[ (1 - z_T) \tilde F_2(1-z_T) - z_T \tilde F_1(1- z_T) \right] = 0,
\end{equation}
where $\tilde F_n(z) = F(1+\alpha_+, 1+\alpha_-, n, z)$, and $z_{T,\Lambda} = 1/(1 + x_{T,\Lambda})$. 
For $e_\alpha\ll c_d\Lambda$, one has $z_\Lambda\ll1$. For small $z$ it is true that
\begin{equation}\label{eq:AppRelFn}
    \tilde F_n(z) = 1 + \frac{(2 - g_\alpha)}{n}z, \quad \tilde F_1(1- z_\Lambda) \approx \frac{1}{z_\Lambda^2}.
\end{equation}
Substitution of Eq. \eqref{eq:AppRelFn} into Eq. \eqref{eq:BigPuddleFullCondition} simplifies the self-consistency condition for $T = T_{c, \alpha}$ to be
\begin{equation}\label{eq:BigPuddleCondition}
    (1 - z_T) \left( \tilde F_1(z_T) + \frac{g_\alpha z_\Lambda^2}{2} \tilde F_2(1-z_T) \right) = z_T \left( \tilde F_2(z_T) + \frac{g_\alpha z_\Lambda^2}{2} \tilde F_1(1 - z_T) \right).
\end{equation}

We first analyze Eq. \eqref{eq:BigPuddleCondition} by considering the regime $g_\alpha \lesssim 1$. 
In this limit, one would expect $x_T \ll 1$, ($z_T \approx 1$), so we can approximate $\tilde F_1(z)$ and $\tilde F_2(z)$ as
\begin{equation}
    \tilde F_1(z) \approx \frac{1}{(1 - z)^2} - \frac{g_\alpha z}{(1 - z)^2}, \quad \tilde F_2(z) \approx \frac{1}{1-z} - \frac{g_\alpha}{1 - z} - \frac{g_\alpha}{z}\log(1 - z).
\end{equation}
so Eq. \eqref{eq:BigPuddleCondition} simplifies greatly, leading to $z_T = 1 - e^{-1/g_\alpha}$. Therefore, the $T_{c, \alpha}$ of the superconducting puddle is
\begin{equation}\label{eq:AppBigPuddleTcSmall}
    T_{c, \alpha} \approx \frac{e_\alpha}{\pi c_d} \left(  \exp \left[\frac{\pi^2 \xi_\alpha^2 e_\alpha}{\bar g^2} \right] - 1 \right)^{-1}, \quad\quad \bar g^2 \lesssim  \pi^2 \xi^2_\alpha e_\alpha.
\end{equation}
Since the bosonic density of states $n(\mathcal{E})$ is constant in the vicinity $\mathcal{E} = 0$, states with arbitrary small $e_\alpha$ appear with non-zero probability. 
For a fixed value of $\bar g^2$, the value of $T_{c,\alpha}$ peaks for $e_\alpha \sim \bar g^2/\xi_\alpha^2$, leading to maximal $T_{c,\alpha}$ to have a scale $T_{c,\alpha} \sim \bar g^2 /c_d\xi_\alpha^2$.

Next, we consider the solution of Eq. \eqref{eq:BigPuddleCondition} in the limit of $g_\alpha \gtrsim 1$.
To estimate the behavior of the hypergeometric functions in such a limit, we use Watson's expansion \cite[Chap.2.3.2, Eq. 17]{bateman1953}:
\begin{equation}
    \tilde F_1(z) = \frac{1}{\sqrt{\pi \lambda}} \frac{\cos \left( \lambda \theta - \frac{\pi}{4} \right)}{\sqrt{\sin \left( \frac{\theta}{2} \right)\cos^5 \left( \frac{\theta}{2} \right)}} \left( 1 + O\left( \frac{1}{\lambda} \right) \right), \quad\quad \tilde F_2(z) = \frac{1}{\sqrt{\pi \lambda} \left( \lambda - \frac{1}{2}\right)} \frac{\sin \left( \lambda \theta - \frac{\pi}{4} \right)}{\sqrt{\sin^3 \left( \frac{\theta}{2} \right)\cos^3 \left( \frac{\theta}{2} \right)}} \left( 1 + O\left( \frac{1}{\lambda} \right) \right),
\end{equation}
where $\theta = \mathrm{Arccos(1 - 2 z)}$ and $\lambda = \sqrt{ g_\alpha}$. 
The approximation works particularly well for $g_\alpha \gtrsim 5$. 
Due to the oscillatory nature of $\tilde F_1(z)$ and $\tilde F_2(z)$, one should expect to find the first solution at high temperatures $T$.
Such temperatures would correspond to $z_T \ll 1$ for large enough $g_\alpha$. 
This allows to approximate $\theta$ as $\theta \approx 2\sqrt{z}$ with an expectation that $\theta \lesssim 1/\lambda$, leading to 
\begin{equation}\label{eq:AppThetaEq}
    \cos\left( \lambda\theta -\frac{\pi}{4} \right) \left[ 1 - \frac{\lambda^2 z_\Lambda^2}{\theta \left( \lambda - \frac{1}{2} \right)} \cos \pi \lambda - 2 \lambda^2 z_\Lambda^2 \sin \pi \lambda \right] = \frac{2 \theta}{\lambda - \frac{1}{2}} \left( 1 - 2 \lambda^2 z_\Lambda^2 \cos \pi \lambda + \frac{\lambda^2 z_\Lambda^2}{2 \theta^2} \sin \pi \lambda \right) \sin\left( \lambda \theta - \frac{\pi}{4}\right).
\end{equation}
In the leading order in $1/\Lambda$, solution of Eq. \eqref{eq:AppThetaEq} is independent of $\Lambda$, leading to
\begin{equation}
    \theta = \frac{3 \pi}{4 \sqrt{g_\alpha}} \left( 1 + O\left( \frac{1}{g_\alpha} \right) + O\left( \frac{g_\alpha^2}{\Lambda^2} \right)\right).
\end{equation}
Therefore, $z_T \approx 9 \pi^2/16 g_\alpha$ for $g_\alpha \gtrsim 1$, which gives the puddle transition temperature
\begin{equation}\label{eq:AppBigPuddleTcBig}
    T_{c, \alpha} = \frac{64 \bar g^2}{9 \pi^5 c_d \xi^2_\alpha}, \quad\quad \bar g^2 \gtrsim \pi^2 \xi_\alpha^2 e_\alpha.
\end{equation}
It is interesting to note that Eq.~\eqref{eq:AppBigPuddleTcSmall} provides a useful interpolation for $T_{c,\alpha}$ across both regimes, up to order-one numerical factors.

The large-puddle approximation is valid if the spatial gradient in Eq. \eqref{eq:gap_puddle} can be neglected.
The order of magnitude of the spatial gradient is $2D/\xi_\alpha^2 = E_{\mathrm{Th}, \alpha}$, and the smallest absolute value of $\varepsilon_n$ appearing in Eq. \eqref{eq:gap_puddle} is $\pi T_{c, \alpha}$.
Then, for the ``large" puddle picture to be valid, it must follow that 

\begin{equation}
    \frac{E_{\mathrm{Th}, \alpha}}{\pi T_{c, \alpha}} \sim  \frac{g_\alpha}{\vartheta} \left( \exp \left[\frac{1}{g_\alpha}\right] -1 \right)  \lesssim 1.
\end{equation}
Therefore, the large puddle picture requires $\vartheta \gtrsim g_\alpha (\exp[g_\alpha^{-1}]  - 1) \gtrsim 1$.
This regime lies outside the weak-inhomogeneity limit $\vartheta\ll1$ emphasized in the main text.


\section{Approximate solution of Eq. \eqref{eq:Ckeq}}
\label{sec:SmallPuddleEigenvalueEq}

To obtain Eq. \eqref{eq:Ckeq}, one substitutes Eq. \eqref{eq:SinglePuddleForm} into Eq. \eqref{eq:gap_puddle}.
Ensuring that Eq. \eqref{eq:gap_puddle} holds for each $C_{\varkappa_m}(\varepsilon_n)$ separately after substitution requires it to satisfy
\begin{equation}\label{eq:CkeqFinT}
    (|\varepsilon_n| - \varkappa_k) C_{\varkappa_k}(\varepsilon_n) = \pi\tilde g^2 T \sum_{\varepsilon_m} \ln \left( \frac{\Lambda}{ \max (|\varepsilon_m|, |\varepsilon_n|)}\right)  C_{\varkappa_k}(\varepsilon_m).
\end{equation}
Applying the Euler-Maclaurin approximation to Eq. \eqref{eq:CkeqFinT} results in Eq. \eqref{eq:Ckeq} of the main text.

Eq. \eqref{eq:Ckeq} is solved in the form
$C_\varkappa(\varepsilon) = C'_\varkappa(\varepsilon) + \delta(\varepsilon - \varkappa)$ for $\varepsilon > 0$, where $C'_\varkappa(\varepsilon)$ obeys the following integral equation:
\begin{equation}\label{eq:CprimeEq}
    (\varepsilon - \varkappa) C'_\varkappa(\varepsilon) = \tilde g^2 \int_0^{\Lambda} d\varepsilon' \ln \left( \frac{\Lambda}{\max(\varepsilon, \varepsilon')}\right) C'_\varkappa(\varepsilon') + \tilde g^2 \ln \left( \frac{\Lambda}{\max(\varepsilon, \varkappa)} \right).
\end{equation}
Eq. \eqref{eq:CprimeEq} can be reduced to a differential equation 
\begin{equation}
    \p_{\varepsilon} \left[ \varepsilon \, \p_{\varepsilon} \left(\left(\varepsilon  - \varkappa\right) C'_{\varkappa}(\varepsilon)\right)  \right] = - \tilde g^2 C'_{\varkappa}(\varepsilon),
\end{equation}
which is solved by
\begin{equation}\label{eq:C_bulk}
    C'_\varkappa(\varepsilon) = A_{\varkappa} F_1\left( \frac{\varepsilon}{\varkappa}\right) + B_{\varkappa} F_2 \left(1 - \frac{ \varepsilon}{\varkappa} \right).
\end{equation}
Functions $F_1$ and $F_2$ can be expressed in terms of a Gauss Hypergeometric function $F(a, b,c; z)$ as
\begin{equation}
    F_1(z) = \mathrm{Re} \, F\left(1 + i \tilde g, 1- i \tilde g ; \,  1 ; \, z\right), \quad\quad
    F_2(z) = \Theta(-z)\mathrm{Re} \, F\left(1 + i \tilde g, 1- i \tilde g ; \,  2 ; \, z\right).
\end{equation}
Function $F_2(1 - \varepsilon/\varkappa)$ is finite for all $\pi T < \varepsilon, \varkappa < \Lambda$, and the $(1 - \varepsilon/\varkappa)^{-1}$ divergence in $F_1(\varepsilon/\varkappa)$ for $\varepsilon \approx \varkappa$ should be understood in the sense of the principal value.

To determine the coefficients $A_\varkappa$ and $B_\varkappa$, we substitute Eq. \eqref{eq:C_bulk} back into Eq. \eqref{eq:CprimeEq}.
Substituting $\varepsilon = \Lambda$ into Eq. \eqref{eq:CprimeEq} gives
$C_\varkappa(\Lambda) = 0$.
With the help of the asymptotic expansion
\begin{align}
    \label{eq:Asymptotic1}
    F(1+i\tilde g, 1-i\tilde g, 1, z) &\sim - e^{\pi \tilde g}\frac{\cos \left( \tilde g \ln z \right)}{2 z} \left( 1 + O\left(\frac{\tilde g}{z}\right) \right), \quad \quad \arg z = +0,\quad  z\gg 1.
    \\\label{eq:Asymptotic2}
    F(1+i \tilde g, 1-i\tilde g, 2, 1-z) &\sim \frac{\sin \left( \tilde g \ln z \right)}{2 z \tilde g} \left( 1 + O\left(\frac{\tilde g}{z}\right) \right), \quad \quad \arg z = +0,\quad z\gg 1,
\end{align}
the relation between $A_\varkappa$ and $B_\varkappa$ in case of $\tilde g \ll 1$ reduces to
\begin{equation}\label{eq:Bplus}
    B_{\varkappa} = - A_\varkappa \frac{F_1( \Lambda/\varkappa)}{F_2(1-\Lambda/\varkappa)} \approx A_\varkappa \,\tilde g e^{- \pi \tilde g} \cot \left( \tilde g \ln \frac{\Lambda}{\varkappa}\right) \approx A_\varkappa \,\tilde g \tan \left( \tilde g \ln \frac{\varkappa}{\pi T_{c0}} \right).
\end{equation}
As before, $\pi T_{c0} = \Lambda e^{- \pi / 2 \tilde g}$.
Substitution of Eq. \eqref{eq:C_bulk} into Eq. \eqref{eq:CprimeEq} results in
\begin{align}\label{eq:CSelfConsEq}
    \tilde g^2 \int_{0}^\Lambda d\varepsilon' \, \ln \left( \frac{\Lambda}{\max(\varepsilon, \varepsilon')} \right) C'_\varkappa(\varepsilon')  + (\varkappa -  \varepsilon) C'_\varkappa(\varepsilon) 
    = 
     B_\varkappa \varkappa \ln \left( \frac{\Lambda}{\max(\varepsilon, \varkappa)} \right) + (\varkappa - \Lambda) C'_\varkappa(\Lambda).
\end{align}
The following integral hypergeometric identities are useful in the derivation of Eq. \eqref{eq:CSelfConsEq}:
\begin{align}
    \int_{\varepsilon_2}^{\varepsilon_1} d  \varepsilon^\prime \, F_1 \left(\frac{ \varepsilon^\prime}{\varkappa} \right) &=  \left. \varepsilon \, F_2 \left(\frac{ \varepsilon}{\varkappa} \right) \right|_{\varepsilon_2}^{\varepsilon_1},
    \\
    \int_{\varepsilon_2}^{\varepsilon_1} d  \varepsilon^\prime \, F_2 \left( 1 - \frac{ \varepsilon^\prime}{\varkappa} \right) &= - \frac{1}{\tilde g^2} \left.   \varepsilon \, F_1 \left( 1- \frac{ \varepsilon}{\varkappa} \right)\right|_{\varepsilon_2}^{\varepsilon_1},
    \\
    \int_{\varepsilon_2}^{\varepsilon_1} d\varepsilon^\prime \, \ln \left( \frac{\Lambda}{\varepsilon^\prime} \right) \, F_1 \left(\frac{ \varepsilon^\prime}{\varkappa} \right) 
    &= \left.\left[\varepsilon \ln \left( \frac{\Lambda}{\varepsilon} \right) F_2 \left( \frac{\varepsilon}{\varkappa} \right) + \frac{1}{\tilde g^2}  (\varkappa -  \varepsilon) F_1 \left( \frac{ \varepsilon}{\varkappa} \right) \right] \right|_{\varepsilon_2}^{\varepsilon_1},
    \\
    \int_{\varepsilon_2}^{\varepsilon_1} d\varepsilon^\prime \, \ln\left( \frac{\Lambda}{\varepsilon^\prime} \right) \, F_2\left(1 -\frac{\varepsilon^\prime}{\varkappa} \right)
    &= - \frac{1}{\tilde g^2} \left[ \varepsilon \ln \left( \frac{\Lambda}{\varepsilon} \right) F_1\left(  1 - \frac{\varepsilon}{\varkappa} \right) + ( \varepsilon - \varkappa) F_2\left(1 - \frac{ \varepsilon}{\varkappa}\right)\right]_{\varepsilon_2}^{\varepsilon_1}.
\end{align}
Comparing Eq. \eqref{eq:CSelfConsEq} to Eq. \eqref{eq:CprimeEq} along with the boundary condition $C_\varkappa(\Lambda) =0$ immediately implies
\begin{equation}
    B_\varkappa = - \frac{\tilde g^2}{\varkappa}, 
    \quad\quad
    A_\varkappa = - \frac{\tilde g}{\varkappa} \cot \left( \tilde g \ln \left( \frac{\varkappa}{\pi T_{c0}} \right) \right),
\end{equation}
which gives the result quoted in the main text.


\section{Solution of Eq. \eqref{eq:PuddleBoundary}}
\label{sec:PuddleBoundaryCond}

Eq. \eqref{eq:PuddleBoundary} can be written as
\begin{equation}\label{eq:PuddleBoundaryLowT}
    \Psi(\varepsilon) = \frac{2\vartheta}{3} \int_{\pi T}^{\Lambda} d\varepsilon' \frac{\Psi(\varepsilon')K_0(k(\varepsilon') \xi_\alpha)}{\max(\varepsilon, \varepsilon') + e_\alpha/c_d} + \Delta_0
\end{equation}
in the limit of small temperature. The $k(\varepsilon) = \sqrt{2 |\varepsilon|/D}$ defines a thermal inverse length scale characteristic of the puddle.
The infinitesimal source $\Delta_0$ is included to fix the normalization of the linear response.
In general, one expects $T\sim e_\alpha/c_d \ll \Lambda$ for most relevant puddle solutions, so Eq. \eqref{eq:PuddleBoundaryLowT} can be approximated as
\begin{equation}\label{eq:PuddleBoundaryLowTMod}
    \Psi(\varepsilon) = \frac{2\vartheta}{3} \int_{\pi T+ e_\alpha/c_d}^{\Lambda} d\varepsilon' \frac{\Psi(\varepsilon')K_0(k(\varepsilon') \xi_\alpha)}{\max(\varepsilon, \varepsilon')}+\Delta_0.
\end{equation}
Assuming $E_{\mathrm{Th},\alpha}$ is large, the modified Bessel function can be approximated by a logarithm as
$K_0(z)\approx - \ln(z/2)$, leading to
\begin{equation}\label{eq:ModPuddleBoundary}
    \Psi(\varepsilon) = \frac{\vartheta}{3} \int_{\pi T'}^{\Lambda} d\varepsilon' \frac{\Psi(\varepsilon')\ln(E_{\mathrm{Th}, \alpha}/\varepsilon')}{\max(\varepsilon, \varepsilon')}+\Delta_0,
\end{equation}
where we introduced
$\pi T' = \pi T+e_\alpha/c_d$
for conciseness.
The replacement of the lower cutoff reflects that the energy term $e_\alpha/c_d$ cuts off the infrared divergence.

The function $\Psi(\varepsilon)$ that satisfies Eq. \eqref{eq:ModPuddleBoundary} also satisfies
\begin{equation}\label{eq:PuddleBoundaryDifEq}
    - \p_\varepsilon \left( \varepsilon^2 \, \p_\varepsilon \Psi(\varepsilon) \right) = \vartheta' \, \Psi(\varepsilon)\ln \left(\frac{E_{\mathrm{Th},\alpha}}{\varepsilon} \right),
\end{equation}
where
$\vartheta' = \vartheta/3$.
Solution of Eq. \eqref{eq:PuddleBoundaryDifEq} can be found exactly:
\begin{align}\label{eq:PuddleBoundarySolution}
    \Psi(\varepsilon) = \sqrt{\frac{E_{\mathrm{Th}, \alpha}}{ \varepsilon}} \left[ a(T') \, \mathrm{Ai} \,(z(\varepsilon)) + b(T') \, \mathrm{Bi} \, (z(\varepsilon)) \right], \quad\quad
    z(\varepsilon) = \frac{1}{\vartheta^{\prime 2/3}} \left( \frac{1}{4} - \vartheta' \ln \left(\frac{E_{\mathrm{Th}, \alpha}}{\varepsilon} \right) \right).
\end{align}
The constants  $a(T')$ and $b(T')$ are determined by substituting Eq. \eqref{eq:PuddleBoundarySolution} back into Eq. \eqref{eq:ModPuddleBoundary}, resulting in
\begin{align}
    a(T') &= 2 \vartheta' \Delta_0 \sqrt{\frac{\Lambda}{E_{\mathrm{Th},\alpha}}} \frac{B_-(z(\pi T'))}{B_-(z(\pi T')) \, A_+(z(\Lambda)) - A_-(z(\pi T')) B_+(z(\Lambda))},
    \\
    b(T') &= - \vartheta' \Delta_0 \sqrt{\frac{\Lambda}{E_{\mathrm{Th},\alpha}}} \frac{A_-(z(\pi T'))}{B_-(z(\pi T')) \, A_+(z(\Lambda)) - A_-(z(\pi T')) B_+(z(\Lambda))},
\end{align}
where $A_{\pm}(z) = \mathrm{Ai}\, (z) \pm 2 \vartheta^{\prime 1/3} \, \mathrm{Ai}^\prime(z)$, and $B_{\pm}(z) = \mathrm{Bi}\, (z) \pm 2 \vartheta^{\prime 1/3} \, \mathrm{Bi}^\prime(z)$.
The pairing amplitude at the distance $|\vec r| = \xi_\alpha$ away from the center of the localized bosonic mode then becomes
\begin{equation}
    \Phi_\mathrm{stat}(r = \xi_\alpha) \sim \int_{\pi T}^{\Lambda^\prime} \frac{d\varepsilon}{2\pi} \Phi(i\varepsilon, r = \xi) = \Delta_0\Lambda \frac{B_-(z(\pi T')) A_-(z(\Lambda)) - A_-(z(\pi T')) B_-(z(\Lambda))}{B_-(z(\pi T')) A_+(z(\Lambda)) - A_-(z(\pi T')) B_+(z(\Lambda))}.
\end{equation}

\bibliographystyle{apsrev4-2_custom}
\bibliography{note}

\begin{thebibliography}{112}%
\makeatletter
\providecommand \@ifxundefined [1]{%
 \@ifx{#1\undefined}
}%
\providecommand \@ifnum [1]{%
 \ifnum #1\expandafter \@firstoftwo
 \else \expandafter \@secondoftwo
 \fi
}%
\providecommand \@ifx [1]{%
 \ifx #1\expandafter \@firstoftwo
 \else \expandafter \@secondoftwo
 \fi
}%
\providecommand \natexlab [1]{#1}%
\providecommand \enquote  [1]{``#1''}%
\providecommand \bibnamefont  [1]{#1}%
\providecommand \bibfnamefont [1]{#1}%
\providecommand \citenamefont [1]{#1}%
\providecommand \href@noop [0]{\@secondoftwo}%
\providecommand \href [0]{\begingroup \@sanitize@url \@href}%
\providecommand \@href[1]{\@@startlink{#1}\@@href}%
\providecommand \@@href[1]{\endgroup#1\@@endlink}%
\providecommand \@sanitize@url [0]{\catcode `\\12\catcode `\$12\catcode
  `\&12\catcode `\#12\catcode `\^12\catcode `\_12\catcode `\%12\relax}%
\providecommand \@@startlink[1]{}%
\providecommand \@@endlink[0]{}%
\providecommand \url  [0]{\begingroup\@sanitize@url \@url }%
\providecommand \@url [1]{\endgroup\@href {#1}{\urlprefix }}%
\providecommand \urlprefix  [0]{URL }%
\providecommand \Eprint [0]{\href }%
\providecommand \doibase [0]{https://doi.org/}%
\providecommand \selectlanguage [0]{\@gobble}%
\providecommand \bibinfo  [0]{\@secondoftwo}%
\providecommand \bibfield  [0]{\@secondoftwo}%
\providecommand \translation [1]{[#1]}%
\providecommand \BibitemOpen [0]{}%
\providecommand \bibitemStop [0]{}%
\providecommand \bibitemNoStop [0]{.\EOS\space}%
\providecommand \EOS [0]{\spacefactor3000\relax}%
\providecommand \BibitemShut  [1]{\csname bibitem#1\endcsname}%
\let\auto@bib@innerbib\@empty
\bibitem [{\citenamefont {Keimer}\ \emph {et~al.}(2015)\citenamefont {Keimer},
  \citenamefont {Kivelson}, \citenamefont {Norman}, \citenamefont {Uchida},\
  and\ \citenamefont {Zaanen}}]{Keimer2015}%
  \BibitemOpen
  \bibfield  {author} {\bibinfo {author} {\bibfnamefont {B.}~\bibnamefont
  {Keimer}}, \bibinfo {author} {\bibfnamefont {S.~A.}\ \bibnamefont
  {Kivelson}}, \bibinfo {author} {\bibfnamefont {M.~R.}\ \bibnamefont
  {Norman}}, \bibinfo {author} {\bibfnamefont {S.}~\bibnamefont {Uchida}},\
  and\ \bibinfo {author} {\bibfnamefont {J.}~\bibnamefont {Zaanen}},\
  }\bibfield  {title} {\emph {\bibinfo {title} {From quantum matter to
  high-temperature superconductivity in copper oxides}},\ }\href
  {https://doi.org/10.1038/nature14165} {\bibfield  {journal} {\bibinfo
  {journal} {Nature}\ }\textbf {\bibinfo {volume} {518}},\ \bibinfo {pages}
  {179} (\bibinfo {year} {2015})}\BibitemShut {NoStop}%
\bibitem [{\citenamefont {Varma}\ \emph {et~al.}(1989)\citenamefont {Varma},
  \citenamefont {Littlewood}, \citenamefont {Schmitt-Rink}, \citenamefont
  {Abrahams},\ and\ \citenamefont {Ruckenstein}}]{MarginalFLPhenomenology}%
  \BibitemOpen
  \bibfield  {author} {\bibinfo {author} {\bibfnamefont {C.~M.}\ \bibnamefont
  {Varma}}, \bibinfo {author} {\bibfnamefont {P.~B.}\ \bibnamefont
  {Littlewood}}, \bibinfo {author} {\bibfnamefont {S.}~\bibnamefont
  {Schmitt-Rink}}, \bibinfo {author} {\bibfnamefont {E.}~\bibnamefont
  {Abrahams}},\ and\ \bibinfo {author} {\bibfnamefont {A.~E.}\ \bibnamefont
  {Ruckenstein}},\ }\bibfield  {title} {\emph {\bibinfo {title} {{Phenomenology
  of the normal state of Cu-O high-temperature superconductors}}},\ }\href
  {https://doi.org/10.1103/PhysRevLett.63.1996} {\bibfield  {journal} {\bibinfo
   {journal} {Phys. Rev. Lett.}\ }\textbf {\bibinfo {volume} {63}},\ \bibinfo
  {pages} {1996} (\bibinfo {year} {1989})}\BibitemShut {NoStop}%
\bibitem [{\citenamefont {Cooper}\ \emph {et~al.}(2009)\citenamefont {Cooper},
  \citenamefont {Wang}, \citenamefont {Vignolle}, \citenamefont {Lipscombe},
  \citenamefont {Hayden}, \citenamefont {Tanabe}, \citenamefont {Adachi},
  \citenamefont {Koike}, \citenamefont {Nohara}, \citenamefont {Takagi},
  \citenamefont {Proust},\ and\ \citenamefont {Hussey}}]{Hussey_foot}%
  \BibitemOpen
  \bibfield  {author} {\bibinfo {author} {\bibfnamefont {R.~A.}\ \bibnamefont
  {Cooper}}, \bibinfo {author} {\bibfnamefont {Y.}~\bibnamefont {Wang}},
  \bibinfo {author} {\bibfnamefont {B.}~\bibnamefont {Vignolle}}, \bibinfo
  {author} {\bibfnamefont {O.~J.}\ \bibnamefont {Lipscombe}}, \bibinfo {author}
  {\bibfnamefont {S.~M.}\ \bibnamefont {Hayden}}, \bibinfo {author}
  {\bibfnamefont {Y.}~\bibnamefont {Tanabe}}, \bibinfo {author} {\bibfnamefont
  {T.}~\bibnamefont {Adachi}}, \bibinfo {author} {\bibfnamefont
  {Y.}~\bibnamefont {Koike}}, \bibinfo {author} {\bibfnamefont
  {M.}~\bibnamefont {Nohara}}, \bibinfo {author} {\bibfnamefont
  {H.}~\bibnamefont {Takagi}}, \bibinfo {author} {\bibfnamefont
  {C.}~\bibnamefont {Proust}},\ and\ \bibinfo {author} {\bibfnamefont {N.~E.}\
  \bibnamefont {Hussey}},\ }\bibfield  {title} {\emph {\bibinfo {title}
  {{Anomalous Criticality in the Electrical Resistivity of
  La$_{2-x}$Sr$_x$CuO$_4$}}},\ }\href {https://doi.org/10.1126/science.1165015}
  {\bibfield  {journal} {\bibinfo  {journal} {Science}\ }\textbf {\bibinfo
  {volume} {323}},\ \bibinfo {pages} {603} (\bibinfo {year}
  {2009})}\BibitemShut {NoStop}%
\bibitem [{\citenamefont {Hussey}\ \emph {et~al.}(2011)\citenamefont {Hussey},
  \citenamefont {Cooper}, \citenamefont {Xu}, \citenamefont {Wang},
  \citenamefont {Mouzopoulou}, \citenamefont {Vignolle},\ and\ \citenamefont
  {Proust}}]{Hussey2011}%
  \BibitemOpen
  \bibfield  {author} {\bibinfo {author} {\bibfnamefont {N.~E.}\ \bibnamefont
  {Hussey}}, \bibinfo {author} {\bibfnamefont {R.~A.}\ \bibnamefont {Cooper}},
  \bibinfo {author} {\bibfnamefont {X.}~\bibnamefont {Xu}}, \bibinfo {author}
  {\bibfnamefont {Y.}~\bibnamefont {Wang}}, \bibinfo {author} {\bibfnamefont
  {I.}~\bibnamefont {Mouzopoulou}}, \bibinfo {author} {\bibfnamefont
  {B.}~\bibnamefont {Vignolle}},\ and\ \bibinfo {author} {\bibfnamefont
  {C.}~\bibnamefont {Proust}},\ }\bibfield  {title} {\emph {\bibinfo {title}
  {Dichotomy in the t-linear resistivity in hole-doped cuprates}},\ }\href
  {https://doi.org/10.1098/rsta.2010.0196} {\bibfield  {journal} {\bibinfo
  {journal} {Philosophical Transactions of the Royal Society A: Mathematical,
  Physical and Engineering Sciences}\ }\textbf {\bibinfo {volume} {369}},\
  \bibinfo {pages} {1626} (\bibinfo {year} {2011})},\ \Eprint
  {https://arxiv.org/abs/https://royalsocietypublishing.org/rsta/article-pdf/369/1941/1626/1346564/rsta.2010.0196.pdf}
  {https://royalsocietypublishing.org/rsta/article-pdf/369/1941/1626/1346564/rsta.2010.0196.pdf}
  \BibitemShut {NoStop}%
\bibitem [{\citenamefont {Kirchner}\ \emph {et~al.}(2020)\citenamefont
  {Kirchner}, \citenamefont {Paschen}, \citenamefont {Chen}, \citenamefont
  {Wirth}, \citenamefont {Feng}, \citenamefont {Thompson},\ and\ \citenamefont
  {Si}}]{Kirchner2020}%
  \BibitemOpen
  \bibfield  {author} {\bibinfo {author} {\bibfnamefont {S.}~\bibnamefont
  {Kirchner}}, \bibinfo {author} {\bibfnamefont {S.}~\bibnamefont {Paschen}},
  \bibinfo {author} {\bibfnamefont {Q.}~\bibnamefont {Chen}}, \bibinfo {author}
  {\bibfnamefont {S.}~\bibnamefont {Wirth}}, \bibinfo {author} {\bibfnamefont
  {D.}~\bibnamefont {Feng}}, \bibinfo {author} {\bibfnamefont {J.~D.}\
  \bibnamefont {Thompson}},\ and\ \bibinfo {author} {\bibfnamefont
  {Q.}~\bibnamefont {Si}},\ }\bibfield  {title} {\emph {\bibinfo {title}
  {Colloquium: Heavy-electron quantum criticality and single-particle
  spectroscopy}},\ }\href {https://doi.org/10.1103/RevModPhys.92.011002}
  {\bibfield  {journal} {\bibinfo  {journal} {Rev. Mod. Phys.}\ }\textbf
  {\bibinfo {volume} {92}},\ \bibinfo {pages} {011002} (\bibinfo {year}
  {2020})}\BibitemShut {NoStop}%
\bibitem [{\citenamefont {Ayres}\ \emph {et~al.}(2021)\citenamefont {Ayres},
  \citenamefont {Berben}, \citenamefont {{\v C}ulo}, \citenamefont {Hsu},
  \citenamefont {van Heumen}, \citenamefont {Huang}, \citenamefont {Zaanen},
  \citenamefont {Kondo}, \citenamefont {Takeuchi}, \citenamefont {Cooper},
  \citenamefont {Putzke}, \citenamefont {Friedemann}, \citenamefont
  {Carrington},\ and\ \citenamefont {Hussey}}]{Ayres2021}%
  \BibitemOpen
  \bibfield  {author} {\bibinfo {author} {\bibfnamefont {J.}~\bibnamefont
  {Ayres}}, \bibinfo {author} {\bibfnamefont {M.}~\bibnamefont {Berben}},
  \bibinfo {author} {\bibfnamefont {M.}~\bibnamefont {{\v C}ulo}}, \bibinfo
  {author} {\bibfnamefont {Y.~T.}\ \bibnamefont {Hsu}}, \bibinfo {author}
  {\bibfnamefont {E.}~\bibnamefont {van Heumen}}, \bibinfo {author}
  {\bibfnamefont {Y.}~\bibnamefont {Huang}}, \bibinfo {author} {\bibfnamefont
  {J.}~\bibnamefont {Zaanen}}, \bibinfo {author} {\bibfnamefont
  {T.}~\bibnamefont {Kondo}}, \bibinfo {author} {\bibfnamefont
  {T.}~\bibnamefont {Takeuchi}}, \bibinfo {author} {\bibfnamefont {J.~R.}\
  \bibnamefont {Cooper}}, \bibinfo {author} {\bibfnamefont {C.}~\bibnamefont
  {Putzke}}, \bibinfo {author} {\bibfnamefont {S.}~\bibnamefont {Friedemann}},
  \bibinfo {author} {\bibfnamefont {A.}~\bibnamefont {Carrington}},\ and\
  \bibinfo {author} {\bibfnamefont {N.~E.}\ \bibnamefont {Hussey}},\ }\bibfield
   {title} {\emph {\bibinfo {title} {Incoherent transport across the
  strange-metal regime of overdoped cuprates}},\ }\href
  {https://doi.org/10.1038/s41586-021-03622-z} {\bibfield  {journal} {\bibinfo
  {journal} {Nature}\ }\textbf {\bibinfo {volume} {595}},\ \bibinfo {pages}
  {661} (\bibinfo {year} {2021})}\BibitemShut {NoStop}%
\bibitem [{\citenamefont {Hartnoll}\ and\ \citenamefont
  {Mackenzie}(2022)}]{Hartnoll2022}%
  \BibitemOpen
  \bibfield  {author} {\bibinfo {author} {\bibfnamefont {S.~A.}\ \bibnamefont
  {Hartnoll}}\ and\ \bibinfo {author} {\bibfnamefont {A.~P.}\ \bibnamefont
  {Mackenzie}},\ }\bibfield  {title} {\emph {\bibinfo {title} {Colloquium:
  Planckian dissipation in metals}},\ }\href
  {https://doi.org/10.1103/RevModPhys.94.041002} {\bibfield  {journal}
  {\bibinfo  {journal} {Rev. Mod. Phys.}\ }\textbf {\bibinfo {volume} {94}},\
  \bibinfo {pages} {041002} (\bibinfo {year} {2022})}\BibitemShut {NoStop}%
\bibitem [{\citenamefont {Sachdev}(2011)}]{Sachdev_2011}%
  \BibitemOpen
  \bibfield  {author} {\bibinfo {author} {\bibfnamefont {S.}~\bibnamefont
  {Sachdev}},\ }\href@noop {} {\emph {\bibinfo {title} {Quantum Phase
  Transitions}}},\ \bibinfo {edition} {2nd}\ ed.\ (\bibinfo  {publisher}
  {Cambridge University Press},\ \bibinfo {year} {2011})\BibitemShut {NoStop}%
\bibitem [{\citenamefont {Sachdev}(2025)}]{Sachdev2025}%
  \BibitemOpen
  \bibfield  {author} {\bibinfo {author} {\bibfnamefont {S.}~\bibnamefont
  {Sachdev}},\ }\bibfield  {title} {\emph {\bibinfo {title} {The foot, the fan,
  and the cuprate phase diagram: Fermi-volume-changing quantum phase
  transitions}},\ }\href
  {https://doi.org/https://doi.org/10.1016/j.physc.2025.1354707} {\bibfield
  {journal} {\bibinfo  {journal} {Physica C: Superconductivity and its
  Applications}\ }\textbf {\bibinfo {volume} {633}},\ \bibinfo {pages}
  {1354707} (\bibinfo {year} {2025})}\BibitemShut {NoStop}%
\bibitem [{\citenamefont {Hertz}(1976)}]{Hertz1976}%
  \BibitemOpen
  \bibfield  {author} {\bibinfo {author} {\bibfnamefont {J.~A.}\ \bibnamefont
  {Hertz}},\ }\bibfield  {title} {\emph {\bibinfo {title} {Quantum critical
  phenomena}},\ }\href {https://doi.org/10.1103/PhysRevB.14.1165} {\bibfield
  {journal} {\bibinfo  {journal} {Phys. Rev. B}\ }\textbf {\bibinfo {volume}
  {14}},\ \bibinfo {pages} {1165} (\bibinfo {year} {1976})}\BibitemShut
  {NoStop}%
\bibitem [{\citenamefont {Moriya}\ and\ \citenamefont
  {Kawabata}(1973)}]{Moriya1973}%
  \BibitemOpen
  \bibfield  {author} {\bibinfo {author} {\bibfnamefont {T.}~\bibnamefont
  {Moriya}}\ and\ \bibinfo {author} {\bibfnamefont {A.}~\bibnamefont
  {Kawabata}},\ }\bibfield  {title} {\emph {\bibinfo {title} {Effect of spin
  fluctuations on itinerant electron ferromagnetism}},\ }\href
  {https://doi.org/10.1143/JPSJ.34.639} {\bibfield  {journal} {\bibinfo
  {journal} {Journal of the Physical Society of Japan}\ }\textbf {\bibinfo
  {volume} {34}},\ \bibinfo {pages} {639} (\bibinfo {year} {1973})}\BibitemShut
  {NoStop}%
\bibitem [{\citenamefont {Millis}(1993)}]{Millis1993}%
  \BibitemOpen
  \bibfield  {author} {\bibinfo {author} {\bibfnamefont {A.~J.}\ \bibnamefont
  {Millis}},\ }\bibfield  {title} {\emph {\bibinfo {title} {Effect of a nonzero
  temperature on quantum critical points in itinerant fermion systems}},\
  }\href {https://doi.org/10.1103/PhysRevB.48.7183} {\bibfield  {journal}
  {\bibinfo  {journal} {Phys. Rev. B}\ }\textbf {\bibinfo {volume} {48}},\
  \bibinfo {pages} {7183} (\bibinfo {year} {1993})}\BibitemShut {NoStop}%
\bibitem [{\citenamefont {Zhang}\ and\ \citenamefont
  {Sachdev}(2020{\natexlab{a}})}]{Zhang2020}%
  \BibitemOpen
  \bibfield  {author} {\bibinfo {author} {\bibfnamefont {Y.-H.}\ \bibnamefont
  {Zhang}}\ and\ \bibinfo {author} {\bibfnamefont {S.}~\bibnamefont
  {Sachdev}},\ }\bibfield  {title} {\emph {\bibinfo {title} {From the pseudogap
  metal to the fermi liquid using ancilla qubits}},\ }\href
  {https://doi.org/10.1103/PhysRevResearch.2.023172} {\bibfield  {journal}
  {\bibinfo  {journal} {Phys. Rev. Res.}\ }\textbf {\bibinfo {volume} {2}},\
  \bibinfo {pages} {023172} (\bibinfo {year} {2020}{\natexlab{a}})}\BibitemShut
  {NoStop}%
\bibitem [{\citenamefont {Zhang}\ and\ \citenamefont
  {Sachdev}(2020{\natexlab{b}})}]{Zhang2020Decon}%
  \BibitemOpen
  \bibfield  {author} {\bibinfo {author} {\bibfnamefont {Y.-H.}\ \bibnamefont
  {Zhang}}\ and\ \bibinfo {author} {\bibfnamefont {S.}~\bibnamefont
  {Sachdev}},\ }\bibfield  {title} {\emph {\bibinfo {title} {Deconfined
  criticality and ghost fermi surfaces at the onset of antiferromagnetism in a
  metal}},\ }\href {https://doi.org/10.1103/PhysRevB.102.155124} {\bibfield
  {journal} {\bibinfo  {journal} {Phys. Rev. B}\ }\textbf {\bibinfo {volume}
  {102}},\ \bibinfo {pages} {155124} (\bibinfo {year}
  {2020}{\natexlab{b}})}\BibitemShut {NoStop}%
\bibitem [{\citenamefont {Mascot}\ \emph {et~al.}(2022)\citenamefont {Mascot},
  \citenamefont {Nikolaenko}, \citenamefont {Tikhanovskaya}, \citenamefont
  {Zhang}, \citenamefont {Morr},\ and\ \citenamefont {Sachdev}}]{Mascot2022}%
  \BibitemOpen
  \bibfield  {author} {\bibinfo {author} {\bibfnamefont {E.}~\bibnamefont
  {Mascot}}, \bibinfo {author} {\bibfnamefont {A.}~\bibnamefont {Nikolaenko}},
  \bibinfo {author} {\bibfnamefont {M.}~\bibnamefont {Tikhanovskaya}}, \bibinfo
  {author} {\bibfnamefont {Y.-H.}\ \bibnamefont {Zhang}}, \bibinfo {author}
  {\bibfnamefont {D.~K.}\ \bibnamefont {Morr}},\ and\ \bibinfo {author}
  {\bibfnamefont {S.}~\bibnamefont {Sachdev}},\ }\bibfield  {title} {\emph
  {\bibinfo {title} {Electronic spectra with paramagnon fractionalization in
  the single-band hubbard model}},\ }\href
  {https://doi.org/10.1103/PhysRevB.105.075146} {\bibfield  {journal} {\bibinfo
   {journal} {Phys. Rev. B}\ }\textbf {\bibinfo {volume} {105}},\ \bibinfo
  {pages} {075146} (\bibinfo {year} {2022})}\BibitemShut {NoStop}%
\bibitem [{\citenamefont {Nikolaenko}\ \emph {et~al.}(2023)\citenamefont
  {Nikolaenko}, \citenamefont {von Milczewski}, \citenamefont {Joshi},\ and\
  \citenamefont {Sachdev}}]{Nikolaenko2023}%
  \BibitemOpen
  \bibfield  {author} {\bibinfo {author} {\bibfnamefont {A.}~\bibnamefont
  {Nikolaenko}}, \bibinfo {author} {\bibfnamefont {J.}~\bibnamefont {von
  Milczewski}}, \bibinfo {author} {\bibfnamefont {D.~G.}\ \bibnamefont
  {Joshi}},\ and\ \bibinfo {author} {\bibfnamefont {S.}~\bibnamefont
  {Sachdev}},\ }\bibfield  {title} {\emph {\bibinfo {title} {Spin density wave,
  fermi liquid, and fractionalized phases in a theory of antiferromagnetic
  metals using paramagnons and bosonic spinons}},\ }\href
  {https://doi.org/10.1103/PhysRevB.108.045123} {\bibfield  {journal} {\bibinfo
   {journal} {Phys. Rev. B}\ }\textbf {\bibinfo {volume} {108}},\ \bibinfo
  {pages} {045123} (\bibinfo {year} {2023})}\BibitemShut {NoStop}%
\bibitem [{\citenamefont {Lunts}\ \emph {et~al.}(2025)\citenamefont {Lunts},
  \citenamefont {Patel},\ and\ \citenamefont {Sachdev}}]{Lunts2025}%
  \BibitemOpen
  \bibfield  {author} {\bibinfo {author} {\bibfnamefont {P.}~\bibnamefont
  {Lunts}}, \bibinfo {author} {\bibfnamefont {A.~A.}\ \bibnamefont {Patel}},\
  and\ \bibinfo {author} {\bibfnamefont {S.}~\bibnamefont {Sachdev}},\
  }\bibfield  {title} {\emph {\bibinfo {title} {Thermopower across
  fermi-volume-changing quantum phase transitions without translational
  symmetry breaking}},\ }\href {https://doi.org/10.1103/3639-byq1} {\bibfield
  {journal} {\bibinfo  {journal} {Phys. Rev. B}\ }\textbf {\bibinfo {volume}
  {111}},\ \bibinfo {pages} {245151} (\bibinfo {year} {2025})}\BibitemShut
  {NoStop}%
\bibitem [{\citenamefont {Bonetti}\ \emph {et~al.}(2026)\citenamefont
  {Bonetti}, \citenamefont {Christos}, \citenamefont {Nikolaenko},
  \citenamefont {Patel},\ and\ \citenamefont {Sachdev}}]{Bonetti_2026}%
  \BibitemOpen
  \bibfield  {author} {\bibinfo {author} {\bibfnamefont {P.~M.}\ \bibnamefont
  {Bonetti}}, \bibinfo {author} {\bibfnamefont {M.}~\bibnamefont {Christos}},
  \bibinfo {author} {\bibfnamefont {A.}~\bibnamefont {Nikolaenko}}, \bibinfo
  {author} {\bibfnamefont {A.~A.}\ \bibnamefont {Patel}},\ and\ \bibinfo
  {author} {\bibfnamefont {S.}~\bibnamefont {Sachdev}},\ }\bibfield  {title}
  {\emph {\bibinfo {title} {Fractionalized fermi liquids and the cuprate phase
  diagram}},\ }\href {https://doi.org/10.1088/1361-6633/ae530d} {\bibfield
  {journal} {\bibinfo  {journal} {Reports on Progress in Physics}\ }\textbf
  {\bibinfo {volume} {89}},\ \bibinfo {pages} {044501} (\bibinfo {year}
  {2026})}\BibitemShut {NoStop}%
\bibitem [{\citenamefont {Bonesteel}\ \emph {et~al.}(1996)\citenamefont
  {Bonesteel}, \citenamefont {McDonald},\ and\ \citenamefont
  {Nayak}}]{Bonesteel1996}%
  \BibitemOpen
  \bibfield  {author} {\bibinfo {author} {\bibfnamefont {N.~E.}\ \bibnamefont
  {Bonesteel}}, \bibinfo {author} {\bibfnamefont {I.~A.}\ \bibnamefont
  {McDonald}},\ and\ \bibinfo {author} {\bibfnamefont {C.}~\bibnamefont
  {Nayak}},\ }\bibfield  {title} {\emph {\bibinfo {title} {Gauge fields and
  pairing in double-layer composite fermion metals}},\ }\href
  {https://doi.org/10.1103/PhysRevLett.77.3009} {\bibfield  {journal} {\bibinfo
   {journal} {Phys. Rev. Lett.}\ }\textbf {\bibinfo {volume} {77}},\ \bibinfo
  {pages} {3009} (\bibinfo {year} {1996})}\BibitemShut {NoStop}%
\bibitem [{\citenamefont {Son}(1999)}]{Son:1998uk}%
  \BibitemOpen
  \bibfield  {author} {\bibinfo {author} {\bibfnamefont {D.~T.}\ \bibnamefont
  {Son}},\ }\bibfield  {title} {\emph {\bibinfo {title} {{Superconductivity by
  long range color magnetic interaction in high density quark matter}}},\
  }\href {https://doi.org/10.1103/PhysRevD.59.094019} {\bibfield  {journal}
  {\bibinfo  {journal} {Phys. Rev. D}\ }\textbf {\bibinfo {volume} {59}},\
  \bibinfo {pages} {094019} (\bibinfo {year} {1999})},\ \Eprint
  {https://arxiv.org/abs/hep-ph/9812287} {arXiv:hep-ph/9812287} \BibitemShut
  {NoStop}%
\bibitem [{\citenamefont {{Abanov}}\ \emph {et~al.}(2001)\citenamefont
  {{Abanov}}, \citenamefont {{Chubukov}},\ and\ \citenamefont
  {{Finkel'stein}}}]{ChubukovFinkelstein}%
  \BibitemOpen
  \bibfield  {author} {\bibinfo {author} {\bibfnamefont {A.}~\bibnamefont
  {{Abanov}}}, \bibinfo {author} {\bibfnamefont {A.~V.}\ \bibnamefont
  {{Chubukov}}},\ and\ \bibinfo {author} {\bibfnamefont {A.~M.}\ \bibnamefont
  {{Finkel'stein}}},\ }\bibfield  {title} {\emph {\bibinfo {title} {{Coherent
  vs. incoherent pairing in 2D systems near magnetic instability}}},\ }\href
  {https://doi.org/10.1209/epl/i2001-00266-0} {\bibfield  {journal} {\bibinfo
  {journal} {EPL (Europhysics Letters)}\ }\textbf {\bibinfo {volume} {54}},\
  \bibinfo {pages} {488} (\bibinfo {year} {2001})},\ \Eprint
  {https://arxiv.org/abs/cond-mat/9911445} {arXiv:cond-mat/9911445
  [cond-mat.supr-con]} \BibitemShut {NoStop}%
\bibitem [{\citenamefont {{Abanov}}\ \emph {et~al.}(2003)\citenamefont
  {{Abanov}}, \citenamefont {{Chubukov}},\ and\ \citenamefont
  {{Schmalian}}}]{Abanov_review}%
  \BibitemOpen
  \bibfield  {author} {\bibinfo {author} {\bibfnamefont {A.}~\bibnamefont
  {{Abanov}}}, \bibinfo {author} {\bibfnamefont {A.~V.}\ \bibnamefont
  {{Chubukov}}},\ and\ \bibinfo {author} {\bibfnamefont {J.}~\bibnamefont
  {{Schmalian}}},\ }\bibfield  {title} {\emph {\bibinfo {title}
  {{Quantum-critical theory of the spin-fermion model and its application to
  cuprates. Normal state analysis}}},\ }\href
  {https://doi.org/10.1080/0001873021000057123} {\bibfield  {journal} {\bibinfo
   {journal} {Advances in Physics}\ }\textbf {\bibinfo {volume} {52}},\
  \bibinfo {pages} {119} (\bibinfo {year} {2003})},\ \Eprint
  {https://arxiv.org/abs/cond-mat/0107421} {arXiv:cond-mat/0107421
  [cond-mat.supr-con]} \BibitemShut {NoStop}%
\bibitem [{\citenamefont {{Moon}}\ and\ \citenamefont
  {{Chubukov}}(2010)}]{Moon2010}%
  \BibitemOpen
  \bibfield  {author} {\bibinfo {author} {\bibfnamefont {E.-G.}\ \bibnamefont
  {{Moon}}}\ and\ \bibinfo {author} {\bibfnamefont {A.}~\bibnamefont
  {{Chubukov}}},\ }\bibfield  {title} {\emph {\bibinfo {title}
  {{Quantum-critical Pairing with Varying Exponents}}},\ }\href
  {https://doi.org/10.1007/s10909-010-0199-y} {\bibfield  {journal} {\bibinfo
  {journal} {Journal of Low Temperature Physics}\ }\textbf {\bibinfo {volume}
  {161}},\ \bibinfo {pages} {263} (\bibinfo {year} {2010})},\ \Eprint
  {https://arxiv.org/abs/1005.0356} {arXiv:1005.0356 [cond-mat.supr-con]}
  \BibitemShut {NoStop}%
\bibitem [{\citenamefont {{Metlitski}}\ and\ \citenamefont
  {{Sachdev}}(2010)}]{Metlitski10}%
  \BibitemOpen
  \bibfield  {author} {\bibinfo {author} {\bibfnamefont {M.~A.}\ \bibnamefont
  {{Metlitski}}}\ and\ \bibinfo {author} {\bibfnamefont {S.}~\bibnamefont
  {{Sachdev}}},\ }\bibfield  {title} {\emph {\bibinfo {title} {{Quantum phase
  transitions of metals in two spatial dimensions. II. Spin density wave
  order}}},\ }\href {https://doi.org/10.1103/PhysRevB.82.075128} {\bibfield
  {journal} {\bibinfo  {journal} {Phys. Rev. B}\ }\textbf {\bibinfo {volume}
  {82}},\ \bibinfo {eid} {075128} (\bibinfo {year} {2010})},\ \Eprint
  {https://arxiv.org/abs/1005.1288} {arXiv:1005.1288 [cond-mat.str-el]}
  \BibitemShut {NoStop}%
\bibitem [{\citenamefont {{Wang}}\ and\ \citenamefont
  {{Chubukov}}(2013)}]{Wang13}%
  \BibitemOpen
  \bibfield  {author} {\bibinfo {author} {\bibfnamefont {Y.}~\bibnamefont
  {{Wang}}}\ and\ \bibinfo {author} {\bibfnamefont {A.~V.}\ \bibnamefont
  {{Chubukov}}},\ }\bibfield  {title} {\emph {\bibinfo {title}
  {{Superconductivity at the Onset of Spin-Density-Wave Order in a Metal}}},\
  }\href {https://doi.org/10.1103/PhysRevLett.110.127001} {\bibfield  {journal}
  {\bibinfo  {journal} {Phys. Rev. Lett.}\ }\textbf {\bibinfo {volume} {110}},\
  \bibinfo {eid} {127001} (\bibinfo {year} {2013})},\ \Eprint
  {https://arxiv.org/abs/1210.2408} {arXiv:1210.2408 [cond-mat.supr-con]}
  \BibitemShut {NoStop}%
\bibitem [{\citenamefont {Metlitski}\ \emph {et~al.}(2015)\citenamefont
  {Metlitski}, \citenamefont {Mross}, \citenamefont {Sachdev},\ and\
  \citenamefont {Senthil}}]{Metlitski15}%
  \BibitemOpen
  \bibfield  {author} {\bibinfo {author} {\bibfnamefont {M.~A.}\ \bibnamefont
  {Metlitski}}, \bibinfo {author} {\bibfnamefont {D.~F.}\ \bibnamefont
  {Mross}}, \bibinfo {author} {\bibfnamefont {S.}~\bibnamefont {Sachdev}},\
  and\ \bibinfo {author} {\bibfnamefont {T.}~\bibnamefont {Senthil}},\
  }\bibfield  {title} {\emph {\bibinfo {title} {{Cooper pairing in non-Fermi
  liquids}}},\ }\href {https://doi.org/10.1103/PhysRevB.91.115111} {\bibfield
  {journal} {\bibinfo  {journal} {Phys. Rev. B}\ }\textbf {\bibinfo {volume}
  {91}},\ \bibinfo {pages} {115111} (\bibinfo {year} {2015})},\ \Eprint
  {https://arxiv.org/abs/1403.3694} {arXiv:1403.3694 [cond-mat.str-el]}
  \BibitemShut {NoStop}%
\bibitem [{\citenamefont {Raghu}\ \emph {et~al.}(2015)\citenamefont {Raghu},
  \citenamefont {Torroba},\ and\ \citenamefont {Wang}}]{Raghu2015}%
  \BibitemOpen
  \bibfield  {author} {\bibinfo {author} {\bibfnamefont {S.}~\bibnamefont
  {Raghu}}, \bibinfo {author} {\bibfnamefont {G.}~\bibnamefont {Torroba}},\
  and\ \bibinfo {author} {\bibfnamefont {H.}~\bibnamefont {Wang}},\ }\bibfield
  {title} {\emph {\bibinfo {title} {Metallic quantum critical points with
  finite bcs couplings}},\ }\href {https://doi.org/10.1103/PhysRevB.92.205104}
  {\bibfield  {journal} {\bibinfo  {journal} {Phys. Rev. B}\ }\textbf {\bibinfo
  {volume} {92}},\ \bibinfo {pages} {205104} (\bibinfo {year}
  {2015})}\BibitemShut {NoStop}%
\bibitem [{\citenamefont {Esterlis}\ and\ \citenamefont
  {Schmalian}(2019)}]{Esterlis:2019ola}%
  \BibitemOpen
  \bibfield  {author} {\bibinfo {author} {\bibfnamefont {I.}~\bibnamefont
  {Esterlis}}\ and\ \bibinfo {author} {\bibfnamefont {J.}~\bibnamefont
  {Schmalian}},\ }\bibfield  {title} {\emph {\bibinfo {title} {{Cooper pairing
  of incoherent electrons: an electron-phonon version of the Sachdev-Ye-Kitaev
  model}}},\ }\href {https://doi.org/10.1103/PhysRevB.100.115132} {\bibfield
  {journal} {\bibinfo  {journal} {Phys. Rev. B}\ }\textbf {\bibinfo {volume}
  {100}},\ \bibinfo {pages} {115132} (\bibinfo {year} {2019})},\ \Eprint
  {https://arxiv.org/abs/1906.04747} {arXiv:1906.04747 [cond-mat.str-el]}
  \BibitemShut {NoStop}%
\bibitem [{\citenamefont {{Chowdhury}}\ and\ \citenamefont
  {{Berg}}(2020)}]{Debanjan20}%
  \BibitemOpen
  \bibfield  {author} {\bibinfo {author} {\bibfnamefont {D.}~\bibnamefont
  {{Chowdhury}}}\ and\ \bibinfo {author} {\bibfnamefont {E.}~\bibnamefont
  {{Berg}}},\ }\bibfield  {title} {\emph {\bibinfo {title} {{The unreasonable
  effectiveness of Eliashberg theory for pairing of non-Fermi liquids}}},\
  }\href {https://doi.org/10.1016/j.aop.2020.168125} {\bibfield  {journal}
  {\bibinfo  {journal} {Annals of Physics}\ }\textbf {\bibinfo {volume}
  {417}},\ \bibinfo {eid} {168125} (\bibinfo {year} {2020})},\ \Eprint
  {https://arxiv.org/abs/1912.07646} {arXiv:1912.07646 [cond-mat.supr-con]}
  \BibitemShut {NoStop}%
\bibitem [{\citenamefont {{Hauck}}\ \emph {et~al.}(2020)\citenamefont
  {{Hauck}}, \citenamefont {{Klug}}, \citenamefont {{Esterlis}},\ and\
  \citenamefont {{Schmalian}}}]{Esterlis20}%
  \BibitemOpen
  \bibfield  {author} {\bibinfo {author} {\bibfnamefont {D.}~\bibnamefont
  {{Hauck}}}, \bibinfo {author} {\bibfnamefont {M.~J.}\ \bibnamefont {{Klug}}},
  \bibinfo {author} {\bibfnamefont {I.}~\bibnamefont {{Esterlis}}},\ and\
  \bibinfo {author} {\bibfnamefont {J.}~\bibnamefont {{Schmalian}}},\
  }\bibfield  {title} {\emph {\bibinfo {title} {{Eliashberg equations for an
  electron-phonon version of the Sachdev-Ye-Kitaev model: Pair breaking in
  non-Fermi liquid superconductors}}},\ }\href
  {https://doi.org/10.1016/j.aop.2020.168120} {\bibfield  {journal} {\bibinfo
  {journal} {Annals of Physics}\ }\textbf {\bibinfo {volume} {417}},\ \bibinfo
  {eid} {168120} (\bibinfo {year} {2020})},\ \Eprint
  {https://arxiv.org/abs/1911.04328} {arXiv:1911.04328 [cond-mat.str-el]}
  \BibitemShut {NoStop}%
\bibitem [{\citenamefont {Chubukov}\ \emph {et~al.}(2020)\citenamefont
  {Chubukov}, \citenamefont {Abanov}, \citenamefont {Wang},\ and\ \citenamefont
  {Wu}}]{Chubukov20}%
  \BibitemOpen
  \bibfield  {author} {\bibinfo {author} {\bibfnamefont {A.~V.}\ \bibnamefont
  {Chubukov}}, \bibinfo {author} {\bibfnamefont {A.}~\bibnamefont {Abanov}},
  \bibinfo {author} {\bibfnamefont {Y.}~\bibnamefont {Wang}},\ and\ \bibinfo
  {author} {\bibfnamefont {Y.-M.}\ \bibnamefont {Wu}},\ }\bibfield  {title}
  {\emph {\bibinfo {title} {{The interplay between superconductivity and
  non-Fermi liquid at a quantum-critical point in a metal}}},\ }\href
  {https://doi.org/https://doi.org/10.1016/j.aop.2020.168142} {\bibfield
  {journal} {\bibinfo  {journal} {Annals of Physics}\ }\textbf {\bibinfo
  {volume} {417}},\ \bibinfo {pages} {168142} (\bibinfo {year} {2020})},\
  \bibinfo {note} {{Eliashberg theory at 60: Strong-coupling superconductivity
  and beyond}}\BibitemShut {NoStop}%
\bibitem [{\citenamefont {Esterlis}\ \emph {et~al.}(2021)\citenamefont
  {Esterlis}, \citenamefont {Guo}, \citenamefont {Patel},\ and\ \citenamefont
  {Sachdev}}]{Esterlis:2021eth}%
  \BibitemOpen
  \bibfield  {author} {\bibinfo {author} {\bibfnamefont {I.}~\bibnamefont
  {Esterlis}}, \bibinfo {author} {\bibfnamefont {H.}~\bibnamefont {Guo}},
  \bibinfo {author} {\bibfnamefont {A.~A.}\ \bibnamefont {Patel}},\ and\
  \bibinfo {author} {\bibfnamefont {S.}~\bibnamefont {Sachdev}},\ }\bibfield
  {title} {\emph {\bibinfo {title} {{Large $N$ theory of critical Fermi
  surfaces}}},\ }\href {https://doi.org/10.1103/PhysRevB.103.235129} {\bibfield
   {journal} {\bibinfo  {journal} {Phys. Rev. B}\ }\textbf {\bibinfo {volume}
  {103}},\ \bibinfo {pages} {235129} (\bibinfo {year} {2021})},\ \Eprint
  {https://arxiv.org/abs/2103.08615} {arXiv:2103.08615 [cond-mat.str-el]}
  \BibitemShut {NoStop}%
\bibitem [{\citenamefont {Inkof}\ \emph {et~al.}(2022)\citenamefont {Inkof},
  \citenamefont {Schalm},\ and\ \citenamefont {Schmalian}}]{Inkof:2021ohk}%
  \BibitemOpen
  \bibfield  {author} {\bibinfo {author} {\bibfnamefont {G.~A.}\ \bibnamefont
  {Inkof}}, \bibinfo {author} {\bibfnamefont {K.}~\bibnamefont {Schalm}},\ and\
  \bibinfo {author} {\bibfnamefont {J.}~\bibnamefont {Schmalian}},\ }\bibfield
  {title} {\emph {\bibinfo {title} {{Quantum critical Eliashberg theory, the
  Sachdev-Ye-Kitaev superconductor and their holographic duals}}},\ }\href
  {https://doi.org/10.1038/s41535-022-00460-8} {\bibfield  {journal} {\bibinfo
  {journal} {npj Quant. Mater.}\ }\textbf {\bibinfo {volume} {7}},\ \bibinfo
  {pages} {56} (\bibinfo {year} {2022})},\ \Eprint
  {https://arxiv.org/abs/2108.11392} {arXiv:2108.11392 [cond-mat.str-el]}
  \BibitemShut {NoStop}%
\bibitem [{\citenamefont {{Esterlis}}\ and\ \citenamefont
  {{Schmalian}}(2025)}]{Esterlis25}%
  \BibitemOpen
  \bibfield  {author} {\bibinfo {author} {\bibfnamefont {I.}~\bibnamefont
  {{Esterlis}}}\ and\ \bibinfo {author} {\bibfnamefont {J.}~\bibnamefont
  {{Schmalian}}},\ }\bibfield  {title} {\emph {\bibinfo {title} {{Quantum
  Critical Eliashberg Theory}}},\ }\href
  {https://doi.org/10.48550/arXiv.2506.11952} {\bibfield  {journal} {\bibinfo
  {journal} {arXiv e-prints}\ ,\ \bibinfo {eid} {arXiv:2506.11952}} (\bibinfo
  {year} {2025})},\ \Eprint {https://arxiv.org/abs/2506.11952}
  {arXiv:2506.11952 [cond-mat.str-el]} \BibitemShut {NoStop}%
\bibitem [{\citenamefont {Chakravarty}\ \emph {et~al.}(1998)\citenamefont
  {Chakravarty}, \citenamefont {Yin},\ and\ \citenamefont
  {Abrahams}}]{Chakravarty1998}%
  \BibitemOpen
  \bibfield  {author} {\bibinfo {author} {\bibfnamefont {S.}~\bibnamefont
  {Chakravarty}}, \bibinfo {author} {\bibfnamefont {L.}~\bibnamefont {Yin}},\
  and\ \bibinfo {author} {\bibfnamefont {E.}~\bibnamefont {Abrahams}},\
  }\bibfield  {title} {\emph {\bibinfo {title} {Interactions and scaling in a
  disordered two-dimensional metal}},\ }\href
  {https://doi.org/10.1103/PhysRevB.58.R559} {\bibfield  {journal} {\bibinfo
  {journal} {Phys. Rev. B}\ }\textbf {\bibinfo {volume} {58}},\ \bibinfo
  {pages} {R559(R)} (\bibinfo {year} {1998})}\BibitemShut {NoStop}%
\bibitem [{\citenamefont {Rosch}(1999)}]{Rosch1999}%
  \BibitemOpen
  \bibfield  {author} {\bibinfo {author} {\bibfnamefont {A.}~\bibnamefont
  {Rosch}},\ }\bibfield  {title} {\emph {\bibinfo {title} {Interplay of
  disorder and spin fluctuations in the resistivity near a quantum critical
  point}},\ }\href {https://doi.org/10.1103/PhysRevLett.82.4280} {\bibfield
  {journal} {\bibinfo  {journal} {Phys. Rev. Lett.}\ }\textbf {\bibinfo
  {volume} {82}},\ \bibinfo {pages} {4280} (\bibinfo {year}
  {1999})}\BibitemShut {NoStop}%
\bibitem [{\citenamefont {Maebashi}\ \emph {et~al.}(2002)\citenamefont
  {Maebashi}, \citenamefont {Miyake},\ and\ \citenamefont
  {Varma}}]{Maebashi2002}%
  \BibitemOpen
  \bibfield  {author} {\bibinfo {author} {\bibfnamefont {H.}~\bibnamefont
  {Maebashi}}, \bibinfo {author} {\bibfnamefont {K.}~\bibnamefont {Miyake}},\
  and\ \bibinfo {author} {\bibfnamefont {C.~M.}\ \bibnamefont {Varma}},\
  }\bibfield  {title} {\emph {\bibinfo {title} {Singular effects of impurities
  near the ferromagnetic quantum-critical point}},\ }\href
  {https://doi.org/10.1103/PhysRevLett.88.226403} {\bibfield  {journal}
  {\bibinfo  {journal} {Phys. Rev. Lett.}\ }\textbf {\bibinfo {volume} {88}},\
  \bibinfo {pages} {226403} (\bibinfo {year} {2002})}\BibitemShut {NoStop}%
\bibitem [{\citenamefont {Paul}\ \emph {et~al.}(2005)\citenamefont {Paul},
  \citenamefont {P\'epin}, \citenamefont {Narozhny},\ and\ \citenamefont
  {Maslov}}]{Paul2005}%
  \BibitemOpen
  \bibfield  {author} {\bibinfo {author} {\bibfnamefont {I.}~\bibnamefont
  {Paul}}, \bibinfo {author} {\bibfnamefont {C.}~\bibnamefont {P\'epin}},
  \bibinfo {author} {\bibfnamefont {B.~N.}\ \bibnamefont {Narozhny}},\ and\
  \bibinfo {author} {\bibfnamefont {D.~L.}\ \bibnamefont {Maslov}},\ }\bibfield
   {title} {\emph {\bibinfo {title} {Quantum correction to conductivity close
  to a ferromagnetic quantum critical point in two dimensions}},\ }\href
  {https://doi.org/10.1103/PhysRevLett.95.017206} {\bibfield  {journal}
  {\bibinfo  {journal} {Phys. Rev. Lett.}\ }\textbf {\bibinfo {volume} {95}},\
  \bibinfo {pages} {017206} (\bibinfo {year} {2005})}\BibitemShut {NoStop}%
\bibitem [{\citenamefont {Nosov}\ \emph {et~al.}(2020)\citenamefont {Nosov},
  \citenamefont {Burmistrov},\ and\ \citenamefont {Raghu}}]{Nosov2020}%
  \BibitemOpen
  \bibfield  {author} {\bibinfo {author} {\bibfnamefont {P.~A.}\ \bibnamefont
  {Nosov}}, \bibinfo {author} {\bibfnamefont {I.~S.}\ \bibnamefont
  {Burmistrov}},\ and\ \bibinfo {author} {\bibfnamefont {S.}~\bibnamefont
  {Raghu}},\ }\bibfield  {title} {\emph {\bibinfo {title} {Interaction-induced
  metallicity in a two-dimensional disordered non-fermi liquid}},\ }\href
  {https://doi.org/10.1103/PhysRevLett.125.256604} {\bibfield  {journal}
  {\bibinfo  {journal} {Phys. Rev. Lett.}\ }\textbf {\bibinfo {volume} {125}},\
  \bibinfo {pages} {256604} (\bibinfo {year} {2020})}\BibitemShut {NoStop}%
\bibitem [{\citenamefont {Halbinger}\ and\ \citenamefont
  {Punk}(2021)}]{Halbinger2021}%
  \BibitemOpen
  \bibfield  {author} {\bibinfo {author} {\bibfnamefont {J.}~\bibnamefont
  {Halbinger}}\ and\ \bibinfo {author} {\bibfnamefont {M.}~\bibnamefont
  {Punk}},\ }\bibfield  {title} {\emph {\bibinfo {title} {Quenched disorder at
  antiferromagnetic quantum critical points in two-dimensional metals}},\
  }\href {https://doi.org/10.1103/PhysRevB.103.235157} {\bibfield  {journal}
  {\bibinfo  {journal} {Phys. Rev. B}\ }\textbf {\bibinfo {volume} {103}},\
  \bibinfo {pages} {235157} (\bibinfo {year} {2021})}\BibitemShut {NoStop}%
\bibitem [{\citenamefont {Wu}\ \emph {et~al.}(2022)\citenamefont {Wu},
  \citenamefont {Liao},\ and\ \citenamefont {Foster}}]{Wu2022}%
  \BibitemOpen
  \bibfield  {author} {\bibinfo {author} {\bibfnamefont {T.~C.}\ \bibnamefont
  {Wu}}, \bibinfo {author} {\bibfnamefont {Y.}~\bibnamefont {Liao}},\ and\
  \bibinfo {author} {\bibfnamefont {M.~S.}\ \bibnamefont {Foster}},\ }\bibfield
   {title} {\emph {\bibinfo {title} {Quantum interference of hydrodynamic modes
  in a dirty marginal fermi liquid}},\ }\href
  {https://doi.org/10.1103/PhysRevB.106.155108} {\bibfield  {journal} {\bibinfo
   {journal} {Phys. Rev. B}\ }\textbf {\bibinfo {volume} {106}},\ \bibinfo
  {pages} {155108} (\bibinfo {year} {2022})}\BibitemShut {NoStop}%
\bibitem [{\citenamefont {{Nosov}}\ \emph {et~al.}(2023)\citenamefont
  {{Nosov}}, \citenamefont {{Burmistrov}},\ and\ \citenamefont
  {{Raghu}}}]{Nosov2023}%
  \BibitemOpen
  \bibfield  {author} {\bibinfo {author} {\bibfnamefont {P.~A.}\ \bibnamefont
  {{Nosov}}}, \bibinfo {author} {\bibfnamefont {I.~S.}\ \bibnamefont
  {{Burmistrov}}},\ and\ \bibinfo {author} {\bibfnamefont {S.}~\bibnamefont
  {{Raghu}}},\ }\bibfield  {title} {\emph {\bibinfo {title} {{Interplay of
  superconductivity and localization near a two-dimensional ferromagnetic
  quantum critical point}}},\ }\href
  {https://doi.org/10.1103/PhysRevB.107.144508} {\bibfield  {journal} {\bibinfo
   {journal} {Phys. Rev. B}\ }\textbf {\bibinfo {volume} {107}},\ \bibinfo
  {eid} {144508} (\bibinfo {year} {2023})},\ \Eprint
  {https://arxiv.org/abs/2211.02668} {arXiv:2211.02668 [cond-mat.str-el]}
  \BibitemShut {NoStop}%
\bibitem [{\citenamefont {Wu}\ \emph {et~al.}(2023)\citenamefont {Wu},
  \citenamefont {Lee},\ and\ \citenamefont {Foster}}]{Wu2023}%
  \BibitemOpen
  \bibfield  {author} {\bibinfo {author} {\bibfnamefont {T.~C.}\ \bibnamefont
  {Wu}}, \bibinfo {author} {\bibfnamefont {P.~A.}\ \bibnamefont {Lee}},\ and\
  \bibinfo {author} {\bibfnamefont {M.~S.}\ \bibnamefont {Foster}},\ }\bibfield
   {title} {\emph {\bibinfo {title} {Enhancement of superconductivity in a
  dirty marginal fermi liquid}},\ }\href
  {https://doi.org/10.1103/PhysRevB.108.214506} {\bibfield  {journal} {\bibinfo
   {journal} {Phys. Rev. B}\ }\textbf {\bibinfo {volume} {108}},\ \bibinfo
  {pages} {214506} (\bibinfo {year} {2023})}\BibitemShut {NoStop}%
\bibitem [{\citenamefont {Kim}\ and\ \citenamefont {Kim}(2024)}]{Kim2024}%
  \BibitemOpen
  \bibfield  {author} {\bibinfo {author} {\bibfnamefont {K.-M.}\ \bibnamefont
  {Kim}}\ and\ \bibinfo {author} {\bibfnamefont {K.-S.}\ \bibnamefont {Kim}},\
  }\bibfield  {title} {\emph {\bibinfo {title} {{Disordered non-Fermi liquid
  fixed point for two-dimensional metals at Ising-nematic quantum critical
  points}}},\ }\href {https://doi.org/10.21468/SciPostPhys.17.2.059} {\bibfield
   {journal} {\bibinfo  {journal} {SciPost Phys.}\ }\textbf {\bibinfo {volume}
  {17}},\ \bibinfo {pages} {059} (\bibinfo {year} {2024})}\BibitemShut
  {NoStop}%
\bibitem [{\citenamefont {Harris}(1974)}]{Harris1974}%
  \BibitemOpen
  \bibfield  {author} {\bibinfo {author} {\bibfnamefont {A.~B.}\ \bibnamefont
  {Harris}},\ }\bibfield  {title} {\emph {\bibinfo {title} {{Effect of random
  defects on the critical behaviour of Ising models}}},\ }\href
  {https://doi.org/10.1088/0022-3719/7/9/009} {\bibfield  {journal} {\bibinfo
  {journal} {Journal of Physics C: Solid State Physics}\ }\textbf {\bibinfo
  {volume} {7}},\ \bibinfo {pages} {1671} (\bibinfo {year} {1974})}\BibitemShut
  {NoStop}%
\bibitem [{\citenamefont {{Aldape}}\ \emph {et~al.}(2022)\citenamefont
  {{Aldape}}, \citenamefont {{Cookmeyer}}, \citenamefont {{Patel}},\ and\
  \citenamefont {{Altman}}}]{Aldape-2022}%
  \BibitemOpen
  \bibfield  {author} {\bibinfo {author} {\bibfnamefont {E.~E.}\ \bibnamefont
  {{Aldape}}}, \bibinfo {author} {\bibfnamefont {T.}~\bibnamefont
  {{Cookmeyer}}}, \bibinfo {author} {\bibfnamefont {A.~A.}\ \bibnamefont
  {{Patel}}},\ and\ \bibinfo {author} {\bibfnamefont {E.}~\bibnamefont
  {{Altman}}},\ }\bibfield  {title} {\emph {\bibinfo {title} {{Solvable theory
  of a strange metal at the breakdown of a heavy Fermi liquid}}},\ }\href
  {https://doi.org/10.1103/PhysRevB.105.235111} {\bibfield  {journal} {\bibinfo
   {journal} {Phys. Rev. B}\ }\textbf {\bibinfo {volume} {105}},\ \bibinfo
  {eid} {235111} (\bibinfo {year} {2022})},\ \Eprint
  {https://arxiv.org/abs/2012.00763} {arXiv:2012.00763 [cond-mat.str-el]}
  \BibitemShut {NoStop}%
\bibitem [{\citenamefont {Hartnoll}\ \emph {et~al.}(2014)\citenamefont
  {Hartnoll}, \citenamefont {Mahajan}, \citenamefont {Punk},\ and\
  \citenamefont {Sachdev}}]{Hartnoll:2014gba}%
  \BibitemOpen
  \bibfield  {author} {\bibinfo {author} {\bibfnamefont {S.~A.}\ \bibnamefont
  {Hartnoll}}, \bibinfo {author} {\bibfnamefont {R.}~\bibnamefont {Mahajan}},
  \bibinfo {author} {\bibfnamefont {M.}~\bibnamefont {Punk}},\ and\ \bibinfo
  {author} {\bibfnamefont {S.}~\bibnamefont {Sachdev}},\ }\bibfield  {title}
  {\emph {\bibinfo {title} {{Transport near the Ising-nematic quantum critical
  point of metals in two dimensions}}},\ }\href
  {https://doi.org/10.1103/PhysRevB.89.155130} {\bibfield  {journal} {\bibinfo
  {journal} {Phys. Rev. B}\ }\textbf {\bibinfo {volume} {89}},\ \bibinfo
  {pages} {155130} (\bibinfo {year} {2014})},\ \Eprint
  {https://arxiv.org/abs/1401.7012} {arXiv:1401.7012 [cond-mat.str-el]}
  \BibitemShut {NoStop}%
\bibitem [{\citenamefont {Patel}\ and\ \citenamefont
  {Sachdev}(2014)}]{Patel:2014jfa}%
  \BibitemOpen
  \bibfield  {author} {\bibinfo {author} {\bibfnamefont {A.~A.}\ \bibnamefont
  {Patel}}\ and\ \bibinfo {author} {\bibfnamefont {S.}~\bibnamefont
  {Sachdev}},\ }\bibfield  {title} {\emph {\bibinfo {title} {{DC resistivity at
  the onset of spin density wave order in two-dimensional metals}}},\ }\href
  {https://doi.org/10.1103/PhysRevB.90.165146} {\bibfield  {journal} {\bibinfo
  {journal} {Phys. Rev. B}\ }\textbf {\bibinfo {volume} {90}},\ \bibinfo
  {pages} {165146} (\bibinfo {year} {2014})},\ \Eprint
  {https://arxiv.org/abs/1408.6549} {arXiv:1408.6549 [cond-mat.str-el]}
  \BibitemShut {NoStop}%
\bibitem [{\citenamefont {Goldman}\ \emph {et~al.}(2020)\citenamefont
  {Goldman}, \citenamefont {Thomson}, \citenamefont {Nie},\ and\ \citenamefont
  {Bi}}]{Goldman2020}%
  \BibitemOpen
  \bibfield  {author} {\bibinfo {author} {\bibfnamefont {H.}~\bibnamefont
  {Goldman}}, \bibinfo {author} {\bibfnamefont {A.}~\bibnamefont {Thomson}},
  \bibinfo {author} {\bibfnamefont {L.}~\bibnamefont {Nie}},\ and\ \bibinfo
  {author} {\bibfnamefont {Z.}~\bibnamefont {Bi}},\ }\bibfield  {title} {\emph
  {\bibinfo {title} {Interplay of interactions and disorder at the
  superfluid-insulator transition: A dirty two-dimensional quantum critical
  point}},\ }\href {https://doi.org/10.1103/PhysRevB.101.144506} {\bibfield
  {journal} {\bibinfo  {journal} {Phys. Rev. B}\ }\textbf {\bibinfo {volume}
  {101}},\ \bibinfo {pages} {144506} (\bibinfo {year} {2020})}\BibitemShut
  {NoStop}%
\bibitem [{\citenamefont {Patel}\ \emph {et~al.}(2023)\citenamefont {Patel},
  \citenamefont {Guo}, \citenamefont {Esterlis},\ and\ \citenamefont
  {Sachdev}}]{Aavishkar2023}%
  \BibitemOpen
  \bibfield  {author} {\bibinfo {author} {\bibfnamefont {A.~A.}\ \bibnamefont
  {Patel}}, \bibinfo {author} {\bibfnamefont {H.}~\bibnamefont {Guo}}, \bibinfo
  {author} {\bibfnamefont {I.}~\bibnamefont {Esterlis}},\ and\ \bibinfo
  {author} {\bibfnamefont {S.}~\bibnamefont {Sachdev}},\ }\bibfield  {title}
  {\emph {\bibinfo {title} {{Universal theory of strange metals from spatially
  random interactions}}},\ }\href {https://doi.org/10.1126/science.abq6011}
  {\bibfield  {journal} {\bibinfo  {journal} {Science}\ }\textbf {\bibinfo
  {volume} {381}},\ \bibinfo {pages} {abq6011} (\bibinfo {year} {2023})},\
  \Eprint {https://arxiv.org/abs/2203.04990} {arXiv:2203.04990
  [cond-mat.str-el]} \BibitemShut {NoStop}%
\bibitem [{\citenamefont {Li}\ \emph {et~al.}(2024)\citenamefont {Li},
  \citenamefont {Valentinis}, \citenamefont {Patel}, \citenamefont {Guo},
  \citenamefont {Schmalian}, \citenamefont {Sachdev},\ and\ \citenamefont
  {Esterlis}}]{Li:2024kxr}%
  \BibitemOpen
  \bibfield  {author} {\bibinfo {author} {\bibfnamefont {C.}~\bibnamefont
  {Li}}, \bibinfo {author} {\bibfnamefont {D.}~\bibnamefont {Valentinis}},
  \bibinfo {author} {\bibfnamefont {A.~A.}\ \bibnamefont {Patel}}, \bibinfo
  {author} {\bibfnamefont {H.}~\bibnamefont {Guo}}, \bibinfo {author}
  {\bibfnamefont {J.}~\bibnamefont {Schmalian}}, \bibinfo {author}
  {\bibfnamefont {S.}~\bibnamefont {Sachdev}},\ and\ \bibinfo {author}
  {\bibfnamefont {I.}~\bibnamefont {Esterlis}},\ }\bibfield  {title} {\emph
  {\bibinfo {title} {{Strange Metal and Superconductor in the Two-Dimensional
  Yukawa-Sachdev-Ye-Kitaev Model}}},\ }\href
  {https://doi.org/10.1103/PhysRevLett.133.186502} {\bibfield  {journal}
  {\bibinfo  {journal} {Phys. Rev. Lett.}\ }\textbf {\bibinfo {volume} {133}},\
  \bibinfo {pages} {186502} (\bibinfo {year} {2024})},\ \Eprint
  {https://arxiv.org/abs/2406.07608} {arXiv:2406.07608 [cond-mat.str-el]}
  \BibitemShut {NoStop}%
\bibitem [{\citenamefont {{Kim}}\ and\ \citenamefont
  {{Chatterjee}}(2026)}]{Chatterjee25}%
  \BibitemOpen
  \bibfield  {author} {\bibinfo {author} {\bibfnamefont {J.}~\bibnamefont
  {{Kim}}}\ and\ \bibinfo {author} {\bibfnamefont {S.}~\bibnamefont
  {{Chatterjee}}},\ }\bibfield  {title} {\emph {\bibinfo {title} {{Theory of
  Linear Magnetoresistance in a Strange Metal}}},\ }\href
  {https://doi.org/10.1103/6vg9-98hp} {\bibfield  {journal} {\bibinfo
  {journal} {Phys. Rev. Lett.}\ }\textbf {\bibinfo {volume} {136}},\ \bibinfo
  {pages} {186301} (\bibinfo {year} {2026})},\ \Eprint
  {https://arxiv.org/abs/2504.01059} {arXiv:2504.01059 [cond-mat.str-el]}
  \BibitemShut {NoStop}%
\bibitem [{\citenamefont {{Sin}}\ and\ \citenamefont
  {{Wang}}(2025)}]{SangJin25}%
  \BibitemOpen
  \bibfield  {author} {\bibinfo {author} {\bibfnamefont {S.-J.}\ \bibnamefont
  {{Sin}}}\ and\ \bibinfo {author} {\bibfnamefont {Y.-L.}\ \bibnamefont
  {{Wang}}},\ }\bibfield  {title} {\emph {\bibinfo {title} {{Linear Resistivity
  from Spatially Random Interactions and the Uniqueness of Yukawa Coupling}}},\
  }\href {https://doi.org/10.48550/arXiv.2507.09442} {\bibfield  {journal}
  {\bibinfo  {journal} {arXiv e-prints}\ ,\ \bibinfo {eid} {arXiv:2507.09442}}
  (\bibinfo {year} {2025})},\ \Eprint {https://arxiv.org/abs/2507.09442}
  {arXiv:2507.09442 [hep-th]} \BibitemShut {NoStop}%
\bibitem [{\citenamefont {{Kleger}}\ \emph {et~al.}(2026)\citenamefont
  {{Kleger}}, \citenamefont {{Gnezdilov}},\ and\ \citenamefont
  {{Boyack}}}]{Boyack26}%
  \BibitemOpen
  \bibfield  {author} {\bibinfo {author} {\bibfnamefont {A.}~\bibnamefont
  {{Kleger}}}, \bibinfo {author} {\bibfnamefont {N.}~\bibnamefont
  {{Gnezdilov}}},\ and\ \bibinfo {author} {\bibfnamefont {R.}~\bibnamefont
  {{Boyack}}},\ }\bibfield  {title} {\emph {\bibinfo {title} {{Universal Theory
  of Incoherent Metals}}},\ }\href@noop {} {\bibfield  {journal} {\bibinfo
  {journal} {arXiv e-prints}\ ,\ \bibinfo {eid} {arXiv:2605.03013}} (\bibinfo
  {year} {2026})},\ \Eprint {https://arxiv.org/abs/2605.03013}
  {arXiv:2605.03013 [cond-mat.str-el]} \BibitemShut {NoStop}%
\bibitem [{\citenamefont {{Patel}}\ \emph {et~al.}(2024)\citenamefont
  {{Patel}}, \citenamefont {{Lunts}},\ and\ \citenamefont
  {{Sachdev}}}]{Aavishkar2024}%
  \BibitemOpen
  \bibfield  {author} {\bibinfo {author} {\bibfnamefont {A.~A.}\ \bibnamefont
  {{Patel}}}, \bibinfo {author} {\bibfnamefont {P.}~\bibnamefont {{Lunts}}},\
  and\ \bibinfo {author} {\bibfnamefont {S.}~\bibnamefont {{Sachdev}}},\
  }\bibfield  {title} {\emph {\bibinfo {title} {{Localization of overdamped
  bosonic modes and transport in strange metals}}},\ }\href
  {https://doi.org/10.1073/pnas.2402052121} {\bibfield  {journal} {\bibinfo
  {journal} {Proceedings of the National Academy of Science}\ }\textbf
  {\bibinfo {volume} {121}},\ \bibinfo {eid} {e2402052121} (\bibinfo {year}
  {2024})},\ \Eprint {https://arxiv.org/abs/2312.06751} {arXiv:2312.06751
  [cond-mat.str-el]} \BibitemShut {NoStop}%
\bibitem [{\citenamefont {{Patel}}\ \emph {et~al.}(2025)\citenamefont
  {{Patel}}, \citenamefont {{Lunts}},\ and\ \citenamefont
  {{Albergo}}}]{patel2024strange}%
  \BibitemOpen
  \bibfield  {author} {\bibinfo {author} {\bibfnamefont {A.~A.}\ \bibnamefont
  {{Patel}}}, \bibinfo {author} {\bibfnamefont {P.}~\bibnamefont {{Lunts}}},\
  and\ \bibinfo {author} {\bibfnamefont {M.~S.}\ \bibnamefont {{Albergo}}},\
  }\bibfield  {title} {\emph {\bibinfo {title} {{Strange Metals and Planckian
  Transport in a Gapless Phase from Spatially Random Interactions}}},\ }\href
  {https://doi.org/10.1103/611k-yxb9} {\bibfield  {journal} {\bibinfo
  {journal} {Physical Review X}\ }\textbf {\bibinfo {volume} {15}},\ \bibinfo
  {eid} {031064} (\bibinfo {year} {2025})},\ \Eprint
  {https://arxiv.org/abs/2410.05365} {arXiv:2410.05365 [cond-mat.str-el]}
  \BibitemShut {NoStop}%
\bibitem [{\citenamefont {{Greene}}\ \emph {et~al.}(2020)\citenamefont
  {{Greene}}, \citenamefont {{Mandal}}, \citenamefont {{Poniatowski}},\ and\
  \citenamefont {{Sarkar}}}]{Greene_rev}%
  \BibitemOpen
  \bibfield  {author} {\bibinfo {author} {\bibfnamefont {R.~L.}\ \bibnamefont
  {{Greene}}}, \bibinfo {author} {\bibfnamefont {P.~R.}\ \bibnamefont
  {{Mandal}}}, \bibinfo {author} {\bibfnamefont {N.~R.}\ \bibnamefont
  {{Poniatowski}}},\ and\ \bibinfo {author} {\bibfnamefont {T.}~\bibnamefont
  {{Sarkar}}},\ }\bibfield  {title} {\emph {\bibinfo {title} {{The Strange
  Metal State of the Electron-Doped Cuprates}}},\ }\href
  {https://doi.org/10.1146/annurev-conmatphys-031119-050558} {\bibfield
  {journal} {\bibinfo  {journal} {Annual Review of Condensed Matter Physics}\
  }\textbf {\bibinfo {volume} {11}},\ \bibinfo {pages} {213} (\bibinfo {year}
  {2020})},\ \Eprint {https://arxiv.org/abs/1905.04998} {arXiv:1905.04998
  [cond-mat.str-el]} \BibitemShut {NoStop}%
\bibitem [{\citenamefont {{Radaelli}}\ \emph {et~al.}(2026)\citenamefont
  {{Radaelli}}, \citenamefont {{Lipscombe}}, \citenamefont {{Zhu}},
  \citenamefont {{Stewart}}, \citenamefont {{Patel}}, \citenamefont
  {{Sachdev}},\ and\ \citenamefont {{Hayden}}}]{Hayden25}%
  \BibitemOpen
  \bibfield  {author} {\bibinfo {author} {\bibfnamefont {J.}~\bibnamefont
  {{Radaelli}}}, \bibinfo {author} {\bibfnamefont {O.~J.}\ \bibnamefont
  {{Lipscombe}}}, \bibinfo {author} {\bibfnamefont {M.}~\bibnamefont {{Zhu}}},
  \bibinfo {author} {\bibfnamefont {J.~R.}\ \bibnamefont {{Stewart}}}, \bibinfo
  {author} {\bibfnamefont {A.~A.}\ \bibnamefont {{Patel}}}, \bibinfo {author}
  {\bibfnamefont {S.}~\bibnamefont {{Sachdev}}},\ and\ \bibinfo {author}
  {\bibfnamefont {S.~M.}\ \bibnamefont {{Hayden}}},\ }\bibfield  {title} {\emph
  {\bibinfo {title} {{Critical spin fluctuations across the superconducting
  dome in La$_{2-x}$Sr$_x$CuO$_4$}}},\ }\href
  {https://doi.org/10.1038/s41467-026-71319-w} {\bibfield  {journal} {\bibinfo
  {journal} {Nature Communications}\ }\textbf {\bibinfo {volume} {17}},\
  \bibinfo {pages} {4564} (\bibinfo {year} {2026})},\ \Eprint
  {https://arxiv.org/abs/2503.13600} {arXiv:2503.13600 [cond-mat.str-el]}
  \BibitemShut {NoStop}%
\bibitem [{\citenamefont {Fisher}\ \emph {et~al.}(1989)\citenamefont {Fisher},
  \citenamefont {Weichman}, \citenamefont {Grinstein},\ and\ \citenamefont
  {Fisher}}]{Fisher1989}%
  \BibitemOpen
  \bibfield  {author} {\bibinfo {author} {\bibfnamefont {M.~P.~A.}\
  \bibnamefont {Fisher}}, \bibinfo {author} {\bibfnamefont {P.~B.}\
  \bibnamefont {Weichman}}, \bibinfo {author} {\bibfnamefont {G.}~\bibnamefont
  {Grinstein}},\ and\ \bibinfo {author} {\bibfnamefont {D.~S.}\ \bibnamefont
  {Fisher}},\ }\bibfield  {title} {\emph {\bibinfo {title} {Boson localization
  and the superfluid-insulator transition}},\ }\href
  {https://doi.org/10.1103/PhysRevB.40.546} {\bibfield  {journal} {\bibinfo
  {journal} {Phys. Rev. B}\ }\textbf {\bibinfo {volume} {40}},\ \bibinfo
  {pages} {546} (\bibinfo {year} {1989})}\BibitemShut {NoStop}%
\bibitem [{\citenamefont {Bulaevskii}\ and\ \citenamefont
  {Sadovskii}(1984)}]{Sadovskii1984}%
  \BibitemOpen
  \bibfield  {author} {\bibinfo {author} {\bibfnamefont {L.~N.}\ \bibnamefont
  {Bulaevskii}}\ and\ \bibinfo {author} {\bibfnamefont {M.~V.}\ \bibnamefont
  {Sadovskii}},\ }\bibfield  {title} {\emph {\bibinfo {title} {Localization and
  superconductivity}},\ }\href {http://jetpletters.ru/ps/0/article_19696.shtml}
  {\bibfield  {journal} {\bibinfo  {journal} {JETP Lett.}\ }\textbf {\bibinfo
  {volume} {39}},\ \bibinfo {pages} {640} (\bibinfo {year} {1984})}\BibitemShut
  {NoStop}%
\bibitem [{\citenamefont {Ma}\ and\ \citenamefont {Lee}(1985)}]{Ma1985}%
  \BibitemOpen
  \bibfield  {author} {\bibinfo {author} {\bibfnamefont {M.}~\bibnamefont
  {Ma}}\ and\ \bibinfo {author} {\bibfnamefont {P.~A.}\ \bibnamefont {Lee}},\
  }\bibfield  {title} {\emph {\bibinfo {title} {{Localized superconductors}}},\
  }\href {https://doi.org/10.1103/PhysRevB.32.5658} {\bibfield  {journal}
  {\bibinfo  {journal} {Phys. Rev. B}\ }\textbf {\bibinfo {volume} {32}},\
  \bibinfo {pages} {5658} (\bibinfo {year} {1985})}\BibitemShut {NoStop}%
\bibitem [{\citenamefont {Kapitulnik}\ and\ \citenamefont
  {Kotliar}(1985)}]{Kapitulnik1985}%
  \BibitemOpen
  \bibfield  {author} {\bibinfo {author} {\bibfnamefont {A.}~\bibnamefont
  {Kapitulnik}}\ and\ \bibinfo {author} {\bibfnamefont {G.}~\bibnamefont
  {Kotliar}},\ }\bibfield  {title} {\emph {\bibinfo {title} {{Anderson
  Localization and the Theory of Dirty Superconductors}}},\ }\href
  {https://doi.org/10.1103/PhysRevLett.54.473} {\bibfield  {journal} {\bibinfo
  {journal} {Phys. Rev. Lett.}\ }\textbf {\bibinfo {volume} {54}},\ \bibinfo
  {pages} {473} (\bibinfo {year} {1985})}\BibitemShut {NoStop}%
\bibitem [{\citenamefont {Kotliar}\ and\ \citenamefont
  {Kapitulnik}(1986)}]{Kapitulnik1986}%
  \BibitemOpen
  \bibfield  {author} {\bibinfo {author} {\bibfnamefont {G.}~\bibnamefont
  {Kotliar}}\ and\ \bibinfo {author} {\bibfnamefont {A.}~\bibnamefont
  {Kapitulnik}},\ }\bibfield  {title} {\emph {\bibinfo {title} {{Anderson
  localization and the theory of dirty superconductors. II}}},\ }\href
  {https://doi.org/10.1103/PhysRevB.33.3146} {\bibfield  {journal} {\bibinfo
  {journal} {Phys. Rev. B}\ }\textbf {\bibinfo {volume} {33}},\ \bibinfo
  {pages} {3146} (\bibinfo {year} {1986})}\BibitemShut {NoStop}%
\bibitem [{\citenamefont {{Feigel'Man}}\ \emph {et~al.}(2007)\citenamefont
  {{Feigel'Man}}, \citenamefont {{Ioffe}}, \citenamefont {{Kravtsov}},\ and\
  \citenamefont {{Yuzbashyan}}}]{Feigelman2007}%
  \BibitemOpen
  \bibfield  {author} {\bibinfo {author} {\bibfnamefont {M.~V.}\ \bibnamefont
  {{Feigel'Man}}}, \bibinfo {author} {\bibfnamefont {L.~B.}\ \bibnamefont
  {{Ioffe}}}, \bibinfo {author} {\bibfnamefont {V.~E.}\ \bibnamefont
  {{Kravtsov}}},\ and\ \bibinfo {author} {\bibfnamefont {E.~A.}\ \bibnamefont
  {{Yuzbashyan}}},\ }\bibfield  {title} {\emph {\bibinfo {title}
  {{Eigenfunction Fractality and Pseudogap State near the
  Superconductor-Insulator Transition}}},\ }\href
  {https://doi.org/10.1103/PhysRevLett.98.027001} {\bibfield  {journal}
  {\bibinfo  {journal} {Phys. Rev. Lett.}\ }\textbf {\bibinfo {volume} {98}},\
  \bibinfo {eid} {027001} (\bibinfo {year} {2007})},\ \Eprint
  {https://arxiv.org/abs/cond-mat/0610554} {arXiv:cond-mat/0610554
  [cond-mat.supr-con]} \BibitemShut {NoStop}%
\bibitem [{\citenamefont {{Feigel'man}}\ \emph {et~al.}(2010)\citenamefont
  {{Feigel'man}}, \citenamefont {{Ioffe}}, \citenamefont {{Kravtsov}},\ and\
  \citenamefont {{Cuevas}}}]{Feigelman2010}%
  \BibitemOpen
  \bibfield  {author} {\bibinfo {author} {\bibfnamefont {M.~V.}\ \bibnamefont
  {{Feigel'man}}}, \bibinfo {author} {\bibfnamefont {L.~B.}\ \bibnamefont
  {{Ioffe}}}, \bibinfo {author} {\bibfnamefont {V.~E.}\ \bibnamefont
  {{Kravtsov}}},\ and\ \bibinfo {author} {\bibfnamefont {E.}~\bibnamefont
  {{Cuevas}}},\ }\bibfield  {title} {\emph {\bibinfo {title} {{Fractal
  superconductivity near localization threshold}}},\ }\href
  {https://doi.org/10.1016/j.aop.2010.04.001} {\bibfield  {journal} {\bibinfo
  {journal} {Annals of Physics}\ }\textbf {\bibinfo {volume} {325}},\ \bibinfo
  {pages} {1390} (\bibinfo {year} {2010})},\ \Eprint
  {https://arxiv.org/abs/1002.0859} {arXiv:1002.0859 [cond-mat.supr-con]}
  \BibitemShut {NoStop}%
\bibitem [{\citenamefont {Burmistrov}\ \emph {et~al.}(2012)\citenamefont
  {Burmistrov}, \citenamefont {Gornyi},\ and\ \citenamefont
  {Mirlin}}]{BGM2012}%
  \BibitemOpen
  \bibfield  {author} {\bibinfo {author} {\bibfnamefont {I.~S.}\ \bibnamefont
  {Burmistrov}}, \bibinfo {author} {\bibfnamefont {I.~V.}\ \bibnamefont
  {Gornyi}},\ and\ \bibinfo {author} {\bibfnamefont {A.~D.}\ \bibnamefont
  {Mirlin}},\ }\bibfield  {title} {\emph {\bibinfo {title} {{Enhancement of the
  Critical Temperature of Superconductors by Anderson Localization}}},\ }\href
  {https://doi.org/10.1103/PhysRevLett.108.017002} {\bibfield  {journal}
  {\bibinfo  {journal} {Phys. Rev. Lett.}\ }\textbf {\bibinfo {volume} {108}},\
  \bibinfo {pages} {017002} (\bibinfo {year} {2012})},\ \Eprint
  {https://arxiv.org/abs/1102.3323} {arXiv:1102.3323 [cond-mat.mes-hall]}
  \BibitemShut {NoStop}%
\bibitem [{\citenamefont {{Burmistrov}}\ \emph {et~al.}(2021)\citenamefont
  {{Burmistrov}}, \citenamefont {{Gornyi}},\ and\ \citenamefont
  {{Mirlin}}}]{BGM2021}%
  \BibitemOpen
  \bibfield  {author} {\bibinfo {author} {\bibfnamefont {I.~S.}\ \bibnamefont
  {{Burmistrov}}}, \bibinfo {author} {\bibfnamefont {I.~V.}\ \bibnamefont
  {{Gornyi}}},\ and\ \bibinfo {author} {\bibfnamefont {A.~D.}\ \bibnamefont
  {{Mirlin}}},\ }\bibfield  {title} {\emph {\bibinfo {title}
  {{Multifractally-enhanced superconductivity in thin films}}},\ }\href
  {https://doi.org/10.1016/j.aop.2021.168499} {\bibfield  {journal} {\bibinfo
  {journal} {Annals of Physics}\ }\textbf {\bibinfo {volume} {435}},\ \bibinfo
  {eid} {168499} (\bibinfo {year} {2021})},\ \Eprint
  {https://arxiv.org/abs/2101.12713} {arXiv:2101.12713 [cond-mat.mes-hall]}
  \BibitemShut {NoStop}%
\bibitem [{\citenamefont {L\"ohneysen}\ \emph {et~al.}(2007)\citenamefont
  {L\"ohneysen}, \citenamefont {Rosch}, \citenamefont {Vojta},\ and\
  \citenamefont {W\"olfle}}]{VojtaInstabilities}%
  \BibitemOpen
  \bibfield  {author} {\bibinfo {author} {\bibfnamefont {H.~v.}\ \bibnamefont
  {L\"ohneysen}}, \bibinfo {author} {\bibfnamefont {A.}~\bibnamefont {Rosch}},
  \bibinfo {author} {\bibfnamefont {M.}~\bibnamefont {Vojta}},\ and\ \bibinfo
  {author} {\bibfnamefont {P.}~\bibnamefont {W\"olfle}},\ }\bibfield  {title}
  {\emph {\bibinfo {title} {Fermi-liquid instabilities at magnetic quantum
  phase transitions}},\ }\href {https://doi.org/10.1103/RevModPhys.79.1015}
  {\bibfield  {journal} {\bibinfo  {journal} {Rev. Mod. Phys.}\ }\textbf
  {\bibinfo {volume} {79}},\ \bibinfo {pages} {1015} (\bibinfo {year}
  {2007})}\BibitemShut {NoStop}%
\bibitem [{\citenamefont {Belitz}\ \emph {et~al.}(2001)\citenamefont {Belitz},
  \citenamefont {Kirkpatrick}, \citenamefont {Mercaldo},\ and\ \citenamefont
  {Sessions}}]{Belitz2001}%
  \BibitemOpen
  \bibfield  {author} {\bibinfo {author} {\bibfnamefont {D.}~\bibnamefont
  {Belitz}}, \bibinfo {author} {\bibfnamefont {T.~R.}\ \bibnamefont
  {Kirkpatrick}}, \bibinfo {author} {\bibfnamefont {M.~T.}\ \bibnamefont
  {Mercaldo}},\ and\ \bibinfo {author} {\bibfnamefont {S.~L.}\ \bibnamefont
  {Sessions}},\ }\bibfield  {title} {\emph {\bibinfo {title} {Local field
  theory for disordered itinerant quantum ferromagnets}},\ }\href
  {https://doi.org/10.1103/PhysRevB.63.174427} {\bibfield  {journal} {\bibinfo
  {journal} {Phys. Rev. B}\ }\textbf {\bibinfo {volume} {63}},\ \bibinfo
  {pages} {174427} (\bibinfo {year} {2001})}\BibitemShut {NoStop}%
\bibitem [{\citenamefont {Brando}\ \emph {et~al.}(2016)\citenamefont {Brando},
  \citenamefont {Belitz}, \citenamefont {Grosche},\ and\ \citenamefont
  {Kirkpatrick}}]{Brando2016}%
  \BibitemOpen
  \bibfield  {author} {\bibinfo {author} {\bibfnamefont {M.}~\bibnamefont
  {Brando}}, \bibinfo {author} {\bibfnamefont {D.}~\bibnamefont {Belitz}},
  \bibinfo {author} {\bibfnamefont {F.~M.}\ \bibnamefont {Grosche}},\ and\
  \bibinfo {author} {\bibfnamefont {T.~R.}\ \bibnamefont {Kirkpatrick}},\
  }\bibfield  {title} {\emph {\bibinfo {title} {Metallic quantum
  ferromagnets}},\ }\href {https://doi.org/10.1103/RevModPhys.88.025006}
  {\bibfield  {journal} {\bibinfo  {journal} {Rev. Mod. Phys.}\ }\textbf
  {\bibinfo {volume} {88}},\ \bibinfo {pages} {025006} (\bibinfo {year}
  {2016})}\BibitemShut {NoStop}%
\bibitem [{\citenamefont {{Del Maestro}}\ \emph {et~al.}(2008)\citenamefont
  {{Del Maestro}}, \citenamefont {{Rosenow}}, \citenamefont {{M{\"u}ller}},\
  and\ \citenamefont {{Sachdev}}}]{Maestro2008}%
  \BibitemOpen
  \bibfield  {author} {\bibinfo {author} {\bibfnamefont {A.}~\bibnamefont {{Del
  Maestro}}}, \bibinfo {author} {\bibfnamefont {B.}~\bibnamefont {{Rosenow}}},
  \bibinfo {author} {\bibfnamefont {M.}~\bibnamefont {{M{\"u}ller}}},\ and\
  \bibinfo {author} {\bibfnamefont {S.}~\bibnamefont {{Sachdev}}},\ }\bibfield
  {title} {\emph {\bibinfo {title} {{Infinite Randomness Fixed Point of the
  Superconductor-Metal Quantum Phase Transition}}},\ }\href
  {https://doi.org/10.1103/PhysRevLett.101.035701} {\bibfield  {journal}
  {\bibinfo  {journal} {Phys. Rev. Lett.}\ }\textbf {\bibinfo {volume} {101}},\
  \bibinfo {eid} {035701} (\bibinfo {year} {2008})},\ \Eprint
  {https://arxiv.org/abs/0802.3900} {arXiv:0802.3900 [cond-mat.dis-nn]}
  \BibitemShut {NoStop}%
\bibitem [{\citenamefont {Vojta}(2006)}]{vojta_rare_2006}%
  \BibitemOpen
  \bibfield  {author} {\bibinfo {author} {\bibfnamefont {T.}~\bibnamefont
  {Vojta}},\ }\bibfield  {title} {\emph {\bibinfo {title} {Rare region effects
  at classical, quantum and nonequilibrium phase transitions}},\ }\href
  {https://doi.org/10.1088/0305-4470/39/22/R01} {\bibfield  {journal} {\bibinfo
   {journal} {Journal of Physics A: Mathematical and General}\ }\textbf
  {\bibinfo {volume} {39}},\ \bibinfo {pages} {R143} (\bibinfo {year}
  {2006})}\BibitemShut {NoStop}%
\bibitem [{\citenamefont {Vojta}(2003)}]{Vojta2003}%
  \BibitemOpen
  \bibfield  {author} {\bibinfo {author} {\bibfnamefont {T.}~\bibnamefont
  {Vojta}},\ }\bibfield  {title} {\emph {\bibinfo {title} {Disorder-induced
  rounding of certain quantum phase transitions}},\ }\href
  {https://doi.org/10.1103/PhysRevLett.90.107202} {\bibfield  {journal}
  {\bibinfo  {journal} {Phys. Rev. Lett.}\ }\textbf {\bibinfo {volume} {90}},\
  \bibinfo {pages} {107202} (\bibinfo {year} {2003})}\BibitemShut {NoStop}%
\bibitem [{\citenamefont {{Hoyos}}\ \emph {et~al.}(2007)\citenamefont
  {{Hoyos}}, \citenamefont {{Kotabage}},\ and\ \citenamefont
  {{Vojta}}}]{Vojta07}%
  \BibitemOpen
  \bibfield  {author} {\bibinfo {author} {\bibfnamefont {J.~A.}\ \bibnamefont
  {{Hoyos}}}, \bibinfo {author} {\bibfnamefont {C.}~\bibnamefont
  {{Kotabage}}},\ and\ \bibinfo {author} {\bibfnamefont {T.}~\bibnamefont
  {{Vojta}}},\ }\bibfield  {title} {\emph {\bibinfo {title} {{Effects of
  Dissipation on a Quantum Critical Point with Disorder}}},\ }\href
  {https://doi.org/10.1103/PhysRevLett.99.230601} {\bibfield  {journal}
  {\bibinfo  {journal} {Phys. Rev. Lett.}\ }\textbf {\bibinfo {volume} {99}},\
  \bibinfo {eid} {230601} (\bibinfo {year} {2007})},\ \Eprint
  {https://arxiv.org/abs/0705.1865} {arXiv:0705.1865 [cond-mat.str-el]}
  \BibitemShut {NoStop}%
\bibitem [{\citenamefont {Vojta}\ \emph {et~al.}(2009)\citenamefont {Vojta},
  \citenamefont {Kotabage},\ and\ \citenamefont {Hoyos}}]{Vojta2009}%
  \BibitemOpen
  \bibfield  {author} {\bibinfo {author} {\bibfnamefont {T.}~\bibnamefont
  {Vojta}}, \bibinfo {author} {\bibfnamefont {C.}~\bibnamefont {Kotabage}},\
  and\ \bibinfo {author} {\bibfnamefont {J.~A.}\ \bibnamefont {Hoyos}},\
  }\bibfield  {title} {\emph {\bibinfo {title} {Infinite-randomness quantum
  critical points induced by dissipation}},\ }\href
  {https://doi.org/10.1103/PhysRevB.79.024401} {\bibfield  {journal} {\bibinfo
  {journal} {Phys. Rev. B}\ }\textbf {\bibinfo {volume} {79}},\ \bibinfo
  {pages} {024401} (\bibinfo {year} {2009})}\BibitemShut {NoStop}%
\bibitem [{\citenamefont {Thill}\ and\ \citenamefont {Huse}(1995)}]{Thill95}%
  \BibitemOpen
  \bibfield  {author} {\bibinfo {author} {\bibfnamefont {M.}~\bibnamefont
  {Thill}}\ and\ \bibinfo {author} {\bibfnamefont {D.}~\bibnamefont {Huse}},\
  }\bibfield  {title} {\emph {\bibinfo {title} {Equilibrium behaviour of
  quantum ising spin glass}},\ }\href
  {https://doi.org/https://doi.org/10.1016/0378-4371(94)00247-Q} {\bibfield
  {journal} {\bibinfo  {journal} {Physica A: Statistical Mechanics and its
  Applications}\ }\textbf {\bibinfo {volume} {214}},\ \bibinfo {pages} {321}
  (\bibinfo {year} {1995})}\BibitemShut {NoStop}%
\bibitem [{\citenamefont {{Read}}\ \emph {et~al.}(1995)\citenamefont {{Read}},
  \citenamefont {{Sachdev}},\ and\ \citenamefont {{Ye}}}]{RSY95}%
  \BibitemOpen
  \bibfield  {author} {\bibinfo {author} {\bibfnamefont {N.}~\bibnamefont
  {{Read}}}, \bibinfo {author} {\bibfnamefont {S.}~\bibnamefont {{Sachdev}}},\
  and\ \bibinfo {author} {\bibfnamefont {J.}~\bibnamefont {{Ye}}},\ }\bibfield
  {title} {\emph {\bibinfo {title} {{Landau theory of quantum spin glasses of
  rotors and Ising spins}}},\ }\href {https://doi.org/10.1103/PhysRevB.52.384}
  {\bibfield  {journal} {\bibinfo  {journal} {Phys. Rev. B}\ }\textbf {\bibinfo
  {volume} {52}},\ \bibinfo {pages} {384} (\bibinfo {year} {1995})},\ \Eprint
  {https://arxiv.org/abs/cond-mat/9412032} {arXiv:cond-mat/9412032 [cond-mat]}
  \BibitemShut {NoStop}%
\bibitem [{\citenamefont {Mermin}\ and\ \citenamefont
  {Wagner}(1966)}]{MerminWagner}%
  \BibitemOpen
  \bibfield  {author} {\bibinfo {author} {\bibfnamefont {N.~D.}\ \bibnamefont
  {Mermin}}\ and\ \bibinfo {author} {\bibfnamefont {H.}~\bibnamefont
  {Wagner}},\ }\bibfield  {title} {\emph {\bibinfo {title} {Absence of
  ferromagnetism or antiferromagnetism in one- or two-dimensional isotropic
  heisenberg models}},\ }\href {https://doi.org/10.1103/PhysRevLett.17.1133}
  {\bibfield  {journal} {\bibinfo  {journal} {Phys. Rev. Lett.}\ }\textbf
  {\bibinfo {volume} {17}},\ \bibinfo {pages} {1133} (\bibinfo {year}
  {1966})}\BibitemShut {NoStop}%
\bibitem [{\citenamefont {Codello}\ and\ \citenamefont
  {D'Odorico}(2013)}]{Codello2013}%
  \BibitemOpen
  \bibfield  {author} {\bibinfo {author} {\bibfnamefont {A.}~\bibnamefont
  {Codello}}\ and\ \bibinfo {author} {\bibfnamefont {G.}~\bibnamefont
  {D'Odorico}},\ }\bibfield  {title} {\emph {\bibinfo {title}
  {$o(n)$-universality classes and the mermin-wagner theorem}},\ }\href
  {https://doi.org/10.1103/PhysRevLett.110.141601} {\bibfield  {journal}
  {\bibinfo  {journal} {Phys. Rev. Lett.}\ }\textbf {\bibinfo {volume} {110}},\
  \bibinfo {pages} {141601} (\bibinfo {year} {2013})}\BibitemShut {NoStop}%
\bibitem [{\citenamefont {Usadel}(1970)}]{Usadel1970}%
  \BibitemOpen
  \bibfield  {author} {\bibinfo {author} {\bibfnamefont {K.~D.}\ \bibnamefont
  {Usadel}},\ }\bibfield  {title} {\emph {\bibinfo {title} {Generalized
  diffusion equation for superconducting alloys}},\ }\href
  {https://doi.org/10.1103/PhysRevLett.25.507} {\bibfield  {journal} {\bibinfo
  {journal} {Phys. Rev. Lett.}\ }\textbf {\bibinfo {volume} {25}},\ \bibinfo
  {pages} {507} (\bibinfo {year} {1970})}\BibitemShut {NoStop}%
\bibitem [{\citenamefont {Meyer}\ and\ \citenamefont
  {Simons}(2001)}]{Meyer2001}%
  \BibitemOpen
  \bibfield  {author} {\bibinfo {author} {\bibfnamefont {J.~S.}\ \bibnamefont
  {Meyer}}\ and\ \bibinfo {author} {\bibfnamefont {B.~D.}\ \bibnamefont
  {Simons}},\ }\bibfield  {title} {\emph {\bibinfo {title} {Gap fluctuations in
  inhomogeneous superconductors}},\ }\href
  {https://doi.org/10.1103/PhysRevB.64.134516} {\bibfield  {journal} {\bibinfo
  {journal} {Phys. Rev. B}\ }\textbf {\bibinfo {volume} {64}},\ \bibinfo
  {pages} {134516} (\bibinfo {year} {2001})}\BibitemShut {NoStop}%
\bibitem [{\citenamefont {Tikhonov}\ and\ \citenamefont
  {Feigel’man}(2020)}]{Tikhonov2020}%
  \BibitemOpen
  \bibfield  {author} {\bibinfo {author} {\bibfnamefont {K.~S.}\ \bibnamefont
  {Tikhonov}}\ and\ \bibinfo {author} {\bibfnamefont {M.~V.}\ \bibnamefont
  {Feigel’man}},\ }\bibfield  {title} {\emph {\bibinfo {title} {Strange metal
  state near quantum superconductor-metal transition in thin films}},\ }\href
  {https://doi.org/https://doi.org/10.1016/j.aop.2020.168138} {\bibfield
  {journal} {\bibinfo  {journal} {Annals of Physics}\ }\textbf {\bibinfo
  {volume} {417}},\ \bibinfo {pages} {168138} (\bibinfo {year} {2020})},\
  \bibinfo {note} {{Eliashberg theory at 60: Strong-coupling superconductivity
  and beyond}}\BibitemShut {NoStop}%
\bibitem [{\citenamefont {Larkin}\ and\ \citenamefont
  {Ovchinnikov}(1972)}]{larkin1972density}%
  \BibitemOpen
  \bibfield  {author} {\bibinfo {author} {\bibfnamefont {A.}~\bibnamefont
  {Larkin}}\ and\ \bibinfo {author} {\bibfnamefont {Y.~N.}\ \bibnamefont
  {Ovchinnikov}},\ }\bibfield  {title} {\emph {\bibinfo {title} {Density of
  states in inhomogeneous superconductors}},\ }\href@noop {} {\bibfield
  {journal} {\bibinfo  {journal} {Sov. Phys. JETP}\ }\textbf {\bibinfo {volume}
  {34}},\ \bibinfo {pages} {1144} (\bibinfo {year} {1972})}\BibitemShut
  {NoStop}%
\bibitem [{\citenamefont {Gomes}\ \emph {et~al.}(2007)\citenamefont {Gomes},
  \citenamefont {Pasupathy}, \citenamefont {Pushp}, \citenamefont {Ono},
  \citenamefont {Ando},\ and\ \citenamefont {Yazdani}}]{Yazdani_puddles_2007}%
  \BibitemOpen
  \bibfield  {author} {\bibinfo {author} {\bibfnamefont {K.~K.}\ \bibnamefont
  {Gomes}}, \bibinfo {author} {\bibfnamefont {A.~N.}\ \bibnamefont
  {Pasupathy}}, \bibinfo {author} {\bibfnamefont {A.}~\bibnamefont {Pushp}},
  \bibinfo {author} {\bibfnamefont {S.}~\bibnamefont {Ono}}, \bibinfo {author}
  {\bibfnamefont {Y.}~\bibnamefont {Ando}},\ and\ \bibinfo {author}
  {\bibfnamefont {A.}~\bibnamefont {Yazdani}},\ }\bibfield  {title} {\emph
  {\bibinfo {title} {{Visualizing pair formation on the atomic scale in the
  high-Tc superconductor
  $\text{Bi}_2\text{Sr}_2\text{CaCu}_2\text{O}_{8+\delta}$}}},\ }\href
  {https://doi.org/10.1038/nature05881} {\bibfield  {journal} {\bibinfo
  {journal} {Nature}\ }\textbf {\bibinfo {volume} {447}},\ \bibinfo {pages}
  {569} (\bibinfo {year} {2007})}\BibitemShut {NoStop}%
\bibitem [{\citenamefont {Lee}\ and\ \citenamefont
  {Ramakrishnan}(1985)}]{Lee1985}%
  \BibitemOpen
  \bibfield  {author} {\bibinfo {author} {\bibfnamefont {P.~A.}\ \bibnamefont
  {Lee}}\ and\ \bibinfo {author} {\bibfnamefont {T.~V.}\ \bibnamefont
  {Ramakrishnan}},\ }\bibfield  {title} {\emph {\bibinfo {title} {Disordered
  electronic systems}},\ }\href {https://doi.org/10.1103/RevModPhys.57.287}
  {\bibfield  {journal} {\bibinfo  {journal} {Rev. Mod. Phys.}\ }\textbf
  {\bibinfo {volume} {57}},\ \bibinfo {pages} {287} (\bibinfo {year}
  {1985})}\BibitemShut {NoStop}%
\bibitem [{\citenamefont {Ioffe}\ and\ \citenamefont
  {Larkin}(1981)}]{Larkin1981}%
  \BibitemOpen
  \bibfield  {author} {\bibinfo {author} {\bibfnamefont {L.~B.}\ \bibnamefont
  {Ioffe}}\ and\ \bibinfo {author} {\bibfnamefont {A.~I.}\ \bibnamefont
  {Larkin}},\ }\bibfield  {title} {\emph {\bibinfo {title} {Properties of
  superconductors with a smeared transition temperature}},\ }\href
  {https://www.osti.gov/biblio/5726409} {\bibfield  {journal} {\bibinfo
  {journal} {Sov. Phys. - JETP (Engl. Transl.)}\ }\textbf {\bibinfo {volume}
  {54:2}} (\bibinfo {year} {1981})}\BibitemShut {NoStop}%
\bibitem [{\citenamefont {Dodaro}\ and\ \citenamefont
  {Kivelson}(2018)}]{Dodaro2018}%
  \BibitemOpen
  \bibfield  {author} {\bibinfo {author} {\bibfnamefont {J.~F.}\ \bibnamefont
  {Dodaro}}\ and\ \bibinfo {author} {\bibfnamefont {S.~A.}\ \bibnamefont
  {Kivelson}},\ }\bibfield  {title} {\emph {\bibinfo {title} {{Generalization
  of Anderson's theorem for disordered superconductors}}},\ }\href
  {https://doi.org/10.1103/PhysRevB.98.174503} {\bibfield  {journal} {\bibinfo
  {journal} {Phys. Rev. B}\ }\textbf {\bibinfo {volume} {98}},\ \bibinfo
  {pages} {174503} (\bibinfo {year} {2018})}\BibitemShut {NoStop}%
\bibitem [{\citenamefont {Atkinson}\ \emph {et~al.}(2000)\citenamefont
  {Atkinson}, \citenamefont {Hirschfeld},\ and\ \citenamefont
  {MacDonald}}]{Atkinson2000}%
  \BibitemOpen
  \bibfield  {author} {\bibinfo {author} {\bibfnamefont {W.~A.}\ \bibnamefont
  {Atkinson}}, \bibinfo {author} {\bibfnamefont {P.~J.}\ \bibnamefont
  {Hirschfeld}},\ and\ \bibinfo {author} {\bibfnamefont {A.~H.}\ \bibnamefont
  {MacDonald}},\ }\bibfield  {title} {\emph {\bibinfo {title} {Gap
  inhomogeneities and the density of states in disordered d-wave
  superconductors}},\ }\href {https://doi.org/10.1103/PhysRevLett.85.3922}
  {\bibfield  {journal} {\bibinfo  {journal} {Phys. Rev. Lett.}\ }\textbf
  {\bibinfo {volume} {85}},\ \bibinfo {pages} {3922} (\bibinfo {year}
  {2000})}\BibitemShut {NoStop}%
\bibitem [{\citenamefont {Nunner}\ \emph {et~al.}(2005)\citenamefont {Nunner},
  \citenamefont {Andersen}, \citenamefont {Melikyan},\ and\ \citenamefont
  {Hirschfeld}}]{Nunner2005}%
  \BibitemOpen
  \bibfield  {author} {\bibinfo {author} {\bibfnamefont {T.~S.}\ \bibnamefont
  {Nunner}}, \bibinfo {author} {\bibfnamefont {B.~M.}\ \bibnamefont
  {Andersen}}, \bibinfo {author} {\bibfnamefont {A.}~\bibnamefont {Melikyan}},\
  and\ \bibinfo {author} {\bibfnamefont {P.~J.}\ \bibnamefont {Hirschfeld}},\
  }\bibfield  {title} {\emph {\bibinfo {title} {Dopant-modulated pair
  interaction in cuprate superconductors}},\ }\href
  {https://doi.org/10.1103/PhysRevLett.95.177003} {\bibfield  {journal}
  {\bibinfo  {journal} {Phys. Rev. Lett.}\ }\textbf {\bibinfo {volume} {95}},\
  \bibinfo {pages} {177003} (\bibinfo {year} {2005})}\BibitemShut {NoStop}%
\bibitem [{\citenamefont {{Li}}\ \emph {et~al.}(2021)\citenamefont {{Li}},
  \citenamefont {{Kivelson}},\ and\ \citenamefont {{Lee}}}]{KivelsonLee21}%
  \BibitemOpen
  \bibfield  {author} {\bibinfo {author} {\bibfnamefont {Z.-X.}\ \bibnamefont
  {{Li}}}, \bibinfo {author} {\bibfnamefont {S.~A.}\ \bibnamefont
  {{Kivelson}}},\ and\ \bibinfo {author} {\bibfnamefont {D.-H.}\ \bibnamefont
  {{Lee}}},\ }\bibfield  {title} {\emph {\bibinfo {title}
  {{Superconductor-to-metal transition in overdoped cuprates}}},\ }\href
  {https://doi.org/10.1038/s41535-021-00335-4} {\bibfield  {journal} {\bibinfo
  {journal} {npj Quantum Materials}\ }\textbf {\bibinfo {volume} {6}},\
  \bibinfo {eid} {36} (\bibinfo {year} {2021})},\ \Eprint
  {https://arxiv.org/abs/2010.06091} {arXiv:2010.06091 [cond-mat.supr-con]}
  \BibitemShut {NoStop}%
\bibitem [{\citenamefont {Bulaevskii}\ \emph {et~al.}(1987)\citenamefont
  {Bulaevskii}, \citenamefont {Panyukov},\ and\ \citenamefont
  {Sadovskii}}]{bulaevskii1987}%
  \BibitemOpen
  \bibfield  {author} {\bibinfo {author} {\bibfnamefont {L.}~\bibnamefont
  {Bulaevskii}}, \bibinfo {author} {\bibfnamefont {S.}~\bibnamefont
  {Panyukov}},\ and\ \bibinfo {author} {\bibfnamefont {M.}~\bibnamefont
  {Sadovskii}},\ }\bibfield  {title} {\emph {\bibinfo {title} {Inhomogeneous
  superconductivity in disordered metals}},\ }\href
  {https://jetp.ras.ru/cgi-bin/e/index/e/65/2/p380?a=list} {\bibfield
  {journal} {\bibinfo  {journal} {Zh. Eksp. Teor. Fiz.}\ }\textbf {\bibinfo
  {volume} {92}},\ \bibinfo {pages} {672} (\bibinfo {year} {1987})}\BibitemShut
  {NoStop}%
\bibitem [{\citenamefont {Gnedenko}\ and\ \citenamefont
  {Kolmogorov}(1968)}]{gnedenko1968limit}%
  \BibitemOpen
  \bibfield  {author} {\bibinfo {author} {\bibfnamefont {B.}~\bibnamefont
  {Gnedenko}}\ and\ \bibinfo {author} {\bibfnamefont {A.}~\bibnamefont
  {Kolmogorov}},\ }\href@noop {} {\emph {\bibinfo {title} {Limit Distributions
  for Sums of Independent Random Variables}}},\ Addison-Wesley Mathematical
  Series\ (\bibinfo  {publisher} {Addison-Wesley},\ \bibinfo {year}
  {1968})\BibitemShut {NoStop}%
\bibitem [{\citenamefont {Abrikosov}\ \emph {et~al.}(2012)\citenamefont
  {Abrikosov}, \citenamefont {Gorkov}, \citenamefont {Dzyaloshinski},\ and\
  \citenamefont {Silverman}}]{AGD}%
  \BibitemOpen
  \bibfield  {author} {\bibinfo {author} {\bibfnamefont {A.}~\bibnamefont
  {Abrikosov}}, \bibinfo {author} {\bibfnamefont {L.}~\bibnamefont {Gorkov}},
  \bibinfo {author} {\bibfnamefont {I.}~\bibnamefont {Dzyaloshinski}},\ and\
  \bibinfo {author} {\bibfnamefont {R.}~\bibnamefont {Silverman}},\ }\href@noop
  {} {\emph {\bibinfo {title} {Methods of Quantum Field Theory in Statistical
  Physics}}},\ Dover Books on Physics\ (\bibinfo  {publisher} {Dover
  Publications},\ \bibinfo {year} {2012})\BibitemShut {NoStop}%
\bibitem [{\citenamefont {Levtov}\ and\ \citenamefont
  {Shytov}(2003)}]{Levitov}%
  \BibitemOpen
  \bibfield  {author} {\bibinfo {author} {\bibfnamefont {L.~S.}\ \bibnamefont
  {Levtov}}\ and\ \bibinfo {author} {\bibfnamefont {A.~V.}\ \bibnamefont
  {Shytov}},\ }\href@noop {} {\emph {\bibinfo {title} {Green's Functions:
  Problems and Solutions}}}\ (\bibinfo  {publisher} {Fizmatlit},\ \bibinfo
  {year} {2003})\BibitemShut {NoStop}%
\bibitem [{\citenamefont {Abanov}\ and\ \citenamefont
  {Chubukov}(2020)}]{Gamma1}%
  \BibitemOpen
  \bibfield  {author} {\bibinfo {author} {\bibfnamefont {A.}~\bibnamefont
  {Abanov}}\ and\ \bibinfo {author} {\bibfnamefont {A.~V.}\ \bibnamefont
  {Chubukov}},\ }\bibfield  {title} {\emph {\bibinfo {title} {{Interplay
  between superconductivity and non-Fermi liquid at a quantum critical point in
  a metal. {I}. The $\ensuremath{\gamma}$ model and its phase diagram at
  $\ensuremath{T}=0$: The case $0\ensuremath{<\gamma<}1$}}},\ }\href
  {https://doi.org/10.1103/PhysRevB.102.024524} {\bibfield  {journal} {\bibinfo
   {journal} {Phys. Rev. B}\ }\textbf {\bibinfo {volume} {102}},\ \bibinfo
  {pages} {024524} (\bibinfo {year} {2020})}\BibitemShut {NoStop}%
\bibitem [{\citenamefont {Wu}\ \emph {et~al.}(2020{\natexlab{a}})\citenamefont
  {Wu}, \citenamefont {Abanov}, \citenamefont {Wang},\ and\ \citenamefont
  {Chubukov}}]{Gamma2}%
  \BibitemOpen
  \bibfield  {author} {\bibinfo {author} {\bibfnamefont {Y.-M.}\ \bibnamefont
  {Wu}}, \bibinfo {author} {\bibfnamefont {A.}~\bibnamefont {Abanov}}, \bibinfo
  {author} {\bibfnamefont {Y.}~\bibnamefont {Wang}},\ and\ \bibinfo {author}
  {\bibfnamefont {A.~V.}\ \bibnamefont {Chubukov}},\ }\bibfield  {title} {\emph
  {\bibinfo {title} {{Interplay between superconductivity and non-Fermi liquid
  at a quantum critical point in a metal. {II}. The $\ensuremath{\gamma}$ model
  at a finite $\ensuremath{T}$ for $0\ensuremath{<\gamma<}1$}}},\ }\href
  {https://doi.org/10.1103/PhysRevB.102.024525} {\bibfield  {journal} {\bibinfo
   {journal} {Phys. Rev. B}\ }\textbf {\bibinfo {volume} {102}},\ \bibinfo
  {pages} {024525} (\bibinfo {year} {2020}{\natexlab{a}})}\BibitemShut
  {NoStop}%
\bibitem [{\citenamefont {Wu}\ \emph {et~al.}(2020{\natexlab{b}})\citenamefont
  {Wu}, \citenamefont {Abanov},\ and\ \citenamefont {Chubukov}}]{Gamma3}%
  \BibitemOpen
  \bibfield  {author} {\bibinfo {author} {\bibfnamefont {Y.-M.}\ \bibnamefont
  {Wu}}, \bibinfo {author} {\bibfnamefont {A.}~\bibnamefont {Abanov}},\ and\
  \bibinfo {author} {\bibfnamefont {A.~V.}\ \bibnamefont {Chubukov}},\
  }\bibfield  {title} {\emph {\bibinfo {title} {{Interplay between
  superconductivity and non-Fermi liquid behavior at a quantum critical point
  in a metal. {III}. The $\ensuremath{\gamma}$ model and its phase diagram
  across $\ensuremath{\gamma}=1$}}},\ }\href
  {https://doi.org/10.1103/PhysRevB.102.094516} {\bibfield  {journal} {\bibinfo
   {journal} {Phys. Rev. B}\ }\textbf {\bibinfo {volume} {102}},\ \bibinfo
  {pages} {094516} (\bibinfo {year} {2020}{\natexlab{b}})}\BibitemShut
  {NoStop}%
\bibitem [{\citenamefont {Wu}\ \emph {et~al.}(2021{\natexlab{a}})\citenamefont
  {Wu}, \citenamefont {Zhang}, \citenamefont {Abanov},\ and\ \citenamefont
  {Chubukov}}]{Gamma4}%
  \BibitemOpen
  \bibfield  {author} {\bibinfo {author} {\bibfnamefont {Y.-M.}\ \bibnamefont
  {Wu}}, \bibinfo {author} {\bibfnamefont {S.-S.}\ \bibnamefont {Zhang}},
  \bibinfo {author} {\bibfnamefont {A.}~\bibnamefont {Abanov}},\ and\ \bibinfo
  {author} {\bibfnamefont {A.~V.}\ \bibnamefont {Chubukov}},\ }\bibfield
  {title} {\emph {\bibinfo {title} {{Interplay between superconductivity and
  non-Fermi liquid at a quantum critical point in a metal. {IV}. The
  $\ensuremath{\gamma}$ model and its phase diagram at
  $1\ensuremath{<\gamma<}2$}}},\ }\href
  {https://doi.org/10.1103/PhysRevB.103.024522} {\bibfield  {journal} {\bibinfo
   {journal} {Phys. Rev. B}\ }\textbf {\bibinfo {volume} {103}},\ \bibinfo
  {pages} {024522} (\bibinfo {year} {2021}{\natexlab{a}})}\BibitemShut
  {NoStop}%
\bibitem [{\citenamefont {Wu}\ \emph {et~al.}(2021{\natexlab{b}})\citenamefont
  {Wu}, \citenamefont {Zhang}, \citenamefont {Abanov},\ and\ \citenamefont
  {Chubukov}}]{Gamma5}%
  \BibitemOpen
  \bibfield  {author} {\bibinfo {author} {\bibfnamefont {Y.-M.}\ \bibnamefont
  {Wu}}, \bibinfo {author} {\bibfnamefont {S.-S.}\ \bibnamefont {Zhang}},
  \bibinfo {author} {\bibfnamefont {A.}~\bibnamefont {Abanov}},\ and\ \bibinfo
  {author} {\bibfnamefont {A.~V.}\ \bibnamefont {Chubukov}},\ }\bibfield
  {title} {\emph {\bibinfo {title} {{Interplay between superconductivity and
  non-Fermi liquid behavior at a quantum-critical point in a metal. {V}. The
  $\ensuremath{\gamma}$ model and its phase diagram: The case
  $\ensuremath{\gamma}=2$}}},\ }\href
  {https://doi.org/10.1103/PhysRevB.103.184508} {\bibfield  {journal} {\bibinfo
   {journal} {Phys. Rev. B}\ }\textbf {\bibinfo {volume} {103}},\ \bibinfo
  {pages} {184508} (\bibinfo {year} {2021}{\natexlab{b}})}\BibitemShut
  {NoStop}%
\bibitem [{\citenamefont {Zhang}\ \emph {et~al.}(2021)\citenamefont {Zhang},
  \citenamefont {Wu}, \citenamefont {Abanov},\ and\ \citenamefont
  {Chubukov}}]{Gamma6}%
  \BibitemOpen
  \bibfield  {author} {\bibinfo {author} {\bibfnamefont {S.-S.}\ \bibnamefont
  {Zhang}}, \bibinfo {author} {\bibfnamefont {Y.-M.}\ \bibnamefont {Wu}},
  \bibinfo {author} {\bibfnamefont {A.}~\bibnamefont {Abanov}},\ and\ \bibinfo
  {author} {\bibfnamefont {A.~V.}\ \bibnamefont {Chubukov}},\ }\bibfield
  {title} {\emph {\bibinfo {title} {{Interplay between superconductivity and
  non-Fermi liquid at a quantum critical point in a metal. {VI}. The
  $\ensuremath{\gamma}$ model and its phase diagram at
  $2<\ensuremath{\gamma}<3$}}},\ }\href
  {https://doi.org/10.1103/PhysRevB.104.144509} {\bibfield  {journal} {\bibinfo
   {journal} {Phys. Rev. B}\ }\textbf {\bibinfo {volume} {104}},\ \bibinfo
  {pages} {144509} (\bibinfo {year} {2021})}\BibitemShut {NoStop}%
\bibitem [{\citenamefont {Apostol}(1999)}]{Apostol01051999}%
  \BibitemOpen
  \bibfield  {author} {\bibinfo {author} {\bibfnamefont {T.~M.}\ \bibnamefont
  {Apostol}},\ }\bibfield  {title} {\emph {\bibinfo {title} {{An Elementary
  View of Euler's Summation Formula}}},\ }\href
  {https://doi.org/10.1080/00029890.1999.12005063} {\bibfield  {journal}
  {\bibinfo  {journal} {The American Mathematical Monthly}\ }\textbf {\bibinfo
  {volume} {106}},\ \bibinfo {pages} {409} (\bibinfo {year}
  {1999})}\BibitemShut {NoStop}%
\bibitem [{\citenamefont {Evers}\ and\ \citenamefont
  {Mirlin}(2008)}]{RevModPhys.80.1355}%
  \BibitemOpen
  \bibfield  {author} {\bibinfo {author} {\bibfnamefont {F.}~\bibnamefont
  {Evers}}\ and\ \bibinfo {author} {\bibfnamefont {A.~D.}\ \bibnamefont
  {Mirlin}},\ }\bibfield  {title} {\emph {\bibinfo {title} {Anderson
  transitions}},\ }\href {https://doi.org/10.1103/RevModPhys.80.1355}
  {\bibfield  {journal} {\bibinfo  {journal} {Rev. Mod. Phys.}\ }\textbf
  {\bibinfo {volume} {80}},\ \bibinfo {pages} {1355} (\bibinfo {year}
  {2008})}\BibitemShut {NoStop}%
\bibitem [{\citenamefont {Cuevas}\ and\ \citenamefont
  {Kravtsov}(2007)}]{Cuevas2007}%
  \BibitemOpen
  \bibfield  {author} {\bibinfo {author} {\bibfnamefont {E.}~\bibnamefont
  {Cuevas}}\ and\ \bibinfo {author} {\bibfnamefont {V.~E.}\ \bibnamefont
  {Kravtsov}},\ }\bibfield  {title} {\emph {\bibinfo {title} {Two-eigenfunction
  correlation in a multifractal metal and insulator}},\ }\href
  {https://doi.org/10.1103/PhysRevB.76.235119} {\bibfield  {journal} {\bibinfo
  {journal} {Phys. Rev. B}\ }\textbf {\bibinfo {volume} {76}},\ \bibinfo
  {pages} {235119} (\bibinfo {year} {2007})}\BibitemShut {NoStop}%
\bibitem [{\citenamefont {Pasupathy}\ \emph {et~al.}(2008)\citenamefont
  {Pasupathy}, \citenamefont {Pushp}, \citenamefont {Gomes}, \citenamefont
  {Parker}, \citenamefont {Wen}, \citenamefont {Xu}, \citenamefont {Gu},
  \citenamefont {Ono}, \citenamefont {Ando},\ and\ \citenamefont
  {Yazdani}}]{Yazdani_puddles_2008}%
  \BibitemOpen
  \bibfield  {author} {\bibinfo {author} {\bibfnamefont {A.~N.}\ \bibnamefont
  {Pasupathy}}, \bibinfo {author} {\bibfnamefont {A.}~\bibnamefont {Pushp}},
  \bibinfo {author} {\bibfnamefont {K.~K.}\ \bibnamefont {Gomes}}, \bibinfo
  {author} {\bibfnamefont {C.~V.}\ \bibnamefont {Parker}}, \bibinfo {author}
  {\bibfnamefont {J.}~\bibnamefont {Wen}}, \bibinfo {author} {\bibfnamefont
  {Z.}~\bibnamefont {Xu}}, \bibinfo {author} {\bibfnamefont {G.}~\bibnamefont
  {Gu}}, \bibinfo {author} {\bibfnamefont {S.}~\bibnamefont {Ono}}, \bibinfo
  {author} {\bibfnamefont {Y.}~\bibnamefont {Ando}},\ and\ \bibinfo {author}
  {\bibfnamefont {A.}~\bibnamefont {Yazdani}},\ }\bibfield  {title} {\emph
  {\bibinfo {title} {{Electronic Origin of the Inhomogeneous Pairing
  Interaction in the High-$T_c$ Superconductor
  Bi$_2$Sr$_2$CaCu$_2$O$_{8+\delta}$}}},\ }\href
  {https://doi.org/10.1126/science.1154700} {\bibfield  {journal} {\bibinfo
  {journal} {Science}\ }\textbf {\bibinfo {volume} {320}},\ \bibinfo {pages}
  {196} (\bibinfo {year} {2008})}\BibitemShut {NoStop}%
\bibitem [{\citenamefont {Eliashberg}(1987)}]{EliashbergLinear}%
  \BibitemOpen
  \bibfield  {author} {\bibinfo {author} {\bibfnamefont {G.}~\bibnamefont
  {Eliashberg}},\ }\bibfield  {title} {\emph {\bibinfo {title} {On a possible
  mechanism for superconductivity and linear resistance}},\ }\href@noop {}
  {\bibfield  {journal} {\bibinfo  {journal} {JETP Letters}\ }\textbf {\bibinfo
  {volume} {46}} (\bibinfo {year} {1987})}\BibitemShut {NoStop}%
\bibitem [{\citenamefont {{Bashan}}\ \emph {et~al.}(2024)\citenamefont
  {{Bashan}}, \citenamefont {{Tulipman}}, \citenamefont {{Schmalian}},\ and\
  \citenamefont {{Berg}}}]{Tulipman24}%
  \BibitemOpen
  \bibfield  {author} {\bibinfo {author} {\bibfnamefont {N.}~\bibnamefont
  {{Bashan}}}, \bibinfo {author} {\bibfnamefont {E.}~\bibnamefont
  {{Tulipman}}}, \bibinfo {author} {\bibfnamefont {J.}~\bibnamefont
  {{Schmalian}}},\ and\ \bibinfo {author} {\bibfnamefont {E.}~\bibnamefont
  {{Berg}}},\ }\bibfield  {title} {\emph {\bibinfo {title} {{Tunable Non-Fermi
  Liquid Phase from Coupling to Two-Level Systems}}},\ }\href
  {https://doi.org/10.1103/PhysRevLett.132.236501} {\bibfield  {journal}
  {\bibinfo  {journal} {Phys. Rev. Lett.}\ }\textbf {\bibinfo {volume} {132}},\
  \bibinfo {eid} {236501} (\bibinfo {year} {2024})},\ \Eprint
  {https://arxiv.org/abs/2310.07768} {arXiv:2310.07768 [cond-mat.str-el]}
  \BibitemShut {NoStop}%
\bibitem [{\citenamefont {{Tulipman}}\ \emph {et~al.}(2024)\citenamefont
  {{Tulipman}}, \citenamefont {{Bashan}}, \citenamefont {{Schmalian}},\ and\
  \citenamefont {{Berg}}}]{Tulipman24B}%
  \BibitemOpen
  \bibfield  {author} {\bibinfo {author} {\bibfnamefont {E.}~\bibnamefont
  {{Tulipman}}}, \bibinfo {author} {\bibfnamefont {N.}~\bibnamefont
  {{Bashan}}}, \bibinfo {author} {\bibfnamefont {J.}~\bibnamefont
  {{Schmalian}}},\ and\ \bibinfo {author} {\bibfnamefont {E.}~\bibnamefont
  {{Berg}}},\ }\bibfield  {title} {\emph {\bibinfo {title} {{Solvable models of
  two-level systems coupled to itinerant electrons: Robust non-Fermi liquid and
  quantum critical pairing}}},\ }\href
  {https://doi.org/10.1103/PhysRevB.110.155118} {\bibfield  {journal} {\bibinfo
   {journal} {Phys. Rev. B}\ }\textbf {\bibinfo {volume} {110}},\ \bibinfo
  {eid} {155118} (\bibinfo {year} {2024})},\ \Eprint
  {https://arxiv.org/abs/2404.06532} {arXiv:2404.06532 [cond-mat.str-el]}
  \BibitemShut {NoStop}%
\bibitem [{\citenamefont {{Bashan}}\ \emph {et~al.}(2026)\citenamefont
  {{Bashan}}, \citenamefont {{Tulipman}}, \citenamefont {{Kivelson}},
  \citenamefont {{Schmalian}},\ and\ \citenamefont {{Berg}}}]{Tulipman26}%
  \BibitemOpen
  \bibfield  {author} {\bibinfo {author} {\bibfnamefont {N.}~\bibnamefont
  {{Bashan}}}, \bibinfo {author} {\bibfnamefont {E.}~\bibnamefont
  {{Tulipman}}}, \bibinfo {author} {\bibfnamefont {S.~A.}\ \bibnamefont
  {{Kivelson}}}, \bibinfo {author} {\bibfnamefont {J.}~\bibnamefont
  {{Schmalian}}},\ and\ \bibinfo {author} {\bibfnamefont {E.}~\bibnamefont
  {{Berg}}},\ }\bibfield  {title} {\emph {\bibinfo {title} {{Extended strange
  metal regime from superconducting puddles}}},\ }\href
  {https://doi.org/10.1103/m23x-96cr} {\bibfield  {journal} {\bibinfo
  {journal} {Phys. Rev. B}\ }\textbf {\bibinfo {volume} {113}},\ \bibinfo {eid}
  {075124} (\bibinfo {year} {2026})},\ \Eprint
  {https://arxiv.org/abs/2502.08699} {arXiv:2502.08699 [cond-mat.str-el]}
  \BibitemShut {NoStop}%
\bibitem [{\citenamefont {Kamenev}(2011)}]{Kamenev}%
  \BibitemOpen
  \bibfield  {author} {\bibinfo {author} {\bibfnamefont {A.}~\bibnamefont
  {Kamenev}},\ }\href@noop {} {\emph {\bibinfo {title} {Field Theory of
  Non-Equilibrium Systems}}}\ (\bibinfo  {publisher} {Cambridge University
  Press},\ \bibinfo {year} {2011})\BibitemShut {NoStop}%
\bibitem [{\citenamefont {Kryhin}\ \emph {et~al.}(2025)\citenamefont {Kryhin},
  \citenamefont {Sachdev},\ and\ \citenamefont {Volkov}}]{Kryhin2025}%
  \BibitemOpen
  \bibfield  {author} {\bibinfo {author} {\bibfnamefont {S.}~\bibnamefont
  {Kryhin}}, \bibinfo {author} {\bibfnamefont {S.}~\bibnamefont {Sachdev}},\
  and\ \bibinfo {author} {\bibfnamefont {P.~A.}\ \bibnamefont {Volkov}},\
  }\bibfield  {title} {\emph {\bibinfo {title} {Strong nonlinear response of
  strange metals}},\ }\href {https://doi.org/10.1103/4jxd-z3pm} {\bibfield
  {journal} {\bibinfo  {journal} {Phys. Rev. Lett.}\ }\textbf {\bibinfo
  {volume} {135}},\ \bibinfo {pages} {016503} (\bibinfo {year}
  {2025})}\BibitemShut {NoStop}%
\bibitem [{\citenamefont {Huang}\ \emph {et~al.}(2024)\citenamefont {Huang},
  \citenamefont {Lunts},\ and\ \citenamefont {Hafezi}}]{Huang2024}%
  \BibitemOpen
  \bibfield  {author} {\bibinfo {author} {\bibfnamefont {T.-S.}\ \bibnamefont
  {Huang}}, \bibinfo {author} {\bibfnamefont {P.}~\bibnamefont {Lunts}},\ and\
  \bibinfo {author} {\bibfnamefont {M.}~\bibnamefont {Hafezi}},\ }\bibfield
  {title} {\emph {\bibinfo {title} {Nonbosonic moir\'e excitons}},\ }\href
  {https://doi.org/10.1103/PhysRevLett.132.186202} {\bibfield  {journal}
  {\bibinfo  {journal} {Phys. Rev. Lett.}\ }\textbf {\bibinfo {volume} {132}},\
  \bibinfo {pages} {186202} (\bibinfo {year} {2024})}\BibitemShut {NoStop}%
\bibitem [{\citenamefont {Bateman}\ and\ \citenamefont
  {Project}(1953)}]{bateman1953}%
  \BibitemOpen
  \bibfield  {author} {\bibinfo {author} {\bibfnamefont {H.}~\bibnamefont
  {Bateman}}\ and\ \bibinfo {author} {\bibfnamefont {B.~M.}\ \bibnamefont
  {Project}},\ }\href@noop {} {\emph {\bibinfo {title} {Higher Transcendental
  Functions [Volumes I-III]}}}\ (\bibinfo  {publisher} {McGraw-Hill Book
  Company},\ \bibinfo {year} {1953})\BibitemShut {NoStop}%
\end{thebibliography}%



\end{document}